\documentclass[11pt,a4paper]{article}

\def\Anonymity{1}%Anonymity = 0 means authors are anonymous.
\def\IddoSys{0}%IddoSys = 1 means working on Iddo local.

\usepackage[dvipsnames]{xcolor}
\usepackage[normalem]{ulem}
\usepackage{lineno} % line numbers for reviewers

\ifnum\IddoSys=0
\usepackage[inline,marginclue,index]{fixme}  %Remove draft to erase all notes, remove in line to erase inline notes.
\fxsetup{theme=color,mode=multiuser}
\else
\usepackage{comment}   % then \excludecomment{comment}
\excludecomment{comment}
\fi 

\usepackage{lmodern}
\usepackage[T1]{fontenc}
\usepackage{setspace}
\usepackage{graphicx}

\usepackage{algorithm,algpseudocode}
\usepackage{xspace}
\usepackage[utf8]{inputenc}
\usepackage[noadjust]{cite}
\usepackage{amsmath, amssymb, amsfonts, amsthm}
\usepackage{enumerate}
\usepackage{mathtools}
\usepackage{bm}
\usepackage{paralist}
\usepackage{textcomp, gensymb}
\usepackage{ebproof}
\usepackage{thmtools}
\usepackage{rotating}
\ifnum\IddoSys=0
\usepackage{tcolorbox}
\fi
\usepackage{multicol}
\usepackage{multirow}
\usepackage{fancyvrb}
\usepackage{mathrsfs}
\usepackage{makecell}
\usepackage{indentfirst}
\usepackage{mleftright}
\usepackage[bb=boondox]{mathalfa}
\usepackage[nottoc]{tocbibind}%add bib in toc
\usepackage{complexity}
\usepackage{tablefootnote}
\usepackage{array}
\usepackage{tabularx}
\usepackage{longtable}
\usepackage{tikz}
\usepackage{xcolor}
\usepackage{forest}
\usetikzlibrary{calc,fit,shapes.geometric}
\usepackage{float}
\usepackage{microtype}

\usepackage[pdfstartview=FitH,pagebackref=true,colorlinks,linkcolor=RoyalBlue,citecolor=CadetBlue,bookmarks,bookmarksopen,bookmarksnumbered,hyperfootnotes=true,hypertexnames=false]{hyperref}
\usepackage[left=0.9in,top=0.9in,right=0.9in,bottom=0.9in]{geometry}

\ifnum\IddoSys=0
\usepackage{zref-clever}
\usepackage{zref-vario}
\zcsetup{cap,nameinlink,noabbrev}
\newcommand{\cref}{\zcref}
\newcommand{\Cref}{\zcref}

\zcRefTypeSetup{claim}{
     Name-sg = Claim ,
     name-sg = claim ,
     Name-pl = Claims ,
     name-pl = claims ,
}
\else
\usepackage[capitalize,nameinlink,noabbrev]{cleveref}
\fi

\newtheorem{theorem}{Theorem}[section]
\newtheorem{lemma}[theorem]{Lemma}
\newtheorem{corollary}[theorem]{Corollary}
\newtheorem{claim}[theorem]{Claim}

\newtheorem{fact}[theorem]{Fact}

\newtheorem{proposition}[theorem]{Proposition}

\theoremstyle{definition}
\newtheorem{definition}[theorem]{Definition}

\newtheorem*{notation*}{Notation}

\theoremstyle{remark}
\newtheorem{remark}[theorem]{Remark}

\newenvironment{claimproof}[1][\proofname]{\begin{proof}[#1]}{\end{proof}}

\algtext*{EndWhile}% Remove "end while" text
\algtext*{EndIf}% Remove "end if" text
\algtext*{EndFor}% Remove "end for" text
\algtext*{EndFunction}% Remove "end function" text

\algnewcommand{\IfThen}[2]% \IfThenElse{<if>}{<then>}{<else>}
{\State \algorithmicif\ #1\ \algorithmicthen\ #2}

\def\moverlay{\mathpalette\mov@rlay}
\def\mov@rlay#1#2{\leavevmode\vtop{%
                \baselineskip\z@skip \lineskiplimit-\maxdimen
                \ialign{\hfil$\m@th#1##$\hfil\cr#2\crcr}}}
\newcommand{\charfusion}[3][\mathord]{
        #1{\ifx#1\mathop\vphantom{#2}\fi
                \mathpalette\mov@rlay{#2\cr#3}
        }
        \ifx#1\mathop\expandafter\displaylimits\fi}
\makeatother

\renewcommand{\poly}{\mathrm{poly}}

\newcommand{\eps}{\epsilon}
\newcommand{\Res}{\mathsf{Res}}

\newcommand{\Enc}{\mathsf{Enc}}

\newlang{\MCSP}{MCSP}
\newlang{\MFSP}{MFSP}
\newlang{\MKtP}{MKtP}
\newlang{\MKTP}{MKTP}
\newlang{\itrMCSP}{itrMCSP}
\newlang{\itrMKTP}{itrMKTP}
\newlang{\itrMINKT}{itrMINKT}
\newlang{\MINKT}{MINKT}
\newlang{\MINK}{MINK}
\newlang{\MINcKT}{MINcKT}
\newlang{\CMD}{CMD}
\newlang{\DCMD}{DCMD}
\newlang{\CGL}{CGL}
\newlang{\PARITY}{PARITY}
\renewlang{\Gap}{Gap}
\newlang{\Empty}{\textsc{Empty}}
\newlang{\Avoid}{\textsc{Avoid}}
\newlang{\Sparsification}{\textsc{Sparsification}}
\newlang{\HamEst}{\mathsf{HammingEst}}
\newlang{\HamHit}{\mathsf{HammingHit}}
\newlang{\CktEval}{\textsc{Circuit-Eval}}
\newlang{\Hard}{\textsc{Hard}}
\newlang{\cHard}{\textsc{cHard}}
\newlang{\CAPP}{CAPP}
\newlang{\GapUNSAT}{GapUNSAT}
\newlang{\OV}{OV}
\newlang{\PRIMES}{PRIMES}
\renewlang{\PCP}{PCP}
\newlang{\PCPP}{PCPP}
\newclass{\Avg}{Avg}
\newclass{\ZPEXP}{ZPEXP}
\newclass{\DLOGTIME}{DLOGTIME}
\newclass{\ALOGTIME}{ALOGTIME}
\newclass{\ATIME}{ATIME}%alternating time
\newclass{\SZKA}{SZKA}
\newclass{\Laconic}{Laconic\text{-}}
\newclass{\APEPP}{APEPP}
\newclass{\SAPEPP}{SAPEPP}
\newclass{\TFSigma}{TF\Sigma}
\newclass{\NTIMEGUESS}{NTIMEGUESS}
\newclass{\VPnc}{VP_{nc}}
\newclass{\VNPnc}{VNP_{nc}}
\newclass{\VBPnc}{VBP_{nc}}
\newclass{\FAC}{FAC}

\newlang{\Formula}{Formula}
\newlang{\THR}{THR}
\newlang{\MAJ}{MAJ}
\newlang{\DOR}{DOR}
\newlang{\ETHR}{ETHR}
\newlang{\Midbit}{Midbit}
\newlang{\LCS}{LCS}
\newlang{\TAUT}{TAUT}

\newclass{\VAC}{VAC}
\newclass{\LA}{LA}
\newcommand{\wRank}{{\mathrm{WRank}}}
\newcommand{\ACZ}{{\mathsf{AC}^0}}
\newcommand{\NCTwo}{{\mathsf{NC}^2}}
\newcommand{\AND}{\mathsf{AND}}
\newcommand{\OR}{\mathsf{OR}}

\newcommand{\NOT}{\mathsf{NOT}}

\newcommand{\calB}{\mathcal{B}}
\newcommand{\calC}{\mathcal{C}}
\newcommand{\calD}{\mathcal{D}}
\newcommand{\calE}{\mathcal{E}}

\newcommand{\calL}{\mathcal{L}}
\newcommand{\calM}{\mathcal{M}}
\newcommand{\calN}{\mathcal{N}}

\newcommand{\calP}{\mathcal{P}}

\newcommand{\calS}{\mathcal{S}}

\newcommand{\N}{\mathbb{N}}

\newcommand{\F}{\mathbb{F}}
\renewcommand{\R}{\mathbb{R}}

\newcommand{\wC}{\tilde{C}}

\newcommand{\FTwo}{\ensuremath{\F_2}}

\newcommand{\Eval}{\mathsf{Eval}}%evaluation algorithm for stack programs

\newcommand{\BASIC}{\mathbf{BASIC}}
\newcommand{\COMP}{\mathbf{COMP}}
\newcommand{\MIN}{\mathbf{MIN}}
\newcommand{\IND}{\mathbf{IND}}

\newcommand{\Krajicek}{Kraj\'{\i}\v{c}ek\xspace}
\newcommand{\Jerabek}{Je\v{r}\'{a}bek\xspace}

\DeclareMathOperator{\Range}{\mathrm{Range}}
\DeclareMathOperator{\OpChar}{\mathrm{char}}%characteristic
\newcommand{\coeff}{\mathsf{coeff}}
\newcommand{\Coeff}{\mathsf{Coeff}}
\newcommand{\rank}{\mathrm{rank}}
\newcommand{\Rank}{\mathsf{Rank}}
\newcommand{\PHP}{\mathsf{PHP}}

\newcommand{\Count}{\mathsf{Count}}

\newcommand{\OT}{\mathsf{OT}}

\newcommand{\ncABP}{\mathrm{ncABP}}

\newcommand{\SA}{\ensuremath{\mathsf{SA}}}%Sherali--Adams
\newcommand{\NS}{\mathsf{NS}}%Nullstellensatz
\newcommand{\PC}{\ensuremath{\mathsf{PC}}}%polynomial calculus
\newcommand{\PCR}{\ensuremath{\mathsf{PCR}}}%polynomial calculus resolution
\newcommand{\SoS}{\ensuremath{\mathsf{SoS}}}%sum of squares

\newcommand{\alg}{\mathrm{alg}}

\newcommand{\subpara}[1]{\medskip\noindent\textit{#1}}

\newcommand{\Q}{{\{0,1\}}}
\newcommand{\aOT}{\OT}
\newcommand{\PMOT}{{\mathsf{PMOT}}}

\newcommand{\BTRank}{{\mathsf{BTRank}}}
\newcommand{\PMRank}{\mathsf{PMRank}}
\newcommand{\simpleBTRank}{\mathsf{BT}^{\bullet}\mathsf{Rank}}
\newcommand{\IRank}{\mathsf{IRank}}
\newcommand{\FPHP}{\mathsf{FPHP}}

\newcommand{\rhocol}{\sigma} % was \rho^{\textit{col}}
\newcommand{\rhorow}{\rho} % was \rho^{\textit{row}}
\newcommand{\boundary}{\partial}
\newcommand{\Deltamin}{{\Delta_{\mathit{min}}}}
\newcommand{\Deltamax}{{\Delta_{\mathit{max}}}}
\DeclareMathOperator{\vars}{vars}
\DeclareMathOperator{\Exp}{\mathbb{E}}

\newcommand{\lb}{\mathsf{lb}}
\newcommand{\ax}{\mathsf{axiom}}

\newcommand{\card}[1]{\lvert#1\rvert}
\newcommand{\bfA}{\mathbf{A}}

\newcommand{\bbR}{\mathbb{R}}

\newcommand{\Log}{\mathsf{Log}}

\newcommand{\V}{\mathbf{V}}

\newcolumntype{C}[1]{>{\centering\arraybackslash}p{#1}}

\newcommand{\groupref}[1]{(\hyperref[#1]{Group~\ref*{#1}})}

\ifnum\IddoSys=0
 \definecolor{color1}{RGB}{46,134,193}
 \FXRegisterAuthor{hanlin}{ahanlin}{\colorbox{color1}{\color{white}Hanlin}}
 \newcommand{\hanlin}[1]{{\hanlinnote{#1}}}
 \definecolor{color2}{RGB}{160, 52, 114}
 \FXRegisterAuthor{iddo}{aiddo}{\colorbox{color2}{\color{white}Iddo}}
 \newcommand{\Iddo}[1]{{\iddonote{#1}}}
 \definecolor{color3}{RGB}{240, 182, 213}
 \FXRegisterAuthor{slava}{aslava}{\colorbox{color3}{\color{white}Slava}}
 \newcommand{\slava}[1]{{\slavanote{#1}}}
 \definecolor{color4}{RGB}{210, 102, 253}
 \FXRegisterAuthor{michal}{amichal}{\colorbox{color4}{\color{white}Michal}}
 \newcommand{\michal}[1]{{\michalnote{#1}}}
 \allowdisplaybreaks
\else
\newcommand{\IddoSysComment}[1]{{\small{\textcolor{green}{~#1}}}}
 \newcommand{\hanlin}[1]{Hanlin: \IddoSysComment{#1}}%{\hanlinnote{#1}}}
 \newcommand{\Iddo}[1]{~Iddo: \IddoSysComment{#1}}%{{\iddonote{#1}}}
 \newcommand{\slava}[1]{Slava: \IddoSysComment{#1}}%{{\slavanote{#1}}}
 \newcommand{\michal}[1]{Michal: \IddoSysComment{#1}}%{\michalnote{#1}}}
\fi

\renewcommand{\subpara}[1]{\smallskip\par\noindent\textbf{#1}}

\ifnum\IddoSys=0
\DeclarePairedDelimiter{\parens}{\lparen}{\rparen}
\DeclarePairedDelimiter{\brackets}{[}{]}
\newcommand{\Prb}[1]{\Pr\brackets*{#1}}
\else
\newcommand{\parens}[1]{\mleft(#1\mright)}
\newcommand{\brackets}[1]{\mleft[#1\mright]}
\newcommand{\Prb}[1]{\ensuremath{\mathrm{Pr}\mleft[{#1}\mright]}}
\fi

\makeatletter
\newcommand{\thankssep}{%
  \hphantom{\@textsuperscript{\normalfont\@thefnmark}}%
  \textsuperscript{,}%
}
\makeatother

\ifnum\IddoSys=0
\tcbuselibrary{breakable}
\tcbset{colourfuldefinitionbox/.style={%
    breakable,%
    colback=gray!10,%
    frame empty,%
    width=\columnwidth,%
    arc=2pt%
}}
\newtcolorbox{colourfuldefinition}{colourfuldefinitionbox}
\fi
\vfuzz \hfuzz
\allowdisplaybreaks

\usepackage{draftwatermark}
 \SetWatermarkText{}
 \SetWatermarkScale{5.4}     % Increase the watermark size (default is 1)
 \SetWatermarkColor[gray]{0.94}  % Set transparency (0 = black, 1 = white)

\begin{document}

% ---------- ------------------------------ ----------

% ---------- ------------------------------ ----------
\title{\textbf{The Weak Rank Principle: \\ Lower Bounds and Applications}\thanks{Preliminary version of this work appears in \textit{58th Annual ACM Symposium on Theory of Computing} (STOC) 2026, Salt Lake City, Utah, June 2026. This project has received funding from the European Research
Council (ERC) under the European Union's Horizon 2020 research and innovation programme (grant agreement No 101002742,
EPRICOT project). It was also supported by the Engineering
and Physical Sciences Research Council (EPSRC) under grant
EP/Z534158/1, Integrated Approach to Computational Complexity:
Structure, Self-Reference and Lower Bounds.}}

%  --------------------------------------------------

\date{}

\ifnum\Anonymity=1
\author{
    Michal Garl\'ik\thanks{Email: \href{mailto:michal.garlik@gmail.com}{\texttt{michal.garlik@gmail.com}}}\\ \small{Imperial College London}
    \and
    Svyatoslav Gryaznov\thanks{Email: \href{mailto:svyatoslav.i.gryaznov@gmail.com}{\texttt{svyatoslav.i.gryaznov@gmail.com}}}\\ \small{Imperial College London}
    \and
    Hanlin Ren\thanks{Email: \href{mailto:h4n1in.r3n@gmail.com}{\texttt{h4n1in.r3n@gmail.com}}}\\ \small{Institute for Advanced Study}\vspace{4pt} 
    \and
    Iddo Tzameret\thanks{Email: \href{mailto:iddo.tzameret@gmail.com}{\texttt{iddo.tzameret@gmail.com}}}\\ \small{Imperial College London}  
}

%\hypersetup{urlcolor=black} % black emails
% \author{
%     Michal Garl\'ik\thanks{Email: \href{mailto:michal.garlik@gmail.com}{\texttt{michal.garlik@gmail.com}}}\rlap{\thankssep{}\thanks{This project has received funding from the European Research
% Council (ERC) under the European Union's Horizon 2020 research and innovation programme (grant agreement No 101002742,
% EPRICOT project).}}\\ \small{Imperial College London}  
%     \and
%     Svyatoslav Gryaznov\thanks{Email: \href{mailto:svyatoslav.i.gryaznov@gmail.com}{\texttt{svyatoslav.i.gryaznov@gmail.com}}}\rlap{\thankssep{}\footnotemark[2]}\\ \small{Imperial College London}
%     \and
%     Hanlin Ren\thanks{Email: \href{mailto:h4n1in.r3n@gmail.com}{\texttt{h4n1in.r3n@gmail.com}}}\\ \small{Institute for Advanced Study}\vspace{4pt} 
%     \and
%     Iddo Tzameret\thanks{Email: \href{mailto:iddo.tzameret@gmail.com}{\texttt{iddo.tzameret@gmail.com}}~This project has received funding from the European Research
% Council (ERC) under the European Union's Horizon 2020 research and innovation programme (grant agreement No 101002742,
% EPRICOT project). It was also supported by the Engineering
% and Physical Sciences Research Council (EPSRC) under grant
% EP/Z534158/1, Integrated Approach to Computational Complexity:
% Structure, Self-Reference and Lower Bounds.}\\ \small{Imperial College London}  
% }
%\hypersetup{linkcolor=magenta} % bring back default color
\fi

%\pagenumbering{gobble}
\maketitle
\thispagestyle{empty}

\vspace{-.59cm}
%!TEX root = main.tex

\begin{abstract}
Given two symbolic matrices $X$ and $Y$ of dimensions $m\times n$ and $n\times m$, respectively, the \emph{rank principle} states that when $m = n+1$ and $A$ is a scalar matrix of rank $n+1$, the equation $XY = A$ is unsatisfiable. 
When $m$ is arbitrarily larger than $n$ and $A$ has rank exceeding $n$, we obtain the \emph{weak rank principle}.
We study this principle as an algebraic generalisation of the weak pigeonhole principle (WPHP), asserting that $m$ pigeons cannot be injected into $n$ holes.
As a strengthening of WPHP, it admits  proof complexity  lower bounds in settings where none are known for WPHP, while still supporting analogous applications.
Using new generalised random restrictions applied to the weak rank principle, which may be of independent interest, we resolve several  open problems in proof complexity: we construct proof complexity generators for Polynomial Calculus Resolution over the two-element field ($\PCR_{\F_2}$), new generators for Sherali--Adams ($\SA$), and hardness results for circuit lower bound statements against $\PCR_{\F_2}$.

\begin{description}
%[labelwidth=!,itemindent=!,labelindent=2.pt,leftmargin=5.8pt] % This doesn't work on my system (Iddo) 
% because I can't upload enumitem with the other packages \itemsep=0pt
\setlength{\leftskip}{-12pt}    % removes indentation from the left
  \setlength{\itemindent}{0pt}  % ensures item label not indented
  \setlength{\labelsep}{0.4em}  % space between bullet and text
  \setlength{\labelwidth}{9pt}

\item[Generators for $\PCR_{\FTwo}$:] We prove exponential size lower bounds for several encodings---both algebraic and CNF---of the weak rank principle in $\PCR$ over $\F_2$, where no such bounds are known for the WPHP in the regime with arbitrarily many pigeons.
In particular, we obtain $2^{\Omega(n)}$ size lower bounds for both algebraic and standard CNF encodings, including the \emph{bamboo-tree encoding}, which is the most relevant for applications  and corresponds to a circuit encoding, as considered by Alekhnovich, Ben-Sasson, Razborov, and Wigderson (\emph{SIAM J.\ Comput.}, 2004) and  Razborov (\emph{Ann.~Math.}, 2015). 
Our bounds hold for every matrix $A$ in $XY = A$, implying that the rank principle forms a proof complexity generator with nearly quadratic stretch. 
Using a standard iteration technique we amplify the stretch to $2^{n^{\Omega(1)}}$, thereby obtaining a function generator.  
This resolves the open problem posed by Alekhnovich~\textit{et al.}~(\emph{SIAM J.\ Comput.}, 2004) and  Razborov (\emph{Ann.~Math.}, 2015) concerning the construction of proof complexity generators with good stretch for $\PCR_{\F_2}$.

\item[Generators for Sherali--Adams:]
Since in $\SA$ even the \emph{strong} pigeonhole principle is easy, we develop a new size lower-bound technique showing that the weak rank principle, encoded as a  bamboo-tree CNF, serves as a proof complexity generator for $\SA$. 
Our method introduces a new relaxed notion of degree and a  corresponding pseudoexpectation tailored specifically to the rank principle (and incompatible with the pigeonhole principle).

\item[Circuit lower bound formulas:]
We show that $\PCR_{\F_2}$ does not admit short proofs of lower-bound statements against Boolean circuits, nor against weak models of algebraic circuits. 
This settles the open problem raised by Razborov (\emph{Ann.~Math.}, 2015) concerning the provability of such lower bounds in $\PCR_{\FTwo}$.

\item[Strength of the weak rank principle:]
Finally, we show that  the  weak rank principle is \emph{necessary} for proving $\NCTwo$ circuit lower bounds and, for odd primes $p$, \emph{sufficient} within the theory corresponding to $\ACZ[p]$ for deriving $\ACZ[p]$ lower bounds. 
%Specifically, we show that the hardness of the weak rank principle implies the hardness of $\NCTwo$ lower-bound statements in any proof system closed under low-degree reductions. 
%Moreover, for odd primes $p$, we show that within weak formal theories corresponding to $\ACZ[p]$-reasoning, the weak rank principle serves as an axiom from which $\ACZ[p]$ circuit lower bounds can be derived.
\end{description}
\end{abstract}

\tableofcontents
%\pagenumbering{arabic}
%!TEX root = main.tex

\section{Introduction}
Proof complexity provides both a concrete and conceptual framework for studying computational lower bounds. On the one hand, it seeks to develop combinatorial and algebraic techniques for proving unconditional lower bounds on the lengths of proofs, with relations to complexity class separations. On the other hand, it offers a setting for formulating and exploring metamathematical questions—for example, which proof systems can efficiently establish which lower bounds in computational complexity theory.

For concrete proof-size lower bound questions, it remains a long-standing open problem to establish superpolynomial size lower bounds against sufficiently strong proof systems, such as textbook propositional logic (i.e., Frege systems). Such lower bounds are unknown even against significantly weaker fragments, such as those operating with $\AC^0[2]$ circuits. Nonetheless, the development of theoretical frameworks and techniques aimed at tackling such lower bounds remains a highly active and vibrant area of research.

\subpara{The Weak Pigeonhole Principle.}
A prominent example illustrating the broad reach of proof complexity is the \emph{pigeonhole principle} (PHP for short), which is the (suitably encoded) statement that $m=n+1$ pigeons cannot be mapped into $n$ holes if each hole can accommodate at most one pigeon. The \emph{weak pigeonhole principle} (WPHP) refers to the case where the number of pigeons is possibly much larger than the number of holes; namely, it is the collection of such statements for all pairs $m>n$, where $m$ may be arbitrarily larger than 
$n$. In this regard, WPHP is logically weaker than PHP where $m=n+1$.

The pigeonhole principle is perhaps the most influential example in proof complexity, serving as a driving force behind many lower bound techniques and frameworks. These include bottleneck counting~\cite{Hak85}, pigeonhole switching lemmas and $k$-evaluations~\cite{Ajt88, PBI93, KPW95}, pigeon dance~\cite{Razb98, IPS99}, pseudo-width~\cite{Raz04, Razb02}, etc. As for upper bounds, already $\TC^0$-Frege admits polynomial-size proofs of PHP (and hence WPHP); see~\cite{CN10,Bus87}. Despite this progress, a notable open problem posed by Razborov~\cite{Razb15-annals} remains unresolved: proving size lower bounds for WPHP against the polynomial calculus resolution (\PCR) proof system (see \Cref{footn:91} and discussion therein).

Beyond concrete lower bounds, the pigeonhole principle plays a central role in the development of theories in bounded arithmetic, which study the computational complexity of the concepts required to prove various statements within formal systems. In bounded arithmetic, the weak pigeonhole principle serves as an axiom from which important results—particularly those related to randomness in computation—can be derived. As early as 1981, Woods~\cite{Woo81} observed that explicit counting of the number of elements in a finite set can often be replaced by applications of the pigeonhole principle for bounded formulas. Building on this idea, Paris, Wilkie, and Woods~\cite{PWW88} introduced the weak pigeonhole principle, and showed that it often serves as a more suitable substitute for  PHP in such contexts and that it is provable in bounded arithmetic $\mathsf{T}_2$. Initial work by Wilkie (unpublished; see~\Krajicek{}~\cite[Theorem 7.3.7]{Kra95}), further developed by Thapen~\cite{Tha02} and systematically pursued by \Jerabek{}~\cite{Jerabek04, Jer05-PhD, Jer07}, established WPHP as a useful axiom for reasoning about randomized computation.% when added to base theories for polynomial-time reasoning ($\mathsf{PV}$).

% ======= MERGE-CLASH
% Beyond concrete lower bounds, the pigeonhole principle plays a central role in the development of theories in bounded arithmetic, which study the computational complexity of the concepts required to prove various statements within formal systems. In bounded arithmetic, the weak pigeonhole principle serves as an axiom from which important results—particularly those related to randomness in computation—can be derived. As early as~\cite{Woo81}, Woods observed that explicit counting of the number of elements in a finite set can often be replaced by applications of the pigeonhole principle for bounded formulas. Building on this idea, Paris, Wilkie, and Woods~\cite{PWW88} introduced the weak pigeonhole principle, and showed that it often serves as a more suitable substitute for full PHP in such contexts and that it is provable in bounded arithmetic $\mathsf{T}_2$. Initial work by Wilkie (unpublished; see~\Krajicek{}~\cite[Theorem 7.3.7]{Kra95}), further developed by Thapen~\cite{Tha02} and systematically pursued by Jeřábek~\cite{Jerabek04, Jer05-PhD, Jer07}, established WPHP as a useful axiom for reasoning about randomized computation.% when added to base theories for polynomial-time reasoning ($\mathsf{PV}$).
% >>>>>>> 25293b2c06c9ea434cbc2a3fc80fb1e46f8507ac
% 

\subpara{WPHP and proof complexity generators.} A concept closely related to the weak pigeonhole principle is that of a proof complexity generator. In fact, it is more directly connected to the \emph{dual} weak pigeonhole principle (denoted dWPHP), which states that when $n$ pigeons are mapped to $2n$ holes, at least one hole must remain empty. The notion of proof complexity generators was introduced independently by Alekhnovich, Ben-Sasson, Razborov, and Wigderson~\cite{ABRW04}, and by \Krajicek{}~\cite{Kra01-Fundamenta, Kra04}; see also the  monograph~\cite{Kra25} for a comprehensive treatment.

A \emph{proof complexity generator} is any encoding of the statement that expresses that a point $b \in \Q^\ell$ is not in the image of a given polynomial time mapping $G\colon \Q^m \to \Q^\ell$, with $m<\ell$.
The proof complexity of such statements depends on $G$ and is sensitive to the encodings; indeed,~\cite{ABRW04} suggested three different encodings (see also~Sokolov~\cite{Sok20-heavy} for a discussion).
We say that a proof complexity generator is \emph{hard} against a proof system when for \emph{every} point $b$ there is no short proof that $b$ is not in the image of $G$. 
 This means that the proof system cannot prove that $G$ is non-surjective. (Notice that even a single $b$ for which the proof complexity generator is easy would constitute a proof of non-surjectivity, hence we insist that no such $b$ exists;  this point is explained further in Alekhnovich~\textit{et~al.}~\cite{ABRW04}.)

The hope is that for strong propositional proof systems, one can establish (at least conditionally) that there are no $\poly(\ell)$-size proofs that $g$ is non-surjective, under the assumption that the mapping $g$ is sufficiently pseudorandom (cf.~\cite{Razb15-annals}). Razborov in~\cite{Razb15-annals} proved the existence of hard  proof complexity generators for systems such as $\PCR$ and $\Res(k)$. However, as Razborov mentions, the important case of generators for $\PCR$ over \FTwo\ is completely open (the importance of this case is explained after stating 
\Cref{thm:intro:Bamboo-lower bound}).

\subpara{WPHP and the provability of circuit lower bounds.} 
WPHP and proof complexity generators also play a central role in the metamathematics of complexity theory—particularly in understanding which systems can efficiently prove circuit lower bounds. This connection was first observed by Razborov~\cite{Razb98}, who showed that WPHP reduces to circuit lower bound statements. Informally, this means that if WPHP is hard for some (``nice'') proof system $\calP$, then $\calP$ cannot prove any circuit lower bound efficiently. 
%WPHP and proof complexity generators are also useful in meta-mathematics, particularly as a way to study which proof systems can efficiently prove lower bound statements (encoded as propositional formulas). This was observed by Razborov~\cite{Razb98}: WPHP, where the number of pigeons is $2^{n^\eps}$ and the number of holes is $n$, for $0<\eps<1$, can be reduced to the statement that a circuit of size approximately $n$ cannot compute  any fixed function. 
This idea was employed by Razborov~\cite{Razb98} and later by Raz~\cite{Raz04}. The former established degree lower bounds for WPHP in polynomial calculus, and the latter established resolution size lower bounds. Both work then leveraged the above reduction to show that the statement ``$\NP\nsubseteq\P/\poly$'', when encoded as a family of CNF formulas, does not admit efficient refutations in those systems.

%This idea was employed by Razborov~\cite{Razb98}, and later by Raz~\cite{Raz04}. They first established lower bounds for WPHP in their respective proof systems—polynomial calculus for Razborov and resolution for Raz—and then used these bounds to show that the statement ``SAT cannot be computed by small Boolean circuits'' when encoded as a family of CNF formulas, does not admit short refutations in those systems.

More generally, one can define the \emph{truth-table generator}, which maps a description of a small Boolean circuit to the truth table of the function it computes.
Establishing that a string $b$ lies outside the image of this generator is equivalent to proving that no small circuit computes the function with truth table $b$.
Hence, the truth-table generator constitutes a proof-complexity generator against a system $\mathcal{P}$ exactly when $\mathcal{P}$ cannot efficiently prove any circuit lower bounds.

%More generally, one can consider the \emph{truth table generator} that maps a description of a small Boolean circuit to the truth table of the function it computes. Proving that a string $b$ lies outside the range of this generator amounts to proving that no small circuit computes the function with truth table $b$. Thus, the truth table generator being a proof complexity generator against a system $\mathcal{P}$ is equivalent to that $\mathcal{P}$ cannot efficiently prove any circuit lower bounds.

%More generally, a proof complexity generator, when viewed as a truth table generator—where the input is a description of a small circuit and the output is the truth table of the function computed by that circuit—can be used to encode a lower bound statement. In particular, proving that a string $b$ is outside the range of the generator amounts to proving that no small circuit computes the function whose truth table is $b$.

\subpara{Linear algebraic instances.}
Beyond simple counting arguments like the pigeonhole principle, linear algebra is a key tool in discrete mathematics. Accordingly,  Cook and Rackoff suggested linear algebraic statements as potential hard instances for strong propositional proof systems. 
Bonet, Buss, and Pitassi~\cite[Sec.~3.1.1]{BBP95} further explored this idea and identified candidates hard tautologies such as the \emph{oddtown principle}, which asserts $XX^T \ne I_m$ for an $m \times n$ variable matrix $X$ with $m > n$, where $I_m$ is the $m \times m$ identity matrix. Soltys and Cook~\cite{SC04} systematically studied the \emph{hard matrix identities} in the context of bounded arithmetic, focusing on statements such as the \emph{inversion principle}, which 
asserts that $XY = I_n$ implies $YX = I_n$ for square $n\times n$ variable matrices $X, Y$, as well as other equivalent formulations of basic linear algebraic facts. Their goal was to identify minimal formal theories capable of proving such 
statements. Soltys and Urquhart~\cite{SU04} showed that the pigeonhole principle is reducible to (hence, not stronger than) these matrix identities.
Later, it was demonstrated by Hrube\v{s} and Tzameret over $\FTwo$~\cite{HT15} and by Tzameret and Cook over the integers~\cite{TC21} that these linear algebraic principles admit quasipolynomial-size propositional (i.e., Frege) proofs and are provable in relatively weak theories of arithmetic. The latter two works  roughly  show that linear algebra sits within ``$\NC^2$-reasoning'' (formally, the theory $\mathsf{VNC}^2$). The hard matrix identities were also considered in the context of the Ideal Proof System (IPS) by Grochow and Pitassi~\cite{GP18}, and Andrews and Forbes~\cite{AF22} who investigated a close unsatisfiable instance $\{\det(X)=0,XY=I\}$ for two square variable matrices $X,Y$.

A related algebraic principle is the (\emph{strong}) rank principle: Let $m=n+1$. Given two variable matrices $X,Y$ of dimension $m\times n$ and $n\times m$, respectively, the product $XY=A$ is unsatisfiable whenever $A$ is an $m\times m$ scalar matrix of rank $m$. This formula  was considered implicitly in \Krajicek{}~\cite{Kra09}, as a way to algebraically express proof complexity generators based on PHP. 

For the specific case where $A$ is the identity matrix $I$,
Soltys and Cook~\cite[Equation (V), page 287]{SC04} showed that the strong rank principle is equivalent over their theories (the base one denoted \LA; hence over Frege) to the hard matrix identities, and specifically the inversion principle. Galesi, Grochow, Pitassi, and She~\cite{GGPS23} formulated the rank principle explicitly, for the special case when $A$ is the identity matrix $I$, and when $m$ is not necessarily $n+1$. Galesi \textit{et al.}  established $\PC$ degree lower bounds for the rank principle, using the same reduction from PHP degree lower bounds that was used in~\cite{SU04}.

%
%Linear algebraic instances and their relation to the pigeonhole principle were considered in prior works (e.g.,~\cite{SC04}). Specifically, PHP was shown to be reducible to the inversion principle $AB=I\implies BA=I$~\cite{SU04}. 
%
%The rank principle was also considered implicitly by \Krajicek{}~\cite{Kra09}.
%He considered PHP as a proof complexity generator, and uses the rank principle to rewrite in an algebraic form his PHP generator, with the intention to derive the hardness of this generator from the \emph{hardness of PHP}.

%Galesi, Grochow, Pitassi and She~\cite{GGPS23} considered the rank principle  XXX

\medskip 
%----------------------------------------
\subpara{The weak rank principle.}
%----------------------------------------
In this work we show the advantage in considering what we call the  \emph{weak rank principle}, {\emph{WRank} for short, as a natural algebraic analogue and augmentation of the weak pigeonhole principle.
Given two variable matrices $X$ and $Y$ of dimensions $m\times n$ and $n\times m$, respectively, WRank states that

\begin{center}
\fbox{%
\parbox{0.9\textwidth}{%
\textbf{WRank.} 
For an arbitrary scalar matrix $A$ of rank at least $n+1$, 
the equation $XY = A$ is unsatisfiable, where $m > n$; 
here we mean that $m$ is \emph{arbitrarily larger} than $n$.
}%
}
\end{center}
%
% \begin{quote}
% for an arbitrary scalar matrix $A$ of rank at least  $n+1$, the equation $XY = A$ is unsatisfiable, where $m>n$; where here we mean that $m$ is \emph{arbitrarily} \emph{larger} than $n$.  
% \end{quote}
Notice that WPHP is (usually) considered in the same regime, namely when $m$ is arbitrarily larger than $n$~\cite{Razb02}.
Accordingly, all our lower bounds are expressed in terms of $n$ and are \emph{independent} of $m$.   
% $$
%  (\wRank) ~~~~~
% XY = A, \qquad \text{for $X$, $Y$ symbolic matrices of dimensions $m \times n$, $n \times m$, resp., where $m > n+1$.}
% $$
%\Iddo{I define WRP for $m> n+1$, is that correct for out purposes?}
% \Iddo{\sout{Decide between WRP against WRank.}}
% \slava{\sout{I'm in favour of WRank.} Also, would saying $m \gg n$ here be too vague?}\Iddo{Okay WRank.  $m\gg  n$ seems too vague indeed.}
% \slava{Why not just $m>n$?} \Iddo{since then it's not the WEAK version}
% \slava{Should it be WPHP and WRank? Although we usually call them just PHP and $\Rank$ even in the weak setting.}\Iddo{Perhaps, but then it's not in the style of "WPHP"}
 
%Our results show how WRank can be used in settings where WPHP fails to, and in partiuclar in constituting a hard tautology for various relatively strong proof systems (up to $\NCTwo$-Frege), a hard proof complexity generator, and as a basic axiom in weak arithmetic theories. %

Our results show that the weak rank principle can be applied in settings where the weak pigeonhole principle fails, and that it provides a framework for both lower and upper bounds in proof complexity.
In particular, we obtain lower bounds and corresponding proof-complexity generators with a good stretch for $\PCR_{\F_2}$ which can also be iterated to a function generator (where none were previously known) and introduce a new generator for the Sherali--Adams system, establish the hardness of circuit lower-bound statements for $\PCR_{\F_2}$ (which was open), and identify upper bounds---namely, feasibly constructive proofs---that are facilitated by the weak rank principle itself.

The proof systems we shall consider in this work are
polynomial calculus resolution \PCR\ and Sherali--Adams \SA.
%polynomial calculus resolution \PCR, sum-of-squares \SoS, Nullstellensatz $\NS$, \ACZ-Frege, and $\ACZ[p]$-Frege.
See \cref{sec:prelims} for the precise definitions and references.

\medskip 
\noindent\textit{What is known about WRank.} 
For the \emph{strong} rank principle (where $m=n+1$, and in fact up to $m\le n^2$), size lower bounds against \PCR\ could be shown over any field, by a reduction to the (strong) pigeonhole principle (with $m$ pigeons; see~\cite{SU04,GGPS23} for this reduction), and then using a size-degree tradeoff argument.  However, for the \emph{weak} rank principle (when $m$ is arbitrary) only \emph{degree} lower bounds are known in polynomial calculus resolution ($\PCR$); again, this follows easily from known degree lower bounds for WPHP. However, the more challenging and meaningful regime is that of \emph{size} lower bounds for $\PCR$ and stronger systems. 
Notably, no $\PCR$ size lower bounds are currently known for WPHP for arbitrary many pigeons, and in fact when the number of pigeons exceeds $n^2$ (see Mik\v{s}a and Nordstr\"{o}m~\cite{MiksaN24} and de Rezende, Nordstr\"{o}m, Risse and Sokolov~\cite{RezendeNR025})\footnotemark.%%
\footnotetext{\label{footn:91} Beyond $n^2$ pigeons, size lower bound techniques against \PCR\  based on degree break down. Note that for \emph{resolution}, using different techniques  by  Raz~\cite{Raz04} and subsequently Razborov~\cite{Razborov:2001:WPHP,Razborov04} as well as de Rezende \textit{et al.}~\cite{RezendeNR025}, the case of WPHP lower bounds with arbitrary many pigeons $m$ is resolved.} This motivates algebraically extending WPHP to WRank, which enables new size lower bounds in settings where WPHP has so far proven insufficient (as the current work shows).

With respect to \textit{upper bounds}, while WPHP admits polynomial-size proofs in $\TC^0$-Frege, the best known upper  bound for WRank is a polynomial-size proof in $\NCTwo$-Frege. This could be shown as follows: for strong enough proof systems closed under low degree reductions WRank is implied by the special case of WRank when $A=I$ (see \Cref{lem:degree-1 reduction for CNF rank formula}), which in turn is implied by the inversion principle by~\cite{SC04}; finally, the inversion principle admits polynomial-size proofs in $\NCTwo$-Frege by the work of  Hrube\v{s} and Tzameret~\cite{HT15}. On the other hand, it is reasonable to conjecture that Frege does not admit polynomial-size proofs of WRank by the fact that linear algebraic statements are expected not to have polynomial-size Frege proofs as noted by Buss \textit{el al.}~\cite{BBP95}. 

%While the strong rank principle, when $A$ is the identity matrix $I$, cannot be proved efficiently in any proof system for which the hard matrix identities are hard {SC04}, it is unclear how the weak rank formulas are precisely related to the hard matrix identities. 

Note that the relation between the weak rank principle and the strong one (and hence,  by~\cite{SC04}, its relation to the hard matrix identities) is not entirely understood, since the latter concerns square matrices, while our focus is on highly non-square matrices (hence, the ``weak'' regime).

% \color{lightgray}%%%%%%%%%%%%%%%%%%%%%%%%%

% \subsection{Notes}

% %  \hanlin{Note: Smolensky's lower bounds, as currently written in this paper, require wRF over GF(3). We can just say that we consider finite fields of constant size, our results are robust w.r.t.~the specific field, and for convenience we mostly focus on GF(2).}
% \slava{Not only because of convenience: for tree-like CNF encodings extension variables must be field elements, so we have to assume $\F_2$. Also,  our random restrictions work for other fields, even for the algebraic encoding. Although, it's plausible they \emph{might} be modified to work over $\F_p$, but don't quote me on this.}

% -- Open problems with respect to wPHP: PCR over \FTwo\.
 
% \color{black} %%%%%%%%%%%%%%%%%%%%%%%%%%%%

%We start by presenting the instances we  work with in a purely algebraic form. Later we shall encode them as CNF formulas. 

% --------------------------------------------
\section{Our Results}
% --------------------------------------------
We present four types of results based on the weak rank principle: Lower bounds and generators for $\PCR$ over $\FTwo$, hereafter denoted  $\PCR_{\FTwo}$ (\Cref{sec:intro:lbandapp}), lower bounds and generators for \SA\ (\Cref{sec:intro:SA}), hardness of circuit lower bound statements for $\PCR_{\FTwo}$ (\Cref{sec:intro:cktlbs}), and upper bounds (namely, feasible constructive proofs) facilitated by the weak rank principle (\Cref{sec:intro:rank-is-axiom}).

% --------------------------------------------
\subsection{Lower Bounds and Generators for \texorpdfstring{$\PCR_{\FTwo}$}{PCR(F2)}}\label{sec:intro:lbandapp}
% --------------------------------------------
% 
We establish lower bounds and construct generators for \PCR\ over \FTwo, under the following encodings of increasing strength:
\begin{itemize}\itemsep=0pt
\setlength{\leftskip}{-12pt}    % removes indentation from the left
  \setlength{\itemindent}{0pt}  % ensures item label not indented
  \setlength{\labelsep}{0.4em}  % space between bullet and text
  \setlength{\labelwidth}{9pt}

  \item \textbf{Algebraic encoding.}    
          This setting provides the most direct algebraic formulation and serves as a foundation for the subsequent encodings.

  \item \textbf{Perfect matching (PM) encoding.}  
          In the PM encoding, we obtain lower bounds for the weak rank principle, when $A=I$, against $\PCR_{\FTwo}$. This encoding is in a CNF. 
          To get a generator in this encoding, we should prove the lower bound for every $A$. 
          Although it is achievable by  similar techniques used in the next item, we do not formally prove this case, since this encoding does not appear to be directly useful   to establish hardness of circuit lower bound  statements. The case $A=I$ is conceptually simpler than the general case of arbitrary $A$, yet serves as an instructive case for the forthcoming generators because those results expand on similar ideas. It requires a different notion of degree from the algebraic encoding, that is useful in the sequel (e.g., for \SA\ lower bounds). 
          %For this purpose we consider another CNF encoding, in the following item.  
          
          %And the stretch is $2mn$ into $m^2$. 

  \item \textbf{Bamboo-tree encoding.}  
  The bamboo-tree encoding yields the strongest construction, achieving the same stretch as the generators above ($2mn$ to $m^2$).  In addition, we are able to prove that this generator is \emph{iterable} (in the sense of~\cite{Kra04, Razb15-annals}) under this encoding, hence we establish a \emph{function} generator (i.e., a generator with a stretch of $2^{n^{\Omega(1)}}$). A notable consequence of this function generator is that $\PCR_{\F_2}$ does not admit short proofs of circuit lower bound statements such as $\NP\nsubseteq \P/\poly$.
  \end{itemize}

We start by providing more technical background to our new generators.

\paragraph{Context and motivation for our generators.} Recall that a proof complexity generator encodes the statement that a point $b \in \Q^\ell$ is not in the image of a polynomial-time map $G\colon \Q^m \to \Q^\ell$, with $m < \ell$.
When the weak rank principle is viewed in this framework, the matrix product $XY$ plays the role of the map $G$, and the point outside its image is the matrix $A$.

The known generators for $\PCR_{\F_2}$ in the literature are based on reductions to the pigeonhole principle.
For the \emph{strong} pigeonhole principle, \Krajicek{}~\cite[Section~1]{Kra09} constructed a generator stretching $n$ bits to $n+1$ bits.
%For the \emph{weak} pigeonhole principle,
%combining the same construction with the best known lower bounds for WPHP over $\PCR_{\F_2}$ 
The same construction can be combined with the best known lower bounds for the \emph{weak} pigeonhole principle over $\PCR_{\F_2}$~\cite{AlekhnovichRazborov01,MiksaN24} to yield a slightly superlinear stretch---from $n^3$ to $n^4$ bits%
\footnote{Given a formula $\PHP_n^m$, \Krajicek{}~\cite{Kra09} builds a function stretching $nm{+}tn$ bits into $tm$, for any parameter $t$; the ratio $tm/(nm{+}tn)$ is maximised at $t=\Theta(n^2)$.
The best known lower bounds for $\PHP_n^m$ against $\PCR$ hold for $m=o(n^2)$.
If we plug these values of $t$ and $m$ into $nm{+}tn$ and $tm$, respectively, we obtain a stretch of $n^3$ to $n^4$.}.
(For comparison, our results achieve a nearly quadratic stretch of $2mn$ to $m^2$ since our lower bounds hold in the regime where $m$ is arbitrarily larger than $n$.)

%--- The generators known for $\PCR_{\FTwo}$ are based on reductions to the pigeonhole principle. For strong PHP the best generator that  can be achieved is $n$ to $ n+1$ bits, as shown by \Krajicek{}~\cite[Section 1]{Kra09}. If  we work with WPHP, then using the same construction from~\cite{Kra09} together with the best known  lower bounds for WPHP over PCR-two from~\cite{AlekhnovichRazborov01,MiksaN24} we can improve the stretch up to slightly superlinear, namely $n^3$ to $n^4$ bits\footnote{Given a $\PHP_n^m$ \Krajicek{}~\cite{Kra09} constructs a function that stretches $nm+tn$ bits into $tm$, for any $t$, and this stretch is maximal when $t=\Theta(n^2)$ (because the ratio $tm/(nm+tn)$ is maximised that way). As mentioned before the best known lower bound for $\PHP_n^m$ against \PCR\ is for $m=o(n^2)$. Note that if we plug this $t$ and $m$ to $nm+tn$ and $tm$, respectively, we get a stretch of $n^3$ to $n^4$.}. We achieve a near quadratic stretch $2mn$ to $m^2$.

In the present work, we aim to base the hardness of such generators on the hardness of WRank instead, which is \emph{stronger} than WPHP.
%
%For instance, in the $\SoS$ proof system, WPHP is easy, whereas WRank
%is plausibly  hard for this system; 
%
For instance, WPHP is easy for the $\SoS$ proof system, but WRank is plausibly hard for it; likewise, $\Res(\log n)$ admits quasipolynomial-size refutations of WPHP~\cite{MPW02}, yet statements based on linear algebra (and in particular WRank)
are expected to be hard for it~\cite{BBP95}.

%- Note that WPHP for arbitray $m$ is open for \PCR\ over any field.If it was known (for arbitray $m$)  then we could have used it both for getting generators with good stretch (by the construction of \Krajicek{}) as well as for applications: using Raz~\cite{Raz04} method we could have shown circuit lower bound statements are unprovable efficiently (e.g., for \PCR), using directly WPHP lower bounds. 

As mentioned above, for arbitrary $m$, the hardness of WPHP against  $\PCR$ over any field remains open.
If such bounds were known, they could yield strong generators through \Krajicek{}'s construction and  further applications. In particular, a sufficiently strong lower bound for the Onto-WPHP would, via Raz's method~\cite{Raz04}, yield that certain circuit lower bound statements for unbounded fan-in circuits cannot be efficiently proven in $\PCR$.

%, following Raz's method~\cite{Raz04}, could potentially show that certain circuit lower-bound statements are not efficiently provable (e.g., in $\PCR$).\footnote{This connection, however, is incomplete: Raz's reduction requires the \emph{onto} version of PHP and produces circuits of unbounded fan-in---later refined by Razborov through a sequence of papers to the bounded fan-in case.} Thus, hardness of WPHP for general $m$ over $\PCR$ (and any field) remains unresolved.

%Note that establishing lower bounds for WPHP for arbitrary $m$ remains an open problem for $\PCR$ over any field. If such bounds were known, they could be used both to construct pseudorandom generators with good stretch (following \Krajicek{}'s construction) and for further applications.
%In particular, a sufficiently strong lower bound for the Onto-WPHP would, via Raz's method~\cite{Raz04}, yield that certain circuit lower bound statements for unbounded fan-in circuits cannot be efficiently proven in $\PCR$.

%\michal{This needs to be rewritten because we wouldn't be able to apply Raz's approach \emph{directly}. The reason is that Raz needs the Onto-PHP lower bound. And even if we succeeded to apply the ideas behind Raz's direct approach, his reduction is to a circuit of unbounded fan-in. People prefer fan-in two circuits; at least this was the motivation of Razborov to write three papers gradually improving the mentioned result of Raz.}\Iddo{Okay, can you please rewrite it?} However, it is not known that WPHP arbitrary $m$ is hard for \PCR\ (over any field). 

This led Alekhnovich \emph{et al.}~\cite{ABRW04} to study hard instances distinct from WPHP on which to base proof-complexity generators and their applications to circuit lower-bound statements.
%They proposed using systems of linear equations over~$\F_2$ defined by an expander graph, giving rise to the so-called \emph{Nisan generator}, a special case of the Nisan--Wigderson generator~\cite{NisanWigderson94}.
They proposed the so-called \emph{Nisan generator}, a special case of the Nisan--Wigderson generator~\cite{NisanWigderson94}, which is based on systems of linear equations over~$\F_2$ defined over expander graphs.
The stretch of this construction alone is not sufficient to yield application to circuit lower bound statements. 

%Given a proof-complexity generator, the goal is to \emph{maximise} its stretch. 
An important research direction is to obtain proof complexity generators with \emph{as large stretch as possible}.
When the stretch is exponential (i.e., $2^{n^{\Omega(1)}}$ in the seed length $n$) the generator is called a \emph{function generator}.
To achieve such amplification, \Krajicek{}~\cite{Kra04} introduced the notion of \emph{$s$-iterability}, designed to boost the stretch of a variant of the Nisan generator~\cite{NisanWigderson94}.
%Building on this, Razborov~\cite{Razb15-annals} showed how to iterate a generator $2^{n^{\Omega(1)}}$ times to obtain a function generator that remains hard for $\PCR_\F$ whenever $\operatorname{char}(\F)\neq 2$.
%
Improving~\cite{ABRW04}, Razborov~\cite{Razb15-annals} relaxed the expander requirements and showed that  linear systems as in~\cite{ABRW04} can be \emph{iterated} to achieve a much larger stretch---about~$2^{n^{\varepsilon}}$.
In particular, the hardness of the base generator transfers to the iterated version, since an iterated expander remains an expander.
However, systems of linear equations over~$\F_2$ are \emph{easy} to refute in~$\PCR_{\F_2}$; therefore, this construction cannot yield generators hard for~$\PCR_{\F_2}$. 

The limitations of existing principles motivate the use of WRank which subsumes both the weak pigeonhole principle and systems of linear equations.
This allows us to obtain lower bounds for WRank (over~$\F_2$) for arbitrary~$m$, thereby producing a hard \emph{base generator}.
The hardness of WRank persists under iteration, which enables us to construct function generators that remain hard and leads to the hardness of corresponding circuit lower-bound statements.

%This motivates  considering the  stronger principle WRank, which generalises both WPHP and systems of linear equations. Because WRank is strictly stronger than WPHP, we are able to  establish lower bounds for WRank (over~$\F_2$) for arbitrary~$m$, providing a hard \emph{base generator}. This generator can then be iterated while preserving hardness, leading directly to the hardness of corresponding circuit lower-bound statements.

Among other results known about generators are Khaniki's Nisan--Wigderson generator against $\ACZ[p]$-Frege~\cite{Kha22} encoded as a (non-CNF) propositional formula. In addition, Sokolov~\cite{Sok20-heavy} improved the lower bounds for a particular generator encoding (functional encoding as in~\cite{ABRW04}).

% 
% We establish lower bounds and generators in encoding of increasing strength and importance:
% 
% Algebraic 
% 
% PMT
% 
% -- We can have a lower bound on WRank here; but the m is slightly weaker than the Bamboo encoding. Still, for PMT we can have iterability. Hence, a function generator. 
% 
% 
% Bamboo
% 
% --- Best generator stretch. However, we do not attempt to carry out iteration for this construction, hence we don't show a function generator based on this. 

% ------------------------------------
\paragraph{Algebraic encoding.} 
The algebraic encoding of the negation of WRank, denoted $\Rank^m_n(A)$, consists of $m^2$ polynomial equations of degree 2, 
\begin{equation}\label{eq:intro:WRank}
\Rank^m_n(A)\text{:}~~~~~~~~~~~ 
\sum_{k=1}^n x_{i,k} y_{k,j} = A_{i,j},
\text{~~for all $i,j\in[m]$}. 
\end{equation}

%\slava{We do not define $\ACZ$ and $\ACZ[p]$-Frege currently.}\Iddo{That's why I say references}

\begin{theorem}[Algebraic encoding lower bound; see \cref{lem: Rank lower bound algebraic,thm: lower bound for iterated algebraic irank}]
\label{thm:intro:Algebraic encoding lower bound}
For every $m>n$ and  $A\in\F_2^{m\times m}$, every   $\PCR_{\F_2}$ refutation of $\Rank^m_n(A)$ requires size $2^{\Omega(n)}$. Consequently, $\Rank^m_n(A)$ is a proof complexity generator stretching $2nm$ bits into $m^2$.
Moreover, the iterated variant of $\Rank^m_n(A)$ is a function generator with stretch $2^{n^{\Omega(1)}}$.  
\end{theorem}
Here, size is measured by the number of distinct monomials appearing in the refutation (see precise definitions in \cref{sec:prelims}).

The algebraic encoding lower bounds are obtained using random restrictions, which yield a size-to-degree reduction and already suffice to derive the hardness of certain circuit lower-bound statements (for example, for non-commutative algebraic branching programs; see \cref{sec: LB-ncABP}).
The argument in this setting is technically simpler than the proofs for the CNF encodings introduced below, as the latter requires translating the algebraic reasoning into a combinatorial framework. The lower bound for the algebraic instance in \cref{thm:intro:Algebraic encoding lower bound} thus forms the basis for the CNF lower bounds presented later and underlies the high-level random-restriction approach  used throughout.

The algebraic encoding, however, is specific to algebraic proof systems and therefore not directly comparable with encodings suited for other systems.
To address this, we encode WRank as a CNF formula based on the perfect matching principle, and later consider the more demanding \emph{bamboo-tree} encoding (see~\Cref{sec: generator bt}).
The choice to work with CNF encodings is natural, as CNF is the standard formulation in proof complexity.
Moreover, to extend lower bounds from algebraic proof systems such as $\PCR$ and $\SoS$ to systems like $\ACZ$-Frege or $\Res(\mathrm{lin})$ (that is, resolution over linear equations~\cite{RT07}), CNF encodings are the appropriate framework, whereas the algebraic formulation---involving degree-2 polynomial equations---lies outside the expressive power of these systems.

% Algebraic encoding lower bounds goes quite smoothly using random restrictions which enables a size-to-degree reduction. This would already be sufficient to yield some circuit lower bound statement lower bounds (for non-commutative algebraic branching program; see in the sequel). However the algebraic encoding is specific to algebraic proof systems, and hence incomparable to other formulas for different proof systems (for example, proof complexity generators against \ACZ-Frege by Khaniki~\cite{Kha22}). For that purpose, we encode WRank as a CNF based on the perfect matching principle (later on we shall consider the more challenging `bamboo-tree' encoding; see XXX). 
% 
% 
% \textcolor[rgb]{0.501961,0,0.25098}{Combine with above perhaps: The lower bound for the algebraic instance in \cref{thm:intro:OT-alg-lower-bound} above serves as the basis of the lower bounds for CNF formulas below and the random restrictions we use. We present two different CNF encodings with two different aims. \textcolor[rgb]{0.752941,0.752941,0.752941}{}\\ 
% COMBINE ALSO WITH: The choice to study CNF encodings is natural given this is the most common formulation in proof complexity. Further, in order to go from \PCR\ and \SoS\ lower bounds towards proof system such as \ACZ-Frege or Res(lin) (i.e., resolution over linear equations, over the integers or a finite field~\cite{RT07}), CNF encoding is the  appropriate one, while the algebraic formulation (of degree 2 polynomials) is beyond the language of these proof systems.
% \\
%}

\vspace{-5pt} 
 
\paragraph{Perfect matching encoding.}
The perfect matching CNF encoding of WRank, denoted $\PMRank^m_n(A)$, encodes each equation $\sum_{k=1}^n x_{i,k} y_{k,j} =A_{i,j}$ of $\Rank^m_n(A)$ by stating there exists  a perfect matching on the satisfied monomials (i.e., those equal $1$) of this equation. To encode such a matching we use extension variables for each pair of monomials in the equation (together with an extra point if $A_{i,j}=1$). See \cref{sec:generator-pm} for the definition.
Similar encodings of parity computations were considered for example in Impagliazzo--Segerlind and Ken~\cite{IS01,ken24}.

As mentioned above, we shall consider only the case $A=I$ and prove lower bounds for this encoding  against $\PCR_{\F_2}$. The case for every $A$, yielding a generator could be achieved with  techniques similar  to the ones in the bamboo-tree encoding. 

%of the form  $w_1+\dots+w_n=0 \bmod 2$ such that a satisfying assignment to the linear equation $w_1+\dots+w_n=0 \bmod 2$, as a  perfect matching between pairs of variables $w_i,w_j$ ($i\neq j$), such that every variable which is 1 under the assignment participates in this matching, and variables which are 0 under the assignment do not participate .

%Similarly to $\OT^m_n(A)$, we denote by $\PMOT^m_n(A)$ %the CNF encoding of the oddtown principle where the %identity matrix $I_m$ is replaced by the matrix $A$. 

\begin{theorem}[Perfect matching encoding lower bound against $\PCR_{\FTwo}$; see \cref{thm: PMRank size bound PCR2}]\label{thm:intro:PMRank size bound PCR2}
For every $m>n$, every $\PCR_{\FTwo}$ refutation of the perfect matching CNF encoding $\PMRank^m_n(I_m)$  requires size $2^{\Omega(n)}$.
\end{theorem}
 
The case of $\PCR_{\FTwo}$ is the most interesting because the perfect matching principle (stating that there is no perfect matching on an odd size set) is in fact easy for $\PCR_{\FTwo}$, hence the hardness of $\PMRank^m_n(I_m)$ does not stem from the encoding per se. This contrasts many  other cases where the hardness follows from the hardness of the perfect matching principle  itself~\cite{Kra95,CleggEI96,IPS99,BGIP01, Gri01, AH19}, as we show for the case of $\PCR_{\F}$ when $\OpChar(\F)\neq 2$ in the next theorem.   

%Further demonstrating the importance of working over $\FTwo$, we  establish WRank as a generators for $\PCR_{\F}$ when $\F\neq 2$. This result is much easier than Thm 2.2. and uses the hardness of the encoding itself, not the hardness of WRank.  

The next result demonstrates the sensitivity of the encoding of the generators. Although we can construct generators for several proof systems using the PM encoding, it is not known how to get hardness of  circuit lower bound statements from these generators.

\begin{theorem}[Perfect matching encoding lower bound over any field;  see \cref{thm: PMOT arb A other fields}]\label{thm:intro:OT-CNF-PM-lower-bound-Other-Fields}
For every $m>n$ and  $A\in\Q^{m\times m}$, any $\PCR_{\F}$ refutation (for a field $\F$ with $\OpChar(\F) \neq 2$) or $\SoS$ refutation of the perfect matching CNF encoding $\PMRank^m_n(A)$  requires size $2^{\Omega(n)}$. Similarly, any $\ACZ$-Frege refutation requires size $2^{n^{\Omega(1)}}$, where the constant in the exponent depends on the circuit depth.
\end{theorem}

%The size measure for $\SoS$ is defined in terms of monomial-size (namely, number of distinct monomials in the refutation), similar to $\PCR$.

%This follows from \Iddo{TBC}

%\begin{corollary}[Informal; see \cref{}]
%\label{thm:intro:SOS-OT-CNF-PM-lower-bound}
%very refutation of the perfect matching CNF encoding of the oddtown principle $\PMOT^m_n$ in SOS over the reals requires XXX monomial-size.
%\end{corollary}
%\Iddo{changed to corollary?}

\paragraph{Bamboo-tree encoding.} 
%
% The first encodes computation in \FTwo\ implicitly using perfect matchings, and the second more explicitly providing essentially extension variables for each mod 2 gate, where mod 2 of $n$ variables is encoded using a bamboo tree  with mod 2 gates on the leaves. %
The main challenge and open problems regarding proof complexity generators and their applications   is predominantly about proving lower bounds for good encodings; where an encoding is good, if it has extension variables that correspond to gates of a circuit, hence could be applied to circuit statement lower bounds~\cite{ABRW04,Razb15-annals}.

The bamboo-tree  CNF encoding of WRank, denoted $\BTRank^m_n(A)$, encodes each equation $\sum_{k=1}^n x_{i,k} y_{k,j} =A_{ij}$ of $\Rank^m_n(A)$ by using extension variables  introduced to sequentially compute all inner products involved.
The extension variables for each inner product are arranged in a totally unbalanced binary tree known as a ``bamboo''. Each extension variable $u_{i,j,\ell}$ encodes the partial sum $\sum_{k=1}^{\ell}x_{i,k}y_{k,j}$ (for $\ell\in[n]$). Moreover, each monomial $x_{i,k}y_{k,j}$ has its own extension variable $z_{i,j,k}$.

\begin{figure}[H]\centering
    % LTeX: enabled=false
    \begin{forest}
        for tree={grow'=east,inner sep=1mm}
        [
            [$u_{i,j,1}$, no edge
                [$u_{i,j,2}$
                    [$\cdots$
                        [$u_{i,j,n-1}$
                            [$u_{i,j,n}$
                                [,no edge]
                                [$z_{i,j,n}$]
                            ]
                            [$z_{i,j,n-1}$]]
                        [, no edge]]
                    [$z_{i,j,2}$]]
                [, no edge]]
            [$z_{i,j,1}$, no edge]]
        \path (!1) -- coordinate[midway] (mid) (!2);
        \node at (mid) {\rotatebox{90}{$=$}};
    \end{forest}
    % LTeX: enabled=true
    \caption{Example of one axiom of $\BTRank^m_n(A)$. Here, $u_{i,j,\ell}$ are the extension variables for the partial sum $\sum_{k=1}^{\ell}x_{i,k}y_{k,j} = \sum_{k=1}^{\ell}z_{i,j,k}$ (for $\ell\in[n]$). See \Cref{def:bt-enc}.}
\end{figure}
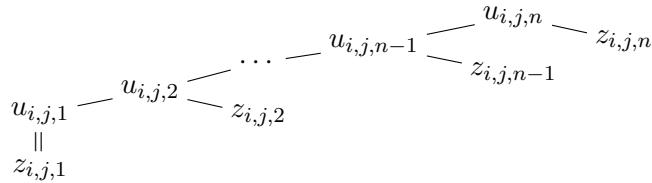

  This natural encoding corresponds to a circuit in the sense of~\cite{ABRW04,Razb15-annals}, and similar encodings have been used repeatedly for parity computations~\cite{SU04,ABRW04,Razb15-annals}.

\begin{theorem}[Bamboo-tree encoding lower bound; see \cref{thm:BTRank_A_lower_bound,thm: lower bound on iterated formula psi}]
\label{thm:intro:Bamboo-lower bound}
For every $m>n$ and  $A\in\F_2^{m\times m}$, every   $\PCR_{\F_2}$ refutation of $\BTRank^m_n(A)$ requires size $2^{\Omega(n)}$. 
Consequently, $\BTRank^m_n(A)$ is a proof complexity generator stretching $2nm$ bits into $m^2$ bits.
Moreover, there is an  iterated variant of $\BTRank^m_n(A)$ that yields a function generator with stretch $2^{n^{\Omega(1)}}$.  
\end{theorem}

This result resolves an open problem raised in~\cite{ABRW04,Razb15-annals} concerning the construction of proof-complexity generators for $\PCR_{\F_2}$.
Razborov's~\cite{Razb15-annals} established  generators against \PCR\  based on the Nisan construction only over fields of characteristic different from~$2$.
% Our result closes this gap by exhibiting hard generators for $\PCR_{\F_2}$.
Lower bounds over $\F_2$ are particularly significant and technically more challenging than those over other fields, as we now explain. 

In general, when considering the weak rank principle, there are two distinct field parameters to consider: the \emph{ground field}, over which the proof system operates, and the \emph{expressed field}, over which the matrix product expressed in the weak rank principle is computed.
Although these two fields may differ, the case where both are $\F_2$---that is, WRank over $\F_2$ and $\PCR$ over $\F_2$---is the most natural and the most difficult.

For example, unsatisfiable linear systems over $\F_2$ are easy for $\PCR_{\F_2}$, and more generally, linear systems over $\F_p$ are easy for $\PCR_{\F_p}$.
When $p \neq 2$, however, such linear systems are non-Boolean and require explicit encoding of non-Boolean field elements, which makes the lower-bound task formally easier but conceptually less meaningful, as the hardness would then usually stem from the encoding rather than from the inherent structure of the weak rank principle.
In contrast, working with both ground and expressed fields equal to $\F_2$ keeps all variables Boolean and ensures that the difficulty arises from the combinatorial and algebraic content of the principle itself.

Furthermore, $\F_2$ is usually the most challenging case for algebraic and semi-algebraic proof systems.
For instance, lower bounds in the Nullstellensatz system or its variants often require substantially more complex designs over $\F_2$ than over larger or characteristic-$0$ fields (cf.~\cite[Theorem 12]{BCEIP98},~\cite{Kra19}).
Likewise, proof systems such as $\Res(\mathrm{lin})$~\cite{RT07} admit lower bounds over characteristic $0$~\cite{PT21}, whereas corresponding results over finite fields---and in particular over $\F_2$ (cf.~\cite{IS14})---remain open.
Thus, establishing hard generators over $\F_2$ advances our understanding precisely in the setting that is both the most natural and the least understood.

\vspace{-5pt}

\paragraph{Lower bounds techniques.}

All our lower bounds follow the same general \emph{random self-reduction scheme}: we first reduce proof size to suitable notions of degree, tailored to the formulas and results under consideration. For example, although total degree is the standard measure for monomials, one may also work with refined degree measures that count only variables from a designated set. This means that we shall start, by way of contradiction, from a small refutation (having only a few distinct monomials) and apply a random restriction or a random substitution  (changing variables to different polynomials in different variables\Iddo{Correct?}) that simplifies terms in the refutation \emph{with respect to the appropriate notion of degree}, reducing the formula to a smaller instance of the same principle. The second step is proving a lower bound on the relevant notion of degree, which will result in a contradiction (namely, no small size refutation exists).  

The novelty and challenge is in finding the appropriate notions of degree  and in devising the right random substitutions which will decrease the corresponding notion of degree.
These notions of degree increase in level of complication along with the complexity of the encoding. Accordingly, the random substitutions increase in level of complication.

One idea that is different from the usual proof complexity lower bounds is that our generalised random restrictions are \emph{random substitutions} of both $0$-$1$ and polynomials \Iddo{Check} in formulas. In PHP lower bounds, the usual argument hinges on applying $0$-$1$ restrictions (representing partial matchings). But in our case partial $0$-$1$ random restrictions are not always sufficient, hence we need to substitute the original variables by both $0$-$1$ and other \emph{variables or their negations}. Our goal will be to get self-reducibility: a smaller version of the principle. In the literature such random substitutions were used in works by, e.g., Pitassi, Rossman, Servedio and Tan~\cite{PRSLY16} and subsequently Hast\aa{}d~\cite{Has20} for $\ACZ$-Frege lower bounds.

\medskip 
\noindent\textit{\underline{Algebraic and bamboo-tree encoding of rank principle.}}
The following explains the lower bounds for both the algebraic encoding $\Rank^m_n(A)$ and the bamboo-tree encoding $\BTRank^m_n(A)$.
We outline some ideas behind the proof of \cref{thm:intro:Bamboo-lower bound}, which establishes a $2^{\Omega(n)}$ size lower bound for $\mathsf{BTRank}^m_n(A)$ in $\PCR_{\F_2}$. The key challenge is that this lower bound must hold independently of the parameter $m$, which precludes the use of standard expander-based techniques. These techniques would typically involve restricting one of the matrices, say $X$, to an expander, zeroing out most of the entries $x_{i,k}$, for $k \in [n]$, in each row $i \in [m]$. However, when the gap between $m$ and $n$ is too large, there are no bipartite expanders with suitable parameters for achieving degree lower bounds.

Our solution hence avoids using expanders. We relax the degree notion, aiming to balance two competing requirements for establishing size lower bounds: (1) the existence of random self-reductions of $\mathsf{BTRank}^m_n(A)$ that effectively reduce the relaxed degree, and (2) the existence of a lower bound on the relaxed notion of degree for refutations of $\mathsf{BTRank}^m_n(A)$. The random self-reduction we design to achieve (1) is highly sensitive to the structure of the extension variables (the $z$-variables and $u$-variables defined above) and its necessarily technical nature leads us to present it in two stages for clarity. In this outline, we focus solely on motivating the notion of the relaxed degree that the self-reduction aims to decrease, and we primarily discuss requirement (2). 

Consider a term $t$ in $x$-, $y$- and $z$-variables of $\mathsf{BTRank}^m_n(A)$, that is, a product of some $x$-, $y$-  and $z$-variables and their negations. For a variable $v$, we write $v^1$ to denote $v$ and $v^0$ to denote its negation, $\overline{v}$. Let $C$ be the set of indices $k$ such that $x^b_{i,k}$, $y^b_{k,j}$ or $z^b_{i,j,k}$ appears in $t$ for some $b \in \Q$ and $i,j \in [m]$. We refer to $C$ as the set of \emph{columns $X$-, $Y^T$- and $Z$-mentioned} in $t$. Now, let $\tau$ be the substitution of $X^TA$ for $Y$, i.e., 
$\tau(x_{i,k}) = x_{i,k}$, $\tau(y_{k,j}) = \sum_{i' \in [m]} x_{i',k}A_{i',j}$, and $\tau(z_{i,j,k}) = \tau(x_{i,k})\tau(y_{k,j})$ for all $i,j \in [m]$ and $k \in [n]$, with $\tau(\overline{v}) = 1+ \tau(v)$ for any variable $v$. We see that if we apply $\tau$ to $t$, 
the columns $X$-mentioned in each term of the polynomial $\tau(t)$ remain in $C$ (note that under $\tau$ the term $t$ becomes a \emph{polynomial}, not necessarily a single  monomial,  in the $X$-variables only). It follows that each term $t'$ of $\tau(t)$ either has degree at most $\card{C}$ (since we are working over the ring of multilinear polynomials; this happens when each column index $k\in C$ contributes a single variable  $x_{i,k}$ to $t'$), or there is $k \in C$ and $i < i' \in [m]$ such that $t'$ is a multiple of $x_{i,k}x_{i',k}$ (by the pigeonhole principle; this happens when the column $k\in C$ contributes more than a single variable $x_{i,k}$ to $t'$). 

Why is the last observation useful? The substitution $\tau$ maps the set of polynomial equations $XY = A$ to the set of polynomial equations $XX^TA = A$, and the latter has a straightforward degree-2 derivation from the oddtown equations $XX^T = I_m$, which in turn have an immediate degree-2 derivation from the algebraic formulation of the functional pigeonhole principle $\mathsf{FPHP}^m_n$. 
For $\mathsf{FPHP}^m_n$, Razborov~\cite{Razb98} established a degree lower bound of $n/2$, which holds even for $\PCR$ proofs operating with multilinear polynomials modulo the ideal generated by the hole axioms ($x_{i,k}x_{i',k}$ for $i<i' \in [m]$ and $k \in [n]$) and the functionality axioms ($x_{i,k}x_{i,k'}$ for $i \in [m]$ and $k<k' \in [n]$). We conclude that there is no refutation of the \emph{algebraic} formulation $\mathsf{Rank}^m_n(A)$ each term of which has less than $n/2$ $X$- or $Y^T$-mentioned columns. If such a refutation existed, then $\tau$, along with the aforementioned degree-2 derivations, would convert it into a refutation of $\mathsf{FPHP}^m_n$, where every term has either degree less than $n/2$ or is a multiple of a hole axiom, contradicting the cited lower bound. This forms the column-degree lower bound part in the proof of \cref{thm:intro:Algebraic encoding lower bound}. Similarly, this argument rules out the existence of a refutation of a variant of $\mathsf{Rank}^m_n(A)$ that involves the $x$-, $y$-, and $z$-variables (but not the $u$-variables), where each term in the refutation has less than $n/2$ $X$-, $Y^T$- or $Z$-mentioned columns. \medskip 

While the argument in the above paragraph works for the algebraic formulation  $\mathsf{Rank}^m_n(A)$, it is not sufficient for the bamboo-tree encoding $\mathsf{BTRank}^m_n(A)$. This is because $\tau$ must be consistently extended to the $u$-variables by $\tau(u_{i,j,k}) = \sum_{\ell \in [k]} \tau(z_{i,j,\ell})$, and the column index $k$ of $u_{i,j,k}$ is not particularly relevant here: when $\tau$ is applied to a product of multiple $u$-variables mentioning only a small number of columns, it leads to a blow up in the $X$-mentioned columns in the resulting terms. Thus, our new notion of degree for the $u$-variables concerns only the first of the three indices of $u_{i,j,k}$. We say that $i$ is $U$-\emph{left-row-mentioned} in a term $t$ if $u^b_{i,j,k}$ appears in $t$ for some $j \in [m]$, $k \in [n]$, and $b \in \Q$. 

To appreciate this definition, let us first calculate $\tau(u^{1-b}_{i, j, k})$ modulo the hole axioms. We have \vspace{-5pt}
\begin{align*}
\tau(u^{1-b}_{i, j, k}) 
&= b + \tau(u_{i, j, k})
= b + \sum_{\ell \in [k]} \tau(z_{i,j,\ell}) 
= b + \sum_{\ell \in [k]} \tau(x_{i,\ell}) \tau(y_{\ell,j}) 
= b + \sum_{\ell \in [k]} x_{i,\ell} \sum_{i' \in [m]} x_{i',\ell}A_{i',j} \\
&= b + \sum_{\ell \in [k]} \sum_{i' \in [m]} x_{i,\ell} x_{i',\ell}A_{i',j} 
= b + \sum_{\ell \in [k]} x_{i,\ell} x_{i,\ell}A_{i,j}
= b + A_{i,j}  \sum_{\ell \in [k]}  x_{i,\ell},
\end{align*}
where the penultimate equality is by the hole axioms and the last equality is by the Boolean axioms. Hence, we see that, modulo the hole axioms, the only remaining row index is $i$. Let $t$ be a term consisting of $u$-variables and their negations and let $R$ be the set of indices $U$-left-row-mentioned in $t$. Based on the above calculation, we observe that each term $t'$ of the polynomial $\tau(t)$ either has degree at most $\card{R}$, or is a multiple of a hole axiom, or (by the pigeonhole principle) is a multiple of a functionality axiom of $\mathsf{FPHP}^m_n$. 

The preceding discussion motivates the definition of the relaxed degree of a general term $t$ in the variables of $\mathsf{BTRank}^m_n(A)$ as the sum of the cardinality of the set of indices $U$-left-row-mentioned in $t$ and the cardinality of the set of $X$-, $Y^T$- or $Z$-mentioned columns in $t$. For a precise definition of random self-reductions and a proof of their effectiveness in reducing the relaxed degree, we refer the reader to the proof of \cref{thm:BTRank_A_lower_bound}.

%We explain here the proof of \cref{thm:intro:OT-alg-lower-bound}. This pure algebraic instance lower bound already contains many of the ideas of the lower bounds against the CNF encodings of the oddtown principle. They become more apparent in the algebraic encoding. In fact, the special random restriction we employ in the CNF formulation are based on, and were found initially, using the algebraic formulation. %\Iddo{"(In general it is common to think first, when dealing with proof systems of a richer language, about hard instances in this richer language,because it is usually easier to prove the lower bound in such way ["because you give less information of the prover"].)"} 

\slava{The following is written with the assumption that we explained something about special notions of degree above. It's also kinda technical, but this is for the reader that would appreciate some details.} %\hanlin{I also feel what's written below assumes a lot...}
\slava{I don't think we can reasonably explain all the details here.}

\medskip 
\noindent\textit{\underline{Iterated rank principles.}}
In order to address the iterated variants of the rank principle, we need additional techniques besides the ones that were sufficient for \cref{thm:intro:Algebraic encoding lower bound}. We start by describing the ideas behind the iterated algebraic principle.
We consider the classical Goldwasser, Goldreich and Micali~\cite{GGM86}  iteration protocol, which was also used by Razborov~\cite{Razb15-annals}. This protocol achieves a generator with exponential stretch (i.e., a function generator)  by iterating the base generator (the way to achieve such a function generator from the weak rank principle is described after \Cref{thm:intro:boolean lb for PCR2} in this section).

We start with the algebraic encoding. Consider the polynomial system $\IRank^m_n(\{A^{(\pi)} : \pi \in \Q^{\le\kappa}\}, \kappa) \coloneq \bigcup_{\pi \in \Q^{\le\kappa}} X^{(\pi)} Y = A^{(\pi)}$. Here, each $X^{(\pi)} Y = A^{(\pi)}$ is a distinct instance of the weak rank principle with parameters $m$ and $n$ and the $x$-variables indexed by the binary string $\pi$, except that the matrices $A^{(\pi)}$ are not necessarily constant: we also allow them to contain  variables from $X^{(\pi')}$ for some strings $\pi'$ with $\card{\pi'} > \card{\pi}$. 
This instance is a generalisation of WRank: letting ``$()$'' be the empty string, when $\kappa=0$ we have that $\IRank^m_n(\{A^{()}\}, 0) = \Rank^m_n(A^{()})$.
Thus, to specify the entries of different $A^{(\pi)}$'s, it is helpful to view the formula  arranged as a binary tree  shown in \cref{fig: irank structure}.
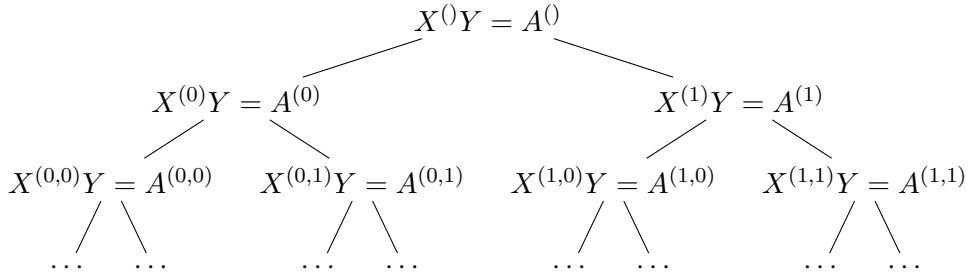
\begin{figure}[H]
\centering
    \begin{forest}
        for tree={grow=south,l sep=0em,s sep=1em,inner sep=1mm}
        [{$X^{()}Y=A^{()}$}
            [{$X^{(0)}Y=A^{(0)}$}
                [{$X^{(0,0)}Y=A^{(0,0)}$}
                    [$\cdots$] [$\cdots$]]
                [{$X^{(0,1)}Y=A^{(0,1)}$}
                    [$\cdots$] [$\cdots$]]]
            [{$X^{(1)}Y=A^{(1)}$}
                [{$X^{(1,0)}Y=A^{(1,0)}$}
                    [$\cdots$] [$\cdots$]]
                [{$X^{(1,1)}Y=A^{(1,1)}$}
                    [$\cdots$] [$\cdots$]]]]
    \end{forest}
    \caption{The  Goldwasser, Goldreich and Micali-style iteration structure of $\IRank$.}
    \label{fig: irank structure}
\end{figure}
In this tree, the entries of each matrix $A^{(\pi)}$ can depend only on the entries of $X^{(\pi')}$ from the corresponding subtree.
This differs from the standard notion of $s$-iterability~\cite{Kra04,Razb15-annals}, which only assumes a flat structure; that is, the iterated copies are arranged as a path and not as a tree. Our lower bound depends on the depth of the arrangement ($\kappa$ in the above notation), and the tree structure will maximise our lower bound. 
The tree structure was also used by Razborov~\cite{Razb15-annals}. Although this construction turns out to be sufficient for applications to circuit lower bounds statements.
Our goal would be to show that any refutation of $\IRank$ with parameters $m$ and $n$ requires size $2^{{(n/\kappa)}^{\Omega(1)}}$.
\smallskip 

We consider two notions of degree for $\IRank$:
\begin{itemize}
\itemsep=0pt
\setlength{\leftskip}{-12pt}    % removes indentation from the left
  \setlength{\itemindent}{0pt}  % ensures item label not indented
  \setlength{\labelsep}{0.4em}  % space between bullet and text
  \setlength{\labelwidth}{9pt}
    \item the \emph{column degree}, which measures the number of different columns of $X^{(\pi)}$ or $Y^T$ the term mentions, where the same columns from different matrices are counted \emph{at most once}; and
    \item the \emph{row degree}, which   measures the number of different rows of $X^{(\pi)}$ or $Y^T$ the term mentions, where the same rows from different matrices \emph{are counted individually}.
\end{itemize}
For example, the term $x^{(0)}_{1,1} x^{(1)}_{1,1} y_{1,1}$ has column degree $1$ and row-degree $3$.
We perform two random self-reductions $\rho$ and $\sigma$. The former reduces the row degree of every term in a refutation to $\le d$ while the latter reduces the column degree to $\le d$ \Iddo{what is d? For some d?}. This implies the total degree does not exceed $2d^2$.

We show that refuting $\IRank^m_n({A^{(\pi)}}, \kappa)$ requires degree $n/2$ by reducing it to a single  copy of the formula $\Rank^m_n(I_m)$ using a reduction that substitutes each $X^{(\pi)}$ with $A^{(\pi)} X$ for a new variable $X$ \Iddo{New variable or variables?}. Since the entries of $A^{(\pi)}$ are not necessarily constant, this reduction can increase the degree of a refutation. However, it can grow by at most a factor of $\kappa+1$, because for each copy the degree grows by at most $1$. This explains the dependency on $\kappa$ we mentioned earlier.% \hanlin{I think we didn't mention the exact bound on $\kappa$ to this point... So the reader may not know what's the ``dependency on $\kappa$ mentioned earlier''}\slava{Yes, we should say that the bound is $2^{{(n/\kappa)}^{\Omega(1)}}$; now it's mentioned}.

\medskip

Up to now we discussed the algebraic encoding of the iterated rank principle. We now turn to the CNF encoding of $\bigcup_{\pi \in \Q^{\le\kappa}} X^{(\pi)} Y^{(\pi)} = A^{(\pi)}$ (note that here the matrix $Y$ is  different in each iteration, unlike before) and again use the iteration protocol from~\cite{GGM86}.
We encode $\bigcup_{\pi \in \Q^{\le\kappa}} X^{(\pi)} Y^{(\pi)} = A^{(\pi)}$ using the extension variables that are similar to the ones from $\BTRank$ and additionally impose some structure on the matrices $Y^{(\pi)}$.
Our goal is again to prove that any $\PCR_{\F_2}$ refutation requires size $2^{{(n/\kappa)}^{\Omega(1)}}$.

% We also want to work with the standard notion of degree here.
The notion of column degree is not well suited for the extension variables.
For instance, given an extension variable $u^{(\pi)}_{i,j,k}$ which semantically encodes the sum $\sum_{\ell \in [k]} x^{(\pi)}_{i,k} y^{(\pi)}_{k,j}$, it is clear that it depends not only on the $k$th column but also on the preceding columns $\{1, \ldots, k-1\}$. To address this, we restrict our matrices $Y^{(\pi)}$ (and hence the extension variables) to a bipartite expander $G$ with a small left-degree $\Delta$. This automatically restricts the column degree of all but $x$-variables to be at most $\Delta$, thus for all extension variables we only need to reduce the row degree.

Overall, the proof strategy is executed as follows: we start by reducing the row-degree (and thus the total degree) in $u$-variables that compute partial inner products. We then reduce the column degree in the $x$- and $z$-variables (recall that the latter compute the binary ANDs: $z^{(\pi)}_{i,j,k} = x^{(\pi)}_{i,k} y^{(\pi)}_{k,j}$). This is performed in the same way as in the algebraic construction. Finally, we reduce the row degree in the $x$-, $y$-, and $z$-variables by using a random substitution similar to the one that reduces the row degree for $\IRank$.
To deal with the $z$-variables we need to further modify the refutation.
This is because of the terms of the form $s = \prod_{j \in J} \overline{z}_{i_0,j,k}$ (or of the symmetric form $\prod_{I \in J} \overline{z}_{i,j_0,k}$), which have the shape of a \emph{star} (when the $x$- and $y$-variables are considered as two sides of a bipartite graph with an edge between $x^{(\pi)}_{i,k}$ and $y^{(\pi)}_{k,j}$ present whenever $\overline {z}^{(\pi)}_{i,j,k}$ appears in $s$). Such stars have common centre $i_0$, and our substitutions do not, in fact, set them to $0$ w.h.p. However, we can show that each such  star with sufficiently many rays---that is, with large enough set $J$---collapses to its centre $\overline{x}_{i_0,k}$ w.h.p.
This property limits the degree of all such star terms.
For the rest of the variables, we can greedily identify a large set of variables in a term that are independently set to $0$ w.h.p., implying that such a term vanishes from the proof w.h.p.
\subsection{Generators for Sherali--Adams}\label{sec:intro:SA}
%%%%%%%%%%%%%%%%%%%%%%%%%%%%%%%%%%%%%%%%%%%%%%%%%%%%%%%%%%%

We address the problem of establishing proof complexity generators for the Sherali--Adams (\SA) semi-algebraic refutation system~\cite{DantchevM13}.

First, note that the Nisan generator construction, denoted $\tau(A, b)$ in~\cite{Razb15-annals}, can be shown to work also for \SA\ and even \SoS.
%
%when $E$ is an expander, $\tau(E, b)$ is hard for $\PCR_{\F}$ with $\OpChar(\F) \neq 2$. 
%
Specifically, its hardness follows from the lower bound on the \emph{Gaussian width} of $\tau(A, b)$ as introduced in~\cite{Ben-SassonI10}, and whose lower bound against $\SoS$ was established in~\cite{Gri01,Schoenebeck08}. Combined with the random restriction from Razborov~\cite{Razb15-annals}, this gives an \SA\  size lower bound for  $\tau(A, b)$.

We provide a different generator for \SA\ than Razborov's by proving that our bamboo-tree encoding of WRank is such a  generator. Note that  WPHP is easy for \SA\ in terms of size. Therefore, WPHP cannot be directly used as a generator. We thus come up with a new technique that does not use the reduction to PHP and instead defines a relaxed notion of degree (row degree, in our case). The lower bounds are thus based on a reduction from size to row degree, followed by a lower bound on row degree. To establish the row degree lower bound we come up with a pseudoexpectation based on a new family of distributions tailored to WRank  (and not to WPHP). 

%The following is the main novel ingredient 

\begin{lemma}[Row degree lower bound for \SA\ (informal); \Cref{lem: SA BTRank rowe degree bound}]
\label{lem:intro:SA BTRank rowe degree bound}
    For every $m > n$ and every $A \in \Q^{m \times m}$, every $\SA$ refutation of (a variant of) $\BTRank^m_n(A)$ requires row degree $n-1$.
\end{lemma}

\begin{theorem}[Size lower bound for \SA\ (informal); see \cref{thm: SA lower bound on simpleBTRank}]
    Assume $m > n $ and $A \in \Q^{m \times m}$ is a Boolean matrix.
    Then any $\SA$ refutation of (a variant of) $\BTRank^m_n(A)$ requires size $2^{\Omega(n)}$.
\end{theorem}

This result fits  in the generator approach to proof complexity  as stated by Razborov~\cite[p.~417]{Razb15-annals} and proposed earlier by \Krajicek{}~\cite{Kra01-Fundamenta, Kra04} and Alekhnovich \textit{et al.}~\cite{ABRW04}. Specifically, the \emph{generator approach} seeks to establish proof complexity generators for as many proof systems as possible, roughly based on the computational hardness of the generator.

\medskip 

%Paragraph about the new pseudoexpectation's construction. \Iddo{TBC}

To rule out low row-degree \SA\ refutations of $\mathsf{BTRank}^m_n(A)$, we define a family of distributions over partial assignments. Each distribution in this family is indexed by a pair $(I,J)$, where $I \subseteq [m]$ and $J \subseteq [m]$ are sets of row indices of $X$ and $Y^T$, respectively, with $|I|+|J| \leq n-2$. The distribution  corresponding to $(I,J)$ is the uniform distribution supported on certain assignments to all variables $x_{i,k}$ with $i \in I, k \in [n]$ and $y_{\ell,j}$ with $\ell \in [n], j \in J$. 

Before specifying these assignments, it is helpful to recall the family of distributions used against the functional pigeonhole principle $\mathsf{FPHP}^m_n$ by~\cite{DantchevMR09}. Each distribution in that family is indexed by a set $I \subseteq [m]$ of at most $n-2$ pigeons and consists of the uniform distribution over all matchings of the pigeons in $I$ to holes. Thus, the distribution indexed by $I$ is supported on all assignments to the variables $x_{i,k} : i \in I, k \in [n]$ that do not violate any axiom of $\mathsf{FPHP}^m_n$. In other words, no condition whatsoever limits these assignments except the axioms involving the pigeons in $I$.

In contrast, for $\mathsf{BTRank}^m_n(A)$, we cannot allow similar freedom in the supports of the distributions in our family. It is straightforward to verify that if the support consisted of all assignments to $x_{i,k} : i \in I, k \in [n]$ and $y_{\ell,j} : \ell \in [n], j \in J$ that do not violate any axiom of $\mathsf{BTRank}^m_n(A)$, the marginal distribution condition---a key ingredient of $\mathsf{SA}$ lower bounds\footnote{In our case, the marginal distribution condition says that marginalising the distribution indexed by $(I \cup \{i_0\},J)$ to the variables $x_{i,k} : i \in I, k \in [n]$ and $y_{\ell,j} : \ell \in [n], j \in J$ coincides with the distribution indexed by $(I,J)$, and similarly for the distribution indexed by $(I,J \cup \{j_0\})$.}---would fail. In other words, limiting the support only by the requirement that the axioms concerning $A_{I,J}$ (the submatrix of $A$ determined by rows $I$ and columns $J$) hold is insufficient. 
Our approach therefore imposes stronger restrictions on the support than those dictated by the axioms of $\mathsf{BTRank}^m_n(A)$ alone. 

Specifically, we require that the row vectors assigned to the rows of $X$ indexed by $I$ be linearly independent, that the column vectors assigned to the columns of $Y$ indexed by $J$ form a linearly independent set of vectors, and that the product of these rows and columns evaluates to $A_{I,J}$. Of these three requirements, only the last is directly dictated by the axioms of $\mathsf{BTRank}^m_n(A)$. We show that the uniform distribution over such assignments satisfies the marginal distribution conditions for members of our family. Notably, these assignments violate the $\mathsf{FPHP}^m_n$ axioms even in the special case $A = I_m$ and $Y = X^T$ (the oddtown case $XX^T=I_m$), highlighting the bespoke nature of our distribution family. 

It is plausible that our method could be extended to the case of \SoS, though we have not pursued this direction.

%%%%%%%%%%%%%%%%%%%%%%%%%%%%%%%%%%%%%%%%%%%%%%%%%%%%%%%%%%%
\subsection{Hardness of Circuit Lower Bounds from the Weak Rank Principle}\label{sec:intro:cktlbs}
%%%%%%%%%%%%%%%%%%%%%%%%%%%%%%%%%%%%%%%%%%%%%%%%%%%%%%%%%%%

Next, we connect the weak rank principle to the provability of circuit lower bounds. We show that the weak rank principle is \emph{necessary} for proving certain circuit lower bound statements, while in the sequel (\Cref{sec:intro:rank-is-axiom}) we show that it is  also \emph{sufficient} for proving some known circuit lower bounds of interest. These results suggest a  fundamental relationship between the weak rank principle and the provability of circuit lower bounds.

%\begin{enumerate}
%\item Can the hardness of the weak rank principle imply \emph{hardness} for circuit lower bound statements in different proof systems? Or in other words, is the weak rank principle \emph{necessary} for proving such statements? 
%\item Conversely, are short proofs of the weak rank principle \emph{sufficient} for proving known circuit lower bounds of interest?
%\end{enumerate}

Our results are inspired by---and, to some extent, mirror---the connection between the weak pigeonhole principles and the provability of complexity lower bounds. Razborov~\cite{Razb98, Razborov04} and Raz~\cite{Raz04} showed the unprovability of circuit lower bounds such as $\NP\nsubseteq\P/\poly$ in weak proof systems via reductions from WPHP. M\"uller and Pich~\cite{MullerP20} formalised a wide range of circuit lower bounds in \Jerabek{}'s theory for approximate counting in bounded arithmetic~\cite{Jerabek04,Jer05-PhD, Jer07}, which includes the dual weak pigeonhole principle as an axiom. More recently, Chen, Li, and Oliveira~\cite{CLO24} showed that certain lower bound statements are \emph{equivalent} to variants of weak pigeonhole principles.

We extend the above line of research in two aspects. First, we present low-degree reductions from the weak rank principle to circuit lower bound sentences, generalising the aforementioned connections by Razborov and Raz~\cite{Razb98, Razborov04, Raz04}. Then, as a concrete example, we show that $\PCR_{\FTwo}$ does not have efficient proofs of circuit lower bound statements such as $\NP\nsubseteq \P/\poly$, by \emph{iterating} the WRank-based proof complexity generators against $\PCR_{\FTwo}$ (as in~\cite{Razb15-annals}). A variant of circuit lower bound statements was recently proved to be hard for \SoS~\cite{AR23}. 
%. These lower bounds are inspired by~\cite{Razb15-annals}, using notions similar to iterability. \hanlin{Need to elaborate a bit on the last sentence}

%%%%%%%%%%%%%%%%%%%%%%
\paragraph{Necessity of the weak rank principle.}
First, we present low-degree reductions showing that WRank is necessary for proving circuit lower bounds:

\begin{theorem}[WRank is necessary for Boolean circuit lower bounds (informal); see \cref{cor: hardness of proving circuit lower bounds from rank principles} and \cref{remark: unprovability of NC2 lower bounds}]\label{informal thm: hardness of circuit LB from rank principles}
    Let $\calP$ be an algebraic proof system closed under low-degree reductions. If $\calP$ cannot prove the weak rank principle efficiently, then for every Boolean function $f$ (represented as a truth table), $\calP$ cannot prove circuit lower bounds for $f$ efficiently.
\end{theorem}

Interestingly, our argument goes through algebraic circuit complexity. Specifically, we show that WRank is necessary for proving lower bounds against non-commutative algebraic branching programs (non-commutative ABPs, or ncABPs for short), which is a fairly weak algebraic circuit model (making our \emph{unprovability} results stronger). Our reduction relies on Nisan's characterisation of $\ncABP$ complexity via matrix rank~\cite{Nis91}. We then establish that lower bounds on $\ncABP$ complexity are themselves necessary for proving lower bounds on (Boolean) circuit complexity.

\paragraph{Concrete lower bounds.} Inspired by the above connection, we use the generators for $\PCR_{\FTwo}$ demonstrated in \cref{sec:intro:lbandapp} to show that this system cannot efficiently prove circuit lower bounds\footnote{Unfortunately, when measured by \emph{size}, $\PCR_{\FTwo}$ is not closed under low-degree reductions. Hence our concrete lower bounds are not direct corollaries of \cref{informal thm: hardness of circuit LB from rank principles} and require more work.}.%For the algebraic result, we formulate the lower bound statement to express lower bounds against non-commutative algebraic branching programs (ncABP). This is a fairly weak algebraic circuit model (note that as the circuit model considered is weaker, the lower bound result gets stronger, since the inability to efficiently prove lower bounds against weak circuit models, informally implies the same for stronger circuit models. Nevertheless, this intuition cannot always be applied since reductions between different circuit models depend on the encoding of the statement).  

%\begin{theorem}[Informal; see \cref{thm: ncABP lb for PCR2}]
%    Let $f$ be any degree-$d$ non-commutative homogeneous polynomial in $n$ variables, and $r \ge 1$ be a size parameter. Then any $\PCR_{\F_2}$ proof of the statement ``$f$ cannot be computed by an $\ncABP$ of width $r$'' (encoded as a system of polynomial equations) requires size $2^{\Omega(\sqrt{r/d})}$. %and denote $\vec{r} = (1, \underbrace{r, r, \dots, r}_{(d-1)\text{ $r$'s}}, 1)$.
%\end{theorem}

%The formula $\lb_\ncABP(f, \vec{r})$ encodes the existence of an ncABP for $f$.

\begin{theorem}[Informal; see \cref{thm: boolean lb for PCR2}]\label{thm:intro:boolean lb for PCR2}
    Let $f\colon \Q^n \to \Q$. 
    For every $s>n^{\Omega(1)}$, any $\PCR_{\F_2}$ refutation of the statement ``$f$ cannot be computed by Boolean circuits of size $s$'' (encoded as a system of polynomial equations) requires size $2^{s^{\Omega(1)} / n^{O(1)}}$.
\end{theorem}

As in the case of Razborov~\cite{Razb15-annals} our Boolean circuits 
are in the basis $\lnot,\land,\lor,\oplus$
where the last three connectives are binary. %$\lb^{\oplus}(f, s)$ encodes the existence of such a circuit for $f$ of size $s$.\hanlin{Should we mention this technical detail (?) here? Even if we want to mention it, we probably just need a footnote...}
%\hanlin{Do we want to say something about the proofs?}\slava{Why did you remove the concrete bound for ncABP?}
The proof idea is similar to~\cite{Razb15-annals}. We work with the iterated rank principle $\bigcup_{\pi \in \Q^{\le n}} X^{(\pi)} Y^{(\pi)} = A^{(\pi)}$ encoded as a CNF using the bamboo-tree encoding. Given $\pi \in \Q^n$, we choose the matrices $A^{(\pi)}$ such that $A^{(\pi)}_{1,1} = f(\pi)$ the rest of the matrices (for $|\pi|<n$) are arranged as
$A^{(\pi)} = \left[ X^{(\pi * 0)}~X^{(\pi * 1)}~{Y^{(\pi * 0)}}^T~{Y^{(\pi * 1)}}^T \right]$.
% $A^{(\pi)} = \begin{bmatrix}
%     X^{(\pi * 0)} & X^{(\pi * 1)} & {Y^{(\pi * 0)}}^T & {Y^{(\pi * 1)}}^T
% \end{bmatrix}$.
That is, given a copy $X^{(\pi)} Y^{(\pi)} = A^{(\pi)}$, we can compute both copies $X^{(\pi * 0)} Y^{(\pi*0)} = A^{(\pi*0)}$ and $X^{(\pi * 1)} Y^{(\pi*1)} = A^{(\pi*1)}$ recursively. This way, starting with the empty string ``$()$'', we can gradually compute $A^{(\pi)}$ for $\pi \in \Q^n$, thus obtaining $f(\pi)$.
This computation can be naturally expressed as a Boolean circuit, using the gates that correspond to the extension variables.

%Note: We already have lower bounds against PC size (not PCR) for ncABP in this file. This is done via the Razborov 1998 degree PC lower bounds (and then using the size-degree relation by IPS'99 for PC [not PCR]).
%(Note that should imply commutative ABP lower bounds; hence, algebraic circuit lower bounds are hard for PC size as well, though the reductions to these cases may not go through efficiently in PC).

%The fact that the weak rank principle is necessary for \NCTwo\ Boolean circuit lower bounds is expected to carry over to the \emph{algebraic circuit} setting, to yield a rather interesting result: ``rank arguments are necessary for algebraic circuit lower bounds'', namely, to prove algebraic circuit lower bounds one needs to be able to prove statements about linear algebra and specifically the weak rank principle (nevertheless, we have not verified the full details of such a reduction to algebraic circuits; only to Boolean \NCTwo\ circuits). 
%\hanlin{What does this mean? We don't have hardness of algebraic circuit lower bounds yet, so perhaps let's just mention Boolean circuit lower bounds?}\Iddo{It means we went towards this result, but didn't (have time yet to) formally finish it. Maybe move to conclusions?} \hanlin{Since we don't have such results, I commented the above}

%%%%%%%%%%%%%%%%%%%%%%%%%%%%%%%%%%%%%%%%%%%%%%%%%%%%%%%%%
\subsection{On the Strength of WRank}
\label{sec:intro:rank-is-axiom}
% is sufficient for circuit lower bounds.}
%%%%%%%%%%%%%%%%%%%%%%%%%%%%%%%%%%%%%%%%%%%%%%%%%%%%%%%%%

%\hanlin{Only partial progress here...}

Finally, we initiate the study of the weak rank principle as an axiom in bounded arithmetic. 
Linear algebra is used extensively throughout combinatorics and complexity theory, resulting in a variety of breakthroughs such as the resolution of the finite field Kakeya conjecture~\cite{dvir2009size}, Cap Set problem~\cite{croot2017progression, ellenberg2017large} and the Sensitivity Conjecture~\cite{huang2019induced}. Indeed, there is a whole book dedicated to linear algebraic methods in combinatorics~\cite{BF1992}.

Soltys and Cook~\cite{SC04} initiated the study of formal theories that ``capture linear algebra'' from the complexity perspective. Specifically, they considered a few formal theories for linear algebraic reasoning and showed that $\forall\mathrm{LA}\mathtt{P}$, the strongest theory considered in~\cite{SC04} which incorporates matrix powering and certain induction schema, proves many linear algebraic identities including the (strong) rank principle. Subsequently, Tzameret and Cook~\cite{TC21} showed that the seemingly weaker theory $\VNC^2$ suffices for linear algebraic reasoning and specifically proving properties of the determinant. We take a somewhat similar route, but instead of powerful concepts such as matrix powering (which is $\mathsf{DET}$-complete~\cite{Coo85} and hence intuitively captures ``full'' linear algebra), we only consider the \emph{weak} rank principle, hence our theories appear to be somewhat weaker.%, and are tailored especially to capture complexity arguments which are not fully exponential as we explain below.   

An important motivation for considering the \emph{weak} rank principle is that they capture arguments in combinatorics and complexity theory that are ``\emph{loose}'' (i.e., not tight). Intuitively, the \emph{weak} rank principle should already suffice when one is satisfied with a bound that is only tight within multiplicative factors. This is the case when we are satisfied with, say, a bound of $c\cdot 2.756^n$ for the Cap Set problem or a bound of $c\cdot 2^{n^{1/2d}}$ for $\AC^0[p]$ circuit complexity, where $c > 0$ is an unspecified constant. This is similar to the case of approximate counting where we only care about the sizes of sets \emph{approximately}, hence the \emph{weak} pigeonhole principle is already applicable~\cite{Jer07}.

In this work, we put forward a theory $\overline{\V^0(p)} + \wRank_p$, which incorporates the two-sorted theory $\overline{\V^0(p)}$ in the style of Cook--Nguyen~\cite{CN10} with an extra axiom $\wRank_p$ expressing the weak rank principle over $\F_p$. We demonstrate the power of the weak rank principle and confirm the above intuition by showing that this theory can prove Smolensky's circuit lower bound against $\AC^0[p]$~\cite{Smolensky87}:
\begin{theorem}[Informal; Smolensky's lower bound  is implied by the weak rank principle; see \Cref{thm:main-smolensk-in-BA}]
    For every prime number $p > 2$, $\overline{\V^0(p)} + \wRank_p$ proves that $\MOD_2$ cannot be computed by $\AC^0[p]$ circuits of depth $d$ and size $2^{o(n^{1/(2d)})}$.
\end{theorem}

Previously, M\"uller and Pich~\cite{MullerP20} formalised Smolensky's $\AC^0[p]$ lower bounds in \Jerabek{}'s theory $\mathbf{APC}_1$~\cite{Jerabek04, Jer07}. 
\Iddo{I\ moved the comparison here, as it seems quite important.}
%
%The difference between our formalisation and theirs %is as follows.
%will be discussed in \cref{remark: comparison with %Muller-Pich}, at the end of \cref{sec: formalisations %of smolensky}.
%
%\begin{remark}[Comparison with {\cite{MullerP20}}]\label{remark: %comparison with Muller-Pich}
 We point out a few differences between our formalisation of Smolensky's lower bounds and the formalisation in~\cite{MullerP20}.

    \begin{enumerate}
        \item M\"uller and Pich proved a lower bound against circuits of size $n^{\log n}$ in the (single-sorted) theory $\mathbf{APC}_1$ with the assumption that $n^{\log^{9d} n} \in \Log$. This allows randomised quasi-polynomial-time reasoning over the purported $\AC^0[p]$ circuit. We prove a lower bound against size $2^{\Omega(n^{1/2d})}$ in the (two-sorted) theory $\overline{\V^0(p)} + \wRank$ where the circuit is represented as an object of string sort. Except for the weak rank principle, this theory only allows (quasi-polynomial-size) $\AC^0[p]$-reasoning over the purported $\AC^0[p]$ circuit.
        \item One drawback in the M\"uller--Pich proof is the requirement that $n^{\log^{9d} n} \in \Log$. In fact, this allows them to prove the required rank principle by \emph{brute force}. Our formalisation explicitly separates the weak rank principles from other $\AC^0[p]$-manipulations.
    \end{enumerate}

We also note that Razborov mentioned in~\cite[Section E.3]{Razb95} that the metamathematics of Razborov--Smolensky lower bounds are left open. Our work addresses this gap by showing that Smolensky's lower bounds can be proved in $\overline{\V^0(p)} + \wRank_p$.
It is also worth mentioning the already Galesi \emph{et al.}~\cite{GGPS23} considered the rank principle with $A=I$ as an axiom added to the polynomial calculus.

%\hanlin{resembles APC1. Maybe say something here}
%\newpage 
% %%%%%%%%%%%%%%%%%%%%%%%%%%%%%%%%%%%%%%%%%%%%%%
 \section{Conclusions and Open Problems}
% %%%%%%%%%%%%%%%%%%%%%%%%%%%%%%%%%%%%%%%%%%%%%%
% 

Our work  introduces the weak rank principle as a new object of study in proof complexity and applies it to advance the \emph{generator approach} to proof complexity, introduced by \Krajicek{}~\cite{Kra01-Fundamenta,Kra04} and Alekhnovich \emph{et al.}~\cite{ABRW04}, and later emphasized by Razborov~\cite[p.~417]{Razb15-annals}.
The goal of this approach is to construct \emph{proof-complexity generators} for as many proof systems as possible, where the hardness of the generator reflects the computational limitations of the corresponding system.
In our case, the hardness originates from linear-algebraic reasoning: the weak rank principle expresses the unsatisfiability of certain linear systems and matrix identities, and its difficulty is directly tied to the inability of a proof system to efficiently formalise arguments involving matrix rank.
Consequently, the weak rank principle is expected to be hard for proof systems that cannot efficiently reason about linear algebra.

\medskip
\noindent\textbf{Future directions.}
Several directions appear promising for further study:
\begin{itemize}\itemsep=0pt
\setlength{\leftskip}{-18pt}    % removes indentation from the left
  \setlength{\itemindent}{0pt}  % ensures item label not indented
  \setlength{\labelsep}{0.4em}  % space between bullet and text
  \setlength{\labelwidth}{9pt}

    \item The weak rank principle provides a useful family of tautologies to study in proof systems for which the pigeonhole principle is easy or for which no proof-complexity generators are known.
The basic instance, $XY=I$, already gives a concrete candidate for proving lower bounds.
Obtaining a lower bound for this instance is the first step toward establishing lower bounds for the more general case $XY=A$, where $A$ ranges over matrices of rank greater than $n$.
Such results would yield a proof-complexity generator based on WRank.
Once such a base generator is obtained, it can be iterated to obtain a \emph{function generator}, showing that the proof system in question cannot efficiently prove certain complexity-class separations such as $\NP\nsubseteq \P/\poly$.

  % \item Extending the generator construction to other proof systems, such as $\SoS$ and $\ACZ[p]$-Frege, where the behavior of algebraic principles like WRank remains largely unexplored.

  \item Develop bounded arithmetic theories that use WRank as axioms, such as the theory $\overline{\V^0(p)} + \wRank_p$ developed in this paper. We speculate that many interesting results in combinatorics and complexity theory can be formalised in such theories and that the \emph{weak} rank principles suffice. Our formalisation of Smolensky's lower bounds~\cite{Smolensky87} can be seen as the first step towards this goal.
  %\item Using WRank as a \emph{basic axiom} to derive or simulate other lower bounds, potentially formalizing rank-based arguments that appear in combinatorics and circuit complexity (as we have shown for the case of  dimension arguments used in Smolensky's lower bounds~\cite{Smolensky87}). \hanlin{deleted ``for majority'' because the lower bounds are for MOD2, not for majority}
  %\item Developing formal frameworks for expressing and proving rank arguments inside weak proof systems for  non-tight lower bounds (while the strong rank principle may be required to formulate tight exponential lower bounds, the weaker WRank may be enough to prove weaker lower bounds).     

\item The weak rank principle provides a natural candidate for lower bounds against $\ACZ[p]$-Frege (for which no lower bounds are currently known) and $\SoS$.
This connection follows from the conjecture that $\mathsf{NC}^2$ captures the computational complexity of linear algebra.
According to the informal correspondence between circuit classes and propositional proof systems whose proof-lines come from those classes, one expects that proofs operating with $\mathsf{NC}^2$-circuits (that is, $\mathsf{NC}^2$-Frege) are exactly those capable of efficiently reasoning about rank-based arguments.
In contrast, the weak pigeonhole principle admits quasipolynomial-size proofs already in $\Res(\log n)$ and thus in $\ACZ$-Frege~\cite{PWW88,MPW02}, indicating that stronger algebraic principles are required to obtain meaningful lower bounds.

\begin{comment}
Indeed, $\PMOT^m_n$---a specific instance of WRank---combines parity-based reasoning with a rank argument, a structure that aligns with known weaknesses of $\ACZ[p]$-Frege.
This makes such formulas natural candidates for hard instances in these systems.\Iddo{check this sentence}
Moreover, the CNF encoding of WRank appears well suited to existing lower-bound techniques and could serve as a starting point for proving strong $\ACZ$-Frege or $\ACZ[p]$-Frege lower bounds beyond those obtainable via Håstad's switching lemma, potentially leading to bounds independent of the depth parameter~$d$ \hanlin{What does ``bounds independent of $d$'' mean?}.
\end{comment}

\end{itemize}

\Iddo{Mention: LST25?}

\if\Anonymity=0
\section*{Acknowledgement}
We are indebted to Robert Andrews for very helpful discussions throughout this project.
\fi

\addtocontents{toc}{\protect\setcounter{tocdepth}{1}}
%!TEX root = main.tex

%\newpage % for submission 

\section{Preliminaries}\label{sec:prelims}
We will use the following common notations throughout the paper.
\begin{itemize}
    \item For all $k>0$, $I_k$ denotes the $k \times k$ identity matrix ${(I_k)}_{i,j} = \delta_{i,j}$ for all $i,j \in [k]$.
    \item For all $k, \ell>0$, $J_{k,\ell}$ denotes the $k \times \ell$ all-ones matrix, i.e., ${(J_{k,\ell})}_{i,j} = 1$ for all $i \in [k], j \in [\ell]$.
\end{itemize}

\subsection{Concentration Bounds}\label{sec: prelims concentration bounds}

Many of our constructions use probabilistic arguments and rely on concentration bounds.
We first recall a standard version of the Chernoff bound, which can be found in many textbooks; see, for example,~\cite[Corollary~4.6]{MU05-book}.
\begin{theorem}[Chernoff bound]\label{thm:chernoff bound}
    Let $X_1, \ldots, X_n$ be independent Bernoulli random variables with $\Prb{X_i = 1} = p_i$ for all $i \in [n]$. Let $X = \sum_{i \in [n]} X_i$ and $\mu = \Exp{X}$. Then, for every $0 < \delta < 1$,
    \[
        \Prb{X \ge (1 + \delta)\mu} \le \exp(-\delta^2 \mu/3)
    \]
    and
    \[
        \Prb{X \le (1 - \delta)\mu} \le \exp(-\delta^2 \mu/2).
    \]
    Consequently,
    \[
        \Prb{\lvert X - \mu \rvert \ge \delta \mu} \le 2\exp(-\delta^2 \mu/3).
    \]
\end{theorem}

A useful special case of the Chernoff bound is sampling with replacement.
Given a positive integer $N$ and a subset $K \subseteq [N]$, we uniformly draw $n$ samples from $[N]$ with replacement. For every $i \in [n]$, let $X_i$ be the indicator variable of the event that the $i$th sample is in $K$. Then $X = \sum_{i \in [n]} X_i$ counts the number of samples that are in $K$. \cref{thm:chernoff bound} gives a concentration bound for $X$ around its expectation $\mu = n\cdot \card{K}/N$.

We also use the analogous bound for sampling without replacement. The following result follows, for example, from Hoeffding's theorem for samples drawn without replacement; see~\cite[Theorem~4]{Hoeffding63}. In this setting, the number of samples belonging to the set $K$ has a hypergeometric distribution.
\begin{theorem}[Hypergeometric Chernoff bound]\label{thm: hoeffding chernoff bound}
    Let $K\subseteq[N]$ be fixed, and let $S$ be an $s$-subset of $[N]$ chosen uniformly at random.
    Let $X=\card{K \cap S}$ and $\mu=\Exp{X}$. Then, for every $0<\delta<1$,
    \[
        \Prb{X \ge (1 + \delta)\mu} \le \exp(-\delta^2 \mu/3)
    \]
    and
    \[
        \Prb{X \le (1 - \delta)\mu} \le \exp(-\delta^2 \mu/2).
    \]
    Consequently,
    \[
        \Prb{\lvert X - \mu \rvert \ge \delta \mu} \le 2\exp(-\delta^2 \mu/3).
    \]
\end{theorem}

\subsection{Proof Systems}\label{sec: prelims proof systems}

Let $\TAUT$ denote the $\coNP$-complete set of propositional tautologies. Cook and Reckhow~\cite{CookReckhow74a} defined \emph{propositional proof systems} as nondeterministic algorithms for $\TAUT$. Equivalently, a propositional proof system $\calP(\phi, \Pi)$ is a deterministic algorithm that takes as input a propositional formula $\phi$ and a purported proof $\Pi$, and satisfies the following properties:
\begin{itemize}
    \item {\bf Completeness:} If $\phi \in \TAUT$, then there exists a proof $\Pi$ such that $\calP(\phi, \Pi)$ accepts.
    \item {\bf Soundness:} If $\phi \notin \TAUT$, then for every $\Pi$, $\calP(\phi, \Pi)$ rejects.
    \item {\bf Efficiency:} $\calP$ runs in time polynomial in $\card{\phi} + \card{\Pi}$.
\end{itemize}

A propositional proof system $\calP$ is \emph{polynomially bounded} (or \emph{p-bounded}) if there is a polynomial $p(\cdot)$ such that every tautology $\phi \in \TAUT$ admits a proof $\Pi$ of length at most $p(\card{\phi})$ for which $\calP(\phi, \Pi)$ accepts. Cook and Reckhow~\cite{CookReckhow74a} showed that p-bounded propositional proof systems exist if and only if $\NP = \coNP$.

We will sometimes equivalently consider \emph{refutation systems}, which are nondeterministic algorithms for the language $\overline{\SAT}$. A DNF $F$ is a tautology if and only if $\lnot F$ (which can be expressed as a CNF) is unsatisfiable. Thus, a \emph{proof} of $F$ can be equivalently viewed as a \emph{refutation} of $\lnot F$.

While traditional proof systems reason about Boolean formulas, \emph{algebraic proof systems} reason about the unsatisfiability of collections of \emph{polynomial equations}. Fix an underlying field $\F$, and let $f_1, f_2, \ldots, f_m \in \F[x_1, x_2, \ldots, x_n]$ be a collection of polynomials, where each $f_i$ is called an \emph{axiom} and represents the equation $f_i(\vec{x}) = 0$. The goal is to prove that these polynomials do not have a common zero over $\F^n$. It is standard in the literature to add the \emph{Boolean axioms} $x_i^2 - x_i$ to the collection to enforce that all variables take values in $\{0, 1\}$; however, we do \emph{not} include these Boolean axioms in our definition of rank principles.

For a polynomial $f$, let $\|f\|$ denote the number of monomials appearing with a non-zero coefficient in $f$.

\begin{definition}
    Let $F = (f_1, f_2, \dots, f_m)$ be an unsatisfiable collection of polynomial equations over a field $\F$.
    \begin{itemize}
        \item A \emph{Nullstellensatz} ($\NS_\F$)~\cite{BIKPP96} refutation of $F$ is a sequence of polynomials $(g_1, g_2, \dots, g_m)$ such that the following polynomial identity holds:
        \[\sum_{i=1}^m f_i\cdot g_i = 1.\]
        The \emph{degree} of the refutation is $\max_{i\in [m]}\{\deg(f_i\cdot g_i)\}$. The \emph{size} of the refutation is $\sum_{i=1}^m \|f_i\|\cdot \|g_i\|$.
        \item \emph{Polynomial Calculus} ($\PC_\F$)~\cite{CleggEI96} is a dynamic proof system where proof lines are polynomials over $\F$. Starting from $(f_1, f_2, \dots, f_m)$ as axioms, this system has two inference rules:
        \[
        \begin{prooftree}
                \hypo{f}
                \hypo{g}
                \infer2{\alpha f + \beta g}
        \end{prooftree}~\text{(for $\alpha, \beta \in \F$)}
        \quad\text{and}\quad
        \begin{prooftree}
            \hypo{f}
            \infer1{x_i\cdot f}
        \end{prooftree}.
        \]
        A $\PC_\F$ \emph{refutation} of $F$ is a derivation of $1$ from $F$. The \emph{degree} of the refutation is the maximum degree of any polynomial appearing in it. The \emph{size} of the refutation is the total number of monomials appearing in it.
        
        \item Suppose $\F = \R$. A \emph{Sum-of-Squares} ($\SoS$)~\cite{GrigorievV01,ODonnellZ13}
        refutation of $F$ is a sequence of polynomials $(g_1, g_2, \dots, g_m)$ and $(h_1, h_2, \dots, h_\ell)$ such that the following polynomial identity holds:
        \[
            \sum_{i=1}^m f_i\cdot g_i + \sum_{j=1}^\ell h_j^2 = -1.
        \]
        The \emph{degree} of the refutation is the maximum of $\max_{i\in [m]}\{\deg(f_i\cdot g_i)\}$ and $2\max_{j\in [\ell]}\{\deg(h_j)\}$. The \emph{size} of the refutation is $\sum_{i\in [m]}\|f_i\|\cdot \|g_i\| + \sum_{j\in [\ell]} \|h_j\|$.
    \end{itemize}
\end{definition}

\begin{definition}
    Let $F = (f_1, f_2, \dots, f_m)$ be an unsatisfiable collection of polynomial equations over $\R$. We enforce that the variables $x_1, x_2, \dots, x_n$ are Boolean by adding every Boolean axiom $x_i^2 - x_i$ to $F$. A \emph{Sherali--Adams} ($\SA$)~\cite{DantchevM13} refutation of $F$ is a sequence of polynomials $(g_1, g_2, \dots, g_m)$ and a conical junta $\mathcal{J}$ such that the following polynomial identity holds.
    \[\sum_{i\in [m]}f_i\cdot g_i + \mathcal{J} = -1.\]
    Here, $\mathcal{J}$ is a \emph{conical junta} if it is a nonnegative linear combination of terms ($\mathcal{J} = \sum_{j=1}^\ell\alpha_j t_j(x)$ for $\alpha_j \ge 0$), and each term is a product of factors of the form $x_i$ and $(1-x_j)$. The \emph{degree} of the refutation is the maximum of $\deg(\mathcal{J})$ and $\max_i\{\deg(f_i) + \deg(g_i)\}$.
    The \emph{size} of the refutation is $\sum_{i\in [m]}\|f_i\|\cdot \|g_i\| + \ell$.
\end{definition}

\begin{definition}
    Let $F = (f_1, f_2, \dots, f_m)$ be an unsatisfiable collection of polynomial equations over a field $\F$ with underlying variables $x_1, x_2, \dots, x_n$. We enforce that the variables are Boolean by adding every Boolean axiom $x_i^2-x_i$ to $F$.  A \emph{Polynomial Calculus Resolution} ($\PCR_\F$)~\cite{ABSRW99} refutation of $F$ is a $\PC_\F$ refutation of $F'$, where $F'$ is $F$ with the \emph{twin variables} $\overline{x}_1, \overline{x}_2, \ldots, \overline{x}_n$ and the axioms $1 - x_i - \overline{x}_i$ and $\overline{x}_i^2-\overline{x}_i$ added.

    Note that $\PC_\F$ and $\PCR_\F$ are equivalent w.r.t.~the degree measure, but a collection of polynomial equations might have much more efficient $\PCR_\F$ proofs than $\PC_\F$ proofs in terms of size.
\end{definition}

\subsection{Substitutions} \label{sec:Substitutions}

In the setting of proof complexity, substitutions are usually defined as follows.
\begin{definition}[Restrictions]
    Given variables $(x_1, \ldots, x_n)$ and their twins $(\overline{x}_1, \ldots, \overline{x}_n)$, the \emph{restriction $\rho$} is a partial mapping from $\{x_1, \ldots, x_n\} \cup \{\overline{x}_1, \ldots, \overline{x}_n\}$ to $\{0,1\}$.
\end{definition}

Here we consider a generalised notion of substitutions in which we can substitute variables with fresh variables, possibly negated; several variables may be mapped to the same fresh variable.
\begin{definition}[Substitutions]\label{def:subs}
    Given variables $(x_1, \ldots, x_n)$ and $(y_1, \ldots, y_m)$ and their twins $(\overline{x}_1, \ldots, \overline{x}_n)$ and $(\overline{y}_1, \ldots, \overline{y}_m)$, the \emph{substitution $\rho$} is a mapping from $\{x_1 \ldots, x_n\} \cup \{\overline{x}_1, \ldots, \overline{x}_n\}$ to $\{0,1\} \cup \{y_1, \ldots, y_m\} \cup \{\overline{y}_1, \ldots, \overline{y}_m\}$, which is consistent w.r.t.~the twin variables: if $\rho(v)=b$ for some $b \in \Q$, then $\rho(\overline{v})=1-b$; if $\rho(v)=u$ for some variable $u$, then $\rho(\overline{v})=\overline{u}$.
\end{definition}
It coincides with the standard definition when $(y_1, \ldots, y_m) = (x_1, \ldots, x_n)$ and for every $v$ with $\rho(v) \notin \{0,1\}$, $\rho(v)=v$.

Given a polynomial $f \in \F[x_1, \ldots, x_n, \overline{x}_1, \ldots, \overline{x}_n]$ and a substitution $\rho$, the application of $\rho$ to $f$, denoted $f \restriction \rho$, is the polynomial $g(y_1, \ldots y_m, \overline{y}_1, \ldots, \overline{y}_m) = f(\rho(x_1), \ldots, \rho(x_n), \rho(\overline{x}_1), \ldots, \rho(\overline{x}_n))$.

Given a proof $\Pi$ (a sequence of polynomials), the application of a substitution $\rho$ to $\Pi$, denoted $\Pi \restriction \rho$, is obtained by applying the $\rho$ to each line $p$ of $\Pi$ and removing any line that becomes the zero polynomial.
Applying a substitution to a $\PCR_{\F}$, $\SA$, or $\SoS$ proof results in a proof with at most a linear increase in size: this is because \Cref{def:subs} substitutes variables and constants for variables; when we replace a variable with a polynomial, size is not necessarily preserved.

%!TEX root = main.tex

\subsection{Reductions in Proof Complexity}\label{sec: reductions}

Let $F$ and $G$ be two collections of polynomial equations. In this paper, we use \emph{low-degree algebraic reductions}~\cite{BGIP01, RezendeGNPR021} to formalise the idea that $F$ is ``easier'' to prove than $G$. Intuitively, this happens when $F$ is a substitution instance of $G$, namely there is a substitution $R$ to $G$ under which $G \circ R$  equals $F$. More precisely, we require that there is an efficient (e.g., low degree) derivation from $F$ of the substitution instance $G\circ R$. Hence, a short proof of $G\circ R$ can be transformed into a short proof of $F$.
For example, to show that \emph{circuit lower bounds} are as hard to prove as the \emph{weak rank principle}, we exhibit a low-degree reduction from the polynomial equations encoding the weak rank principle to those encoding circuit lower bounds.

%This section is of preliminary character: it introduces the notion of algebraic reductions and discusses the proof systems to which algebraic reductions are applicable. We remark that this notion is not new and has already been considered in previous works.

\begin{definition}[{Algebraic reductions;~\cite[Def.~8.2]{RezendeGNPR021}}]\label{def: algebraic reduction}
        Let $F = (f_1, \dots, f_m)$ be a collection of polynomials over variables $(x_1, \ldots, x_m')$, and $G = (g_1, \dots, g_n)$ be a collection of polynomials over variables $(y_{1}, \ldots, y_{n'})$. A \emph{degree-$d$ algebraic reduction}, denoted $F\le^\alg_d G$, consists of the following:
        \begin{itemize}
                \item [\textbf{(Reduction)}] a mapping $R\colon \F^{m'} \to \F^{n'}$ such that each output element $R_i\in \F[x_1,\ldots,x_{m'}]$, $i \in [n']$, is a polynomial of degree at most $d$ in $(x_1, \ldots, x_{m'})$;
                \item [\textbf{(Axioms)}] for every $i\in[n]$, the polynomial $g_i\circ R$ has an $\NS_\F$-proof from $F$ of degree $d\cdot \deg(g_i)$. That is, there exist polynomials $(q_{i,1}, \dots, q_{i,m})$ such that
                \[g_i\circ R = \sum_{k=1}^m f_kq_{i,k},\]
                and for every $k \in [m]$, $\deg(f_k) + \deg(q_{i,k}) \le d\cdot \deg(g_i)$.
        \end{itemize}
        % Finally, we say that the reduction is \emph{linear} (denoted as $F\le^\lin G$) if each $R_i$ is a degree-$1$ polynomial and each $q_{i,j}$ is a constant in $\F$.
        %\slava{Removed the mention to linear reductions since we call them ``degree-$1$'' consistently.} \hanlin{Sounds good!}
\end{definition}

% \begin{remark}
%    \hanlin{Note that the ``axiom'' definition in~\cite{RezendeGNPR021} needs that the \emph{multilinearisation} of $g_i\circ R$ has an $\NS$ proof. Here we're working in a purely algebraic setting (variables represent field elements, not bits) so multilinearisation doesn't make sense...}

%     \slava{The one in~\cite{BGIP01} doesn't require Boolean axioms. So it's actually closer to ours. The only difference between theirs and ours is that they require $\PC$ proofs and we only need $\NS$.}

%     \slava{\sout{Why are you defining only degree-$d$ and not degree-$(d_1, d_2)$. For CNFs we actually need the latter, since the final encoding is $(1,O(1))$-algebraic reduction, but this is still okay for the applications.}}
%     \slava{\sout{Maybe define it for $(d_1,d_2)$ and say that we call them degree-$d$ when $d_1=d_2=d$?}}
%     \hanlin{\sout{This is indeed a nice suggestion but it seems that we don't need it in this paper}}
% \end{remark}

It is a standard and useful fact that algebraic reductions are closed under composition.

\begin{fact}\label{fact: composition of algebraic reductions}
        Let $F = (f_1, \dots, f_m)$, $G = (g_1, \dots, g_n)$, and $H = (h_1, \dots, h_\ell)$ be systems of polynomial equations. If $F\le^\alg_{d_1} G$ and $G\le^\alg_{d_2} H$, then $F\le^\alg_{d_1d_2} H$.
\end{fact}
\begin{proof}
        Suppose that $F$ is over variables $(x_1, \ldots, x_{m'})$, $G$ is over variables $(y_1, \ldots, y_{n'})$, and $H$ is over variables $(z_1, \ldots, z_{\ell'})$. Let $R^1: \F^{m'} \to \F^{n'}$ be a degree-$d_1$ reduction from $F$ to $G$ and $R^2: \F^{n'}\to \F^{\ell'}$ be a degree-$d_2$ reduction from $G$ to $H$. We claim that $R^2\circ R^1: \F^{m'}\to \F^{\ell'}$ is a degree-$d_1d_2$ reduction from $F$ to $H$. Clearly, every output element of $R^2\circ R^1$ is a polynomial of degree at most $d_1d_2$ in $(x_1, \ldots, x_{m'})$.
        
        Now consider an axiom $h_i$ from $H$. We want to show that $h_i\circ R^2\circ R^1$ has an $\NS$-proof from $F$ of degree $d_1d_2\cdot \deg(h_i)$. We have
        \[h_i\circ R^2 = \sum_{k=1}^n g_k q_{i,k}\]
        for some  polynomials $(q_{i,1}, \dots, q_{i,n})$ such that for every $k \in [n]$, $\deg(g_k) + \deg(q_{i,k}) \le d_2\cdot \deg(h_i)$. For each $k\in[n]$, we also have
        \[g_k\circ R^1 = \sum_{j=1}^m f_j p_{k,j}\]
        for some polynomials $(p_{k,1}, \dots, p_{k,m})$ such that for every $j \in [m]$, $\deg(f_j) + \deg(p_{k,j}) \le d_1\cdot \deg(g_k)$. Hence
        \begin{equation}\label{eq: combined NS proof}
                h_i\circ R^2\circ R^1 =\sum_{k=1}^n (g_k\circ R^1)(q_{i,k}\circ R^1)
                % =\sum_{k=1}^n \sum_{j=1}^m f_jp_{kj}\cdot (q_{ik}\circ R^1)
                =\sum_{j=1}^m f_j \cdot \underbrace{\mleft(\sum_{k=1}^n p_{k,j}\cdot (q_{ik}\circ R^1)\mright)}_{r_{i,j}}.
        \end{equation}
        Note that $\deg(f_j) + \deg(r_{i,j}) \le \max_k\{\deg(f_j) + \deg(p_{k,j}) + \deg(q_{i,k})\cdot d_1\} \le \max_k\{d_1\cdot (\deg(g_k) + \deg(q_{i,k}))\} \le d_1d_2\cdot \deg(h_i)$. Hence, \eqref{eq: combined NS proof} is an $\NS$-proof of $h_i\circ R^2\circ R^1$ from $F$ that has degree $d_1d_2\cdot \deg(h_i)$.
        
        In conclusion, $R^2\circ R^1$ is a degree-$d_1d_2$ algebraic reduction from $F$ to $H$.
\end{proof}

%\hanlin{Define reductions among CNFs.}

%\subsection{Proof Systems Closed Under Algebraic Reductions}

Algebraic proof systems such as Nullstellensatz and Polynomial Calculus are closed under low-degree algebraic reductions when the efficiency measure is degree~\cite[Lemma 8.3]{RezendeGNPR021}. This paper exhibits many low-degree algebraic reductions; for example, \cref{cor: hardness of proving circuit lower bounds from rank principles} shows a low-degree algebraic reduction from the weak rank principles to circuit lower bound statements. Such a result can be interpreted as follows: for any algebraic proof system $\calP$ closed under low-degree algebraic reductions, if $\calP$ cannot prove the weak rank principles, then $\calP$ cannot prove any circuit lower bounds.

%$\Res(\PC_{d, \F_p})$

%\hanlin{Res(lin)? Res(PC-d)? AC0[p]-Frege? Other proof systems?}

\subsection{The Pigeonhole Principle}

We define the \emph{algebraic} encoding of the pigeonhole principle, first considered by Razborov in~\cite{Razb98}---this version is sometimes explicitly called \emph{functional} because of the presence of the Functionality Axioms below.
Fix positive integers $m$ and $n$ with $m > n$.
$\FPHP^m_n$ is defined over $mn$ variables arranged as a Boolean matrix $X = {(x_{i,k})}_{i \in [m], k \in [n]}$. It consists of the following polynomial equations:

\textbf{Pigeon Axioms:} For $i \in [m]$, the polynomial
\begin{equation}\label{eq:php_pigeon_ax}
    1 - \sum_{k \in [n]} x_{i,k}.
\end{equation}

\textbf{Hole Axioms:} For $i,j \in [m], i \neq j, k \in [n]$, the polynomial
\begin{equation}\label{eq:php_hole_ax} 
    x_{i,k}x_{j,k}.
\end{equation}

\textbf{Functionality Axioms:} For $i \in [m], k, \ell \in [n], k \neq \ell$, the polynomial
\begin{equation}\label{eq:php_functionality_ax}
    x_{i,k}x_{i,\ell}.
\end{equation}

\textbf{Boolean Axioms:} For $i \in [m], k \in [n]$, the polynomial
\begin{equation}\label{eq:php_boolean_axioms}
    x_{i,k}^2-x_{i,k}.
\end{equation}

We comment that the standard (non-functional) PHP that maps $m$ pigeons to $n$ holes is defined the same as $\FPHP^m_n$, only excluding the Functionality Axioms~\eqref{eq:php_functionality_ax}.

Razborov~\cite{Razb98} proved a $\PC_\F$ degree lower bound on $\FPHP^m_n$ for arbitrary $m>n$.

\begin{theorem}[\cite{Razb98}] \label{thm:razborov_php_degree}
For every field $\F$, every $\PC_{\F}$ refutation of $\FPHP^m_n$ has a term that mentions at least $n/2+1$ pigeons and at least $n/2+1$ holes.\footnote{In fact, this term encodes a matching of size $n/2+1$.}
\end{theorem}

Although not stated explicitly, the lower bound in~\cite{Razb98} holds when the refutation is considered modulo the ideal generated by the Functionality and Hole Axioms. This implies that every refutation of $\FPHP^m_n$ contains a term $\prod_{\nu \in [\ell]} x_{i_\nu,k_\nu}$, where $\ell \ge n/2+1$, all $i_\nu$ are distinct and all $k_\nu$ are distinct.

\begin{comment}

\begin{theorem}[\cite{Razb98}] \label{old:thm:razborov_php_degree}
For every field $\F$, every $\PC_{\F}$ refutation of $\FPHP^m_n$ has degree at least $n/2+1$.
\end{theorem}

\slava{Rewrite the next corollary such that we also count holes, not only pigeons.}
\slava{We can also just city Razborov directly: he works with the ideal over func.+hole axioms so it's clear that his result implies this.}
\begin{corollary} \label{cor:row_degree_lb_for_php}
For every field $\F$, every $\PCR_{\F}$ refutation of $\FPHP^m_n$ contains a term that mentions at least $n/2$ pigeons. 
\end{corollary}
\begin{proof}
Assume for contradiction that there is a $\PCR_{\F}$ refutation $\Pi$ of $\FPHP^m_n$ in which each term mentions less than $n/2$ pigeons. Substitute every variable $\overline{x}_{i,k}$ in $\Pi$ by $1 - x_{i,k}$. This does not increase the maximum number of pigeons mentioned by a term and transforms the complementarity axioms to $0$. Next, delete from the resulting refutation all terms that mention a row more than once. Then, for each line $p$ that is an incorrect conclusion of the multiplication rule---which can only happen if $p$ is missing some terms each of degree less than $n/2+1$ and having exactly one row mentioned twice---insert the correct conclusion $p'$ of the rule right before $p$ and use the functionality axioms \eqref{eq:php_functionality_ax} to derive $p$ from $p'$. Each term in the resulting $\PC_{\F}$ refutation has degree less than $n/2+1$. This contradicts \cref{thm:razborov_php_degree}. 
\end{proof}

\end{comment}

There is a similar degree lower bound for $\SA$. It was originally stated for $\FPHP^{n+1}_n$, but exactly the same proof applies to $\FPHP^m_n$.
\begin{theorem}[\cite{DantchevMR09}]\label{thm:dantchev_php_degree}
Every $\SA$ refutation of $\FPHP^m_n$ contains a term that mentions at least $n-1$ pigeons.
\end{theorem}

\subsection{The Perfect Matching Principle}

The perfect matching principle, denoted $\Count^2_n$, states that it is possible to find a perfect matching on $n$ elements when only when $n$ is even (\cite{PW85, Ajt90}). Formally, $\Count^2_n$ is defined as a CNF over the variables $y_{\{k,\ell\}} : \{k, \ell\} \in \binom{[n]}{2}$.
These variables give a partition of the elements into pairs. Each variable $y_{\{k,\ell\}}$ is true if and only if the elements $k$ and $\ell$ are matched together.
We assert that each element is matched with some other element by the following axioms:
\begin{align*}
    & \bigvee_{\ell \in [n] \setminus \{k\}} y_{\{k, \ell\}}, & k \in [n].
\end{align*}
We also assert that no element is matched with more than one other element by the following axioms:
\begin{align*}
    & \lnot y_{\{k,\ell\}} \lor \lnot y_{\{k,\ell'\}}, &  k \in [n], \ell, \ell' \in [n] \setminus \{k\}, \ell \neq \ell'.
\end{align*}

When $n$ is odd, it is clear that this formula is unsatisfiable.

\subsection{Lossless Expanders}

Let $G$ be an $m \times n$ bipartite graph with parts $L$ and $R$.

\begin{definition}
    Given $i \in L$, the \emph{neighbourhood} of $i$ in $G$, denoted $\calN_G(i)$, is the set of all $k \in R$ that are connected to $i$ in $G$.
\end{definition}

\begin{definition}
    Given a set $I \subseteq L$, the \emph{boundary} of $I$, denoted $\boundary_G(I)$, is the set of all unique neighbours of $I$:
    \begin{equation*}
        \boundary_G(I) = \{ k \in R : \text{there is exactly one $i \in I$ satisfying $k \in \calN_G(i)$} \}.
    \end{equation*}
\end{definition}

\begin{definition}
    $G$ is an \emph{$(r,c)$-boundary expander} if, for every set $I \subseteq L$ with $\card{I} \le r$,
    \begin{equation*}
        \card{\boundary_G(I)} \ge c \card{I}.
    \end{equation*}
\end{definition}

\begin{definition}
    $G$ is an \emph{$(r,d)$-lossless expander} if, for every set $I \subseteq L$ with $\card{I} \le r$,
    \begin{equation}\label{eq:lossless expander property}
        \sum_{i \in I} \card{\calN_G(i)} - \card{\boundary_G(I)} \le d \card{I}.
    \end{equation}
\end{definition}
This is called a \emph{lossless} expander because 
%informally, it “loses” very few edges when mapping from the left set of vertices to their unique neighbours on the right.
%Namely,
the expansion from the left side to the right side preserves (almost) all edges without significant loss of distinctness.
The following fact immediately follows from this definition.

\begin{fact}
    Let $\Deltamin$ be the minimum left-degree of the vertices in $G$. If $G$ is an $(r,d)$-lossless expander, then it is also an $(r,c)$-boundary expander with $c = \Deltamin - d$.
\end{fact}

\begin{restatable}{lemma}{goodexpanderlemma}\label{lem:good expander exists}
    Let $C > 1$. There exist constants $\delta, d > 0$ such that for every large enough positive integers $m,n$ with $n < m < Cn$, there is an $m \times n$ bipartite graph $G$ satisfying the following:
    \begin{enumerate}
        \item the minimum left-degree of $G$ is at least $\Deltamin = 50\log n$,
        \item the maximum left-degree of $G$ is at most $\Deltamax = 150\log n$,
        \item $G$ is an $(n^\delta,d)$-lossless expander.
    \end{enumerate}
    In particular, $G$ is an $(n^\delta, \Deltamin - d)$-boundary expander.
\end{restatable}

Let $G$ be a bipartite graph with parts $L$ and $R$ and edges $E$. Given $K \subseteq R$, we define $G \setminus K$ as the bipartite graph with parts $L$ and $R \setminus K$ and edges $\{ (i,k) \in E : i \in L, k \in R \setminus K \}$. Observe that for every $i \in L$, $\calN_{G \setminus K}(i) = \calN_G(i) \setminus K$.

\begin{restatable}{lemma}{remainsgoodexpanderclaim}\label{lem:remains good expander}
    Let $m = O(n)$ and let $G$ be an $m \times n$ $(n^\delta,d)$-lossless expander with parts $L$ and $R$, minimum left-degree at least $\Deltamin = 50\log n$ and maximum left-degree at most $\Deltamax = 150\log n$. Let $K$ be a random subset of $R$ such that every $j \in R$ is added to $K$ independently with probability $2/3$. With probability $1-o(1)$, the graph $G \setminus K$ is an $(n^\delta, d)$-lossless expander with a minimum left-degree $\Deltamin / 6$. In particular, $G \setminus K$ is an $(n^\delta, c)$-boundary expander, where $c = \Deltamin / 6 - d$.
\end{restatable}

The proofs of these lemmas can be found in \cref{sec:missing proofs}.

\subsection{The Rank Principle and Oddtown Principle}\label{sec:wRF-and-OT}
In this section, we introduce our main instance. Here we focus on their purely algebraic formulation, as it is the simplest and most transparent. Later, we will also consider its CNF encodings.

Fix a field $\F$ and positive integers $m$ and $n$ with $m > n$. We denote by $\Rank^m_n(A)$ the system of degree-$2$ polynomial equations stating that an $m \times m$ matrix $A$ can be factorised as the product of an $m \times n$ matrix and an $n \times m$ matrix over $\F$. More precisely, the \emph{rank principle} $\Rank^m_n(A)$ is defined over $2mn$ variables arranged into two matrices $X\in \F^{m\times n}$ and $Y\in \F^{n\times m}$. For every $i, j\in [m]$, there is an equation in $\Rank^m_n(A)$ stating that the $(i, j)$th entry of the product $X Y$ is equal to $A_{i,j}$. That is:
\begin{align}
    A_{i,j}-\sum_{k \in [n]} x_{i,k} y_{k,j}&, & i, j \in [m].
\end{align}
Every equation in $\Rank^m_n$ has degree $2$ in the variables of $X$ and $Y$. It follows from basic linear algebra that when the rank of $A$ exceeds $n$, $\Rank^m_n(A)$ is unsatisfiable.

When $A = I_m$, the $m \times m$ identity matrix, we may omit the matrix from the notation and simply write $\Rank^m_n$.

% We often consider the case where $A = I_m$, the $m \times m$ identity matrix. When it is clear from the context, we omit explicitly specifying the matrix in the notation and simply write $\Rank^m_n$. It encodes the following set of the degree-$2$ polynomial equations:
% \begin{align}
%     \sum_{k \in [n]} x_{i,k} y_{k,j}&, & i, j \in [m], i \neq j, \\
%     1 - \sum_{k \in [n]} x_{i,k} y_{k,i}&, & i \in [m].
% \end{align}

We also need the \emph{oddtown principle} $\OT^m_n(A)$, which is the special case of $\Rank^m_n(A)$ in which $Y = X^T$. That is, given a symmetric matrix $A \in \F^{m \times m}$, $\OT^m_n(A)$ is defined over $mn$ variables arranged as a matrix $X \in \F^{m \times n}$, and for every $i, j\in [m]$, there is an equation in $\OT^m_n$ stating that the $(i, j)$th entry of $XX^T$ is equal to $A_{i, j}$:
\begin{align}
    A_{i,j}-\sum_{k \in [n]} x_{i,k} x_{j,k}&, & i, j \in [m], i \le j.
\end{align}

\addtocontents{toc}{\protect\setcounter{tocdepth}{3}}
%!TEX root = main.tex

\section{Proof Complexity Generators for \texorpdfstring{$\PCR_{\F_2}$}{PCR\_F2}}\label{sec:generator}

In this section, we study the following generator and show that, under various encodings, it is a proof complexity generator hard for $\PCR_{\F_2}$.

\ifnum\IddoSys=0
\begin{colourfuldefinition}
\fi
    \begin{center}\textbf{The Rank Generator $g$}\end{center}
    Let $m > n$.

    \underline{\textbf{Input:}} Two matrices $X \in \F^{m\times n}$ and $Y \in \F^{n\times m}$.

    \underline{\textbf{Output:}} Their product $XY \in \F^{m \times m}$.
\ifnum\IddoSys=0
\end{colourfuldefinition}
\else
\medskip
\fi

Note that $g\colon \F^{2mn} \to \F^{m^2}$, so this function is stretching whenever $n < \frac{m}{2}$.
This function was considered by \Krajicek{}~\cite{Kra09}  as a  proof complexity generator for algebraic proof systems based on the pigeonhole principle.
\Iddo{Michal will add a distinguishing note here; perhaps?}
It was also considered as a candidate hitting-set generator in algebraic complexity and polynomial identity testing by Andrews and Forbes~\cite{AF22, Andrews22}.
 
Given an $m \times m$ matrix $A$ over $\F_2$, we consider different encodings of the statement ``$A\notin\Range(g)$''.

In \cref{sec: generator algebraic}, we show the hardness of the algebraic encoding $\Rank^m_n(A)$ for $\PCR_{\F_2}$.

In \cref{sec:generator-pm,sec: generator bt}, we consider two Boolean encodings of the statement ``$A\notin\Range(g)$''.
The first encoding, $\PMRank^m_n(A)$, employs the perfect matching principle to encode parities.
The second encoding, $\BTRank^{m}_{n}(A)$, uses extension variables arranged in unbalanced binary trees (bamboos).
For the perfect matching encoding, when working over $\FTwo$ and specifically  $\PCR_{\F_2}$, we consider only the case $A=I$ and establish its hardness for $\PCR_{\F_2}$. For $\PCR_\F$ with $\OpChar(\F) \neq 2$, $\SoS$, and $\ACZ$-Frege, we consider $\PMRank^m_n(A)$ with any matrix $A$ and reduce its complexity to the perfect matching principle $\Count^2_n$, which is hard for these proof systems.
For the bamboo-tree encoding $\BTRank^m_n(A)$, we establish hardness for $\PCR_{\F_2}$ for arbitrary matrices $A$.

In \cref{sec: iterated generators}, we show that certain iterated variants of $\Rank$ and $\BTRank$ are hard for $\PCR_{\F_2}$. Iteration links multiple copies of the rank generator by allowing the outputs of one copy to be used as inputs to copies at later levels. This amplifies the stretch while preserving the structure of the original rank principle.

%Namely, it will be  sufficient  in \cref{sec: iterability gives ckt lbs}, to apply the lower bounds in the \textcolor{red}{current section} on the iterated formulas and establish lower bounds on the circuit lower bounds statements.

In this section, we consider $\PCR$ in which every proof line is a multilinear polynomial. Thus the multiplication rule consists of multiplying by a variable followed by multilinearisation. Equivalently, we work in the quotient ring $\F_2[\vec{x}] / (\vec{x}^{\,2}-\vec{x})$ factorised over the Boolean axioms for all variables (including the twin variables). This representation increases neither the monomial-size nor the degree of $\PCR$ refutations; see~\cite{CleggEI96}.
 
%!TEX root = main.tex

\subsection{Algebraic Rank Principle}\label{sec: generator algebraic}

In this section, we use a special notion of degree, called \emph{column degree}, which counts the number of distinct \emph{columns} mentioned by a term in a refutation of \(\Rank^m_n(A)\). This is formally defined as follows.

\begin{definition}[Column-mentioned variables]
    Let $t$ be a term in the variables of $\Rank^m_n(A)$.
    We say that $k \in [n]$ is \emph{$X$-column-mentioned in $t$} if there is $i \in [m]$ such that $x_{i,k}$ or $\overline{x}_{i,k}$ appears in $t$.
    We say that $k \in [n]$ is \emph{$Y^T$-column-mentioned in $t$} if there is $j \in [m]$ such that $y_{k,j}$ or $\overline{y}_{k,j}$ appears in $t$.
\end{definition}

\begin{definition}[Column degree]
    Let $t$ be a term in the variables of $\Rank^m_n(A)$.
    Let $K$ be the set of all $k \in [n]$ that are either $X$-column-mentioned or $Y^T$-column-mentioned in $t$.
    We say that $t$ \emph{mentions} columns $K$. 
    The \emph{column degree} of $t$ is defined as $\card{K}$.
\end{definition}

% \Iddo{We start (\Cref{lem: Rank reduction to PHP}) with $\Rank^m_n$ \sout{(in which the right-hand side is the identity matrix $I$)}, and then (\Cref{lem: Rank lower bound algebraic}) consider symmetric matrices with the all-one diagonal in the right-hand side.}

We start by proving a lower bound on the column degree of $\PCR_{\F_2}$ refutations of $\Rank^m_n(A)$.

\begin{lemma}\label{lem: Rank reduction to PHP}
    Suppose that $m > n > 0$ and $A \in \Q^{m \times m}$ is an $m \times m$ Boolean matrix.
    Then any $\PCR_{\F_2}$ refutation of $\Rank^m_n(A)$ requires column degree at least $n/2+1$.
\end{lemma}
\begin{proof}
We reduce the functional pigeonhole principle $\FPHP^m_n$ to this instance (in the sense of \Cref{def: algebraic reduction}); namely, we show that there is a substitution $\tau$ under which $\Rank^m_n(A) \restriction \tau$ is efficiently derivable from the axioms of $\FPHP^m_n$. The reduction $\tau$ is defined as follows.

Let $P = {(p_{i,k})}_{i \in [m], k \in [n]}$ be fresh variables. It is useful to think of $P$ as an $m \times n$ variable matrix.
    \begin{enumerate}
        \item For each $i \in [m]$ and $k \in [n]$, set
        \[
            \tau(x_{i,k}) = {(AP)}_{i,k} = \sum_{s \in [m]} A_{i,s} p_{s,k}.
        \]
        \item For each $j \in [m]$ and $k \in [n]$, set
        \[
            \tau(y_{k,j}) = p_{j,k}.
        \]
    \end{enumerate}
    Observe that $\tau(x_{i,k})$ and $\tau(y_{k,j})$ only use the variables of the form $p_{s,k} : s \in [m]$.

    Let $i, j \in [m]$. The $(i,j)$th axiom of $\Rank^m_n(A)$ under $\tau$ transforms as follows (recall that $A_{i,j}=-A_{i,j}$ over $\FTwo$):
    \begin{align*}
        \left( \sum_{k \in [n]} x_{i,k} y_{k,j} + A_{i,j} \right) \restriction \tau &=
        \sum_{k \in [n]} \left( \sum_{s \in [m]} A_{i,s} p_{s,k} \right) p_{j,k} + A_{i,j} \\
        &= \sum_{s \in [m]} A_{i,s} \left( \sum_{k \in [n]} p_{s,k} p_{j,k} \right) + A_{i,j} \\
        &= \sum_{s \in [m] \setminus \{j\}} A_{i,s} \left( \sum_{k \in [n]} p_{s,k} p_{j,k} \right) +
        A_{i,j} \left( \sum_{k \in [n]} p_{j,k} + 1 \right),
    \end{align*}
    which is a degree-$2$ $\NS_{\FTwo}$ derivation from the Hole Axioms $p_{s,k} p_{j,k} : s \in [m] \setminus \{j\}$, and the Pigeon Axiom $\sum_{k \in [n]} p_{j,k} + 1$.

Let $\Pi$ be a $\PCR_{\F_2}$ refutation of $\Rank^m_n(A)$ of column degree $d$ and let $\Pi' = \Pi \restriction \tau$.

Under the substitution
\[
\tau(x_{i,k})=\sum_{s\in[m]}A_{i,s}p_{s,k}
\quad\text{and}\quad
\tau(y_{k,j})=p_{j,k},
\]
the image of any variable that either $X$- or $Y^T$-column mentions some column $k\in[n]$ depends only on the pigeonhole variables $p_{s,k} : s\in[m]$ associated with hole $k$. Thus each column $k$ in the rank principle corresponds exactly to hole $k$ in the pigeonhole principle. Hence, if a monomial mentions a set $K\subseteq[n]$ of columns, its image under $\tau$ involves only variables corresponding to the holes in $K$.
Consequently, every term in $\Pi'$ mentions at most $d$ holes.

The refutation $\Pi'$ can be extended to a refutation $\Pi''$ of $\FPHP^m_n$ such that each term in $\Pi''$ still mentions at most $d$ holes, in a straightforward manner (we omit the details).
Combined with the bound from \cref{thm:razborov_php_degree}, we conclude that the column degree of $\Pi$ must be at least $n/2+1$. 
\end{proof}

%\sout{Therefore, any $\PCR_{\F_2}$ refutation of $\Rank^m_n(A)$ of column degree $d$ would, after applying $\tau$, yield a $\FPHP^m_n$ refutation of degree $d$ (this is because \PCR\ refutations are closed under substitutions for degree measure by \Cref{XXX}). \slava{No references since we need it to be closed wrt to the column/hole degree here.} \Iddo{I don't follow what's the problem. } Since Razborov's lower bound shows that every $\FPHP^m_n$ refutation requires degree at least $n/2$, \slava{Should say thay it requirees hole degree at least $n/2$ or explain why the degree bound is enough.} it follows that every $\PCR_{\F_2}$ refutation of $\Rank^m_n(A)$ requires column degree at least $n/2$.}

We continue by demonstrating a trade-off between the size and the \emph{column degree} of refutations of $\Rank^m_n(A)$.
We construct a random substitution $\sigma$ that w.h.p.~turns any small refutation $\Pi$ of $\Rank^m_n(A)$ into a refutation $\Rank^m_{\tilde{n}}(A)$ with $\tilde{n} = \Omega(n)$ while simultaneously reducing the column degree of $\Pi$.
The proof of the resulting size lower bound is relatively straightforward, since there are no extension variables to handle. This substitution will later be used to prove lower bounds for CNF encodings.

\begin{definition}[Substitution $\sigma$]\label{def: alg rank substitution}
    Assume $m > n \ge 2$ and $A \in \Q^{m \times m}$ is an $m \times m$ Boolean matrix.
    A random substitution $\sigma$ is chosen as follows.
    \begin{enumerate}
        \item For each $k \in [n-1]$, uniformly and independently sample $c_k \in \{*,1,0\}$. Then set $c_n = \card{\{ k \in [n-1]: c_k = 1\}} \bmod 2$.
        This guarantees that $\card{\{ k \in [n]: c_k = 1\}}$ is even.
        We will call $c_k$ the \emph{column type} of $k$.
        \item Let $\tilde{n} = \card{\{ k \in [n]: c_k = *\}}$ and let $\tilde{X} = {(\tilde{x}_{i,k})}_{i \in [m], k \in [\tilde{n}]}$ and $\tilde{Y} = {(\tilde{y}_{k,j})}_{k \in [\tilde{n}], j \in [m]}$ be fresh variables.
        \item Let $\iota$ be a bijection between $\{ k \in [ n]: c_k = * \}$ and $[\tilde{n}]$.
        \item For each $i \in [m]$ and $k \in [n]$ set
        \[
            \sigma(x_{i,k}) = \begin{cases}
                \tilde{x}_{i, \iota(k)} & \text{if } c_k = *, \\
                c_k & \text{if } c_k \in \Q.
            \end{cases}
        \]
        \item For each $j \in [m]$ and $k \in [n]$ set
        \[
            \sigma(y_{k,j}) = \begin{cases}
                \tilde{y}_{\iota(k),j} & \text{if } c_k = *, \\
                c_k & \text{if } c_k \in \Q.
            \end{cases}
        \]
        \item For any variable $\theta$ mapped to $0$ or $1$ by $\sigma$, set $\sigma(\overline{\theta}) = 1 - \sigma(\theta)$.
        \item For any variable $\theta$ not mapped to a Boolean constant by $\sigma$, set $\sigma(\overline{\theta}) = \overline{\sigma(\theta)}$ (namely, the negated variable of $\sigma(\theta)$).
    \end{enumerate}
\end{definition}

Note that, independently for each $k \in [n-1]$, the substitution $\sigma$ either keeps column $k$ as a fresh variable column, sets it to the all-one column, or sets it to the all-zero column, each with probability $1/3$. The final column is set to a constant in order to preserve the parity of the number of all-one columns.
The last column is unconditionally set to a constant, and thus we do not need to consider it in the analysis of the column degree.
The resulting matrices $X$ and $Y^T$ are thus shrunk under the restriction to fewer columns, where the original variables are replaced with the new $\tilde{x}$- and $\tilde{y}$-variables.

\begin{lemma}\label{lem: PCR2 alg Rank eliminates terms whp}
    Assume $m > n \ge 2$ and $A \in \Q^{m \times m}$ is a Boolean matrix. Let $t$ be a term that does not mention the last $n$th column and mentions columns $K \subseteq [n-1]$. Then $\Prb{t \restriction \sigma \neq 0} \le {(2/3)}^{\card{K}}$. Furthermore, $\Prb{\tilde{n} < (n-1)/6} \le e^{-\frac{n-1}{24}}$.
\end{lemma}
\begin{proof}
    For the first statement, choose for each column $k$ from the set $K$ a single variable $\gamma_k$ in $t$ that mentions column $k$. Let $\Gamma_k$ be the event that $(\gamma_k \restriction \sigma \neq 0)$. Then $\{\Gamma_k : k \in K\}$ is a set of $\card{K}$ mutually independent events and each of them happens with probability at most $2/3$. The term $t$ is non-zero under $\sigma$ only if all events $\Gamma_k$ happen, and thus
    \[
        \Prb{t \restriction \sigma \neq 0} \le {(2/3)}^{\card{K}}.
    \]

    For the second statement in the lemma, define for each $k \in [n-1]$ the random variable
    \begin{equation*}
        z_k = \begin{cases}
            1 & \text{if } c_k = *, \\
            0 & \text{otherwise}.
        \end{cases}
    \end{equation*}
    Then $\tilde{n} = \sum_{k \in [n-1]} z_k$.
    We have $z_k=1$ with probability $1/3$, and thus $\mu=\Exp \tilde{n} = (n-1)/3$.
    Therefore, by \cref{thm:chernoff bound}, applied with $\delta=1/2$,
    \[
        \Prb{\tilde{n} < \frac{n-1}{6}} \le e^{-\frac{n-1}{24}}.
    \]
\end{proof}

The substitution $\sigma$ transforms any refutation of $\Rank^m_n(A)$ into a refutation of $\Rank^m_{\tilde{n}}(A)$.
\begin{lemma}\label{lem: PCR2 alg Rank ref remains ref}
    Assume $m > n \ge 2$ and $A \in \Q^{m \times m}$ is a Boolean matrix.
    Then for every $\PCR_{\F_2}$ refutation $\Pi$ of $\Rank^m_n(A)$, $\Pi \restriction \sigma$ is a refutation of $\Rank^m_{\tilde{n}}(A)$.
\end{lemma}
\begin{proof}
    For every $i, j \in [m]$, the $(i,j)$th axiom of $\Rank^m_n(A)$ transforms under $\sigma$ as follows:
    \[
        \left( A_{i,j}+\sum_{k=1}^n x_{i,k} y_{k,j} \right) \restriction \sigma = A_{i,j} + \sum_{k \in [n], c_k=*} \tilde{x}_{i,\iota(k)} \tilde{y}_{\iota(k),j} + \card{\{k \in [n] : c_k=1\}} = A_{i,j}+\sum_{\ell=1}^{\tilde{n}} \tilde{x}_{i,\ell} \tilde{y}_{\ell,j},
    \]
    since $\card{\{k \in [n] : c_k=1\}}$ is even and $\iota$ is a bijection between $\{k \in [n] : c_k=*\}$ and $[\tilde{n}]$.
\end{proof}

\begin{theorem}\label{lem: Rank lower bound algebraic}
    Suppose that $m > n \ge 2$ and $A \in \Q^{m \times m}$ is an $m \times m$ Boolean matrix.
    Then any $\PCR_{\F_2}$ refutation of $\Rank^m_n(A)$ requires size $2^{\Omega(n)}$.
\end{theorem}
\begin{proof}

Assume that $\Rank^m_n(A)$ has a refutation $\Pi$ of size less than ${(3/2)}^{\frac{n-1}{12}}(1 - e^{-\frac{n-1}{24}}) = 2^{\Theta(n)}$. Then, by the union bound and \cref{lem: PCR2 alg Rank eliminates terms whp,lem: PCR2 alg Rank ref remains ref} there exists a restriction $\sigma$ such that $\Pi \restriction \sigma$ is a refutation of $\Rank^m_{\tilde{n}}(A)$ with $\tilde{n} \ge \frac{n-1}{6}$ and column degree less than $\frac{n-1}{12} \le \frac{\tilde{n}}{2}$. This would contradict \cref{lem: Rank reduction to PHP}.
\end{proof}

%!TEX root = main.tex

\subsection{Perfect Matching Encoding of the Rank Principle}\label{sec:generator-pm}

To encode the rank formula $\Rank^m_n(A)$ as a CNF, we consider an encoding that uses the \emph{perfect matching principle} to encode the parities appearing in the formula. Similar encodings were used in~\cite{IS01,ken24}. For every $i,j \in [m]$, we encode the parity $\sum_{k \in [n]} x_{i,k} y_{k,j} = A_{i,j}$ by exhibiting a perfect matching on the terms $x_{i,k} y_{k,j}$ that evaluate to one (together with an extra point when $A_{i,j} = 1$, added so that the set being matched has even cardinality).

% \begin{mdframed}[hidealllines=true,backgroundcolor=gray!10,skipabove=0.3em,skipbelow=-0.4em,innertopmargin=0.4em]
    % \begin{center}\textbf{The CNF Encoding $\PMRank^m_n(A)$}\end{center}
\begin{definition}[CNF encoding $\PMRank^m_n(A)$]
    Assume $m > n > 0$ and $A \in \Q^{m \times m}$.

    \underline{\textbf{Variables:}} $X = {(x_{i,k})}_{i\in [m],k\in [n]}$, $Y = {(y_{k,j})}_{k \in [n], j \in [m]}$, for each $i,j \in [m]$, $\{z_{i,j,\{k,\ell\}} : \{k,\ell\} \in \binom{[n]}{2}\}$ if $A_{i,j}=0$ and $\{w_{i,j,\{k,\ell\}} : \{k,\ell\} \in \binom{[n+1]}{2}\}$ if $A_{i,j}=1$.

    \underline{\textbf{Even Axioms:}} For every $i,j \in [m]$ with $A_{i,j}=0$, we use the following clauses:
    \begin{align}
        \label{eq:even_matched} & \neg x_{i,k} \lor \neg y_{k,j} \lor \bigvee_{\ell \in [n] \setminus \{k\}} z_{i,j,\{k, \ell\}}, & k \in [n],\\
        \label{eq:even_edge_endp_both_sat_or_none1} & \neg x_{i,k} \lor \neg y_{k,j} \lor \neg z_{i,j,\{k, \ell\}} \lor x_{i,\ell}, & k \in [n], \ell \in [n] \setminus \{k\}, \\
        \label{eq:even_edge_endp_both_sat_or_none2} & \neg x_{i,k} \lor \neg y_{k,j} \lor \neg z_{i,j,\{k, \ell\}} \lor y_{\ell,j}, & k \in [n], \ell \in [n] \setminus \{k\}, \\
        \label{eq:even_no_conflict} & \neg x_{i,k} \lor \neg y_{k,j} \lor \neg z_{i,j,\{k,\ell\}} \lor \neg z_{i,j,\{k,\ell'\}}, & k \in [n], \ell, \ell' \in [n] \setminus \{k\}, \ell \neq \ell'.
    \end{align}

    \underline{\textbf{Odd Axioms:}} For every $i,j \in [m]$ with $A_{i,j}=1$, we use the following clauses:
    \begin{align}
        \label{eq:odd_matched} & \neg x_{i,k} \lor \neg y_{k,j} \lor \bigvee_{\ell \in [n+1] \setminus \{k\}} w_{i,j,\{k,\ell\}}, & k \in [n],\\
        \label{eq:odd_matched_extrapoint} & \bigvee_{\ell \in [n]} w_{i,j,\{n+1,\ell\}}, &  \\
        \label{eq:odd_edge_endp_sat1} & \neg x_{i,k} \lor \neg y_{k,j} \lor \neg w_{i,j,\{k,\ell\}} \lor x_{i,\ell}, & k \in [n], \ell \in [n]\setminus \{k\}, \\
        \label{eq:odd_edge_endp_sat2} & \neg x_{i,k} \lor \neg y_{k,j} \lor \neg w_{i,j,\{k,\ell\}} \lor y_{\ell,j}, & k \in [n], \ell \in [n]\setminus \{k\}, \\
        \label{eq:odd_edge_endp_sat_extrapoint1} & \neg w_{i,j,\{n+1,\ell\}} \lor x_{i,\ell}, & \ell \in [n], \\
        \label{eq:odd_edge_endp_sat_extrapoint2} & \neg w_{i,j,\{n+1,\ell\}} \lor y_{\ell,j}, & \ell \in [n], \\
        \label{eq:odd_no_conflict} & \neg x_{i,k} \lor \neg y_{k,j} \lor \neg w_{i,j,\{k,\ell\}} \lor \neg w_{i,j,\{k,\ell'\}}, & k \in [n], \ell, \ell' \in [n+1] \setminus \{k\}, \ell \neq \ell', \\
        \label{eq:odd_no_conflict_extrapoint} & \neg w_{i,j,\{n+1,\ell\}} \lor \neg w_{i,j,\{n+1,\ell'\}}, &  \ell \in [n], \ell' \in [n] \setminus \{\ell\}.
    \end{align}
% \end{mdframed}
\end{definition}

\medskip

In words, for each polynomial equation $\sum_{k \in [n]} x_{i,k} y_{k,j} = A_{i,j}$ in $\Rank^m_n(A)$, the statement that the equation is satisfied is encoded by exhibiting a perfect matching on the satisfied monomials appearing in the equation (namely, those monomials that evaluate to $1$ under a given assignment).
In particular, clauses \eqref{eq:even_matched}--\eqref{eq:even_no_conflict} express that each polynomial equation $\sum_{k \in [n]} x_{i,k} y_{k,j} = 0$ is satisfied by exhibiting a perfect matching on the set consisting of the satisfied monomials among $x_{i,k} y_{k,j}$, with $k \in [n]$: each monomial $x_{i,k}y_{k,j}$ with $x_{i,k}=y_{k,j}=1$ is matched with some other monomial $x_{i,\ell}y_{\ell,j}$ (that also satisfies $x_{i,\ell}=y_{\ell,j}=1$) by the variable $z_{i,j,\{k,\ell\}}$.
Similarly, clauses \eqref{eq:odd_matched}--\eqref{eq:odd_no_conflict_extrapoint} express $\sum_{k \in [n]} x_{i,k} y_{k,j} = 1$ using an extra point: each monomial $x_{i,k}y_{k,j}$ with $x_{i,k}=y_{k,j}=1$ can be either matched to another such monomial or to the extra point $n+1$ (by setting $w_{i,j,\{k,n+1\}}=1$). The axioms about the extra point (\eqref{eq:odd_matched_extrapoint}, \eqref{eq:odd_edge_endp_sat_extrapoint1}, \eqref{eq:odd_edge_endp_sat_extrapoint2} and \eqref{eq:odd_no_conflict_extrapoint}) are stated separately since they have a simpler formulation than their counterparts \eqref{eq:odd_matched}, \eqref{eq:odd_edge_endp_sat1}, \eqref{eq:odd_edge_endp_sat2} and \eqref{eq:odd_no_conflict}.

\smallskip
To simplify the reasoning and highlight the main proof ideas, we focus on the case where $A$ is the identity matrix $I_m$. The general case requires additional arguments, which we present in the next section (\cref{sec: generator bt}) for the \emph{bamboo-tree} encoding, as it is more useful in the context of this paper. A similar approach to that in \cref{sec: generator bt} can be applied to $\PMRank^m_n(A)$, but this would require repeating many parts of the proofs from \cref{sec: generator bt}.

Our goal is to prove a lower bound for $\PMRank^m_n(I_m)$. It is enough to prove such a lower bound after the substitution $Y=X^T$, that is, after replacing each variable $y_{k,j}$ by $x_{j,k}$. Indeed, applying this substitution to any \PCR\ refutation of $\PMRank^m_n(I_m)$ gives a \PCR\ refutation of the restricted formula, with no increase in size. Thus, a size lower bound for the restricted formula implies the same lower bound for $\PMRank^m_n(I_m)$.

%After the substitution $Y=X^T$, the encoded equations become
%$
%\sum_{k\in[n]} x_{i,k}x_{j,k} = \delta_{i,j}.
%$
%Thus the diagonal equations have right-hand side $1$, while the off-diagonal equations (i.e., when $i\neq j$) have right-hand side $0$. We now introduce a simplified notation for this restricted formula. For the diagonal equations, we write $w_{i,\{k,\ell\}}$ instead of $w_{i,i,\{k,\ell\}}$. For the off-diagonal equations, the equations indexed by $(i,j)$ and $(j,i)$ are identical (after substituting n $Y=X^T$), so in the simplified formula we keep only one copy of the corresponding matching variables and denote it by $z_{\{i,j\},\{k,\ell\}}$. This gives the following set of axioms:

%In the case $A = I_m$, we can consider a weaker variant of $\PMRank^m_n(I_m)$, in which  $Y = X^T$. 
%It is enough to prove the lower bound for this restricted formula since any \PCR\ refutation of the original formula can be restricted by setting $Y=X^T$ and identifying the corresponding matching variables, yielding
%a refutation of the formula below.

After the substitution $Y=X^T$, the encoded equations become
$
\sum_{k\in[n]} x_{i,k}x_{j,k} = \delta_{i,j}.
$
Since the ones lie only on the diagonal we can replace each $w$-variable $w_{i,i,\{k,\ell\}}$ with $w_{i,\{k,\ell\}}$, where $i \in [m]$. For $i,j \in [m]$ with $i \neq j$, the axioms for $A_{i,j}$ and $A_{j,i}$ become identical, thus we unify the variables $z_{i,j,\{k,\ell\}}$ and $z_{j,i,\{k,\ell\}}$ into a single variable $z_{\{i,j\},\{k,\ell\}}$. This leads to the following simplified set of axioms:

\begin{definition}[CNF for perfect matching encoding of Odd Town principle $\PMOT^m_n$]
\mbox{}

\underline{\textbf{Even Axioms:}} For every $i,j \in [m]$ with $i \neq j$:
\begin{align}
    \label{eq:even_matched Identity} & \neg x_{i,k} \lor \neg x_{j,k} \lor \bigvee_{\ell \in [n] \setminus \{k\}} z_{\{i,j\},\{k, \ell\}}, & k \in [n],\\
    \label{eq:even_edge_endp_both_sat_or_none Identity} & \neg x_{i,k} \lor \neg x_{j,k} \lor \neg z_{\{i,j\},\{k, \ell\}} \lor x_{i,\ell}, & k \in [n], \ell \in [n] \setminus \{k\}, \\
    \label{eq:even_no_conflict Identity} & \neg x_{i,k} \lor \neg x_{j,k} \lor \neg z_{\{i,j\},\{k,\ell\}} \lor \neg z_{\{i,j\},\{k,\ell'\}}, & k \in [n], \ell, \ell' \in [n] \setminus \{k\}, \ell \neq \ell'.
\end{align}

\underline{\textbf{Odd Axioms:}} For every $i \in [m]$:
\begin{align}
    \label{eq:odd_matched Identity} & \neg x_{i,k} \lor \bigvee_{\ell \in [n+1] \setminus \{k\}} w_{i,\{k,\ell\}}, & k \in [n],\\
    \label{eq:odd_matched_extrapoint Identity} & \bigvee_{\ell \in [n]} w_{i,\{n+1,\ell\}}, &  \\
    \label{eq:odd_edge_endp_sat Identity} & \neg x_{i,k} \lor \neg w_{i,\{k,\ell\}} \lor x_{i,\ell}, & k \in [n], \ell \in [n]\setminus \{k\}, \\
    \label{eq:odd_edge_endp_sat_extrapoint Identity} & \neg w_{i,\{n+1,\ell\}} \lor x_{i,\ell}, & \ell \in [n], \\
    \label{eq:odd_no_conflict Identity} & \neg x_{i,k} \lor \neg w_{i,\{k,\ell\}} \lor \neg w_{i,\{k,\ell'\}}, & k \in [n], \ell, \ell' \in [n+1] \setminus \{k\}, \ell \neq \ell', \\
    \label{eq:odd_no_conflict_extrapoint Identity} & \neg w_{i,\{n+1,\ell\}} \lor \neg w_{i,\{n+1,\ell'\}}, &  \ell \in [n], \ell' \in [n] \setminus \{\ell\}.
\end{align}
We call the formula consisting of the above axioms $\PMOT^m_n$ (since it encodes the negation of the \emph{oddtown principle}, namely $XX^T=I$).
\end{definition}

\begin{theorem}\label{thm: PMRank size bound PCR2}
    Let $m > n \ge 15$  with $n$ divisible by $5$ and $n/5$ odd. Then every $\PCR_{\F_2}$ refutation of $\PMOT^m_n$ (and thus $\PMRank^m_n(I_m)$) requires size at least ${(10/9)}^{n/20}$.  
\end{theorem}

The proof of this theorem follows a similar approach to the lower-bound proof for the algebraic encoding in \cref{sec: generator algebraic}, but uses row degree instead of column degree. Furthermore, due to the presence of the extension variables (the $z$- and $w$-variables), our substitution needs to be more structured: a simple restriction like the one used in \cref{sec: generator algebraic} would not suffice.

\emph{Row degree} measures the number of different rows that a term mentions. Every $x$- and $w$-variable mentions exactly one row, while every $z$-variable mentions two distinct rows. The latter complicates the analysis: previously we relied on the fact that every variable mentions exactly one column, which allowed us to find many independent events that a term is eliminated by the random restriction. In contrast, some terms in the $z$-variables may not yield independent events. In particular, star-shaped terms like $\prod_{j \in J} \overline{z}_{\{i,j\},\{k_j,\ell_j\}}$ for some fixed $i$ may have high row degree, but all variables in this term simultaneously depend on the same row $i$.

The proof of \cref{thm: PMRank size bound PCR2} consists of three main components.
\begin{itemize}
    \item We define a random restriction $\rho$ that independently assigns each row one of five \emph{types}, and then substitutes the $x$-variables according to the chosen type. Depending on the $x$-variables, the $z$- and $w$-variables are substituted in a way that is consistent with the perfect matching axioms according to fixed matchings, and set arbitrarily otherwise. The structure of the row types guarantees that the latter case happens often enough to eliminate terms of high row degree with high probability.
    \item We show that under $\rho$, any restricted formula $\PMOT^m_n \restriction \rho$ can be further transformed into a smaller formula $\PMOT^{\tilde{m}}_{n/5-1}$ for some $m/5 \le \tilde{m} \le m$. This is achieved by a reduction $\sigma$ that preserves the row degree.
    \item We show the hardness of $\PMOT^{\tilde{m}}_{n/5-1}$ based on the reduction from the pigeonhole principle, similarly to \cref{lem: Rank reduction to PHP}.
\end{itemize}
If there is a purported refutation of $\PMOT^m_n$ of small size, the union bound implies the existence of a restriction $\rho$ that eliminates all terms of high row degree, and thus gives a refutation of $\PMOT^{\tilde{m}}_{n/5-1}$ of small row degree, which is impossible.

% We prove this result by applying a random restriction that employs a different degree measure than the one used in \cref{sec: generator algebraic}. Previously, we counted how many times each \emph{column} is mentioned in a term. While this approach is effective for $x$- and $y$-variables, it becomes less applicable for $z$- and $w$-variables. Instead, we are interested in the \emph{row degree}, that is, the number of different rows a term mentions.  Formally, it is defined as follows.

We start with the definition of row degree.

\begin{definition}[Row-mentioned variables for the perfect matching encoding]
    Let $t$ be a term in the variables of $\PMOT^m_n$.
    We say that $i \in [m]$ is \emph{$X$-row-mentioned in $t$} if there is $k \in [n]$ such that $x_{i,k}$ or $\overline{x}_{i,k}$ appears in $t$.
    We say that $i \in [m]$ is \emph{$W$-row-mentioned in $t$} if there is $\{k,\ell\} \in \binom{[n+1]}{2}$ such that $w_{i,\{k,\ell\}}$ or $\overline{w}_{i,\{k,\ell\}}$ appears in $t$.
    We say that $i \in [m]$ is \emph{$Z$-row-mentioned in $t$} if there are $j \in [m]$ and $\{k,\ell\} \in \binom{[n]}{2}$ such that $z_{\{i,j\},\{k,\ell\}}$ or $\overline{z}_{\{i,j\},\{k,\ell\}}$ appears in $t$.
\end{definition}

\begin{definition}[Row degree]
    Let $t$ be a term in the variables of $\PMOT^m_n$.
    Let $I$ be the set of all $i \in [m]$ that are either $X$-row-mentioned, $W$-row-mentioned or $Z$-row-mentioned in $t$.
    The \emph{row degree} of $t$ is defined as $\card{I}$.
    We also say that $t$ \emph{mentions} each $i \in I$.
\end{definition}

As before, we construct a random substitution $\rho$ such that, for every fixed term of high row degree, the probability that the term is not eliminated by $\rho$ is exponentially small. The construction is based on the matrix $\calB_n$, which we define below.

Recall that $J_{\ell,\ell'}$ denotes the $\ell\times\ell'$ all-ones matrix. We use \(\otimes\) for the Kronecker (tensor) product. For two matrices \(P\) and \(Q\), the matrix \(P \otimes Q\) is obtained by replacing each entry \(P_{i,j}\) by the block \(P_{i,j} Q\). In particular, if \(Q = J_{1,t}\) is the all-ones row vector of size $t$, then \(P \otimes J_{1,t}\) is obtained by repeating each column of \(P\) into a block of \(t\) identical columns.

Let $n$ be a positive integer divisible by $5$ such that $\tilde{n} = n/5$ is odd.
Partition the set of columns $[n]$ into five consecutive blocks of size $\tilde{n}$ each.
Specifically, we have five column blocks $A_1, A_2, A_3, A_4, A_5$, where $A_r = \{(r-1)\tilde{n}+1, (r-1)\tilde{n}+2, \ldots, r\tilde{n}\}$ for $r \in [5]$.
For each row $i \in [m]$, we independently choose \emph{$r_i \in [5]$}, which is called its \emph{row type} and determines the block configuration of the $i$th row of $X$.

Let $\calB$ be the $5 \times 5$ matrix
\begin{equation}\label{eq:matrix B_n for PMOT}
\calB =
\begin{pmatrix}
1 & * & 0 & 0 & 0 \\
0 & 1 & * & 0 & 0 \\
0 & 0 & 1 & * & 0 \\
0 & 0 & 0 & 1 & * \\
* & 0 & 0 & 0 & 1
\end{pmatrix}.
\end{equation}

We require the following properties from $\calB$:
\begin{itemize}
    \item every row $\calB_i$ of $\calB$ has exactly one $*$-entry;
    \item for every two distinct rows $\calB_i$ and $\calB_j$ there is at most one index $k \in [5]$ satisfying $(\calB_{i,k}, \calB_{j,k}) \in \{(*, 1), (1, *)\}$, and the remaining indices $\ell$ have either $\calB_{i,\ell} = 0$ or $\calB_{j,\ell} = 0$;
    \item every column of $\calB$ contains at least one $0$ and at least one $1$;
    \item for every pair $\{k,\ell\} \in \binom{[5]}{2}$, there is at least one row $\calB_i$ of $\calB$ such that $\calB_{i,k} = \calB_{i,\ell} = 0$.
\end{itemize}

We will use the $5 \times n$ matrix $\calB_n = \calB \otimes J_{1,n/5}$. For example, 
\[
\calB_{15} = \calB \otimes J_{1,3} =
\left(
\begin{array}{ccccccccccccccc}
1 & 1 & 1 & * & * & * & 0 & 0 & 0 & 0 & 0 & 0 & 0 & 0 & 0 \\
0 & 0 & 0 & 1 & 1 & 1 & * & * & * & 0 & 0 & 0 & 0 & 0 & 0 \\
0 & 0 & 0 & 0 & 0 & 0 & 1 & 1 & 1 & * & * & * & 0 & 0 & 0 \\
0 & 0 & 0 & 0 & 0 & 0 & 0 & 0 & 0 & 1 & 1 & 1 & * & * & * \\
* & * & * & 0 & 0 & 0 & 0 & 0 & 0 & 0 & 0 & 0 & 1 & 1 & 1
\end{array}
\right).
\]

The rows of $\calB_n$ define five different row types.
This matrix serves as a template for the matrix $X$ in $\PMOT^m_n$: the $i$th row of $X$ is assigned according to the $r_i$th row of $\calB_n$: the constants are assigned as is, while the $*$ entries remain unassigned.
Given a row type $r \in [5]$ and $s \in \{*,1,0\}$, let
\begin{equation}
    C^s_n(r) = \{k \in [n] : {(\calB_n)}_{r, k} = s\}
\end{equation}
be the set of all the column indices of the occurrences of $s$ in the $r$th row of $\calB_n$.
For example, $C^1_n(r) = \{(r-1)n/5 +1, (r-1)n/5 +2, \ldots, rn/5\}$ for all $r \in [5]$.
The sets $C^s_n(r)$ determine the column configuration of the $x$-variables in the $i$th row of $X$ according to the row type $r_i$. In particular, the variables from $C^*_n(r_i)$ remain unassigned, while the remaining $x$-variables are substituted with constants.

The first property of $\calB$ guarantees that there is exactly one block that is left unassigned for each row type $r \in [5]$: for a block $A$ we either have $C^*_n(r) = A$ or $A \cap C_n^*(r) = \varnothing$. Hence the column sets $C_n^*(r)$ are disjoint for different $r \in [5]$.
The second property of $\calB$ implies that for every two distinct row types $r_i$ and $r_j$, there is at most one block $A$ satisfying $C^*_n(r_i) \cap C^1_n(r_j) = A$ or $C^1_n(r_i) \cap C^*_n(r_j) = A$, and for all other blocks $A'$ we have $A' \subseteq C_n^0(r_i)$ or $A' \subseteq C_n^0(r_j)$.
The third property of $\calB$ ensures that for every $k \in [n]$, each variable $x_{i,k}$ has a constant probability of being assigned to $0$ or $1$.
Finally, the third and fourth properties of $\calB$ guarantee that for every two distinct columns $k$ and $\ell$, there is at least one row type $r_i$ such that both $x_{i,k}$ and $x_{i,\ell}$ are assigned to $0$. This allows us to set the corresponding $w$- and $z$-variables arbitrarily, which is crucial for eliminating terms of high row degree.

To process the additional perfect matching variables, we fix a perfect matching $M(r)$ on the set $C^1_n(r) \cup \{n+1\}$ for each $r \in [5]$. This determines how the corresponding $w$-variables are substituted when the $x$-variables are assigned to $1$ according to the row type $r$.

The interaction between two rows $i$ and $j$ depends on the relation between their row types $r_i$ and $r_j$. In particular, if $r_i = r_j$, then the set of axioms for the inner product of the $i$th and $j$th rows is transformed into a perfect matching encoding on the surviving block with $\tilde{n}$ columns. The substitution $\sigma$ flips these parities, yielding the usual oddtown constraints on one fewer column.
If the two row types $r_i$ and $r_j$ are adjacent, then their axiom set is transformed into a perfect matching encoding on the inner product of the $i$th row with itself or the inner product of the $j$th row with itself, depending on the direction of adjacency. This does not add any new constraints.
To restore the structure of this new axiom set, we fix an arbitrary index $k(r)$ in $C^1_n(r)$ for each $r \in [5]$ that plays the role of the new \emph{extra point} and use another perfect matching $M'(r)$ on the set $C^1_n(r) \setminus \{k(r)\}$ that pairs all the remaining elements of $C^1_n(r)$ except for $k(r)$.
Finally, in all other cases the corresponding axioms are satisfied.

% \Iddo{I think there should be a more approachable intuitive discussion on what the effect of $\rho$ is doing, and how it reduces whp the instance into a smaller instance. without such an explanation it would be hard to understand it. }
% \Iddo{In general more intuitions should be given in each part here, as is done nicely, I think, in Sec. 5.3 about the bamboo-tree encoding.}
% \Iddo{In the below defn it would be helpful to explain in words what each part does. This was done  in Sec. 5.3.}
% \sout{\Iddo{Mention the Y becomes X transpose under random assignment.}}
% \sout{\Iddo{There is a lot of intuition here that would make the definition simpler to understand: e.g., part 1 in definition below simply restricts row i in X and column i in Y to the randomly chosen template in "B tensor J". This should be stated in words.}}

Observe that the axioms of $\PMOT^m_n$ allow variables $w_{i,\{k,\ell\}}$ to be assigned arbitrarily when $\{k, \ell\} \subseteq C^0_n(r_i)$. These axioms are satisfied because we assign $x_{i,k}=x_{i,\ell}=0$ for such $k$ and $\ell$.
Similarly, the value of $z_{\{i,j\},\{k,\ell\}}$ can be chosen arbitrarily when $\{k, \ell\} \subseteq C^0_n(r_i) \cup C^0_n(r_j)$: the axioms it appears in are satisfied due to $x_{i,k} x_{j,k}=x_{i,\ell} x_{j,\ell}=0$. 

\begin{definition}[Substitution $\rho$ for $\PMOT^m_n$]\label{def: PMOT substitution}
Assume that $n$ is divisible by $5$ and $\tilde{n} = n/5$ is odd.
For each $r \in [5]$ fix an arbitrary perfect matching $M(r)$ on the set $C^1_n(r) \cup \{n+1\}$. For each $r \in [5]$ fix $k(r)$ to be the first (for definiteness) integer in $C^1_n(r)$ and fix an arbitrary perfect matching $M'(r)$ on the set $C^1_n(r) \setminus \{k(r)\}$. $k(r)$ will play the role of the new \emph{extra point}.
Define a random substitution $\rho$ as a random partial map from the variables of $\PMOT^m_n$ and their twin variables to $\{0,1\}$ by the following random experiment:
\begin{enumerate}
\item\label{item:def_rho_choose_ri_and_set_xij} Independently for each $i \in [m]$, choose uniformly at random an integer $r_i \in [5]$. Then, for each $i \in [m]$ and $k \in [n] \setminus C^{*}_n(r_i)$, set $\rho(x_{i,k}) = {(\calB_n)}_{r_i, k}$. This assigns all the constant ($0$ or $1$) entries of ${(\calB_n)}_{r_i, k}$ in the $i$th row of $X$. The remaining variables $x_{i,k}$, where $i \in [m]$ and $k \in C^{*}_n(r_i)$, are left unassigned.

\item\label{item:def_rho_ii} Independently for each $i \in [m]$ and each $\{k, \ell\} \subseteq C^0_n(r_i)$, $\rho$ maps $w_{i,\{k, \ell\}}$ to $0$ or $1$ each with probability $1/2$. These variables can be assigned to arbitrary values since all the axioms they appear in (\eqref{eq:odd_matched Identity}, \eqref{eq:odd_edge_endp_sat Identity} and \eqref{eq:odd_no_conflict Identity}) also contain $\neg x_{i,k}$ or $\neg x_{i,\ell}$, which we set to $1$ in \cref{item:def_rho_choose_ri_and_set_xij}, satisfying these axioms.

Next, set $\rho(w_{i,\{k, \ell\}}) = 1$ for every $\{k, \ell\} \in M(r_i)$, and $\rho(w_{i,\{k, \ell\}}) = 0$ for all variables $w_{i,\{k, \ell\}}$ that are not yet mapped, except for those with $\{k, \ell\} \subseteq C^{*}_n(r_i)$. This assignment sets all the remaining variables $\rho(w_{i,\{k, \ell\}})$ according to $M(r_i)$, leaving those with $\{k, \ell\} \subseteq C^{*}_n(r_i)$ unassigned.

Observe that, since $(n+1)$ is matched in $M(r_i)$, we always set all $w_{i,\{n+1, k\}}$, $k \in [n]$, to constants (exactly one of these variables to $1$ and the rest to $0$).

\item\label{item:def_rho_i<j_endpoints_at_zeros} Independently for each $\{i,j\} \in \binom{[m]}{2}$ and each $\{k, \ell\} \subseteq C^0_n(r_i) \cup C^0_n(r_j)$, $\rho$ maps $z_{\{i,j\},\{k, \ell\}}$ to $0$ or $1$ each with probability $1/2$. This is done similarly to \cref{item:def_rho_ii} since every axiom containing $z_{\{i,j\},\{k, \ell\}}$ (\eqref{eq:even_matched Identity}, \eqref{eq:even_edge_endp_both_sat_or_none Identity} and \eqref{eq:even_no_conflict Identity}) also contains either $\neg x_{i,k} \lor \neg x_{j,k}$ or $\neg x_{i,\ell} \lor \neg x_{j,\ell}$ and both are satisfied under $\rho$.

\item\label{item:def_rho_i<j_ri_is_rj}  For each $\{i,j\} \in \binom{[m]}{2}$ with $r_i = r_j$ and each $\{k, \ell\} \in M'(r_i)$, set $\rho(z_{\{i,j\},\{k, \ell\}}) = 1$. Set $\rho(z_{\{i,j\},\{k, \ell\}}) = 0$ for all variables with $i \neq j$ and $r_i = r_j$ that are not yet mapped, except for those with $\{k, \ell\} \subseteq C^{*}_n(r_i) \cup \{k(r_i)\}$.

\item\label{item:def_rho_i<j_ri_not_rj} For each $\{i,j\} \in \binom{[m]}{2}$ with $r_i \neq r_j$ and each $\{k, \ell\} \in \binom{[n]}{2}$ with $|\{k, \ell\} \cap (C^0_n(r_i) \cup C^0_n(r_j))| = 1$, set $\rho(z_{\{i,j\},\{k, \ell\}}) = 0$.

\item For any variable $\theta$ mapped to $0$ or $1$ by $\rho$ in \cref{item:def_rho_choose_ri_and_set_xij,item:def_rho_ii,item:def_rho_i<j_endpoints_at_zeros,item:def_rho_i<j_ri_is_rj,item:def_rho_i<j_ri_not_rj}, define $\rho(\overline{\theta}) = 1 - \rho(\theta)$.

\end{enumerate}
\end{definition}

\Iddo{\sout{I stopped reading here this section, awaiting more intuitive explanations for the definitions and proofs. I shall jump to Sec 5.3 now.}}

% For $i \in [m], \{k,\ell\} \in \binom{[n+1]}{2}$, we say that each of the variables $x_{i,k}, z_{i,\{k,\ell\}}$ \emph{mentions} row $i$. For $i<j \in [m], \{k,\ell\} \in \binom{[n]}{2}$, the variable $y_{i,j,\{k,\ell\}}$ \emph{mentions} both row $i$ and row $j$. A twin variable $\overline{v}$ \emph{mentions} row $i \in [m]$ if and only if $v$ does. A term $t$ \emph{mentions} row $i \in [m]$ if and only if a variable or its twin occurring in $t$ mentions row $i$.

Recall that $\PMOT^m_n$ semantically encodes the following statements:
\begin{align*}
    & \sum_{k=1}^n x_{i,k} = 1, & i \in [m], \\
    & \sum_{k=1}^n x_{i,k} x_{j,k} = 0, & \{i, j\} \in \binom{[m]}{2}.
\end{align*}
Under $\rho$, the former is transformed into
\begin{align*}
    & \sum_{k \in C^*_n(r_i)} x_{i,k} = 0, & i \in [m].
\end{align*}
The latter depends on $r_i$ and $r_j$. Let $\{i, j\} \in \binom{[m]}{2}$. Then it transforms into
\begin{align*}
    \sum_{k \in C^*_n(r_i)} x_{i,k} x_{j,k} &= 1 & \text{if } r_i = r_j, \\
    \sum_{k \in C^*_n(r_i)} x_{i,k} &= 0 & \text{if } r_j - r_i \equiv 1 \pmod 5, \\
    \sum_{k \in C^*_n(r_j)} x_{j,k} &= 0 & \text{if } r_j - r_i \equiv -1 \pmod 5, \\
    0 &= 0 & \text{otherwise}.
\end{align*}

It follows from the above definition that for every variable $\theta$ of $\PMOT^m_n$ except the $w$-variables involving the extra point $n+1$, we have $\Prb{\theta \restriction \rho = 0} \ge 1/10$ and $\Prb{\overline{\theta} \restriction \rho = 0} \ge 1/10$. Moreover, these events depend only on the rows that $\theta$ mentions.

The following lemma shows that a term that mentions many rows is eliminated with high probability by $\rho$.
We exclude the $w$-variables involving the extra point $n+1$ in the lemma because their assignments are correlated within each row; all of them are nevertheless assigned constants by $\rho$.
\begin{lemma}\label{lem: PMOT term elimination}
    Suppose that $n$ is divisible by $5$ and $\tilde{n} = n/5$ is a positive odd integer.
    Let $t$ be a term in the variables of $\PMOT^m_n$ and let $t'$ be a subterm of $t$ obtained by removing all variables $w_{i,\{k, n+1\}}, i \in [m], k \in [n]$, and their twins.
    Let $I$ be the set of all $i \in [m]$ that are $X$-, $W$- or $Z$-row mentioned by $t'$.
    Then $\Prb{t \restriction \rho \neq 0} \leq {(9/10)}^{\card{I}/2}$.
\end{lemma}

\begin{proof}
% We may assume without loss of generality that no variables from $w_{i,\{k, n+1\}}, i \in [m], k \in [n]$ or their twins appear in $t$.
Given $q \in \{X,W,Z\}$, let $I(q)$ be the set of rows $i \in [m]$ that are $q$-row-mentioned in $t'$. For example, $I(X) = \{i \in [m] : \exists k \in [n], x_{i,k} \text{ or } \overline{x}_{i,k} \text{ occurs in } t'\}$.

% By \cref{item:def_rho_choose_ri_and_set_xij}, for each $i \in [m]$ and $k \in [n]$, we have $\Prb{x_{i,k} \restriction \rho = 0} = 3/5$ and $\Prb{\overline{x}_{i,k} \restriction \rho = 0} = 1/5$.
% For each $i \in [m]$ and $\{k,\ell\} \in \binom{[n]}{2}$, we have $\Prb{\{k, \ell\} \subseteq C^{0}_n(r_i)} \geq 1/5$ (by \cref{item:def_rho_choose_ri_and_set_xij}), and therefore $\Prb{w_{i,\{k,\ell\}} \restriction \rho = 0} \geq 1/10$ and $\Prb{\overline{w}_{i,\{k,\ell\}} \restriction \rho = 0} \geq 1/10$ (by \cref{item:def_rho_ii}).
% For $i\neq j \in [m]$ and $\{k,\ell\} \in \binom{[n]}{2}$, it follows from either of the two estimates $\Prb{\{k, \ell\} \subseteq C^{0}_n(r_i)} \geq 1/5$, $\Prb{\{k, \ell\} \subseteq C^{0}_n(r_j)} \geq 1/5$ (both by \cref{item:def_rho_choose_ri_and_set_xij}) that $\Prb{z_{\{i,j\},\{k,\ell\}} \restriction \rho = 0} \geq 1/10$ and $\Prb{\overline{z}_{\{i,j\},\{k,\ell\}} \restriction \rho = 0} \geq 1/10$ (by \cref{item:def_rho_i<j_endpoints_at_zeros}).

There is a set $A \subseteq I(Z)$ of at least half of the elements $I(Z)$
and a function $f$ from $A$ to
\[
    \bigcup_{\substack{\{i,j\} \in \binom{[m]}{2}, \\ \{k,\ell\} \in \binom{[n]}{2}}} \{ z_{\{i,j\},\{k,\ell\}}, \overline{z}_{\{i,j\},\{k,\ell\}} \}
\]
such that $f$ is injective\footnote{We require $f$ to injectively map $A$ to the pairs of row indices $\{i,j\} \in \binom{[m]}{2}$.} and for each $i \in A$, $f(i)$ mentions $i$ and $f(i)$ occurs in $t'$. To see this, consider a graph $G$ with the vertex set $I(Z)$ and the edge set $\{\{i,j\} \in \binom{I(Z)}{2} : \text{for some } \{k,\ell\} \in \binom{[n]}{2},\ z_{\{i,j\},\{k,\ell\}} \text{ or } \overline{z}_{\{i,j\},\{k,\ell\}} \text{ occurs in } t'\}$.
%For each isolated vertex $i \in V$ of $G$, set $f(i)$ to any $z$-variable that mentions $i$ and either occurs in $t'$ or its twin occurs in $t'$.
For each connected component of $G$, fix a spanning tree and a root vertex. For each vertex $i$ in the component other than the root, let $j$ be the neighbour of $i$ in the tree that is closer to the root (i.e., $j$ is the parent of $i$ in the rooted tree). We define $f(i)$ to be a $z$-variable (or its twin) that mentions both $i$ and $j$ and occurs in $t'$. This finishes the definition of $A$ and $f$; they clearly have the required properties. The size of $A$ is at least the size of $I(Z)$ minus the number of connected components of $G$, which is at least $\card{I(Z)}/2$ since each component has at least two vertices.

We can further extend $f$ to an injection $f'$ defined on $A \cup I(X) \cup I(W)$ by setting $f'(i)$ for $i \in I(X) \setminus A$ to some $x$-variable (or its twin) that occurs in $t$ and mentions $i$ and by defining $f'(i)$ for $i \in I(W) \setminus (A \cup I(X))$ to be a $w$-variable (or its twin) that occurs in $t$ and mentions $i$.

For every $i \in I(X) \setminus A$, define the event $\Gamma_i = \left( f'(i) = 0 \right)$. By \cref{item:def_rho_choose_ri_and_set_xij} in \cref{def: PMOT substitution}, we have $\Prb{\Gamma_i} \ge 1/5$.

For every $i \in I(W) \setminus (A \cup I(X))$, we have $f'(i) = w_{i,\{k,\ell\}}$ or $f'(i) = \overline{w}_{i,\{k,\ell\}}$ and define the event $\Delta_i = \left( \{k, \ell\} \subseteq C^0_n(r_i) \land f'(i) = 0 \right)$. It follows from the estimate $\Prb{\{k, \ell\} \subseteq C^{0}_n(r_i)} \geq 1/5$ (by \cref{item:def_rho_choose_ri_and_set_xij}) that $\Prb{\Delta_i} \geq 1/10$ (by \cref{item:def_rho_ii}).

For every $i \in A$, we have either $f'(i) = z_{\{i,j\},\{k,\ell\}}$ or $f'(i) = \overline{z}_{\{i,j\},\{k,\ell\}}$. Define the event $\Xi_i = \left( \{k, \ell\} \subseteq C^0_n(r_i) \land f'(i) = 0 \right)$. It follows from the estimate $\Prb{\{k, \ell\} \subseteq C^{0}_n(r_i)} \geq 1/5$ (by \cref{item:def_rho_choose_ri_and_set_xij}) that $\Prb{\Xi_i} \geq 1/10$ (by \cref{item:def_rho_i<j_endpoints_at_zeros}). Note that $\Xi_i$ depends only on the choice of $r_i$ and the value of $f'(i)$.

Since all the events $\{\Gamma_i : i \in I(X) \setminus A\}$, $\{\Delta_i : i \in I(W) \setminus (I(X) \cup A)\}$ and $\{\Xi_i : i \in A\}$ are independent, we have
\begin{equation*}
    \Prb{t \restriction \rho \neq 0} \leq (4/5)^{|I(X) \setminus A|} \cdot (9/10)^{|I(W) \setminus (A \cup I(X))|} \cdot (9/10)^{|I(Z)|/2} \leq (9/10)^{\card{I}/2}. \qedhere
\end{equation*}
\end{proof}

% Now, we demonstrate that a refutation of $\PMOT^m_n \restriction \rho$ can be further restricted by a substitution $\sigma$ that transforms it into a refutation of $\PMOT^{\tilde{m}}_{n/5-1}$ for some $\tilde{m} \ge m/5$.

Since the sets $C^*_n(r)$ are disjoint for distinct $r \in [5]$, the above formulas can be interpreted as a union of five disjoint copies of the same formula, up to the renaming of the variables. Within each row set $I(r)$, they form a copy of the same principle on the set $C^*_n(r)$, while the constraints between adjacent classes duplicate its diagonal constraints.
Notice that the resulting formula asserts that the inner product of the row with itself is $0$ while the inner product of two distinct rows of the same type is $1$. This is the opposite of the original formula. This is why we need to further restrict the formula to invert these parities. Luckily, this can be done by a simple substitution that replaces the last column of each variable block $C_n^*(r)$ with the all-ones column that flips the signs, but reduces the number of columns only by one. This column matches the extra point in the even case and serves as a new extra point in the odd case. The following lemma formalizes this argument.

\begin{lemma}\label{lem: PMOT smaller instance}
Let $m > n \ge 15$ with $n$ divisible by $5$ and $\tilde{n} = n/5$ odd. Let $\rho$ be an element of the support of the distribution from \cref{def: PMOT substitution}. Then, there is an integer $\tilde{m}$ with $m/5 \leq \tilde{m} \leq m$ and a map $\sigma$ from the variables of $\PMOT^m_n \restriction \rho$ and their twins to the variables of $\PMOT^{\tilde{m}}_{\tilde{n}-1}$, their twins, and the constants $0, 1$ such that $(\PMOT^m_n \restriction \rho) \restriction \sigma =  \PMOT^{\tilde{m}}_{\tilde{n}-1}$.
Moreover, there is a function $f : [m] \to [\tilde{m}]$ such that every variable of $\PMOT^m_n \restriction \rho$ that mentions the rows $i, j \in [m]$ ($i$ may be equal to $j$) is mapped by $\sigma$ to a variable that can only mention rows from $\{f(i), f(j)\}$.
\end{lemma}
\begin{proof}
Given $r \in [5]$, define
\[
    I(r) = \{i : r_i = r\}.
\]
$I(r)$ gives a partition of $[m]$ into five disjoint sets. Let $\tilde{m} = \max\{\card{I(r)} : r \in [5]\}$. Then $m/5 \le \tilde{m} \le m$.

The surviving variables of $\PMOT^m_n \restriction \rho$ are the following variables.
\begin{itemize}
    \item For each $i \in [m]$, the variables $x_{i,k}$ with $k \in C^*_n(r_i)$.
    \item For each $i \in [m]$, the variables $w_{i,\{k,\ell\}}$ with $\{k, \ell\} \subseteq C^*_n(r_i)$.
    \item For each $\{i,j\} \in \binom{[m]}{2}$ with $r_i = r_j$, the variables $z_{\{i,j\},\{k,\ell\}}$ with $\{k, \ell\} \subseteq C^*_n(r_i) \cup \{k(r_i)\}$.
    \item For each $\{i,j\} \in \binom{[m]}{2}$ with $r_j - r_i \equiv 1 \pmod 5$, the variables $z_{\{i,j\},\{k,\ell\}}$ with $\{k, \ell\} \subseteq C^*_n(r_i)$.
    \item For each $\{i,j\} \in \binom{[m]}{2}$ with $r_i - r_j \equiv 1 \pmod 5$, the variables $z_{\{i,j\},\{k,\ell\}}$ with $\{k, \ell\} \subseteq C^*_n(r_j)$.
    \item The twin variables of all the above variables.
\end{itemize}

Recall that for every $r \in [5]$, we have $r\tilde{n} \in C^*_n(r)$.
Then for each $r \in [5]$, let $f_r \colon I(r) \to [\tilde{m}]$ be an arbitrary injection and $g_r \colon C_n^*(r) \to [\tilde{n}]$ be any bijection with $g_r(r\tilde{n}) = \tilde{n}$.
The function $f$ from the statement of the lemma is defined as $f(i) = f_r(i)$ for $i \in I(r)$.

For each $r \in [5]$, $\sigma$ replaces the column $r\tilde{n}$ with the all-ones column, which flips the parities of the inner products. For the diagonal inner products, this column plays the role of the extra point, while for the off-diagonal inner products, it is matched with the existing extra point $k(r)$ to satisfy the corresponding axioms. The functions $f_r$ and $g_r$ are used to rename the surviving variables of $\PMOT^m_n \restriction \rho$ to the variables of $\PMOT^{\tilde{m}}_{\tilde{n}-1}$.
Formally, it is defined as follows.
\begin{enumerate}
    \item For every $i \in [m]$ and $k \in C^*_n(r_i)$, set
    \[
        \sigma(x_{i,k}) = \begin{cases}
            x_{f_{r_i}(i), g_{r_i}(k)}, & k \neq r_i \tilde{n}, \\
            1, & k = r_i \tilde{n}.
        \end{cases}
    \]
    \item For every $i \in [m]$ and $\{k, \ell\} \subseteq C^*_n(r_i)$, set
    \[
        \sigma(w_{i,\{k,\ell\}}) = w_{f_{r_i}(i), \{g_{r_i}(k), g_{r_i}(\ell)\}}.
    \]
    \item For every $\{i,j\} \in \binom{[m]}{2}$ with $r_i = r_j$ and $\{k, \ell\} \subseteq C^*_n(r_i) \setminus \{r_i \tilde{n}\}$, set
    \[
        \sigma(z_{\{i,j\},\{k,\ell\}}) = z_{\{f_{r_i}(i), f_{r_i}(j)\}, \{g_{r_i}(k), g_{r_i}(\ell)\}}.
    \]
    \item For every $\{i,j\} \in \binom{[m]}{2}$ with $r_i = r_j$, set
    \[
        \sigma(z_{\{i,j\},\{k(r_i),r_i \tilde{n}\}}) = 1.
    \]
    \item For every $\{i,j\} \in \binom{[m]}{2}$ with $r_i = r_j$, $k \in C^*_n(r_i) \setminus \{r_i \tilde{n}\}$, and $\ell \in \{k(r_i), r_i \tilde{n}\}$, set
    \[
        \sigma(z_{\{i,j\},\{k,\ell\}}) = 0.
    \]

    \item For every $\{i,j\} \in \binom{[m]}{2}$ with $r_j - r_i \equiv 1 \pmod 5$ and $\{k, \ell\} \subseteq C^*_n(r_i)$, set
    \[
        \sigma(z_{\{i,j\},\{k,\ell\}}) = w_{f_{r_i}(i), \{g_{r_i}(k), g_{r_i}(\ell)\}}.
    \]

    \item For every $\{i,j\} \in \binom{[m]}{2}$ with $r_i - r_j \equiv 1 \pmod 5$ and $\{k, \ell\} \subseteq C^*_n(r_j)$, set
    \[
        \sigma(z_{\{i,j\},\{k,\ell\}}) = w_{f_{r_j}(j), \{g_{r_j}(k), g_{r_j}(\ell)\}}.
    \]

    \item For any variable $\theta$ mapped to $0$ or $1$ by $\sigma$, set $\sigma(\overline{\theta}) = 1 - \sigma(\theta)$.
    \item For any variable $\theta$ not mapped to a Boolean constant by $\sigma$, set $\sigma(\overline{\theta}) = \overline{\sigma(\theta)}$.
\end{enumerate}

By the description of the restricted axiom sets above, $\sigma$ maps the diagonal, same-type off-diagonal, and adjacent-type axiom sets to the corresponding axiom sets of $\PMOT^{\tilde{m}}_{\tilde{n}-1}$, while all remaining axiom sets are already satisfied. Moreover, for any $r\in[5]$ with $\card{I(r)}=\tilde{m}$, the map $f_r$ is a bijection onto $[\tilde{m}]$, and hence all axiom sets of $\PMOT^{\tilde{m}}_{\tilde{n}-1}$ are obtained.

Observe that our choice of $f_r$ ensures that the row degree does not increase under $\sigma$.
\end{proof}

Finally, we show a row degree lower bound on $\PMOT^m_n$ by constructing a reduction from $\FPHP^m_n$ to it.
\begin{lemma}\label{lem: PMOT reduction to PHP}
    Every $\PCR_{\F_2}$ refutation of $\PMOT^m_n$ has a term $t$ that mentions at least $n/2+1$ rows.
\end{lemma}
\begin{proof}
    We will reduce $\FPHP^m_n$ to $\PMOT^m_n$. Let $p_{i,k} : i \in [m], k \in [n]$ be fresh variables.

    We construct a reduction $\tau$ that maps the $x$-variables of $\PMOT^m_n$ to the $p$-variables of $\FPHP^m_n$ naturally: $\tau$ replaces $x_{i,k}$ with $p_{i,k}$.
    The hole axioms of $\FPHP$ ensure that all terms appearing in the inner product of two distinct rows are mapped to $0$, so we can set all the $z$-variables to $0$.
    The extra point is used to ensure that at least one variable $x_{i,k}$ is set to $1$. Functional axioms ensure that at most one such variable is set to $1$, so we can set all the $w$-variables involving the extra point to the corresponding $p$-variable -- $\tau$ replaces $w_{i,\{n+1,k\}}$ with $p_{i,k}$ -- and the remaining $w$-variables to $0$.

    Formally, we define $\tau$ as follows:
    \begin{enumerate}
        \item For every $i \in [m]$ and $k \in [n]$, set $\tau(x_{i,k}) = p_{i,k}$ and $\tau(\overline{x}_{i,k}) = \overline{p}_{i,k}$.
        
        \item For every $i \neq j \in [m]$ and $\{k, \ell\} \in \binom{[n]}{2}$, set $\tau(z_{\{i,j\},\{k,\ell\}}) = 0$ and $\tau(\overline{z}_{\{i,j\},\{k,\ell\}}) = 1$.

        \item For every $i \in [m]$ and $\{k, \ell\} \in \binom{[n]}{2}$, set $\tau(w_{i,\{k,\ell\}}) = 0$ and $\tau(\overline{w}_{i,\{k,\ell\}}) = 1$.

        \item For every $i \in [m]$ and $k \in [n]$, set $\tau(w_{i,\{n+1,k\}}) = p_{i,k}$ and $\tau(\overline{w}_{i,\{n+1,k\}}) = \overline{p}_{i,k}$.
    \end{enumerate}

    Let $\Pi$ be a $\PCR_{\F_2}$ refutation of $\PMOT^m_n$ such that every term in $\Pi$ mentions at most $d$ rows. Then every term in $\Pi' = \Pi \restriction \tau$ mentions at most $d$ pigeons.
    Now we show that each polynomial in $\PMOT^m_n \restriction \tau$ has a simple derivation from $\FPHP^m_n$.

    \begin{itemize}
        \item The axioms \eqref{eq:even_edge_endp_both_sat_or_none Identity} and \eqref{eq:even_no_conflict Identity} are satisfied since we set all $z$-variables to zero.

        \item The axiom \eqref{eq:even_matched Identity} becomes $p_{i,k} p_{j,k}$, which is a Hole Axiom of $\FPHP^m_n$~\eqref{eq:php_hole_ax}.

        \item The axioms \eqref{eq:odd_matched Identity} and \eqref{eq:odd_edge_endp_sat_extrapoint Identity} become $p_{i,k} \overline{p}_{i,k}$ and thus trivially follow from the properties of twin variables.

        \item The axiom \eqref{eq:odd_matched_extrapoint Identity} becomes $\prod_{\ell \in [n]} \overline{p}_{i,\ell}$ that semantically follows from the Pigeon Axioms~\eqref{eq:php_pigeon_ax} and has a simple derivation from it, which mentions only pigeon $i$:
        we can multiply the Pigeon Axiom $\sum_{k \in [n]} p_{i,k} = 1$ by $\prod_{\ell \in [n]} \overline{p}_{i,\ell}$ and use the fact that $p_{i,k} \overline{p}_{i,k} = 0$ to obtain $\prod_{\ell \in [n]} \overline{p}_{i,\ell} = 0$.

        \item The axioms \eqref{eq:odd_edge_endp_sat Identity} and \eqref{eq:odd_no_conflict Identity} are satisfied.

        \item The axiom \eqref{eq:odd_no_conflict_extrapoint Identity} becomes $p_{i,\ell} p_{i,\ell'}$, which is a Functionality Axiom of $\FPHP^m_n$~\eqref{eq:php_functionality_ax}.
    \end{itemize}

    Thus, $\Pi'$ can be further extended to a refutation $\Pi''$ of $\FPHP^m_n$ such that each term in $\Pi''$ mentions at most $d$ pigeons.

    The lemma follows from the lower bound on the number of pigeons from \cref{thm:razborov_php_degree}.
\end{proof}

Combining \cref{lem: PMOT term elimination,lem: PMOT smaller instance,lem: PMOT reduction to PHP}, we proceed by proving \cref{thm: PMRank size bound PCR2}.
 
\begin{proof}[Proof of \cref{thm: PMRank size bound PCR2}]
    Assume for the sake of contradiction that $\PMOT^m_n$ has a refutation $\Pi$ of size less than ${(10/9)}^{n/20} = {(10/9)}^{\tilde{n}/4}$.
    We start by sampling a random substitution $\rho$ from \cref{def: PMOT substitution}.
    It follows from \cref{lem: PMOT term elimination} that any term that mentions more than $\tilde{n}/2$ rows (excluding the $w$-variables involving the extra point $n+1$) is not set to $0$ by $\rho$ with probability at most ${(9/10)}^{\tilde{n}/4}$. By the union bound, there exists a substitution $\rho$ such that every term in $\Pi \restriction \rho$ mentions at most $\tilde{n}/2$ different rows.

    By \cref{lem: PMOT smaller instance} we can further restrict $\Pi \restriction \rho$ and obtain a refutation $\Pi' = \Pi \restriction \rho \restriction \sigma$ of $\PMOT^{\tilde{m}}_{\tilde{n}-1}$ with $m/5 \le \tilde{m} \le m$. Moreover, every term in $\Pi'$ still mentions at most $\tilde{n}/2$ rows.

    Applying \cref{lem: PMOT reduction to PHP} to $\Pi'$, we obtain a refutation of $\FPHP^{\tilde{m}}_{\tilde{n}-1}$ in which every term mentions at most $\tilde{n}/2$ pigeons.
    This contradicts the row degree bound from \cref{lem: PMOT reduction to PHP}.
\end{proof}

The fact that we focused on the identity matrix $I_m$ and not the general case of an arbitrary matrix $A$ simplifies the argument since we only need one substitution $\rho$ to reduce the degree of the terms. It is also helpful that we can use the natural notion of row degree in this case.
We consider the general case for a different encoding -- which encodes the parities using a balanced-tree structure -- in \cref{sec: generator bt}. There, we use a series of substitutions to reduce the appropriate notion of degree.
Nevertheless, a similar approach can also be used to show the hardness of $\PMRank^m_n(A)$, but it would require redoing many intermediate steps, which is why we do not present it here.

\medskip

As we mentioned earlier, for proof systems in which the perfect matching principle is already hard, the hardness of $\PMRank^m_n(A)$ follows from the hardness of $\Count^2_n$. We begin by describing this reduction. We describe it for the general case $\PMRank^m_n(A)$, where the matrix $A$ is arbitrary.

\begin{lemma}\label{lem: PMOT A to Count2 reduction for specific matrices}
    Let $A$ be an arbitrary $m \times m$ Boolean matrix that is distinct from matrices of all ones or all zeros.
    Then there exists a reduction $\rho$ such that $\PMRank^m_n(A) \restriction \rho$ coincides with $\Count^2_n \land \Count^2_{n+1}$.
\end{lemma}
\begin{proof}

    Let $u_{\{k,\ell\}} : \{k,\ell\} \in \binom{[n]}{2}$ and $v_{\{k,\ell\}} : \{k,\ell\} \in \binom{[n+1]}{2}$ be fresh variables and consider the reduction $\rho$ defined as follows:
    \begin{enumerate}
        \item For every $i \in [m]$ and $k \in [n]$ set $\rho(x_{i,k}) = 1$ and $\rho(\overline{x}_{i,k}) = 0$.
        \item For every $j \in [m]$ and $k \in [n]$ set $\rho(y_{k,j}) = 1$ and $\rho(\overline{y}_{k,j}) = 0$.
        \item For every $i,j \in [m]$ with $A_{i,j}=0$ and every $\{k,\ell\} \in \binom{[n]}{2}$, set $\rho(z_{i,j,\{k,\ell\}}) = u_{\{k,\ell\}}$.
        \item For every $i,j \in [m]$ with $A_{i,j}=1$ and every $\{k,\ell\} \in \binom{[n+1]}{2}$, set $\rho(w_{i,j,\{k,\ell\}}) = v_{\{k,\ell\}}$.
        \item For every variable $\theta$, if $\rho(\theta)$ is not a Boolean constant, set $\rho(\overline{\theta}) = \overline{\rho(\theta)}$.
    \end{enumerate} 
    This restriction substitutes $X$ with $J_{m,n}$, $Y$ with $J_{n,m}$, and unifies all isomorphic instances of $\Count^2_n$ and $\Count^2_{n+1}$.
    
    We show that $\PMRank^m_n(A) \restriction \rho$ becomes a union of disjoint instances of $\Count^2_n$ and $\Count^2_{n+1}$. Let $i,j \in [m]$. We consider the following cases.
    \begin{itemize}
        \item Suppose $A_{i,j}=0$. In this case, we use the axioms \eqref{eq:even_matched}--\eqref{eq:even_no_conflict}, which under $\rho$ become an instance of $\Count^2_n$ in the variables $u_{\{k,\ell\}} : \{k,\ell\} \in \binom{[n]}{2}$.
        \item Suppose $A_{i,j}=1$. In this case, we use the axioms \eqref{eq:odd_matched}--\eqref{eq:odd_no_conflict_extrapoint}, which under $\rho$ become an instance of $\Count^2_{n+1}$ in the variables $v_{\{k,\ell\}} : \{k,\ell\} \in \binom{[n+1]}{2}$.
    \end{itemize}

    Since $A$ contains both ones and zeros, both formulas $\Count^2_n$ and $\Count^2_{n+1}$ appear in \mbox{$\PMRank^m_n(A)\restriction\rho$}. Clearly, exactly one of them is unsatisfiable.
\end{proof}

% For an arbitrary symmetric matrix $A$, we apply the same idea we used to show \autoref{thm: Oddtown size-row-degree tradeoff for all A}.

This lemma immediately implies the hardness of $\PMRank^m_n(A)$ for proof systems, in which $\Count^2_n$ is hard.

% For every $m>n$ and  $A\in\Q^{m\times m}$, any $\PCR_{\F}$ refutation (for a field $\F$ with $\OpChar(\F) \neq 2$) or $\SoS$ refutation of the perfect matching CNF encoding $\PMRank^m_n(A)$  requires size $2^{\Omega(n)}$. Similarly, any $\ACZ$-Frege refutation requires size $2^{n^{\Omega(1)}}$, where the constant in the exponent depends on the circuit depth.

\begin{theorem}\label{thm: PMOT arb A other fields}
    Let $m>n>0$ be positive integers.
    Let $A$ be an arbitrary $m \times m$ Boolean matrix.
    Then any $\PCR_{\F}$ refutation (for a field $\F$ with $\OpChar(\F) \neq 2$) or $\SoS$ refutation of the perfect matching CNF encoding $\PMRank^m_n(A)$ requires size $2^{\Omega(n)}$. Similarly, any $\ACZ$-Frege refutation requires size $2^{n^{\Omega(1)}}$, where the constant in the exponent depends on the circuit depth.
\end{theorem}
\begin{proof}
    If all entries of $A$ are all ones or all zeros, the formula is satisfiable and the theorem holds trivially.
    Otherwise, by applying \cref{lem: PMOT A to Count2 reduction for specific matrices}, we conclude that there is a reduction from $\Count^2_{n} \land \Count^2_{n+1}$ to $\PMRank^m_n(A)$.
    Depending on the parity of $n$ exactly one of these instances is unsatisfiable. They are defined on disjoint sets of variables, hence we can satisfy the other.

    $\Count^2_k$ requires size $2^{\Omega(k)}$ in $\PCR_{\F}$ when $\OpChar(\F) \neq 2$, by combining the degree lower bounds from~\cite{BGIP01} with the size--degree trade-offs from~\cite{CleggEI96,IPS99}. The analogous lower bound for $\SoS$ follows from~\cite{Gri01,Schoenebeck08,AH19}. Similarly, the lower bound for $\ACZ$-Frege follows from~\cite{Kra95}.
\end{proof}

%!TEX root = main.tex

\subsection{Bamboo-Tree Encoding of the Rank Principle}\label{sec: generator bt}

To encode the rank formula $XY=A$ as a CNF, extension variables are introduced to sequentially compute all inner products involved. The extension variables for each inner product are arranged in a totally unbalanced binary tree known as a ``bamboo''. This natural encoding corresponds to a circuit in the sense of~\cite{ABRW04,Razb15-annals}, and similar encodings have been used repeatedly for parity computations~\cite{SU04,ABRW04,Razb15-annals}.

% \begin{mdframed}[hidealllines=true,backgroundcolor=gray!10,skipabove=0.3em,skipbelow=-0.4em,innertopmargin=0.4em]
% \begin{center}\textbf{The CNF Encoding $\BTRank^m_n(A)$}\end{center}
\begin{definition}[The CNF encoding $\BTRank^m_n(A)$]\label{def:bt-enc}
Let $m > n$ be positive integers and $A \in \Q^{m \times m}$ an $m$ by $m$ Boolean matrix.
The bamboo-tree CNF encoding, denoted $\BTRank^m_n(A)$, uses input variables $x_{i,k}, y_{k,i}: i \in [m], k \in [n]$ together with extension variables $z_{i,j,k}, u_{i,j,k} : i,j \in [m], k \in [n]$, and consists of the following axioms. 

\textbf{Output Axioms:} For every $i,j \in [m]$, the clause $u_{i,j,n}$ if $A_{i,j} = 1$ and the clause $\neg u_{i,j,n}$ if $A_{i,j} = 0$.

\textbf{Binary AND Axioms:} For every $i,j \in [m]$ and $k \in [n]$, $z_{i,j,k} = x_{i,k}y_{k,j}$, encoded by the clauses $x_{i,k} \lor \neg z_{i,j,k}$, $y_{k,j} \lor \neg z_{i,j,k}$, $z_{i,j,k} \lor \neg x_{i,k} \lor \neg y_{k,j}$.

\textbf{Summation Base Axioms:} For every $i,j \in [m]$, $u_{i,j,1} = z_{i,j,1}$, encoded by the clauses $u_{i,j,1} \lor \neg z_{i,j,1}$, $z_{i,j,1} \lor \neg u_{i,j,1}$.

\textbf{Summation Axioms:} For every $i,j \in [m]$, and $k\in \{2,\ldots,n\}$, $u_{i,j,k} = u_{i, j, k-1} + z_{i,j,k}$, encoded by the following four clauses:
    \begin{align*}
        & z_{i,j,k} \lor \neg u_{i,j,k-1} \lor u_{i,j,k}, \\
        & z_{i,j,k} \lor u_{i,j,k-1} \lor \neg u_{i,j,k}, \\
        & \neg z_{i,j,k}  \lor \neg u_{i,j,k-1} \lor \neg u_{i,j,k}, \\
        & \neg z_{i,j,k}  \lor u_{i,j,k-1} \lor u_{i,j,k}. 
    \end{align*}
\end{definition}
% \end{mdframed}

We focus on the regime where $m$ is arbitrarily larger than $n$. As with related formulas like $\PHP^m_n$, obtaining size lower bounds in this regime is challenging across various proof systems, since standard measures such as resolution width or $\PC$ degree do not appear to be effective, at least when applied directly. Our technique applies to any matrix $A$ and provides a $\PCR_{\F_2}$ size lower bound of $2^{\Omega(n)}$. This makes $\BTRank^m_n(A)$ a proof complexity generator, stretching $2mn$ input bits into $m^2$ output bits.

\begin{theorem} \label{thm:BTRank_A_lower_bound}
Suppose that $m > n \geq 16$ are integers, $8$ divides $n$, and $A \in \Q^{m \times m}$ is an $m$ by $m$ matrix over $\F_2$. Then any $\PCR_{\F_2}$ refutation of $\BTRank^m_n(A)$ has size at least $2^{\frac{\log e}{512}n - 2}$.
\end{theorem}

Recall that the basic strategy for the simpler algebraic encoding of the rank principle is to use a random restriction that, with high
probability, transforms a small-size refutation into a small-degree refutation. This then yields a contradiction with the $\PCR$ degree lower bounds for the pigeonhole principle.
Accordingly, our goal here is to show that if $\BTRank^m_n(A)$ admits a small $\PCR_{\F_2}$ refutation, then it can be converted into one in which a certain \emph{relaxed} form of degree is small. This relaxed notion of degree is carefully tailored to the structure of our formula, with distinct definitions for each variable sort. Moreover, it appears that if we require a random substitution to serve as a meaningful self-reduction (i.e., converting $\BTRank^m_n(A)$ to a smaller instance of the same principle), then this relaxed degree measure is close to the limit of what can be effectively decreased under such a random transformation. While the definition can be made slightly stricter, any substantial move toward the standard notion of degree (i.e., the total degree of a term) does not seem to yield a measure that decreases adequately under random self-reductions. This limitation seems to stem from the distinct nature of the (``extension'')  $u$-variables, whose behaviour differs from that of the other variables in the formula. Fortunately, our relaxed degree is still strong enough to yield, after an additional substitution and some proof manipulations, a $\PCR_{\F_2}$ refutation of $\FPHP^m_{n'}$ of degree less than $n'/2$, which is ruled out by a theorem of Razborov~\cite{Razb98}. 
%\Iddo{There is \textbf{no explicit definition of a new degree measure in Section 5.3}}

The initial conversion of a purported small refutation of $\BTRank^m_n(A)$ begins by applying two distinct random substitutions, $\rho$ and later $\sigma$. The random substitution  $\rho$ converts (with probability $1$) a refutation of $\BTRank^m_n(A)$ into a refutation of an instance with fewer columns $\BTRank^m_{\tilde{n}}(A)$, for some $\tilde n<n$. 
Additionally, with high probability, this substitution simplifies the occurrences of $u$-variables in the terms of the refutation.

We now define the random substitution $\rho$ (not to be confused with the substitution with the same notation in \Cref{sec:generator-pm}). 
This  substitution randomly compresses the matrices $X$ and $Y$ from $n$ columns to about $n/4$ columns, row by row, in a way that preserves the bamboo structure and makes different rows behave independently. 

\begin{definition}[Substitution $\rho$ for $\BTRank^m_n(A)$] \label{def:first_restriction_BTRank}
Assume that $m > n \geq 8$ and $n$ is divisible by $4$. Let $\tilde{n} = (n-4)/4$ and $\{\tilde{x}_{i,\ell} : i \in [m], \ell \in [\tilde{n}]\} 
\cup \{\tilde{y}_{\ell,i} : i \in [m], \ell \in [\tilde{n}]\}
\cup \{\tilde{z}_{i,j,\ell} : i,j \in [m], \ell \in [\tilde{n}]\} 
\cup \{\tilde{u}_{i,j,\ell} : i,j \in [m], \ell \in [\tilde{n}]\}$ be a set of fresh variables.  The random substitution $\rho$ is defined as follows.
Independently and uniformly at random, choose $4m$ bits $\{p_{i,0}, p_{i,1}, q_{i,0}, q_{i,1} : i \in [m]\} \in \Q^{4m}$.
These bits are used as follows: $p_{i,0}$ chooses how the boundary columns of row $i$ in $X$ are hard-fixed,
$p_{i,1}$ chooses which half of the interior columns of row $i$ carry $\tilde{x}$-variables (the other half become $0$),
$q_{j,0}$ similarly fixes boundary entries for column $j$ in $Y$,
$q_{j,1}$ chooses which interior blocks of column $j$ carry $\tilde{y}$-variables (the rest become $0$). See \Cref{fig:rho-step1-X-Y} and
\Cref{fig:rho-step2-X-Y}.
We extend  $\rho$ to the $u$- and $z$-variables so that it agrees with what $\rho$ already does to $X$ and $Y$, and so that $\BTRank^m_n(A)$ restricted by $\rho$ becomes an instance of $\BTRank^m_{\tilde n}(A)$ (with the same matrix $A$). Formally, this is done as follows.

\begin{enumerate}
                
        \item \label{item:first_restriction_BTRank_values_on_x_and_y} For each $i \in [m]$ and $k \in [n]$, set
        \[
            \rho(x_{i,k}) = \begin{cases}
                p_{i,0}  & \text{if } k \in \{1,n-1\}, \\
                1 & \text{if } k \in \{2,n\}, \\
                0 & \text{if } k \in \{3,\ldots, n-2\}, \lfloor \frac{k-3}{\tilde{n}}\rfloor \in \Q \text{ and }p_{i,1} = 0, \\ 
                 \tilde{x}_{i, ((k-3) \bmod{\tilde{n}})+1} & \text{if } k \in \{3,\ldots, n-2\}, \lfloor \frac{k-3}{\tilde{n}}\rfloor \in \{2,3\} \text{ and }p_{i,1} = 0, \\ 
                 \tilde{x}_{i, ((k-3) \bmod{\tilde{n}})+1} & \text{if } k \in \{3,\ldots, n-2\}, \lfloor \frac{k-3}{\tilde{n}}\rfloor \in \Q \text{ and }p_{i,1} = 1, \\ 
                 0 & \text{if } k \in \{3,\ldots, n-2\}, \lfloor \frac{k-3}{\tilde{n}}\rfloor \in \{2,3\} \text{ and }p_{i,1} = 1, \\ 
            
            \end{cases} 
          \]
          and similarly, for each $k \in [n]$ and $j \in [m]$, set  
            \[
            \rho(y_{k,j}) = \begin{cases}
                1 & \text{if } k \in \{1,n-1\}, \\
                q_{j,0} & \text{if } k \in \{2,n\}, \\
                0 & \text{if } k \in \{3,\ldots, n-2\}, \lfloor \frac{k-3}{\tilde{n}}\rfloor \in \{0,2\} \text{ and }q_{j,1} = 0, \\ 
                 \tilde{y}_{((k-3) \bmod{\tilde{n}})+1, j} & \text{if } k \in \{3,\ldots, n-2\}, \lfloor \frac{k-3}{\tilde{n}}\rfloor \in \{1,3\} \text{ and }q_{j,1} = 0, \\ 
                 \tilde{y}_{((k-3) \bmod{\tilde{n}})+1, j} & \text{if } k \in \{3,\ldots, n-2\}, \lfloor \frac{k-3}{\tilde{n}}\rfloor \in \{0,2\} \text{ and }q_{j,1} = 1, \\ 
                 0 & \text{if } k \in \{3,\ldots, n-2\}, \lfloor \frac{k-3}{\tilde{n}}\rfloor \in \{1,3\} \text{ and }q_{j,1} = 1. \\ 
            \end{cases} 
          \]
          In other words, the first two variables $(x_{i,1}, x_{i,2})$ of the row $(x_{i,1}, \ldots, x_{i,n})$ are set to $(0,1)$ if $p_{i,0} = 0$, and to $(1,1)$ if $p_{i,0} = 1$, and the same is true for the last two variables $(x_{i,n-1}, x_{i,n})$. The block of remaining variables $(x_{i,3}, \ldots, x_{i,n-2})$ is set to $(0,0,1,1) \otimes (\tilde{x}_{i,1}, \ldots, \tilde{x}_{i,\tilde{n}})$ if $p_{i,1} = 0$, and to $(1,1,0,0) \otimes (\tilde{x}_{i,1}, \ldots, \tilde{x}_{i,\tilde{n}})$ if $p_{i,1} = 1$. Similarly, for $(y_{1,j}, \ldots, y_{n,j})$, the first two (as well as the last two) variables are set to $(1,0)$ or $(1,1)$ depending on the value of $q_{j,0}$, and the remaining variables $(y_{3,j}, \ldots, y_{n-2,j})$ are set to $(0,1,0,1) \otimes (\tilde{y}_{1,j}, \ldots, \tilde{y}_{\tilde{n},j})$ or $(1,0,1,0) \otimes (\tilde{y}_{1,j}, \ldots, \tilde{y}_{\tilde{n},j})$ depending on the value of $q_{j,1}$.
          
          \item \label{item:first_restriction_BTRank_values_on_conjunction_vars} For each $i,j \in [m]$ and $k \in [n]$, set $z_{i,j,k}$ to the correct value as determined by the values of $\rho(x_{i,k})$ and $\rho(y_{k,j})$ just defined and by Binary AND Axioms. That is, for $k \in \{1,n-1\}$ we define $\rho(z_{i,j,k}) = p_{i,0}$, for $k \in \{2,n\}$ we set $\rho(z_{i,j,k}) = q_{j,0}$, and for $k \in \{3,\ldots, n-2\}$ we have four cases based on the values of $p_{i,1}, q_{j,1}$. First, let $r_{i,j} \in \{0,1,2,3\}$ be defined by $r_{i,j} = 3 - 2p_{i,1} - q_{j,1}$. It follows from \cref{item:first_restriction_BTRank_values_on_x_and_y} that the following definition is consistent with the Binary AND Axioms:            
         \[
            \rho(z_{i,j,k}) = \begin{cases}
                \tilde{z}_{i,j, ((k-3) \bmod{\tilde{n}})+1}  & \text{if } k \in \{r_{i,j}\tilde{n}+3, \ldots, (r_{i,j}+1)\tilde{n}+2\}, \\
                0 & \text{if } k \in \{3, \ldots, n-2\} \setminus \{r_{i,j}\tilde{n}+3, \ldots, (r_{i,j}+1)\tilde{n}+2\}.            
            \end{cases} 
        \]
        In other words, for $k \in \{3,\ldots, n-2\}$ we define $\rho(z_{i,j,k}) = 0$ if either $\rho(x_{i,k})=0$ or $\rho(y_{k,j})=0$; otherwise, it follows from the previous item that $\rho(x_{i,k}) = \tilde{x}_{i,\ell}$ and $\rho(y_{k,j}) = \tilde{y}_{\ell,j}$, where $\ell = ((k-3) \bmod{\tilde{n}})+1$, and in this case we set  $\rho(z_{i,j,k}) = \tilde{z}_{i,j,\ell}$.
          
          \item \label{item:first_restriction_BTRank_values_on_parity_vars} The $u$-variables are assigned by $\rho$ consistently not only with Summation Base Axioms and Summation Axioms, but also with Output Axioms, taking into account that after applying $\rho$ to $\BTRank^m_n(A)$ we want Output Axioms to hold for the new variables $\tilde{u}_{i,j,\tilde{n}}$. Given $i \in [m]$ and $j \in [m]$, the definition of $\rho(u_{i,j,1}), \ldots, \rho(u_{i,j,n})$ splits into cases based on the values $p_{i,0}, p_{i,1}, q_{j,0}, q_{j,1}$. It follows from \cref{item:first_restriction_BTRank_values_on_x_and_y} that the values of the first two variables are constant, namely $\rho(u_{i,j,1}) = p_{i,0}$ and $\rho(u_{i,j,2}) = p_{i,0} \oplus q_{j,0}$. We now define the values $\rho(u_{i,j,3}), \ldots, \rho(u_{i,j,n-2})$ with the help of the integer $r_{i,j}$ defined in \cref{item:first_restriction_BTRank_values_on_conjunction_vars}. 
         \[
            \rho(u_{i,j,k}) = \begin{cases}
                \tilde{u}_{i,j, ((k-3) \bmod{\tilde{n}})+1}  & \text{if } k \in \{r_{i,j}\tilde{n}+3, \ldots, (r_{i,j}+1)\tilde{n}+2\} \text{ and } p_{i,0} \oplus q_{j,0} = 0, \\
                \overline{\tilde{u}}_{i,j, ((k-3) \bmod{\tilde{n}})+1}  & \text{if } k \in \{r_{i,j}\tilde{n}+3, \ldots, (r_{i,j}+1)\tilde{n}+2\} \text{ and } p_{i,0} \oplus q_{j,0} = 1, \\
                 p_{i,0} \oplus q_{j,0} & \text{if } k \in \{3, \ldots, r_{i,j}\tilde{n}+2\}, \\
                 p_{i,0} \oplus q_{j,0} \oplus A_{i,j} & \text{if } k \in \{(r_{i,j}+1)\tilde{n}+3, \ldots, n-2\}.     
            \end{cases} 
          \]
To define $\rho$ consistently at the last two variables, $u_{i,j,n-1}$ and $u_{i,j,n}$, we need to set 
$\rho(u_{i,j,n-1}) = (p_{i,0} \oplus q_{j,0} \oplus A_{i,j}) \oplus p_{i,0} = q_{j,0} \oplus A_{i,j}$ and 
$\rho(u_{i,j,n}) = (q_{j,0} \oplus A_{i,j}) \oplus q_{j,0} = A_{i,j}$. 

        \item For any variable $\theta$ mapped to $0$ or $1$ by $\rho$, set $\rho(\overline{\theta}) = 1 - \rho(\theta)$.
        
        \item For any variable $\theta$ not mapped to a Boolean constant by $\rho$, set $\rho(\overline{\theta}) = \overline{\rho(\theta)}$.
\end{enumerate}
\end{definition}

% In preamble:
% \usepackage{tikz}
% \usetikzlibrary{positioning,calc}
% \usepackage{xcolor}

% Colors (tune if you like)
\definecolor{btgreen}{RGB}{220,245,220}
\definecolor{btred}{RGB}{245,220,220}

% --- TikZ helpers (fixed) ---
\newcommand{\BTcell}[6]{% x,y,w,h,fill,text
    \draw[thick,fill=#5] (#1,#2) rectangle node[font=\scriptsize, text depth=0pt] {#6} ++(#3,#4);
}

% text depth=0pt is very important here since it aligns the text correctly

\newcommand{\BTphaseLabel}[4]{% x,y,w,label
    \node[font=\scriptsize] at (#1+0.5*#3,#2) {#4};
}

\begin{figure}[H]
\centering

% =======================
% Top: X rows (p_{i,1}=0 and p_{i,1}=1)
% =======================
\begin{tikzpicture}[x=.6cm,y=.6cm]

% Geometry parameters (tune sizes here)
\def\h{0.9}        % height of a row strip
\def\wS{1.1}       % width of small boundary cell
\def\wP{3.5}       % width of each phase block

% Common texts
\def\Xkeep{$\tilde x_{i,1}\;\cdots\;\tilde x_{i,\tilde n}$}

% ----- Row 1: p_{i,1}=0 -----
\node[font=\bfseries] at (8.2,3.10){Action of $\rho$ on a fixed row of $X$ when $p_{i,1}=0$};

% Left boundary
\BTcell{0}{1.4}{\wS}{\h}{white}{$p_{i,0}$}
\BTcell{\wS}{1.4}{\wS}{\h}{white}{$1$}

% Phases 0..3
\BTcell{2*\wS}{1.4}{\wP}{\h}{btred}{$0$}
\BTcell{2*\wS+\wP}{1.4}{\wP}{\h}{btred}{$0$}
\BTcell{2*\wS+2*\wP}{1.4}{\wP}{\h}{btgreen}{\Xkeep}
\BTcell{2*\wS+3*\wP}{1.4}{\wP}{\h}{btgreen}{\Xkeep}

% Right boundary
\BTcell{2*\wS+4*\wP}{1.4}{\wS}{\h}{white}{$p_{i,0}$}
\BTcell{3*\wS+4*\wP}{1.4}{\wS}{\h}{white}{$1$}

% Phase labels above
\BTphaseLabel{2*\wS}{2.55}{\wP}{block 0}
\BTphaseLabel{2*\wS+\wP}{2.55}{\wP}{block 1}
\BTphaseLabel{2*\wS+2*\wP}{2.55}{\wP}{block 2}
\BTphaseLabel{2*\wS+3*\wP}{2.55}{\wP}{block 3}

\node[font=\small] at (8.2,0.95){Green = keeps $\tilde x$ variables (one per column inside the block), Red = hard-fixes to $0$};

\end{tikzpicture}

\

\begin{tikzpicture}[x=.6cm,y=.6cm]
% Geometry parameters (tune sizes here)
\def\h{0.9}        % height of a row strip
\def\wS{1.1}       % width of small boundary cell
\def\wP{3.5}       % width of each phase block

% Common texts
\def\Xkeep{$\tilde x_{i,1}\;\cdots\;\tilde x_{i,\tilde n}$}

% ----- Row 2: p_{i,1}=1 -----
\node[font=\bfseries] at (8.2,0.20){Action of $\rho$ on a fixed row of $X$ when $p_{i,1}=1$};

% Left boundary
\BTcell{0}{-1.55}{\wS}{\h}{white}{$p_{i,0}$}
\BTcell{\wS}{-1.55}{\wS}{\h}{white}{$1$}

% Phases 0..3 (swap colors)
\BTcell{2*\wS}{-1.55}{\wP}{\h}{btgreen}{\Xkeep}
\BTcell{2*\wS+\wP}{-1.55}{\wP}{\h}{btgreen}{\Xkeep}
\BTcell{2*\wS+2*\wP}{-1.55}{\wP}{\h}{btred}{$0$}
\BTcell{2*\wS+3*\wP}{-1.55}{\wP}{\h}{btred}{$0$}

% Right boundary
\BTcell{2*\wS+4*\wP}{-1.55}{\wS}{\h}{white}{$p_{i,0}$}
\BTcell{3*\wS+4*\wP}{-1.55}{\wS}{\h}{white}{$1$}

% Phase labels above
\BTphaseLabel{2*\wS}{-0.4}{\wP}{block 0}
\BTphaseLabel{2*\wS+\wP}{-0.4}{\wP}{block 1}
\BTphaseLabel{2*\wS+2*\wP}{-0.4}{\wP}{block 2}
\BTphaseLabel{2*\wS+3*\wP}{-0.4}{\wP}{block 3}

\node[font=\small] at (8.2,-2.0){Green = keeps $\tilde x$ variables (one per column inside the block), Red = hard-fixes to $0$};

\end{tikzpicture}

\caption{Action of $\rho$ on a fixed row of $X$ (two cases: $p_{i,1}=0$ and $p_{i,1}=1$).}
\label{fig:rho-step1-X-Y}
\end{figure}

% In preamble:
% \usepackage{tikz}
% \usetikzlibrary{positioning,calc}
% \usepackage{xcolor}

% Colors (tune if you like)
\definecolor{btgreen}{RGB}{220,245,220}
\definecolor{btred}{RGB}{245,220,220}

% --- TikZ helpers (fixed) ---
\renewcommand{\BTcell}[6]{% x,y,w,h,fill,text
    \draw[thick,fill=#5] (#1,#2) rectangle node[font=\scriptsize, text depth=0pt] {#6} ++(#3,#4);
}

\renewcommand{\BTphaseLabel}[4]{% x,y,w,label
    \node[font=\scriptsize] at (#1+0.5*#3,#2) {#4};
}

% Preamble needs:
% \usepackage{tikz}
% \usetikzlibrary{calc}
% \usepackage{graphicx} % for \rotatebox
% \usepackage{xcolor}

\definecolor{btgreen}{RGB}{220,245,220}
\definecolor{btred}{RGB}{245,220,220}

% --- Small text size used inside cells ---
% \newcommand{\Ycellfont}{\fontsize{6}{7}\selectfont}

% --- Cell drawing (non-rotated) ---
\newcommand{\Ycell}[6]{% x,y,w,h,fill,text
    % \path[fill=#5] (#1,#2) rectangle ++(#3,#4);
    % \draw[thick] (#1,#2) rectangle ++(#3,#4);
    % \node[align=center] at (#1+0.5*#3,#2+0.5*#4) {{\Ycellfont #6}};
    \draw[thick,fill=#5] (#1,#2) rectangle node[font=\scriptsize] {#6} ++(#3,#4);
}
% no text depth here to preserve the vertical alignment

% --- Cell drawing (rotated content) ---
% \newcommand{\YcellRot}[6]{% x,y,w,h,fill,text
%   \path[fill=#5] (#1,#2) rectangle ++(#3,#4);
%   \draw[thick] (#1,#2) rectangle ++(#3,#4);
%   \node[align=center] at (#1+0.5*#3,#2+0.5*#4)
%     {\rotatebox{90}{{\Ycellfont #6}}};
% }

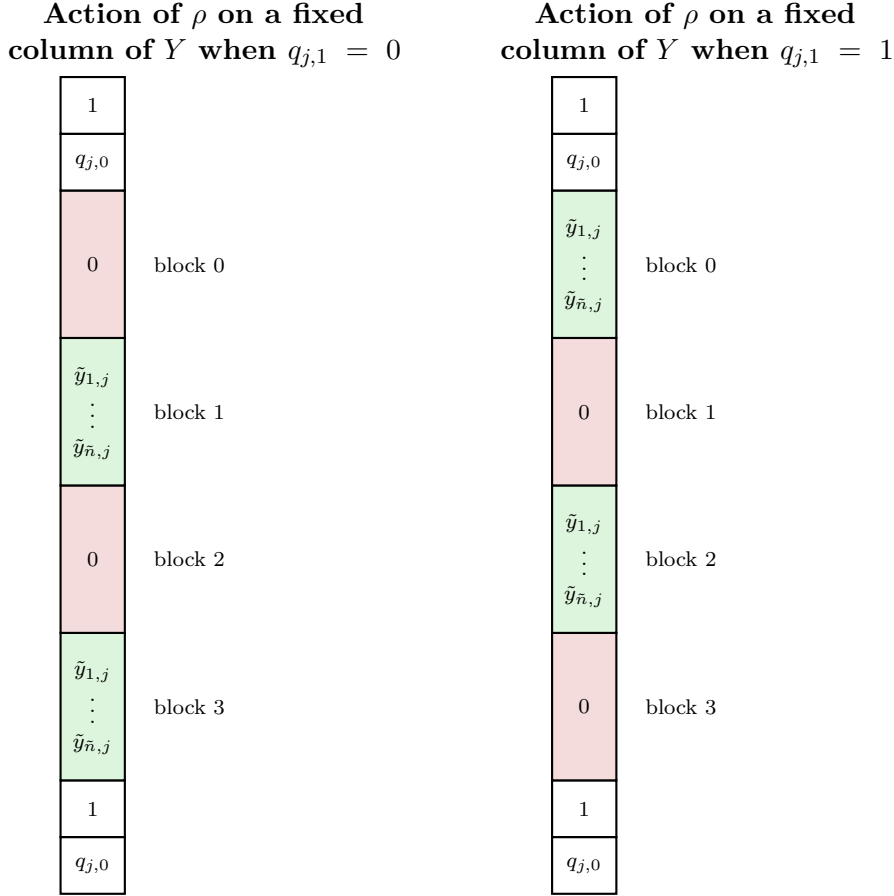
\begin{figure}[H]
\centering

\vspace{8pt} 
% =======================
% Bottom: Y columns (q_{j,1}=0 and q_{j,1}=1)
% =======================
\begin{tikzpicture}[x=1cm,y=1cm]

% Geometry
\def\wC{0.85}
\def\hS{0.75}
\def\hP{1.95}

\def\Ykeep{\begin{tabular}{c}$\tilde y_{1,j}$ \\ \vdots \\ $\tilde y_{\tilde n,j}$\end{tabular}}

% Positions
\def\xL{0.8}
\def\xR{7.3}
\def\yB{0.0}

% Column top and title height
\pgfmathsetmacro{\yTop}{\yB + 3*\hS + 4*\hP}  % top of the column
\pgfmathsetmacro{\yTitle}{\yTop + 0.8}        % push titles up (increase if you want)

% --------- Draw LEFT column first (no titles yet) ---------
% bottom boundary
\Ycell{\xL}{\yB}{\wC}{\hS}{white}{$q_{j,0}$}
\Ycell{\xL}{\yB+\hS}{\wC}{\hS}{white}{$1$}

% phases (from bottom: 3,2,1,0), q_{j,1}=0 keeps 1 and 3
\Ycell{\xL}{\yB+2*\hS}{\wC}{\hP}{btgreen}{\Ykeep}          % block 3
\Ycell{\xL}{\yB+2*\hS+\hP}{\wC}{\hP}{btred}{$0$}              % block 2
\Ycell{\xL}{\yB+2*\hS+2*\hP}{\wC}{\hP}{btgreen}{\Ykeep}    % block 1
\Ycell{\xL}{\yB+2*\hS+3*\hP}{\wC}{\hP}{btred}{$0$}            % block 0

% top boundary
\Ycell{\xL}{\yB+2*\hS+4*\hP}{\wC}{\hS}{white}{$q_{j,0}$}
\Ycell{\xL}{\yB+3*\hS+4*\hP}{\wC}{\hS}{white}{$1$}

% block labels
\node[font=\scriptsize, anchor=west] at (\xL+\wC+0.25, \yB+2*\hS+0.5*\hP) {block 3};
\node[font=\scriptsize, anchor=west] at (\xL+\wC+0.25, \yB+2*\hS+1.5*\hP) {block 2};
\node[font=\scriptsize, anchor=west] at (\xL+\wC+0.25, \yB+2*\hS+2.5*\hP) {block 1};
\node[font=\scriptsize, anchor=west] at (\xL+\wC+0.25, \yB+2*\hS+3.5*\hP) {block 0};

% --------- Draw RIGHT column (no titles yet) ---------
\Ycell{\xR}{\yB}{\wC}{\hS}{white}{$q_{j,0}$}
\Ycell{\xR}{\yB+\hS}{\wC}{\hS}{white}{$1$}

% q_{j,1}=1 keeps 0 and 2
\Ycell{\xR}{\yB+2*\hS}{\wC}{\hP}{btred}{$0$}                  % block 3
\Ycell{\xR}{\yB+2*\hS+\hP}{\wC}{\hP}{btgreen}{\Ykeep}       % block 2
\Ycell{\xR}{\yB+2*\hS+2*\hP}{\wC}{\hP}{btred}{$0$}            % block 1
\Ycell{\xR}{\yB+2*\hS+3*\hP}{\wC}{\hP}{btgreen}{\Ykeep}     % block 0

\Ycell{\xR}{\yB+2*\hS+4*\hP}{\wC}{\hS}{white}{$q_{j,0}$}
\Ycell{\xR}{\yB+3*\hS+4*\hP}{\wC}{\hS}{white}{$1$}

\node[font=\scriptsize, anchor=west] at (\xR+\wC+0.25, \yB+2*\hS+0.5*\hP) {block 3};
\node[font=\scriptsize, anchor=west] at (\xR+\wC+0.25, \yB+2*\hS+1.5*\hP) {block 2};
\node[font=\scriptsize, anchor=west] at (\xR+\wC+0.25, \yB+2*\hS+2.5*\hP) {block 1};
\node[font=\scriptsize, anchor=west] at (\xR+\wC+0.25, \yB+2*\hS+3.5*\hP) {block 0};

% --------- NOW draw the titles LAST, with white background ---------
\node[font=\bfseries, anchor=south, align=center, text width=7.2cm,
      fill=white, inner sep=2pt]
    at (\xL+1.9,\yTitle)
    {Action of $\rho$ on a fixed \mbox{column} of $Y$ when $q_{j,1}=0$};

\node[font=\bfseries, anchor=south, align=center, text width=7.2cm,
      fill=white, inner sep=2pt]
    at (\xR+1.9,\yTitle)
    {Action of $\rho$ on a fixed \mbox{column} of $Y$ when $q_{j,1}=1$};

% Explanatory text under columns (optional)
\node[font=\scriptsize] at (\xL+3.6,\yB-0.8){Green = keeps $\tilde y$ variables, Red = hard-fixes to 0};
% \node[font=\scriptsize] at (\xL+1.9,\yB-0.8)
    % {~~~~~~~~~~~~~~~~~~~~~~~~~~~~~~~~Green = keeps $\tilde y$ variables, Red = hard-fixes to 0};
% \node[font=\scriptsize] at (\xR+1.9,\yB-0.8)
%   {};

\end{tikzpicture}

\caption{Action of $\rho$ on a fixed column of $Y$ (two cases: $q_{j,1}=0$ and $q_{j,1}=1$).}
\label{fig:rho-step2-X-Y}
\end{figure}

First, we demonstrate the desired effect of $\rho$ on the $u$-variables occurring in terms. 

\begin{definition}[Row and column indices mentioned by $u$-variables inside a term]\label{def:column-mentioned u}
Let $t$ be a term in the variables of $\BTRank^m_n(A)$. We say that $i \in [m]$ is \emph{$U$-left-row-mentioned in} $t$ if there is $j \in [m], k \in [n]$, and $b \in \Q$ such that $u^b_{i,j,k}$ appears in $t$. We say that $j \in [m]$ is \emph{$U$-right-row-mentioned in} $t$ if there is $i \in [m], k \in [n]$, and $b \in \Q$ such that $u^b_{i,j,k}$ appears in $t$. We say that $k \in [n]$ is \emph{$U$-column-mentioned in} $t$ if there are $i,j \in [m]$ and $b \in \Q$ such that $u^b_{i,j,k}$ appears in $t$.
\end{definition}

Note that row indices are those that originally range over $[m]$ while column indices originally range over $[n]$.

\begin{lemma}[Random substitution $\rho$ kills terms that mention many $u$-rows]
\label{lem:first_restriction_BTRank_simplifies_extension_vars}
Suppose that $m > n \geq 8$ are integers, $4$ divides $n$, and $A \in \Q^{m \times m}$ is an $m$ by $m$ matrix over $\F_2$. Suppose that $t$ is a term in the variables of $\BTRank^m_n(A)$, $I \subseteq [m]$ is a set and $t'$ is a subterm of $t$ such that every $i \in I$ is $U$-left-row-mentioned in $t'$ and no $k \in \{1,2,n-1,n\}$ is $U$-column-mentioned in $t'$. Then $\Pr[t \restriction \rho \neq 0] \leq (3/4)^{|I|}$.
\end{lemma}
\begin{proof}
Let $r = |I|$. By the assumption of the lemma, $t$ contains a subterm of the form $u^{b_1}_{i_1,j_1,k_1} u^{b_2}_{i_2,j_2,k_2} \cdots u^{b_r}_{i_r,j_r,k_r}$, where $1 \leq i_1 < i_2 < \cdots < i_r \leq m$ and for each $s \in [r]$ we have $b_s \in \Q$, $ j_s \in [m]$ and $k_s \in \{3, \ldots, n-2\}$. 
It follows from the definition of $\rho$ that for each row $i$ that actually appears in the term (via some $u_{i,j,k}$ with $k\in\{3,\ldots,n-2\}$), there is a $1/4$ chance that $\rho$ sets that $u$-factor to the ``opposite''  constant (so the whole product becomes $0$), and these events are independent across the $r$ different rows; hence the term survives with probability at most ${(3/4)}^r$.

More formally, by the definition of $\rho$  for each ${(a_s)}_{s \in [r]} \in \Q^r$, for each ${(a_{j,0}, a_{j,1})}_{j \in \{j_1, \ldots, j_r\}} \in \Q^{2 \cdot |\{j_1, \ldots, j_r\}|}$, and for each $h \in [r]$, we have
\[
    \Prb{\rho(u_{i_h,j_h,k_h}) = a_h \;\vert\; \text{for all } s \in [r], q_{j_s,0} = a_{j_s,0} \text{ and } q_{j_s,1} = a_{j_s,1}} \geq 1/4.
\]
For each $h \in [r]$, at least one of the two possible values of $a_h$ evaluates $t$ to $0$. Since all $i_1, \ldots, i_r$ are distinct, these $r$ conditional events are mutually independent, and hence
\[
    \Prb{t\restriction\rho \neq 0 \;\vert\; \text{for all } s \in [r], q_{j_s,0} = a_{j_s,0} \text{ and } q_{j_s,1} = a_{j_s,1}}
    \le
    (3/4)^r.
\]
By averaging, the lemma follows.
% Moreover, these $r$ conditional events are, for any choice of ${(a_s)}_{s \in [r]} \in \Q^r$ and ${(a_{j,0}, a_{j,1})}_{j \in \{j_1, \ldots, j_r\}} \in \Q^{2 \cdot |\{j_1, \ldots, j_r\}|}$, mutually independent. The lemma follows.
\end{proof}

\smallskip

\Iddo{I think this discussion and \Cref{lem:first_restriction_BTRank_self_reduces} should come before \Cref{lem:first_restriction_BTRank_simplifies_extension_vars}.}
Next, we verify that $\rho$ transforms a refutation of $\BTRank^m_n(A)$ into a refutation of $\BTRank^m_{\tilde{n}}(A)$. Namely, we reduce the number of columns in $X$ and, accordingly, the number of rows in $Y$. This is a structural consequence of the way the substitution is defined, and occurs with probability $1$: 
the $k$-indices in the input variables $x_{i,k}, y_{k,i}: i \in [m], k \in [n]$  (excluding the four boundary ones) are partitioned into four consecutive blocks of equal size, and for each fixed row $i$ the rule for $\rho$ keeps exactly two of these blocks of $x_{i,k}$ as live variables while fixing the other two blocks to constants; similarly, for each fixed column $j$ the rule for $\rho$ keeps exactly two blocks of $y_{k,j}$ (see \Cref{fig:rho-step2-X-Y,fig:rho-step1-X-Y}). These two surviving blocks are deliberately misaligned, so that for every pair $(i,j)$ their intersection is contained in a \emph{single} block.
Since $z_{i,j,k}$ can only remain potentially non-zero when both $x_{i,k}$ and $y_{k,j}$ survive (based on the definition of $\rho$, \Cref{item:first_restriction_BTRank_values_on_conjunction_vars}), it follows that after applying $\rho$ the surviving $z$-variables (hence the relevant $k$-columns) for each $i$ and $j$ are confined to one block, i.e., to at most $(n-4)/4$ columns. Formally, we have the following.

\begin{lemma}\label{lem:first_restriction_BTRank_self_reduces}
Suppose that $m > n \geq 8$ are integers, $4$ divides $n$, and $A \in \Q^{m \times m}$ is an $m$ by $m$ matrix over $\F_2$. Let $\Pi$ be a $\PCR_{\F_2}$ refutation of $\BTRank^m_n(A)$. Then $\Pi \restriction \rho$ is a $\PCR_{\F_2}$ refutation of $\BTRank^m_{(n-4)/4}(A)$.
\end{lemma}
\begin{proof}
\Iddo{\sout{We need to explain that PCR proofs are closed under substitutions, with respect to size, I think. It's the same addition we made in the algebraic WRank lower bound.}}
\michal{We mention this fact in Preliminaries, at the end of \cref{sec:Substitutions}.}

We only need to check how $\rho$ affects (the translations into polynomials of) the clauses of $\BTRank^m_n(A)$. By \cref{item:first_restriction_BTRank_values_on_conjunction_vars,item:first_restriction_BTRank_values_on_parity_vars}, the clauses of each Output Axiom and each Summation Base Axiom are satisfied by $\rho$. Equivalently, $\rho$ converts the translation of each such clause into the zero polynomial (which can be removed from $\Pi \restriction \rho$). Further, by \cref{item:first_restriction_BTRank_values_on_conjunction_vars}, each Binary AND Axiom is either satisfied by $\rho$ or turned into the corresponding  axiom of $\BTRank^m_{\tilde{n}}(A)$. Next, by \cref{item:first_restriction_BTRank_values_on_conjunction_vars,item:first_restriction_BTRank_values_on_parity_vars}, Summation Axioms for $k \in \{2, n\}$ are satisfied by $\rho$. 

Consider now a Summation Axiom for $k = n-1$, i.e., the CNF encoding of $u_{i,j,n-1} = u_{i,j,n-2}  + z_{i,j,n-1}$. By the definition of $\rho$, we have $\rho(z_{i,j,n-1}) = p_{i,0}$, and $\rho(u_{i,j,n-1}) = q_{j,0} \oplus A_{i,j}$. 
If $r_{i,j} = 3$, we have $\rho(u_{i,j,n-2}) = \tilde{u}^{p_{i,0} \oplus q_{j,0} \oplus 1}_{i,j,\tilde{n}}$, which means that each of the four clauses encoding the axiom is either satisfied or is transformed into the corresponding Output Axiom clause $\tilde{u}^{A_{i,j}}_{i,j,\tilde{n}}$. 
If $r_{i,j} \neq 3$, then $\rho(u_{i,j,n-2}) = p_{i,0} \oplus q_{j,0} \oplus A_{i,j}$, and so all four clauses encoding the Summation Axiom are satisfied.  

Finally, consider Summation Axioms for $k \in \{3, \ldots, n-2\}$. For $k = r_{i,j}\tilde{n}+3$, the axiom turns into the corresponding Summation Base Axiom (for either value of $p_{i,0} \oplus q_{j,0}$). For $k \in \{ r_{i,j}\tilde{n}+4, \ldots, (r_{i,j}+1)\tilde{n} + 2\}$, the axiom turns into a Summation Axiom (again, for either value of $p_{i,0} \oplus q_{j,0}$).
For $k = (r_{i,j}+1)\tilde{n}+3$, when $r_{i,j} \neq 3$, we have $\rho(u_{i,j,(r_{i,j}+1)\tilde{n}+2}) = \tilde{u}^{p_{i,0} \oplus q_{j,0} \oplus A_{i,j}}_{i,j,\tilde{n}}$, $\rho(u_{i,j,(r_{i,j}+1)\tilde{n}+3}) = p_{i,0} \oplus q_{j,0} \oplus A_{i,j}$, and $\rho(z_{i,j,(r_{i,j}+1)\tilde{n}+3}) = 0$, so the axiom turns into the corresponding Output Axiom $\tilde{u}^{A_{i,j}}_{i,j,\tilde{n}}$. When $r_{i,j}=3$, we have $k=n-1$, which is already covered above.
For $k \in \{3, \ldots, r_{i,j}\tilde{n}+2\} \cup \{(r_{i,j}+1)\tilde{n}+4, \ldots, n-2\}$, the Summation Axiom is satisfied by $\rho$.
\end{proof}
\smallskip

The second random substitution, denoted by $\sigma$, is intended to simplify the $x$-, $y$- and $z$-variable occurrences  in terms while, again, turning $\BTRank^m_n(A)$ into a smaller instance. In contrast to the substitution $\rho$, which aims, with high probability, to reduce the number of \emph{row} indices (of $X$) appearing through the \emph{$u$-variables} in each term (as well as to reduce the instance from $n$ to $\tilde n$ columns), the goal of $\sigma$ is to reduce the number of \emph{column} indices (of $X$ and $Y^T$) involved via the \emph{$x$-, $y$- and $z$-variables}. \Iddo{Why are we saying here ``of \emph{column} indices (of $X$ and $Y^T$)'' while above we said  ``\emph{row} indices (of $X$)'' only? Shouldn't we add also "of X \underline{and $Y^T$?}} \michal{Because before (i.e., in \cref{lem:first_restriction_BTRank_simplifies_extension_vars}) we were indeed interested only in zeroing terms with many \emph{$U$-left-row-mentioned} indices, where "left" refers to the matrix $X$ (cf. \cref{def:column-mentioned}). We were not trying to do the same with the second indices of $u$-variables, which indicate rows in $Y^T$. In contrast, $\sigma$ aims at all column indices appearing through \emph{$x$-, $y$- and $z$-variables}, cf. \cref{def:column-mentioned}, where an $X$-column-mentioned index refers to a column of $X$, a $Y^T$-column-mentioned index refers to a column of $Y^T$, and a $Z$-column-mentioned index can be thought of as referring both to a column of $X$ and the same column of $Y^T$.} 

Applying $\rho$ reduces the number of columns of the matrices $X$ and $Y^T$ in the original instance from $n$ to $(n-4)/4$ (\cref{lem:first_restriction_BTRank_self_reduces}). However, this does not substantially decrease the number of column indices mentioned by a given term: the term may still mention many columns within the surviving block via the $x$-, $y$-, and $z$-variables. The purpose of $\sigma$ is to significantly reduce, for each term, the number of column indices involved through these variables.

\begin{definition}[Substitution $\sigma$ for $\BTRank^m_n(A)$]\label{def:second_restriction_BTRank}
Assume that $m > n \geq 3$ and $n-1$ is divisible by $2$. A random substitution $\sigma$ is chosen as follows. 
\begin{enumerate}
        \item \label{item:second_restriction_BTRank_picks_some_bits} Choose a subset $S \subseteq [n-1]$ uniformly at random from all $(n-1)/2$-element subsets of $[n-1]$. For each $k \in [n-1] \setminus S$, independently choose $q_k \in \Q$ at random with $\Pr[q_k = 0] = \Pr[q_k = 1] = 1/2$. For each $k \in S$, define $q_k = *$, and set $q_n = |\{\ell \in [n-1] : q_{\ell} = 1\}| \bmod{2}$. Note that this choice of $q_n$ makes the total number of ones even, i.e., if for $k \in [n]$ we let 
$o_k = |\{\ell \in [k] : q_{\ell} = 1\}| \bmod{2}$, then we have $o_n =0$. Denote by $s_k = |S \cap [k]|$ the number of stars among $q_1, \ldots, q_k$. 
        \item \label{item:second_restriction_BTRank_values_on_vars} For each $i,j \in [m]$ and $k \in [n]$, define
         \begin{align*}
            \sigma(x_{i,k})  =  \begin{cases}
                q_k & \text{if } q_k \in \Q, \\
                \tilde{x}_{i,s_k} & \text{if } q_k = *,         
            \end{cases} & &
            \sigma(y_{k,j}) = \begin{cases}
                q_k & \text{if } q_k \in \Q, \\
                \tilde{y}_{s_k,j} & \text{if } q_k = *, 
            \end{cases}
          \end{align*} 
and extend $\sigma$ consistently to the $z$- and $u$-variables by
          \begin{align*}
            \sigma(z_{i,j,k})  & =  \begin{cases}
                q_k & \text{if } q_k \in \Q, \\
                \tilde{z}_{i,j,s_k} & \text{if } q_k = *,         
            \end{cases} \\
            \sigma(u_{i,j,k}) & = \begin{cases}
                o_k & \text{if } k \in \{1, \ldots, \min(S)-1\}, \\
                \tilde{u}^{o_k \oplus 1}_{i,j,s_k} & \text{if } k \in \{\min(S), \ldots, \max(S)\}, \\
                A_{i,j} \oplus o_k & \text{if }  k \in \{\max(S)+1, \ldots, n\}.
            \end{cases}
          \end{align*}
          
         \item For any variable $\theta$ mapped to $0$ or $1$ by $\sigma$, set $\sigma(\overline{\theta}) = 1 - \sigma(\theta)$.
        
        \item For any variable $\theta$ not mapped to a Boolean constant by $\sigma$, set $\sigma(\overline{\theta}) = \overline{\sigma(\theta)}$.
\end{enumerate}
\end{definition}

% -----------------
 
\definecolor{btgray}{RGB}{242,242,242}

\begin{figure}[H]
\centering
\begin{tikzpicture}[x=.70cm,y=.70cm]

% cell geometry
\def\h{0.95}
\def\w{1.20}
\def\x0{2.0}

% row y-levels
\def\yk{0.0}
\def\yq{-1.25}
\def\yo{-2.50}
\def\ys{-3.75}

% title
\node[font=\bfseries] at (7.5,2.05)
{Example for \Cref{def:second_restriction_BTRank}: choosing $S$ and defining $q_k,o_k,s_k$ (here $n=9$)};

% row labels
\node[font=\small, anchor=east] at (\x0-0.25,\yk+0.48) {$k$};
\node[font=\small, anchor=east] at (\x0-0.25,\yq+0.48) {$q_k$};
\node[font=\small, anchor=east] at (\x0-0.25,\yo+0.48) {$o_k$};
\node[font=\small, anchor=east] at (\x0-0.25,\ys+0.48) {$s_k$};

% k row
\BTcell{\x0+0*\w}{\yk}{\w}{\h}{white}{$1$}
\BTcell{\x0+1*\w}{\yk}{\w}{\h}{white}{$2$}
\BTcell{\x0+2*\w}{\yk}{\w}{\h}{white}{$3$}
\BTcell{\x0+3*\w}{\yk}{\w}{\h}{white}{$4$}
\BTcell{\x0+4*\w}{\yk}{\w}{\h}{white}{$5$}
\BTcell{\x0+5*\w}{\yk}{\w}{\h}{white}{$6$}
\BTcell{\x0+6*\w}{\yk}{\w}{\h}{white}{$7$}
\BTcell{\x0+7*\w}{\yk}{\w}{\h}{white}{$8$}
\BTcell{\x0+8*\w}{\yk}{\w}{\h}{white}{$9$}

% q row (stars at S={2,4,6,8})
\BTcell{\x0+0*\w}{\yq}{\w}{\h}{btgray}{$1$}
\BTcell{\x0+1*\w}{\yq}{\w}{\h}{btgreen}{$\ast$}
\BTcell{\x0+2*\w}{\yq}{\w}{\h}{btgray}{$0$}
\BTcell{\x0+3*\w}{\yq}{\w}{\h}{btgreen}{$\ast$}
\BTcell{\x0+4*\w}{\yq}{\w}{\h}{btgray}{$1$}
\BTcell{\x0+5*\w}{\yq}{\w}{\h}{btgreen}{$\ast$}
\BTcell{\x0+6*\w}{\yq}{\w}{\h}{btgray}{$0$}
\BTcell{\x0+7*\w}{\yq}{\w}{\h}{btgreen}{$\ast$}
\BTcell{\x0+8*\w}{\yq}{\w}{\h}{btgray}{$0$}

% o row (prefix parity of 1s in q_1..q_k, treating * as "not counted")
\BTcell{\x0+0*\w}{\yo}{\w}{\h}{btgray}{$1$}
\BTcell{\x0+1*\w}{\yo}{\w}{\h}{btgray}{$1$}
\BTcell{\x0+2*\w}{\yo}{\w}{\h}{btgray}{$1$}
\BTcell{\x0+3*\w}{\yo}{\w}{\h}{btgray}{$1$}
\BTcell{\x0+4*\w}{\yo}{\w}{\h}{btgray}{$0$}
\BTcell{\x0+5*\w}{\yo}{\w}{\h}{btgray}{$0$}
\BTcell{\x0+6*\w}{\yo}{\w}{\h}{btgray}{$0$}
\BTcell{\x0+7*\w}{\yo}{\w}{\h}{btgray}{$0$}
\BTcell{\x0+8*\w}{\yo}{\w}{\h}{btgray}{$0$}

% s row (number of stars among 1..k)
\BTcell{\x0+0*\w}{\ys}{\w}{\h}{btgray}{$0$}
\BTcell{\x0+1*\w}{\ys}{\w}{\h}{btgray}{$1$}
\BTcell{\x0+2*\w}{\ys}{\w}{\h}{btgray}{$1$}
\BTcell{\x0+3*\w}{\ys}{\w}{\h}{btgray}{$2$}
\BTcell{\x0+4*\w}{\ys}{\w}{\h}{btgray}{$2$}
\BTcell{\x0+5*\w}{\ys}{\w}{\h}{btgray}{$3$}
\BTcell{\x0+6*\w}{\ys}{\w}{\h}{btgray}{$3$}
\BTcell{\x0+7*\w}{\ys}{\w}{\h}{btgray}{$4$}
\BTcell{\x0+8*\w}{\ys}{\w}{\h}{btgray}{$4$}

% legend
\node[font=\small, anchor=west] at (\x0, -4.85)
{Green = $k\in S$ (so $q_k=\ast$ survives), Gray = constant bits (fixed by $\sigma$).};

\node[font=\small, anchor=west] at (\x0, -5.35)
{Here $S=\{2,4,6,8\}$ and $q_9$ is chosen so that $o_9=0$.};

\end{tikzpicture}
\caption{A concrete example of the random data in \Cref{def:second_restriction_BTRank}: the set $S$, the values $(q_k$, and the derived sequences $(o_k$ and $s_k=|S\cap[k]|$.}\label{fig:sigma-qs-example}
\end{figure}

% ------------------

\begin{figure}[H]
\centering
\begin{tikzpicture}[x=.70cm,y=.70cm]

\def\cellH{0.95}
\def\cellW{1.25}
\def\xZ{2.0}
\def\yRowK{0.0}
\def\yRowSig{-1.25}

\node[font=\bfseries] at (7.5,1.60)
{Example: action of $\sigma$ on $(x_{i,1},\ldots,x_{i,9})$};

\node[font=\small, anchor=east] at (\xZ-0.25,\yRowK+0.48) {$k$};
\node[font=\small, anchor=east] at (\xZ-0.25,\yRowSig+0.48) {$\sigma(x_{i,k})$};

% k row
\BTcell{\xZ+0*\cellW}{\yRowK}{\cellW}{\cellH}{white}{$1$}
\BTcell{\xZ+1*\cellW}{\yRowK}{\cellW}{\cellH}{white}{$2$}
\BTcell{\xZ+2*\cellW}{\yRowK}{\cellW}{\cellH}{white}{$3$}
\BTcell{\xZ+3*\cellW}{\yRowK}{\cellW}{\cellH}{white}{$4$}
\BTcell{\xZ+4*\cellW}{\yRowK}{\cellW}{\cellH}{white}{$5$}
\BTcell{\xZ+5*\cellW}{\yRowK}{\cellW}{\cellH}{white}{$6$}
\BTcell{\xZ+6*\cellW}{\yRowK}{\cellW}{\cellH}{white}{$7$}
\BTcell{\xZ+7*\cellW}{\yRowK}{\cellW}{\cellH}{white}{$8$}
\BTcell{\xZ+8*\cellW}{\yRowK}{\cellW}{\cellH}{white}{$9$}

% Example choice: n=9, S={2,4,6,8} so s_2=1,s_4=2,s_6=3,s_8=4
\BTcell{\xZ+0*\cellW}{\yRowSig}{\cellW}{\cellH}{btgray}{$q_1$}
\BTcell{\xZ+1*\cellW}{\yRowSig}{\cellW}{\cellH}{btgreen}{$\tilde x_{i,1}$}
\BTcell{\xZ+2*\cellW}{\yRowSig}{\cellW}{\cellH}{btgray}{$q_3$}
\BTcell{\xZ+3*\cellW}{\yRowSig}{\cellW}{\cellH}{btgreen}{$\tilde x_{i,2}$}
\BTcell{\xZ+4*\cellW}{\yRowSig}{\cellW}{\cellH}{btgray}{$q_5$}
\BTcell{\xZ+5*\cellW}{\yRowSig}{\cellW}{\cellH}{btgreen}{$\tilde x_{i,3}$}
\BTcell{\xZ+6*\cellW}{\yRowSig}{\cellW}{\cellH}{btgray}{$q_7$}
\BTcell{\xZ+7*\cellW}{\yRowSig}{\cellW}{\cellH}{btgreen}{$\tilde x_{i,4}$}
\BTcell{\xZ+8*\cellW}{\yRowSig}{\cellW}{\cellH}{btgray}{$q_9$}

\node[font=\small, anchor=west] at (\xZ, -2.55)
{Rule: $\sigma(x_{i,k})=q_k$ if $q_k\in\Q$, and $\sigma(x_{i,k})=\tilde x_{i,s_k}$ if $q_k=\ast$.};

\end{tikzpicture}
\caption{Example of \(\sigma\) on the \(x\)-variables.}
\label{fig:sigma-x-example}
\end{figure}

\begin{figure}[H]
\centering
\begin{tikzpicture}[x=.70cm,y=.70cm]

\def\cellH{0.95}
\def\cellW{1.25}
\def\xZ{2.0}
\def\yRowK{0.0}
\def\yRowSig{-1.25}

\node[font=\bfseries] at (7.5,1.50)
{Example: action of $\sigma$ on $(y_{1,j},\ldots,y_{9,j})$};

\node[font=\small, anchor=east] at (\xZ-0.25,\yRowK+0.48) {$k$};
\node[font=\small, anchor=east] at (\xZ-0.25,\yRowSig+0.48) {$\sigma(y_{k,j})$};

% k row
\BTcell{\xZ+0*\cellW}{\yRowK}{\cellW}{\cellH}{white}{$1$}
\BTcell{\xZ+1*\cellW}{\yRowK}{\cellW}{\cellH}{white}{$2$}
\BTcell{\xZ+2*\cellW}{\yRowK}{\cellW}{\cellH}{white}{$3$}
\BTcell{\xZ+3*\cellW}{\yRowK}{\cellW}{\cellH}{white}{$4$}
\BTcell{\xZ+4*\cellW}{\yRowK}{\cellW}{\cellH}{white}{$5$}
\BTcell{\xZ+5*\cellW}{\yRowK}{\cellW}{\cellH}{white}{$6$}
\BTcell{\xZ+6*\cellW}{\yRowK}{\cellW}{\cellH}{white}{$7$}
\BTcell{\xZ+7*\cellW}{\yRowK}{\cellW}{\cellH}{white}{$8$}
\BTcell{\xZ+8*\cellW}{\yRowK}{\cellW}{\cellH}{white}{$9$}

% Example choice: n=9, S={2,4,6,8}
\BTcell{\xZ+0*\cellW}{\yRowSig}{\cellW}{\cellH}{btgray}{$q_1$}
\BTcell{\xZ+1*\cellW}{\yRowSig}{\cellW}{\cellH}{btgreen}{$\tilde y_{1,j}$}
\BTcell{\xZ+2*\cellW}{\yRowSig}{\cellW}{\cellH}{btgray}{$q_3$}
\BTcell{\xZ+3*\cellW}{\yRowSig}{\cellW}{\cellH}{btgreen}{$\tilde y_{2,j}$}
\BTcell{\xZ+4*\cellW}{\yRowSig}{\cellW}{\cellH}{btgray}{$q_5$}
\BTcell{\xZ+5*\cellW}{\yRowSig}{\cellW}{\cellH}{btgreen}{$\tilde y_{3,j}$}
\BTcell{\xZ+6*\cellW}{\yRowSig}{\cellW}{\cellH}{btgray}{$q_7$}
\BTcell{\xZ+7*\cellW}{\yRowSig}{\cellW}{\cellH}{btgreen}{$\tilde y_{4,j}$}
\BTcell{\xZ+8*\cellW}{\yRowSig}{\cellW}{\cellH}{btgray}{$q_9$}

\node[font=\small, anchor=west] at (\xZ, -2.55)
{Rule: $\sigma(y_{k,j})=q_k$ if $q_k\in\Q$, and $\sigma(y_{k,j})=\tilde y_{s_k,j}$ if $q_k=\ast$.};

\end{tikzpicture}
\caption{Example of \(\sigma\) on the \(y\)-variables.}
\label{fig:sigma-y-example}
\end{figure}

\begin{figure}[H]
\centering
\begin{tikzpicture}[x=.70cm,y=.70cm]

\def\cellH{0.95}
\def\cellW{1.90}
\def\xZ{2.0}
\def\yRowK{0.0}
\def\yRowSig{-1.25}

\node[font=\bfseries] at (7.9,1.50)
{Example: action of $\sigma$ on $(u_{i,j,1},\ldots,u_{i,j,9})$ (here $a=2$, $b=8$)};

\node[font=\small, anchor=east] at (\xZ-0.25,\yRowK+0.48) {$k$};
\node[font=\small, anchor=east] at (\xZ-0.25,\yRowSig+0.48) {$\sigma(u_{i,j,k})$};

% k row
\BTcell{\xZ+0*\cellW}{\yRowK}{\cellW}{\cellH}{white}{$1$}
\BTcell{\xZ+1*\cellW}{\yRowK}{\cellW}{\cellH}{white}{$2$}
\BTcell{\xZ+2*\cellW}{\yRowK}{\cellW}{\cellH}{white}{$3$}
\BTcell{\xZ+3*\cellW}{\yRowK}{\cellW}{\cellH}{white}{$4$}
\BTcell{\xZ+4*\cellW}{\yRowK}{\cellW}{\cellH}{white}{$5$}
\BTcell{\xZ+5*\cellW}{\yRowK}{\cellW}{\cellH}{white}{$6$}
\BTcell{\xZ+6*\cellW}{\yRowK}{\cellW}{\cellH}{white}{$7$}
\BTcell{\xZ+7*\cellW}{\yRowK}{\cellW}{\cellH}{white}{$8$}
\BTcell{\xZ+8*\cellW}{\yRowK}{\cellW}{\cellH}{white}{$9$}

% sigma(u_{i,j,k}) row:
% k<a : o_k (here only k=1)
\BTcell{\xZ+0*\cellW}{\yRowSig}{\cellW}{\cellH}{btgray}{$o_1$}

% a<=k<=b : \tilde u^{o_k \oplus 1}_{i,j,s_k} (k=2..8)
\BTcell{\xZ+1*\cellW}{\yRowSig}{\cellW}{\cellH}{btgreen}{$\tilde u_{i,j,s_2}^{\,o_2\oplus 1}$}
\BTcell{\xZ+2*\cellW}{\yRowSig}{\cellW}{\cellH}{btgreen}{$\tilde u_{i,j,s_3}^{\,o_3\oplus 1}$}
\BTcell{\xZ+3*\cellW}{\yRowSig}{\cellW}{\cellH}{btgreen}{$\tilde u_{i,j,s_4}^{\,o_4\oplus 1}$}
\BTcell{\xZ+4*\cellW}{\yRowSig}{\cellW}{\cellH}{btgreen}{$\tilde u_{i,j,s_5}^{\,o_5\oplus 1}$}
\BTcell{\xZ+5*\cellW}{\yRowSig}{\cellW}{\cellH}{btgreen}{$\tilde u_{i,j,s_6}^{\,o_6\oplus 1}$}
\BTcell{\xZ+6*\cellW}{\yRowSig}{\cellW}{\cellH}{btgreen}{$\tilde u_{i,j,s_7}^{\,o_7\oplus 1}$}
\BTcell{\xZ+7*\cellW}{\yRowSig}{\cellW}{\cellH}{btgreen}{$\tilde u_{i,j,s_8}^{\,o_8\oplus 1}$}

% k>b : A_{i,j} \oplus o_k (here only k=9)
\BTcell{\xZ+8*\cellW}{\yRowSig}{\cellW}{\cellH}{btgray}{$A_{i,j}\oplus o_9$}

% thick region outlines
\draw[very thick] (\xZ+0*\cellW,\yRowSig) rectangle ++(\cellW,\cellH);        % k<a
\draw[very thick] (\xZ+1*\cellW,\yRowSig) rectangle ++(7*\cellW,\cellH);     % a..b
\draw[very thick] (\xZ+8*\cellW,\yRowSig) rectangle ++(\cellW,\cellH);       % k>b

\node[font=\small, anchor=west] at (\xZ, -2.65)
{Rule: $\sigma(u_{i,j,k})=o_k$ for $k<a$;\quad $\sigma(u_{i,j,k})=\tilde u_{i,j,s_k}^{\,o_k\oplus 1}$ for $a\le k\le b$;\quad $\sigma(u_{i,j,k})=A_{i,j}\oplus o_k$ for $k>b$.};

\end{tikzpicture}
\caption{Example of \(\sigma\) on the \(u\)-chain (three regions determined by \(a=\min(S)\) and \(b=\max(S)\)).}
\label{fig:sigma-u-example}
\end{figure}

\begin{definition}[Column indices mentioned in a term by $x$-, $y$- and $z$-variables]\label{def:column-mentioned xyz}
Let $t$ be a term in the variables of $\BTRank^m_n(A)$ and let $k \in [n]$. We say that $k$ is \emph{$X$-column-mentioned in} $t$ if there are $i \in [m]$ and $b \in \Q$ such that $x^b_{i,k}$ appears in $t$. We say that $k$ is \emph{$Y^T$-column-mentioned in} $t$ if there are $j \in [m]$ and $b \in \Q$ such that $y^b_{k,j}$ appears in $t$. We say that $k$ is \emph{$Z$-column-mentioned in} $t$ if there are $i,j \in [m]$ and $b \in \Q$ such that $z^b_{i,j,k}$ appears in $t$.
\end{definition}

\begin{lemma}\label{lem:second_restriction_BTRank_simplifies_extension_vars}
Suppose that $m > n \geq 3$ are integers, $2$ divides $n-1$, and $A \in \Q^{m \times m}$ is an $m$ by $m$ matrix over $\F_2$. Suppose that $t$ is a term in the variables of $\BTRank^m_n(A)$ and $K \subseteq [n-1]$ is a set such that every $k \in K$ is either $X$-column-mentioned or $Y^T$-column-mentioned or $Z$-column-mentioned in $t$. 
Then $\Pr[t \restriction \sigma \neq 0] \leq \exp(-\card{K}/16) + 2^{-\card{K}/4}$.
\end{lemma}
\begin{proof}
    Let $\overline{S} = [n-1] \setminus S$ be the complement of $S$ in $[n-1]$. Since $S$ is chosen uniformly at random from all $(n-1)/2$-element subsets of $[n-1]$, the set $\overline{S}$ is also chosen uniformly at random from all $(n-1)/2$-element subsets of $[n-1]$.
    Define the random variable $X$ by
    \[
        X = \card{K \setminus S} = \card{K \cap \overline{S}}
    \]
    and $\mu = \Exp{X} = \card{K}/2$. Then \cref{thm: hoeffding chernoff bound} (with $\delta=1/2$) implies that
    \[
        \Prb{X \le \mu/2} \le \exp(-\mu/8) = \exp(-\card{K}/16).
    \]

    Now fix any choice of $S$ such that $\card{K \setminus S} = X > \card{K}/4$.
    We pick for each $k \in K \setminus S$ a single variable $\gamma_k$ from $t$ witnessing that $k$ is $X$-column-mentioned or $Y^T$-column-mentioned or $Z$-column-mentioned. Notice that $\sigma$ evaluates these variables independently to $0$ or $1$, each with probability $1/2$. At most one of the $2^X \ge 2^{\card{K}/4}$ assignments to these variables does not zero out $t$, so $\Pr[t \restriction \sigma \neq 0 \;\vert\; S \text{ with } X > \card{K}/4] \le 2^{-\card{K}/4}$.

    The union bound then gives $\Pr[t \restriction \sigma \neq 0] \leq \exp(-\card{K}/16) + 2^{-\card{K}/4}$.
\end{proof}

\begin{lemma}\label{lem:second_restriction_BTRank_self_reduces}
Suppose that $m > n \geq 3$ are integers, $2$ divides $n-1$, and $A \in \Q^{m \times m}$ is an $m$ by $m$ matrix over $\F_2$. Let $\Pi$ be a $\PCR_{\F_2}$ refutation of $\BTRank^m_n(A)$. Then $\Pi \restriction \sigma$ is a $\PCR_{\F_2}$ refutation of $\BTRank^m_{(n-1)/2}(A)$. 
\end{lemma}  
\begin{proof}
We inspect how $\sigma$ acts on (the translations into polynomials of) the clauses of $\BTRank^m_n(A)$. First, consider the Output Axioms. We know by \cref{item:second_restriction_BTRank_picks_some_bits} that $o_n =0$, and that $n \in \{\max(S)+1, \ldots, n\}$ for any choice of $S$. Hence by \cref{item:second_restriction_BTRank_values_on_vars}, $\sigma(u_{i,j,n}) = A_{i,j}$, which means that Output Axioms are satisfied. 
Next, it is immediate by \cref{item:second_restriction_BTRank_values_on_vars} that the Binary AND Axiom for $k \in [n]$ is either satisfied by $\sigma$ (if $q_k \in \Q$) or turned into the Binary AND Axiom $\tilde{z}_{i,j,s_k} = \tilde{x}_{i,s_k}\tilde{y}_{s_k,j}$ of $\BTRank^m_{(n-1)/2}(A)$. 
Further, the Summation Base Axiom is satisfied if $1 \notin S$, since $\sigma(u_{i,j,1}) = o_1 = q_1 = \sigma(z_{i,j,1})$; otherwise $s_1=1$ and it turns into a Summation Base Axiom of $\BTRank^m_{(n-1)/2}(A)$. 

Next, consider Summation Axioms. For $k \in \{2, \ldots, \min(S)-1\}$ the axiom is satisfied by $\sigma$. If $2 \leq k = \min(S)$, then each clause of the axiom is either satisfied or turned into a clause of the Summation Base Axiom of $\BTRank^m_{(n-1)/2}(A)$ (for either value of $o_k$). 
For $k \in \{\min(S), \ldots, \max(S)\} \setminus S$, each clause of the axiom is either satisfied or turned into $u_{i,j,s_k} \lor \neg u_{i,j,s_k}$. For $k \in S \setminus \{\min(S)\}$, the axiom turns into the Summation Axiom for $s_k$ of $\BTRank^m_{(n-1)/2}(A)$ (for either value of $o_k$). For $k = \max(S)+1$, each clause of the axiom is either satisfied or turned into an Output Axiom of $\BTRank^m_{(n-1)/2}(A)$. Finally, for $k \in \{\max(S)+2, \ldots, n\}$, the Summation Axiom is satisfied by $\sigma$. 
\end{proof}

The next lemma combines \cref{lem:first_restriction_BTRank_simplifies_extension_vars,lem:first_restriction_BTRank_self_reduces,lem:second_restriction_BTRank_simplifies_extension_vars,lem:second_restriction_BTRank_self_reduces}.

\begin{lemma} \label{lem:BTRank_size_to_degree_self_reduction}
Suppose that $m > n \geq 16$ are integers, $8$ divides $n$, and $A \in \Q^{m \times m}$ is an $m$ by $m$ matrix over $\F_2$. If $\Pi$ is a $\PCR_{\F_2}$ refutation of $\BTRank^m_n(A)$ of size $s$, then there exists a $\PCR_{\F_2}$ refutation  $\Pi'$ of $\BTRank^m_{(n-8)/8}(A)$ of size at most $s$ such that, letting $d = \frac{16}{\log e}(\log s + 1)$, every term $t$ in $\Pi'$ satisfies the following: 
        \begin{enumerate}
                \item \label{item:left_row_bound_for_terms} the number of indices $i \in [m]$ that are $U$-left-row-mentioned in $t$ is less than $d$, and 
                \item \label{item:column_bound_for_terms} the number of indices $k \in [\frac{n-8}{8}]$ that are $X$-column-mentioned or $Y^T$-column-mentioned or $Z$-column-mentioned in $t$ is less than $d$. 
        \end{enumerate}
\end{lemma}

\begin{proof}
Applying \cref{lem:first_restriction_BTRank_self_reduces,lem:second_restriction_BTRank_self_reduces}, let $\Pi_1 = \Pi \restriction \rho$ be a refutation of $\BTRank^m_{(n-4)/4}(A)$ and $\Pi_2 = \Pi_1 \restriction \sigma$ be a refutation of $\BTRank^m_{(n-8)/8}(A)$. 
A term $t$ violating \cref{item:left_row_bound_for_terms} can only appear in $\Pi_2$ if  $t = (\hat{t} \restriction \rho) \restriction \sigma$ for some term $\hat{t}$ in $\Pi$ such that there exist a set $I \subseteq [m]$ and a subterm $t'$ of $\hat{t}$ with the following properties: $|I| \geq d$, every $i \in I$ is $U$-left-row-mentioned in $t'$, and no $k \in \{1,2,n-1,n\}$ is $U$-column-mentioned in $t'$. (The $u$-variables whose column indices belong to $\{1,2,n-1,n\}$ can be excluded from $t'$ since they are set to a constant by $\rho$ with probability $1$ and hence cannot contribute to a violation of \cref{item:left_row_bound_for_terms} by $t$.) However, for such a term $\hat{t}$ we have $\Pr[\hat{t} \restriction \rho \neq 0] \leq (3/4)^{|I|} \leq (3/4)^{d}$ by \cref{lem:first_restriction_BTRank_simplifies_extension_vars}. 

Similarly, a term $t$ violating \cref{item:column_bound_for_terms} can only appear in $\Pi_2$ if $t = \tilde{t} \restriction \sigma$ for some term $\tilde{t}$ in $\Pi_1$ such that there is a set $K \subseteq [\frac{n-4}{4} - 1]$ with $|K| \geq d$ and every $k \in K$ is either $X$-column-mentioned or $Y^T$-column-mentioned or $Z$-column-mentioned in $\tilde{t}$. (The $x$-, $y$-, and $z$-variables that mention the last column
index $\frac{n-4}{4}$ cannot contribute to a violation of \cref{item:column_bound_for_terms} by $t$ as they are always evaluated to a constant by $\sigma$.) For such a term $\tilde{t}$, 
\cref{lem:second_restriction_BTRank_simplifies_extension_vars} gives $\Pr[\tilde{t} \restriction \sigma \neq 0] \leq \exp(-\card{K}/16) + 2^{-\card{K}/4} \leq \exp(-d/16) + 2^{-d/4}$. 

Denote by $T_r$ the set of all terms $\hat{t}$ in $\Pi$ defined as above. In the event that all of these terms are zeroed out by $\rho$, there are at most $s - |T_r|$ terms $\tilde{t}$ in $\Pi_1$ defined as above. This is because $\Pi_1 = \Pi \restriction \rho$ has no more terms than $\Pi$. 
By the union bound, the probability that there is a term $t$ in $\Pi_2$ that violates \cref{item:left_row_bound_for_terms} or \cref{item:column_bound_for_terms} is bounded above by 
$|T_r| \cdot (3/4)^{d} + (s - |T_r|) \cdot (\exp(-d/16) + 2^{-d/4}) 
\leq s \cdot (\exp(-d/16) + 2^{-d/4}) 
< s \cdot 2^{-\frac{\log e}{16}d+1} 
= 1$ for $d = \frac{16}{\log e}(\log s + 1)$. Thus, we can fix choices of $\rho$ and $\sigma$ such that $\Pi' = (\Pi \restriction \rho) \restriction \sigma$ has the required properties. 
\end{proof}

\begin{proof}[Proof of \autoref{thm:BTRank_A_lower_bound}]
Assume, for contradiction, that there is a $\PCR_{\F_2}$ refutation $\Pi$ of size $s <  2^{\frac{\log e}{512}n - 2}$ witnesses that the theorem fails. We let $\tilde{n} = (n-8)/8$ and construct a $\PCR_{\F_2}$ refutation of $\FPHP^m_{\tilde{n}}$ of degree less than $\tilde{n}/2$, thereby contradicting a theorem of Razborov~\cite{Razb98}. By \cref{lem:BTRank_size_to_degree_self_reduction}, there is a $\PCR_{\F_2}$ refutation $\Pi_1$ of $\BTRank^m_{\tilde{n}}(A)$ such that \cref{item:left_row_bound_for_terms,item:column_bound_for_terms} hold with $d = \frac{16}{\log e}(\log s + 1) <  \frac{n}{32} - \frac{16}{\log e} = \frac{8\tilde{n} + 8}{32} - \frac{16}{\log e} < \frac{\tilde{n}}{4} - 10$. 

Let $\tau$ be the following substitution on the variables of $\BTRank^m_{\tilde{n}}(A)$: $\tau(x_{i,k}) = x_{i,k}$, $\tau(y_{k,j}) = \sum_{i \in [m]} x_{i,k}A_{i,j}$ (that is, $\tau$ substitutes $X^TA$ for $Y$), $\tau(z_{i,j,k}) = \tau(x_{i,k})\tau(y_{k,j})$, and $\tau(u_{i,j,k}) = \sum_{\ell \in [k]} \tau(z_{i,j,\ell})$. Extend $\tau$ to negations by $\tau(\overline{\theta}) = 1+ \tau(\theta)$ for any variable $\theta$ of $\BTRank^m_{\tilde{n}}(A)$. Let $\Pi_2 = \Pi_1 \restriction \tau$. 

Consider a term $t$ occurring in a proof line of $\Pi_1$. We can write $t = t_1t_2$, where $t_1$ consists of all $x$-, $y$- and $z$-variables and their twins that appear in $t$, and $t_2$ consists of all $u$-variables and their twins appearing in $t$. 

\begin{claim} Every term of the polynomial $t_1 \restriction \tau$ either has a degree less than $d$, or is a multiple of a Hole Axiom of $\FPHP^m_{\tilde{n}}$.
\end{claim}

\begin{claimproof}
Let $C$ denote the set of all indices $k \in [\tilde{n}]$ that are $X$-column-mentioned or $Y^T$-column-mentioned or $Z$-column-mentioned in $t_1$. By the choice of $\Pi_1$, we have $|C| < d$. Further, it is immediate from the definition of $\tau$ that the set 
$\{k \in [\tilde{n}] : k \text{ is } X\text{-column-mentioned in some term of the polynomial } t_1 \restriction \tau \}$ is a subset of $C$. Thus, if a term $t'$ in $t_1 \restriction \tau$ has degree at least $d$, then, by the pigeonhole principle, $t'$ must be a multiple of a Hole Axiom of $\FPHP^m_{\tilde{n}}$. 
\end{claimproof}

\begin{claim} Every term of the polynomial $t_2 \restriction \tau$ either has a degree less than $d$, or is a multiple of either a Hole Axiom or a Functionality Axiom of $\FPHP^m_{\tilde{n}}$.
\end{claim}

\begin{claimproof}
Denote by $R$ the set of all indices $i \in [m]$ that are $U$-left-row-mentioned in $t_2$. By the choice of $\Pi_1$, $|R| < d$. Assume that $u^{1-b}_{i, j, k}$ appears in $t_2$ for some $i \in R, j \in [m], k \in [\tilde{n}]$ and $b \in \Q$. Then 
\begin{align*}
\tau(u^{1-b}_{i, j, k}) 
&= b + \sum_{\ell \in [k]} \tau(z_{i,j,\ell}) 
= b + \sum_{\ell \in [k]} \tau(x_{i,\ell}) \tau(y_{\ell,j}) 
= b + \sum_{\ell \in [k]} x_{i,\ell} \sum_{i' \in [m]} x_{i',\ell}A_{i',j} \\
&= b + \sum_{\ell \in [k]} \sum_{i' \in [m]} x_{i,\ell} x_{i',\ell}A_{i',j}. 
\end{align*}
For every $\ell \in [k]$ and every $i' \in [m]$ such that $i' \neq i$ and $A_{i',j} \neq 0$, the summand $x_{i,\ell} x_{i',\ell}A_{i',j}$ in the last expression is a Hole Axiom. Since such summands only give rise to terms in $t_2 \restriction \tau$ that are multiples of a Hole Axiom, we can continue the above calculation modulo Hole Axioms, obtaining
$b + \sum_{\ell \in [k]} x_{i,\ell} x_{i,\ell}A_{i,j} = b + A_{i,j} \sum_{\ell \in [k]}  x_{i,\ell}$.
Thus, modulo Hole Axioms, $t_2 \restriction \tau$ is either $0$ or can be written as a product of degree-$1$ polynomials of the form
\begin{equation*} \label{eqn:product_of_parities_expressing_t_2_tau}
b + \sum_{\ell \in [k]}  x_{i,\ell},
\end{equation*}
where $b \in \Q, k \in [\tilde{n}]$, and $i \in R$. Whenever this product gives rise to a term of degree at least $d$, it follows from $|R| < d$ by the pigeonhole principle that such a term is a multiple of a Functionality Axiom of $\FPHP^m_{\tilde{n}}$. 
\end{claimproof}

Combining the two claims, we see that every term of the polynomial $t \restriction \tau = (t_1 \restriction \tau)(t_2 \restriction \tau)$ either has a degree at most $2d$, or is a multiple of either a Hole Axiom or a Functionality Axiom of $\FPHP^m_{\tilde{n}}$. The same conclusion holds for every term appearing in $\Pi_2 = \Pi_1 \restriction \tau$, since $t$ was an arbitrary term occurring in $\Pi_1$. 

The next step is to transform the sequence of polynomials $\Pi_2$ into a $\PC_{\F_2}$ refutation of $\FPHP^m_{\tilde{n}}$.  

First, we inspect the effect of $\tau$ on the axioms of $\Pi_1$. By definition, $\tau$ satisfies all Complementarity Axioms. Additionally, $\tau$ satisfies the polynomial translation of every clause in the CNF encoding of Binary AND Axioms, Summation Base Axioms, and Summation Axioms. This can be verified by noting that each such clause is a semantic consequence of the algebraic version of the corresponding Binary AND Axiom (i.e., $z_{i,j,k} = x_{i,k}y_{k,j}$ for some $i,j \in [m]$ and $k \in [n]$), Summation Base Axiom (i.e., $u_{i,j,1} = z_{i,j,1}$ for some $i,j \in [m]$), or Summation Axiom (i.e., $u_{i,j,k} = u_{i, j, k-1} + z_{i,j,k}$ for some $i,j \in [m]$ and $k\in \{2,\ldots,n\}$), and all of these are clearly satisfied by $\tau$.

Each Output Axiom in $\Pi_1$ is converted by $\tau$ into one of the axioms algebraically encoding $XX^TA = A$. Each of the latter axioms is a sum of some axioms of $XX^T = I_m$, which is the algebraic oddtown formula $\aOT^m_{\tilde{n}}$. Each of the axioms of $\aOT^m_{\tilde{n}}$ is a sum of some axioms of $\FPHP^m_{\tilde{n}}$. We append the degree-2 derivation just described to each axiom in $\Pi_2$ that is the result of an application of $\tau$ to an Output Axiom in $\Pi_1$. 

Next, we examine the non-axiom proof lines of $\Pi_1$. The addition rule is not affected by $\tau$, since if a polynomial $p$ was inferred in $\Pi_1$ as a sum of $q$ and $r$, then $p \restriction \tau$ is a sum of $q \restriction \tau$ and $r \restriction \tau$. Regarding the multiplication rule, suppose that $p$ and $q$ are proof lines in $\Pi_1$, and let $p$ be inferred from $q$ by multiplying $q$ by $\theta$, where $\theta$ is a variable or the negation of a variable of $\BTRank^m_{\tilde{n}}(A)$. Then $p \restriction \tau = (q \restriction \tau)(\theta \restriction \tau)$, and $\theta \restriction \tau$ is a polynomial of degree at most $2$. We add to $\Pi_2$ a derivation of $p \restriction \tau$ from $q \restriction \tau$ in which each term is a term of $q \restriction \tau$ multiplied by at most two variables.

Let $\Pi_3$ denote the sequence of polynomials resulting from $\Pi_2$ by the modifications introduced in the last two paragraphs. It follows that $\Pi_3$ is a $\PC_{\F_2}$ refutation of $\FPHP^m_{\tilde{n}}$. Moreover, every term appearing in $\Pi_3$ either has a degree at most $2d+2$, or is a multiple of either a Hole Axiom or a Functionality Axiom of $\FPHP^m_{\tilde{n}}$. 

Finally, using Hole Axioms and Functionality Axioms of $\FPHP^m_{\tilde{n}}$, we transform $\Pi_3$ into a $\PC_{\F_2}$ refutation of $\FPHP^m_{\tilde{n}}$ in which each term has a degree at most $2d+3 \leq \frac{\tilde{n}}{2} - 17 < \frac{\tilde{n}}{2}$, which is the desired contradiction. This can be done as follows. Delete from $\Pi_3$ each term that is a multiple of either a Hole Axiom or a Functionality Axiom. Then, for each line $p$ that is an incorrect conclusion of the multiplication rule---which can only happen if $p$ is missing some terms each of degree at most $2d+3$ and having either exactly one row mentioned twice or exactly one column mentioned twice (both of which may occur simultaneously)---insert the correct conclusion $p'$ of the rule before $p$ and use Functionality Axioms and Hole Axioms to derive $p$ from $p'$.
\end{proof}

\subsection{Iterated Generators}\label{sec: iterated generators}

In the previous sections, we proved that the rank generator $g \colon \F_2^{2mn} \to \F_2^{m^2}$ is a hard proof complexity generator against $\PCR_{\F_2}$. However, even when $m$ is arbitrarily large compared to $n$, the stretch of this generator is at most quadratic. In this section, we \emph{iterate} $g$ (in the sense of~\cite{Kra04, Razb15-annals}) to obtain a proof complexity generator of arbitrarily large stretch that is still hard against $\PCR_{\F_2}$.

The iteration process we consider is a weaker notion than $s$-iterability introduced by \Krajicek{}~\cite{Kra04} to boost the stretch of a variant of the Nisan generator~\cite{NisanWigderson94}. Namely, we use a tree-based construction following Razborov~\cite{Razb15-annals}, in which the copies of the rank generator are arranged in a $q$-ary tree. This notion of iterability is sufficient for our applications in \cref{sec: iterability gives ckt lbs}.

At a high level, the iterated process is executed as follows.\footnote{The actual constructions below differ slightly, but the underlying idea is the same.}
Let $\Sigma$ be a finite alphabet of size $q$. We consider the complete $q$-ary tree of depth $\kappa$, whose nodes are indexed by strings $\pi$ of length at most $\kappa$ over the alphabet $\Sigma$. Each node $\pi$ is associated with a copy of the principle
\[
    X^{(\pi)} Y^{(\pi)} = A^{(\pi)},
\]
where $X^{(\pi)}$ and $Y^{(\pi)}$ are the input matrices.

We restrict the entries of $A^{(\pi)}$ to be either constants or variables from the input matrices of copies of $g$ with indices $\pi'$ with $\card{\pi'} > \card{\pi}$.
For the applications in \cref{sec: iterability gives ckt lbs}, we consider a concrete iteration procedure, where each $A^{(\pi)}$ is constant when $\card{\pi} = \kappa$ and is formed from the matrices $X^{(\pi * j)}$ and $Y^{(\pi * j)}$ for $j \in \Sigma$ when $\card{\pi} < \kappa$.

When $\kappa=0$, the iterated formula is simply the original rank principle.
Increasing $\kappa$ increases the stretch of the generator, but also makes the resulting formulas easier to refute. Consequently, proving lower bounds for the iterated formulas is harder than proving them for the original weak rank principle.

%!TEX root = main.tex

\subsubsection{Iterability of \texorpdfstring{$\Rank^m_n(A)$}{Rank}}\label{sec: iterability of Rank}

Here we iterate the \textit{algebraic} version of the rank principle. 
We will use a more limited notion of iterability than $s$-iterability as in~\cite{Kra04,Razb15-annals}, which is sufficient for our purposes. We shall follow the general argument used by Razborov in~\cite{Razb15-annals}.

We denote by $\Sigma^{\le\kappa}$ the set of strings of length at most $\kappa$ over the alphabet $\Sigma$. Similarly, $\Sigma^{<\kappa}$ denotes the strings of length less than $\kappa$ over the alphabet $\Sigma$. We use $()$ to denote the empty string.

% \begin{mdframed}[hidealllines=true,backgroundcolor=gray!10,skipabove=0.3em,skipbelow=-0.4em,innertopmargin=0.4em]
    % \begin{center}\textbf{The Iterated Formula $\IRank^m_n(\bfA,\kappa,\Sigma)$}\end{center}

\begin{definition}[The iterated formula $\IRank^m_n(\bfA,\kappa,\Sigma)$]\label{def: iterated formula}
    Let $m > n > 0$ and $\kappa \ge 0$ be integers and $\Sigma$ a finite alphabet.
    For every string $\pi \in \Sigma^{\le\kappa}$, let $X^{(\pi)} = {(x^{(\pi)}_{i,k})}_{i \in [m], k \in [n]}$ be a variable matrix. Let $Y = {(y_{k,i})}_{k \in [n], i \in [m]}$ be a variable matrix.
    Let $\bfA = \{ A^{(\pi)} : \pi \in \Sigma^{\le\kappa} \}$ be a collection of $m \times m$ matrices, where, for every $\pi \in \Sigma^{\le\kappa}$ and $i,j \in [m]$, the entry $A^{(\pi)}_{i,j}$ is either a Boolean constant or a variable from $X^{(\pi')}$, for some $\pi' \in \Sigma^{\le\kappa}$ with $\card{\pi'} > \card{\pi}$. In particular, all entries of $A^{(\pi)}$, where $\pi \in \Sigma^{\kappa}$, are constant.
    The \emph{iterated rank formula $\IRank^m_n(\bfA, \kappa, \Sigma)$} is the polynomial system $\bigcup_{\pi \in \Sigma^{\le\kappa}} X^{(\pi)} Y = A^{(\pi)}$.
\end{definition}
% \end{mdframed}

Note that the formula $\IRank^m_n(\bfA, 0, \Sigma)$, where $\bfA = \{A^{()}\}$, is precisely the (non-iterated) formula $\Rank^m_n(A^{()})$. Our lower bound will depend on the number of iterations $\kappa$, and for the application, we need $\Sigma$ with $\card{\Sigma} = n^{\Omega(1)}$.

% \Iddo{State (and explain) that this instance is unsatisfiable. I did it:}\slava{This is not true. It is possible that all matrices $A^{(\pi)}$ have small rank but it's just not possible to satisfy all the constraints. It does not matter whether it is satisfiable or not, since we will reduce the l.b. formula to it, which is clearly unsatisfiable.}
% \Iddo{\sout{The iterated rank formula asserts that all matrices $A^{(\pi)}$, each of rank greater than $n$, can be simultaneously expressed as  $X^{(\pi)}Y$ with the same matrix $Y$. This is impossible because the column space of $Y$ has dimension at most~$n$.}}
 
We need the following measure, which describes how many rows are mentioned in a term.

\begin{definition}
    Let $t$ be a term in the variables of $\IRank^m_n(\bfA, \kappa, \Sigma)$.
    We say that $i \in [m]$ is \emph{$X^{(\pi)}$-row-mentioned in $t$} if there is $k \in [n]$ such that $x^{(\pi)}_{i,k}$ or $\overline{x}^{(\pi)}_{i,k}$ appears in $t$.
    We say that $j \in [m]$ is \emph{$Y^T$-row-mentioned in $t$} if there is $k \in [n]$ such that $y_{k,j}$ or $\overline{y}_{k,j}$ appears in $t$.
\end{definition}

\begin{definition}[Row degree]
    Let $t$ be a term in the variables of $\IRank^m_n(\bfA, \kappa, \Sigma)$.
    For every $\pi \in \Sigma^{\le\kappa}$, let $I^{(\pi)}$ be the set of all $i \in [m]$ that are $X^{(\pi)}$-row-mentioned in $t$.
    Similarly, let $J$ be the set of all $j \in [m]$ that are $Y^T$-row-mentioned in $t$.
    In this case we shall denote the mentioned rows as a pair of row-sets,  and say that $t$ mentions rows $(\{I^{(\pi)} : \pi \in \Sigma^{\le\kappa}\},J)$.
    The \emph{row degree} of $t$ is defined as the sum $\sum_{\pi \in \Sigma^{\le\kappa}} \card{I^{(\pi)}} + \card{J}$.
\end{definition}

% \begin{definition}[Row degree]
%     We say that a $\PCR_{\mathbb{F}_2}$ refutation $\Pi$ of the iterated rank formula
%     $\bigcup_{\pi\in\Sigma^{\le\kappa}}(X^{(\pi)}Y=A^{(\pi)})$
%     has \emph{row–column degree}
%     $c+\sum_{\pi\in\Sigma^{\le\kappa}} r^{(\pi)}$
%     if it contains a term that mentions $r^{(\pi)}$ rows of each $X^{(\pi)}$ and $c$ columns of $Y$.
%     The \emph{row–column degree} of $\Pi$ is the maximum row–column degree across all terms  in $\Pi$. We use \emph{rc-degree} to abbreviate row-column degree. 
% \end{definition}

We will construct the restriction that w.h.p.~sets to zero any term of large row degree.
We need the following auxiliary matrices.
\begin{equation*}
    \calD = \begin{pmatrix}
        1 & * & 1 & 0 & 0 & * & * & * \\
        0 & 1 & * & 1 & * & * & * & 0 \\
        * & * & 0 & * & 1 & 0 & 1 & * \\
        * & 0 & * & * & * & 1 & 0 & 1
    \end{pmatrix},
    \quad
    \calE = \begin{pmatrix}
        1 & 0 & 0 & 1 & * & * & * & * \\
        0 & * & 1 & * & 1 & * & 0 & * \\
        * & 1 & * & 0 & * & 0 & * & 1 \\
        * & * & * & * & 0 & 1 & 1 & 0
    \end{pmatrix}.
\end{equation*}
The properties of $\calD$ and $\calE$ that we will use are the following.
\begin{enumerate}
    \item Every pair of rows $\calD_i$ of $\calD$ and $\calE_j$ of $\calE$ satisfies the following:
        \begin{itemize}
            \item there is an odd number of indices $k \in [8]$ such that $\calD_{i,k} = \calE_{j,k} = *$;
            \item there is an even number of indices $k \in [8]$ such that  $\calD_{i,k} = *$ and $\calE_{j,k} = 1$;
            \item there is an even number of indices $k \in [8]$ such that  $\calD_{i,k} = 1$ and $\calE_{j,k} = *$.
        \end{itemize}
    \item Each column of $\calD$ and $\calE$ contains at least one $0$ and at least one $1$.
\end{enumerate}

We will construct a random substitution $\rhorow$ to reduce the row degree, and then follow with a substitution $\rhocol$ to reduce the total degree.

Intuitively, the substitution $\rhorow$ partitions an $m\times n$ matrix into eight blocks. We randomly choose some row type $r_i^{(\pi)}$ defined by $\calD$ for each row $i\in[m]$ of $X^{(\pi)}$. Then we apply this row type to the row $i$: the value of the first entry of the row type $r_i^{(\pi)}$ is assigned to the first block, the value of the second entry to the second block, etc. The assignment for $Y^T$ is similar. This results in matrices that have $0$, $1$ and $*$ entries. The $*$ entries are then replaced with fresh variables with the appropriate indices, forming $m\times \tilde n $ matrices.

\begin{definition}[Substitution $\rhorow$]
    Assume $m > n \ge 16$ and $n$ is divisible by $16$.
    Let $\tilde{n} = n/8$ and then let $\tilde{X}^{(\pi)}={(\tilde{x}^{(\pi)}_{i,k})}_{i \in [m], k \in [\tilde{n}]}$, where $\pi \in \Sigma^{\le\kappa}$, and $\tilde{Y}={(\tilde{y}_{k,i})}_{k \in [\tilde{n}], i \in [m]}$ be fresh variables.
    A random substitution $\rhorow$ is chosen as follows.
    \begin{enumerate}
        \item For each $i \in [m]$ uniformly and independently sample  $s_i \in [4]$ and  $r^{(\pi)}_i \in [4]$, for all $\pi \in \Sigma^{\le\kappa}$.
        \item For each $\pi \in \Sigma^{\le\kappa}$, $i \in [m]$ and $k \in [n]$, let $\nu = \lfloor \frac{k-1}{\tilde{n}} \rfloor + 1$ and set
        \[
            \rhorow(x^{(\pi)}_{i,k}) = \begin{cases}
                \tilde{x}^{(\pi)}_{i, ((k-1) \bmod \tilde{n})+1} & \text{if } {\calD}_{r^{(\pi)}_i,\nu} = *, \\
                {\calD}_{r^{(\pi)}_i,\nu} & \text{if } {\calD}_{r^{(\pi)}_i,\nu} \in \Q.
            \end{cases}
        \]
        \item For each $j \in [m]$ and $k \in [n]$, let $\nu = \lfloor \frac{k-1}{\tilde{n}} \rfloor + 1$ as before and set
        \[
            \rhorow(y_{k,j}) = \begin{cases}
                \tilde{y}_{((k-1) \bmod \tilde{n})+1,j} & \text{if } {\calE}_{s_j,\nu} = *,\\
                {\calE}_{s_j,\nu} & \text{if } {\calE}_{s_j,\nu} \in \Q.
            \end{cases}
        \]
        \item For any variable $\theta$ mapped to $0$ or $1$ by $\rhorow$, set $\rhorow(\overline{\theta}) = 1 - \rhorow(\theta)$.
        \item For any variable $\theta$ not mapped to a Boolean constant by $\rhorow$, set $\rhorow(\overline{\theta}) = \overline{\rhorow(\theta)}$.
    \end{enumerate}
\end{definition}

\begin{lemma}\label{lem: row PCR2 iterated alg Rank eliminates terms whp} % was: clm: iter rc-degree becomes smaller
    Assume $\kappa \ge 0$, $m > n \ge 16$, $n$ divisible by $16$ and $\bfA = \{ A^{(\pi)} : \pi \in \Sigma^{\le\kappa} \}$ is a collection of $m \times m$ matrices as in \cref{def: iterated formula}.
    Let $t$ be a term in the variables of $\IRank^m_n(\bfA, \kappa, \Sigma)$ that mentions rows $(\{I^{(\pi)} : \pi \in \Sigma^{\le\kappa}\},J)$.
    Then $\Prb{t \restriction \rhorow \neq 0} \le {(3/4)}^{\sum_{\pi \in \Sigma^{\le\kappa}}\card{I^{(\pi)}}+\card{J}}$.
\end{lemma}
\begin{proof}
For each $i$ from the set of rows $I^{(\pi)}$ choose a single variable $\gamma^{(\pi)}_i$ in $t$ that mentions row $i$, and for each $j$ from the set $J$ choose a single variable $\delta_j$ in $t$. Let $\Gamma^{(\pi)}_i$ be the event that $(\gamma^{(\pi)}_i \restriction \rhorow \neq 0)$ and $\Delta_j$ the event that $(\delta_j \restriction \rhorow \neq 0)$. Then $\bigcup_{\pi \in \Sigma^{\le\kappa}} \{\Gamma^{(\pi)}_i : i \in I^{(\pi)}\} \cup \{\Delta_j : j \in J\}$ is a set of mutually independent events and each of them happens with probability at most $3/4$.
\end{proof}

The substitution $\rhorow$ transforms any refutation of $\IRank^m_n(\bfA, \kappa, \Sigma)$ into a refutation of $\IRank^m_{\tilde{n}}(\bfA_1, \kappa, \Sigma)$, where $\bfA_1 = \bfA \restriction \rhorow$.
\begin{lemma}\label{lem: row PCR2 iterated alg Rank ref remains ref}
    Assume $\kappa \ge 0$, $m > n \ge 16$, $n$ is divisible by $16$ and $\bfA = \{ A^{(\pi)} : \pi \in \Sigma^{\le\kappa} \}$ is a collection of $m \times m$ matrices as in \cref{def: iterated formula}.
    Let $\bfA_1$ be a collection of matrices defined as $\bfA_1 = \{ A^{(\pi)} \restriction \rhorow : A^{(\pi)} \in \bfA \}$.
    Then for every $\PCR_{\F_2}$ refutation $\Pi$ of $\IRank^m_n(\bfA, \kappa, \Sigma)$, $\Pi \restriction \rhorow$ is a refutation of $\IRank^m_{n/8}(\bfA_1, \kappa, \Sigma)$.
\end{lemma}
\begin{proof}
    The proof is straightforward: each axiom $\sum_{k \in [n]} x^{(\pi)}_{i,k} y_{k,j} = A_{i,j}$ of $\IRank^m_n(\bfA, \kappa, \Sigma)$ transforms into an axiom $\sum_{k \in [n/8]} \tilde{x}^{(\pi)}_{i,k} \tilde{y}_{k,j} = \tilde{A}_{i,j}$ of $\IRank^m_{n/8}(\bfA_1, \kappa, \Sigma)$.
    Consider the axiom
    \[
        \sum_{k \in [n]} x^{(\pi)}_{i,k} y_{k,j} = A^{(\pi)}_{i,j}.
    \]
    By our choice of $\calD$ and $\calE$, we have an odd number of $(*,*)$-blocks, that are replaced with $\sum_{k \in [n/8]} \tilde{x}^{(\pi)}_{i,k} \tilde{y}_{k,j}$, and an even number of $(1,*)$-blocks and $(*,1)$-blocks, which cancel out. There is no shift in the constant term, since each $(1,1)$-block contributes $\sum_{k \in [n/8]} 1 = 0$ due to our choice of $n$. The remaining blocks contribute only zeroes. Thus, the axiom transforms into
    \[
        \sum_{k \in [n/8]} \tilde{x}^{(\pi)}_{i,k} \tilde{y}_{k,j} = \rhorow(A^{(\pi)}_{i,j}).
    \]
\end{proof}

We now construct the substitution $\rhocol$ that reduces the total degree of any term with small row degree. The construction is similar to \cref{def: alg rank substitution}.

\begin{definition}[Substitution $\rhocol$]
    Assume $\kappa \ge 0$, $m > n \ge 2$ and $\bfA = \{ A^{(\pi)} : \pi \in \Sigma^{\le\kappa} \}$ is a collection of $m \times m$ matrices as in \cref{def: iterated formula}.
    A random substitution $\rhocol$ is chosen as follows.
    \begin{enumerate}
        \item For each $k \in [n-1]$, uniformly and independently sample $c_k \in \{*,1,0\}$. Then set $c_n = \card{\{ k \in [n-1]: c_k = 1\}} \bmod 2$. This guarantees that $\card{\{ k \in [n]: c_k = 1\}}$ is even.
        \item Let $\tilde{n} = \card{\{ k \in [n]: c_k = *\}}$ and let $\tilde{X}^{(\pi)} = {(\tilde{x}^{(\pi)}_{i,k})}_{i \in [m], k \in [\tilde{n}]}$, where $\pi \in \Sigma^{\le\kappa}$, and $\tilde{Y} = {(\tilde{y}_{k,j})}_{k \in [\tilde{n}], j \in [m]}$ be fresh variables.
        \item Let $\iota$ be a bijection between $\{ k \in [n]: c_k = * \}$ and $[\tilde{n}]$.
        \item For each $\pi \in \Sigma^{\le\kappa}$, $i \in [m]$ and $k \in [n]$ set
        \[
            \rhocol(x^{(\pi)}_{i,k}) = \begin{cases}
                \tilde{x}^{(\pi)}_{i, \iota(k)} & \text{if } c_k = *, \\
                c_k & \text{if } c_k \in \Q.
            \end{cases}
        \]
        \item For each $j \in [m]$ and $k \in [n]$ set
        \[
            \rhocol(y_{k,j}) = \begin{cases}
                \tilde{y}_{\iota(k),j} & \text{if } c_k = *, \\
                c_k & \text{if } c_k \in \Q.
            \end{cases}
        \]
        \item For any variable $\theta$ mapped to $0$ or $1$ by $\rhocol$, set $\rhocol(\overline{\theta}) = 1 - \rhocol(\theta)$.
        \item For any variable $\theta$ not mapped to a Boolean constant by $\rhocol$, set $\rhocol(\overline{\theta}) = \overline{\rhocol(\theta)}$.
    \end{enumerate}
\end{definition}

$\rhocol$ behaves similarly to the substitution from \cref{def: alg rank substitution}. We have the following analogue of \cref{lem: PCR2 alg Rank eliminates terms whp}.
\begin{lemma}\label{lem: column PCR2 iterated alg Rank eliminates terms whp}
    Assume $\kappa \ge 0$, $m > n \ge 2$ and $\bfA = \{ A^{(\pi)} : \pi \in \Sigma^{\le\kappa} \}$ is a collection of $m \times m$ matrices as in \cref{def: iterated formula}.
    Let $t$ be a term in the variables of $\IRank^m_n(\bfA, \kappa, \Sigma)$ of degree $d$ that does not mention the last column and mentions rows $(\{I^{(\pi)} : \pi \in \Sigma^{\le\kappa}\},J)$.
    Then $\Prb{t \restriction \rhocol \neq 0} \le {(2/3)}^{\frac{d}{\left( \sum_{\pi \in \Sigma^{\le\kappa}} \card{I^{(\pi)}} + \card{J} \right)}}$.
    Furthermore, $\Prb{\tilde{n} < (n-1)/6)} \le e^{-\frac{n-1}{24}}$.
\end{lemma}
%\hanlin{Minor question: Should we subtract $1$ in the exponent, so RHS becomes ${(2/3)}^{\frac{d}{\left( \sum_{\pi \in \Sigma^{\le\kappa}} \card{I^{(\pi)}} + \card{J} \right)}{\color{red}-1}}$? The $n$-th column is not independent from other columns...} \slava{It is always substituted with constants, so we should not consider such variables when defining the indices that are mentioned. At lease that what we do in other places. I've clarified the wording of the lemma by stating that we do not count include such variables.}
\begin{proof}
    Let $K^{(\pi)}$ be the set of all $k \in [n]$ such that for some $i \in [m]$, either $x^{(\pi)}_{i,k}$ or $\overline{x}^{(\pi)}_{i,k}$ appears in $t$.
    Let $L$ be the set of all $\ell \in [n]$ such that for some $j \in [m]$, either $y_{\ell,j}$ or $\overline{y}_{\ell,j}$ appears in $t$.
    
    The degree of $t$ is then upper-bounded by
    \[
        \left( \sum_{\pi \in \Sigma^{\le\kappa}} \card{I^{(\pi)}} \cdot \card{K^{(\pi)}} + \card{J} \cdot \card{L} \right) \le \left( \sum_{\pi \in \Sigma^{\le\kappa}} \card{I^{(\pi)}} + \card{J} \right) \cdot \max\left(\{\card{K^{(\pi)}} : \pi \in \Sigma^{\le\kappa}\} \cup \{\card{L}\}\right).
    \]
    By the pigeonhole principle, one of the sets $\{K^{(\pi)} : \pi \in \Sigma^{\le\kappa}\} \cup \{L\}$ must have size at least $d' = \frac{d}{\left(\sum_{\pi \in \Sigma^{\le\kappa}} \card{I^{(\pi)}} + \card{J} \right)}$.
    Let $S$ be such a set and $\{ \gamma_s : s \in S \}$ some variables in $t$ corresponding to $S$. For every $s \in S$, let $\Gamma_s$ be the event that $(\gamma_s \restriction \rhocol \neq 0)$. Then $\{\Gamma_s : s \in S\}$ is a set of at least $d'$ mutually independent events, each of which happens with probability at most $2/3$.

    The proof of the latter statement is the same as the proof of \cref{lem: PCR2 alg Rank eliminates terms whp}.
\end{proof}

The substitution $\rhocol$ transforms any refutation of $\IRank^m_n(\bfA, \kappa, \Sigma)$ into a refutation of $\IRank^m_{\tilde{n}}(\bfA_2, \kappa, \Sigma)$, where $\bfA_2 = \bfA \restriction \rhocol$.
\begin{lemma}\label{lem: column PCR2 iterated alg Rank ref remains ref}
    Assume $\kappa \ge 0$, $m > n \ge 2$ and $\bfA = \{ A^{(\pi)} : \pi \in \Sigma^{\le\kappa} \}$ is a collection of $m \times m$ matrices as in \cref{def: iterated formula}.
    Let $\bfA_2$ be a collection of matrices defined as $\bfA_2 = \{ A^{(\pi)} \restriction \rhocol : \pi \in \Sigma^{\le\kappa} \}$.
    Then for every $\PCR_{\F_2}$ refutation $\Pi$ of $\IRank^m_n(\bfA, \kappa, \Sigma)$, $\Pi \restriction \rhocol$ is a refutation of $\IRank^m_{\tilde{n}}(\bfA_2, \kappa, \Sigma)$.
\end{lemma}
\begin{proof}
    The proof is the same as the proof of \cref{lem: PCR2 alg Rank ref remains ref}.
\end{proof}

We establish a lower bound on the degree of $\PCR_{\F_2}$ refutations of $\IRank^m_n(\bfA, \kappa, \Sigma)$ by constructing a reduction to a single copy of $\Rank^m_n(I_m)$.

\begin{definition}[Reduction $\tau$]
    Assume $\kappa \ge 0$, $m > n > 0$.
    Let $X = {(x_{i,k})}_{i \in [m], k \in [n]}$ be fresh variables.
    A substitution $\tau$ is defined inductively on $\card{\pi}$ in decreasing order.
    \begin{enumerate}
        \item For each $i \in [m]$ and $k \in [n]$ set
        \[
            \begin{aligned}
                \tau(x^{(\pi)}_{i,k}) &= ((A^{(\pi)} \restriction \tau) X)_{i,k} = \sum_{s \in [m]} \tau(A^{(\pi)}_{i,s}) x_{s,k}, \\
                \tau(y_{k,i}) &= y_{k,i}.
            \end{aligned}
        \]
        \item For any variable $\theta$, set $\tau(\overline{\theta}) = 1 - \tau(\theta)$.\footnote{The proof size may increase after applying this substitution; the degree increase is controlled in \cref{lem: PCR2 iterated degree reduction to one copy}. The resulting refutation will still be a $\PC$ refutation.}
    \end{enumerate}
    Note that $A^{(\pi)} \restriction \tau = A^{(\pi)}$ for every $\pi \in \Sigma^{\kappa}$ since the entries of $A^{(\pi)}$ can only be constant.
\end{definition}

\begin{lemma}\label{lem: PCR2 iterated degree reduction to one copy}
    Assume $\kappa \ge 0$, $m > n > 0$ and $\bfA = \{ A^{(\pi)} : \pi \in \Sigma^{\le\kappa} \}$ is a collection of $m \times m$ matrices as in \cref{def: iterated formula}.
    Then for any $\PCR_{\F_2}$ refutation $\Pi$ of $\IRank^m_n(\bfA, \kappa, \Sigma)$ with degree $d$, $\Pi \restriction \tau$ can be transformed into a refutation of $\Rank^m_n(I_m)$ with degree at most $d(\kappa+2)$.
\end{lemma}
\begin{proof}
    Let $\Pi$ be a $\PCR_{\F_2}$ refutation of $\IRank^m_n(\bfA, \kappa, \Sigma)$.

    It follows by induction on $\card{\pi}$, that for every $\pi \in \Sigma^{\le\kappa}$, $i \in [m]$, and $k \in [n]$, $\tau(x^{(\pi)}_{i,k})$ (and thus $\tau(\overline{x^{(\pi)}_{i,k}})$) is a polynomial of degree at most $\kappa-\card{\pi}+1$. In particular, the entries of $A^{(\pi)} \restriction \tau$ are polynomials of degree at most $\kappa - \card{\pi}$.

    It implies that all polynomials from $\IRank^m_n(\bfA, \kappa, \Sigma) \restriction \tau$ can be easily derived from the axioms of $\Rank^m_n(I_m)$.
    \begin{claim}
        Every polynomial from $\IRank^m_n(\bfA, \kappa, \Sigma) \restriction \tau$ has a degree-$(\kappa+2)$ proof from $\Rank^m_n(I_m)$.
    \end{claim}
    \begin{claimproof}
        Fix $\pi \in \Sigma^{\le\kappa}$.
        As shown above, all the entries of $A^{(\pi)} \restriction \tau$ are polynomials of degree at most $\kappa - \card{\pi}$.

        Let $i, j \in [m]$. The $(i,j)$th axiom of $\IRank^m_n(\bfA, \kappa, \Sigma)$ under $\tau$ transforms as follows
        \begin{align*}
            \left( \sum_{k \in [n]} x^{(\pi)}_{i,k} y_{k,j} + A^{(\pi)}_{i,j} \right) \restriction \tau &=
            \sum_{k \in [n]} \left( \sum_{s \in [m]} \tau(A^{(\pi)}_{i,s}) x_{s,k} \right) y_{k,j} + \tau(A^{(\pi)}_{i,j}) \\
            &= \sum_{s \in [m]} \tau(A^{(\pi)}_{i,s}) \left( \sum_{k \in [n]} x_{s,k} y_{k,j} \right) + \tau(A^{(\pi)}_{i,j}) \\
            &= \sum_{s \in [m] \setminus \{j\}} \tau(A^{(\pi)}_{i,s}) \left( \sum_{k \in [n]} x_{s,k} y_{k,j} \right) +
            \tau(A^{(\pi)}_{i,j}) \left( \sum_{k \in [n]} x_{j,k}y_{k,j} + 1 \right),
        \end{align*}
        which is trivially derivable in degree at most $\kappa-\card{\pi}+2$ from $\Rank^m_n(I_m)$.
    \end{claimproof}

    Thus, we can extend $\Pi \restriction \tau$ to a refutation of $\Rank^m_n(I_m)$ of degree at most $d(\kappa+2)$.
\end{proof}

We now have all the components for proving the size lower bound on the iterated formula $\IRank^m_n(\bfA, \kappa, \Sigma)$.

\begin{theorem}\label{thm: lower bound for iterated algebraic irank}
    Assume $\kappa>0$, $m > n \ge 16+192(\kappa+2)$, $(n-16)$ is divisible by $192(\kappa+2)$ and $\bfA = \{ A^{(\pi)} : \pi \in \Sigma^{\le\kappa} \}$ is a collection of $m \times m$ matrices as in \cref{def: iterated formula}.
    Then any $\PCR_{\F_2}$ refutation of $\IRank^m_n(\bfA, \kappa, \Sigma)$ requires size $2^{\Omega(\sqrt{n/\kappa})}$.
\end{theorem}
\begin{proof}

% \hanlin{If I'm correct, the properties of $P$ and $Q$ that we need are follows: (1) For each row $r^P$ of $P$ and each row $r^Q$ of $Q$, there are an odd number of indices $i\in [8]$ such that $r^P_i = r^Q_i = *$ (this preserves the rank principle); (2) each row/column of $P$ or $Q$ contains a $1/4$ fraction of zeros and a $1/4$ fraction of ones (this kills PCR terms with large rc/cr-degree).}
% \slava{We also need the number of $\{r^P_i, r^Q_i\} = \{*, 1\}$ to be even, so that they cancel out when we multiply $r^P$ and $r^Q$. The $1/4$ fraction is not important; any constant fraction would suffice.}\Iddo{It would be good to add these observations to the text to help the reader.}

Let
\begin{align*}
    d_1 &= \frac{n-16}{192(\kappa+2)}, \\
    d_2 &= \sqrt{d_1} = \sqrt{\frac{n-16}{192(\kappa+2)}}.
\end{align*}

Assume, for the sake of contradiction, that $\IRank^m_n(\bfA, \kappa, \Sigma)$ has a refutation $\Pi$ of size less than ${(4/3)}^{d_2} = 2^{\Theta(\sqrt{n/\kappa})}$.

By the union bound and \cref{lem: row PCR2 iterated alg Rank eliminates terms whp,lem: row PCR2 iterated alg Rank ref remains ref}, there exists a substitution $\rhorow$ such that $\Pi\restriction \rhorow$ has \emph{row degree} less than $d_2$, and it is a refutation of $\IRank^m_{n/8}(\bfA \restriction \rhorow, \kappa, \Sigma)$. Further, extend $\rhorow$ to $\rhorow'$ by also fixing the last $(n/8)$th column in each $X^{(\pi)}$ and $Y^T$ to an all-zero column, thus obtaining $\IRank^m_{n/8-1}(\bfA \restriction \rhorow', \kappa, \Sigma)$.\footnote{This step is needed because we require $16 \mid n$ in order to apply $\rhorow$.}

Since after applying $\rhorow'$ every surviving term has row degree less than $d_2$, \cref{lem: column PCR2 iterated alg Rank eliminates terms whp} implies that any term of degree at least $2d_1$ survives $\rhocol$ with probability at most $(2/3)^{2d_1/d_2}=(2/3)^{2d_2}$. Hence, by the union bound, there is a choice of $\rhocol$ for which all terms of degree at least $d_1$ vanish and $\tilde{n} \ge \frac{n/8-2}{6} = \frac{n-16}{48}$. By \cref{lem: column PCR2 iterated alg Rank ref remains ref}, $\Pi \restriction \rhorow' \restriction \rhocol$ is a refutation of $\IRank^m_{\tilde{n}}(\bfA \restriction \rhorow' \restriction \rhocol, \kappa, \Sigma)$.

By \cref{lem: PCR2 iterated degree reduction to one copy}, we can extend $\Pi \restriction \rhorow' \restriction \rhocol \restriction \tau$ to a refutation of $\Rank^m_{\tilde{n}}(I_m)$ with degree less than $2d_1(\kappa+2) \le \tilde{n}/2$, which contradicts \cref{lem: Rank reduction to PHP}.
\end{proof}

%!TEX root = main.tex

\subsubsection{Iterability of \texorpdfstring{$\BTRank^m_n(A)$}{BTRank}}\label{sec: iterability of BTRank}

We start by defining all the formulas and constructions used in this section, followed by a high-level overview of the proof.

\begin{notation*}
    Given two sequences $(a_1, \ldots, a_n)$ and $(b_1, \ldots, b_m)$, we write $a \cup b$ to denote their concatenation $(a_1, \ldots, a_n, b_1, \ldots, b_m)$.
\end{notation*}

In this section, we make use of several specific variants of the rank principle in which the structure of the matrix $Y$ is determined by a bipartite graph. When this graph has small left-degree (in particular, when it is an expander), each axiom $\sum_k x_{i,k} y_{k,j}$ only depends on a few variables and therefore its CNF encoding does not require too many extension variables.
Unlike \cref{sec: iterability of Rank}, the iteration structure is over the alphabet $\Sigma = \Q$.
We will use the following formulas.

% \begin{mdframed}[hidealllines=true,backgroundcolor=gray!10,skipabove=0.3em,skipbelow=-0.4em,innertopmargin=0.4em]
    % \begin{center}\textbf{The iterated formula $\psi_G(\bfA, \kappa)$}\end{center}
\begin{definition}[The iterated formula $\psi_G(\bfA, \kappa)$]\label{def: iterated cnf psi}
    Let $m, n > 0$ and $\kappa \ge 0$ be integers and $G$ an $m \times n$ bipartite graph with parts $[m]$ and $[n]$. For convenience assume that $1 \in \calN_G(i)$ for every $i \in [m]$.
    We define the formula $\psi_G(\bfA, \kappa)$ that encodes the statement of the form $\bigcup_{\pi \in \Q^{\le\kappa}} X^{(\pi)} Y^{(\pi)} = A^{(\pi)}$, using extension variables that form a binary tree to compute partial parities. The structure of each matrix $Y^{(\pi)}$ is based on $G$.

    The formula uses the following variables:
    \begin{align*}
        \bigcup_{\pi \in \Q^{\le\kappa}} \Biggl[& \begin{aligned}[t]
            \bigcup_{\mu \in [4]} \biggl(& \{ x^{(\pi,\mu)}_{i,k} : i \in [m], k \in [n] \} \\
            & \cup \{ y^{(\pi,\mu)}_{k,j} : j \in [m], k \in \calN_G(j) \} \\
            & \cup \{ z^{(\pi,\mu)}_{i,j,k} : i,j \in [m], k \in \calN_G(j) \} \\
            & \cup \{ u^{(\pi,\mu)}_{i,j,\Sigma} : i,j \in [m], \Sigma \text{ is a non-empty initial segment of } \calN_G(j) \} \biggr)
        \end{aligned} \\
        & \cup \{ w^{(\pi)}_{i,j,\ell} : i,j \in [m], \ell \in \{0,1,2\} \}\Biggr].
    \end{align*}

    Let $\bfA = \{ A^{(\pi)} : \pi \in \Q^{\le\kappa} \}$ be a collection of $m \times m$ matrices, where, for every $\pi \in \Q^{\le\kappa}$ and $i,j \in [m]$, the entry $A^{(\pi)}_{i,j}$ is either a Boolean constant or a variable of the form $x^{(\pi',\mu')}_{i',k'}$ or $y^{(\pi',\mu')}_{k',j'}$ for some $i',j' \in [m]$, $k' \in [n]$, $\mu' \in [4]$ and $\pi' \in \Q^{\le\kappa}$ with $\card{\pi'} > \card{\pi}$. In particular, all entries of $A^{(\pi)}$ are constant for $\pi \in \Q^{\kappa}$.

    The \emph{iterated formula $\psi_G(\bfA, \kappa)$} is a polynomial system consisting of the following axioms.

    \textbf{Binary AND Axiom:} For every $\pi \in \Q^{\le\kappa}$, $\mu \in [4]$, $i,j \in [m]$, $k \in \calN_G(j)$, the polynomial $z^{(\pi,\mu)}_{i,j,k} = x^{(\pi,\mu)}_{i,k} y^{(\pi,\mu)}_{k,j}$.

    \textbf{Summation Base Axiom:} For every $\pi \in \Q^{\le\kappa}$, $\mu \in [4]$, $i,j \in [m]$, the polynomial $u^{(\pi,\mu)}_{i,j,(1)} = z^{(\pi,\mu)}_{i,j,1}$.

    \textbf{Summation Axiom:} For every $\pi \in \Q^{\le\kappa}$, $\mu \in [4]$, $i,j \in [m]$ and every initial segment $\Sigma \cup (k)$ of $\calN_G(j)$ with $\Sigma \neq \varnothing$, the polynomial
    \[
        u^{(\pi,\mu)}_{i,j,\Sigma \cup (k)} = u^{(\pi,\mu)}_{i,j,\Sigma} + z^{(\pi,\mu)}_{i,j,k}.
    \]

    Graphically, the last two types of axioms can be represented as follows. Suppose that $\calN_G(j) = (k_1, \ldots, k_\Delta)$. For every $\ell \in [\Delta]$ define $\Sigma_\ell = (k_1, \ldots, k_\ell)$. In particular, for each $\ell \in \{2, \ldots, \Delta\}$, we have $\Sigma_{\ell} = \Sigma_{\ell-1} \cup (k_\ell)$. Then the $u$- and $z$-variables form a bamboo tree as shown in \cref{fig:iter rank bamboo part}.
    \begin{figure}[H]\centering
        % LTeX: enabled=false
        \begin{forest}
            for tree={grow=south,l sep=0em,s sep=6em,inner sep=1mm}
            [$u^{(\pi,\mu)}_{i,j,\Sigma_\Delta}$
                [$u^{(\pi,\mu)}_{i,j,\Sigma_{\Delta-1}}$
                    [$\cdots$
                        [$u^{(\pi,\mu)}_{i,j,\Sigma_2}$
                            [$u^{(\pi,\mu)}_{i,j,\Sigma_1}$
                                [$z^{(\pi,\mu)}_{i,j,k_1}$, no edge, edge label={node[midway]{\rotatebox{90}{$=$}}}]
                            ]
                            [$z^{(\pi,\mu)}_{i,j,k_2}$]]
                        [, no edge]]
                    [$z^{(\pi,\mu)}_{i,j,k_{\Delta-1}}$]]
                [$z^{(\pi,\mu)}_{i,j,k_\Delta}$]]
        \end{forest}
        % LTeX: enabled=true
        \caption{The bamboo-tree structure of the $u$- and $z$-variables.}\label{fig:iter rank bamboo part}
    \end{figure}
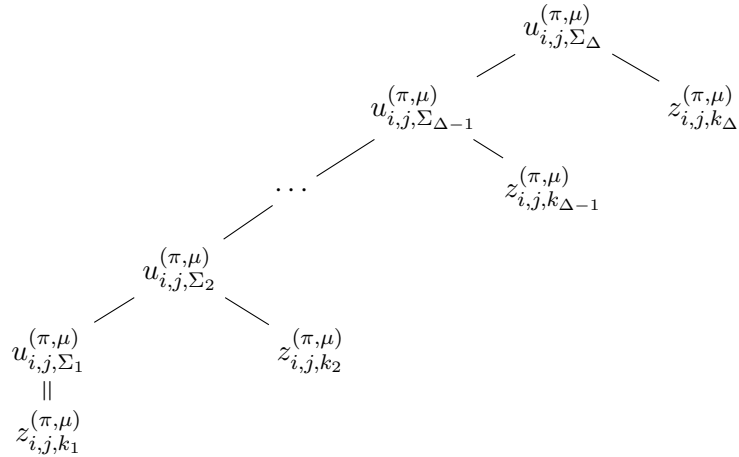

    \textbf{Output Axioms:} For every $\pi \in \Q^{\le\kappa}$, $i,j \in [m]$, the following four axioms:
    \begin{align*}
        & w^{(\pi)}_{i,j,1} = u^{(\pi,1)}_{i,j,\calN_G(j)} + u^{(\pi,2)}_{i,j,\calN_G(j)}, \\
        & w^{(\pi)}_{i,j,2} = u^{(\pi,3)}_{i,j,\calN_G(j)} + u^{(\pi,4)}_{i,j,\calN_G(j)}, \\
        & w^{(\pi)}_{i,j,0} = w^{(\pi)}_{i,j,1} + w^{(\pi)}_{i,j,2}, \\
        & w^{(\pi)}_{i,j,0} = A^{(\pi)}_{i,j}.
    \end{align*}
    Graphically, these axioms can be represented as shown in \cref{fig:iter rank balanced part}.
    \begin{figure}[H]\centering
        % LTeX: enabled=false
        \begin{forest}
            for tree={grow=south,inner sep=1mm} % calign=fixed edge angles
            [$A^{(\pi)}_{i,j}$
                [$w^{(\pi)}_{i,j,0}$, no edge, edge label={node[midway]{\rotatebox{90}{$=$}}}
                    [$w^{(\pi)}_{i,j,1}$
                        [$u^{(\pi,1)}_{i,j,\calN_G(j)}$]
                        [$u^{(\pi,2)}_{i,j,\calN_G(j)}$]]
                    [$w^{(\pi)}_{i,j,2}$
                        [$u^{(\pi,3)}_{i,j,\calN_G(j)}$]
                        [$u^{(\pi,4)}_{i,j,\calN_G(j)}$]]]]
        \end{forest}
        % LTeX: enabled=true
        \caption{The balanced tree structure of the $w$-variables.}\label{fig:iter rank balanced part}
    \end{figure}
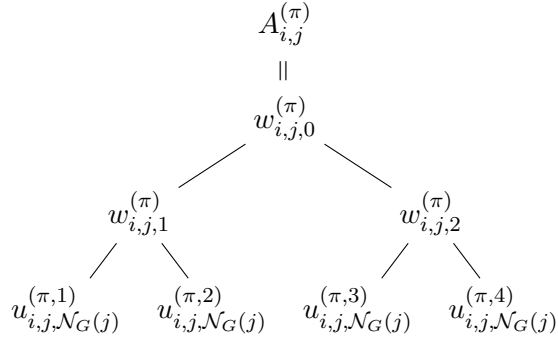

    Although $\psi_G(\bfA, \kappa)$ is defined as a polynomial system, it can be easily encoded as a CNF since each axiom of $\psi_G(\bfA, \kappa)$ depends only on a constant number of variables.
\end{definition}
% \end{mdframed}

The following formula $\phi_G(\bfA, \kappa)$ has a simpler structure, and in particular does not involve the parameter $\mu$ or the $w$-variables. It can be viewed as a substitution instance of $\psi_G(\bfA, \kappa)$.

\begin{definition}[Formula $\phi_G(\bfA, \kappa)$]\label{def: iterated cnf phi}
    Let $m, n > 0$ and $\kappa \ge 0$ be integers and let $G$ be an $m \times n$ bipartite graph.
    Consider the following variables:
    \begin{align*}
        \bigcup_{\pi \in \Q^{\le\kappa}} \Biggl[& \{ x^{(\pi)}_{i,k} : i \in [m], k \in [n] \} \\
        & \cup \{ y^{(\pi)}_{k,j} : j \in [m], k \in \calN_G(j) \} \\
        & \cup \{ z^{(\pi)}_{i,j,k} : i,j \in [m], k \in \calN_G(j) \} \\
        & \cup \{ u^{(\pi)}_{i,j,\Sigma} : i,j \in [m], \Sigma \text{ is a non-empty initial segment of } \calN_G(j) \} \Biggr].
    \end{align*}

    Let $\bfA = \{ A^{(\pi)} : \pi \in \Q^{\le\kappa} \}$ be a collection of $m \times m$ matrices, where, for every $\pi \in \Q^{\le\kappa}$ and $i,j \in [m]$, the entry $A^{(\pi)}_{i,j}$ is either a Boolean constant or a variable of the form $x^{(\pi')}_{i',k'}$ or $y^{(\pi')}_{k',j'}$ for some $i',j' \in [m]$, $k' \in [n]$ and $\pi' \in \Q^{\le\kappa}$ with $\card{\pi'} > \card{\pi}$.

    % $\phi_G(\bfA, \kappa)$ uses the variables $\{ x^{(\pi)}_{i,k} : i \in [m], k \in [n] \}$, $\{ y^{(\pi)}_{k,j} : j \in [m], k \in \calN_{G}(j) \}$, $\{ z^{(\pi)}_{i,j,k} : i, j \in [m], k \in \calN_{G}(j) \}$ and $\{ u^{(\pi)}_{i,j,\Sigma} : i,j \in [m], \Sigma \text{ is a non-empty initial segment of } \calN_{G}(j) \}$.
    % It consists of the following polynomial axioms.

    \textbf{Binary AND Axiom:} For every $\pi \in \Q^{\le\kappa}$, $i,j \in [m]$, $k \in \calN_{G}(j)$, the polynomial $z^{(\pi)}_{i,j,k} = x^{(\pi)}_{i,k} y^{(\pi)}_{k,j}$.

    \textbf{Summation Base Axiom:} For every $\pi \in \Q^{\le\kappa}$, $i,j \in [m]$, the polynomial $u^{(\pi)}_{i,j,(1)} = z^{(\pi)}_{i,j,1}$.

    \textbf{Summation Axiom:} For every $\pi \in \Q^{\le\kappa}$, $i,j \in [m]$ and every non-empty initial segment $\Sigma \cup (k)$ of $\calN_{G}(j)$, the polynomial $u^{(\pi)}_{i,j,\Sigma \cup (k)} = u^{(\pi)}_{i,j,\Sigma} + z^{(\pi)}_{i,j,k}$.

    \textbf{Output Axiom:} For every $\pi \in \Q^{\le\kappa}$, $i,j \in [m]$, the polynomial $u^{(\pi)}_{i,j,\calN_{G}(j)} = A^{(\pi)}_{i,j}$.
\end{definition}

We also consider an even simpler formula $\widetilde{\phi}_G(\bfA, \kappa)$, which is the algebraic analogue of $\phi_G(\bfA, \kappa)$. %That is, $\widetilde{\phi}_G(\bfA, \kappa)$ is the polynomial system
\begin{definition}[Formula $\widetilde{\phi}_G(\bfA, \kappa)$]\label{def: iterated algebraic phi tilde}
    Let $m, n > 0$ and $\kappa \ge 0$ be integers and let $G$ be an $m \times n$ bipartite graph.
    Consider the following variables:
    \[
        \bigcup_{\pi \in \Q^{\le\kappa}} \Biggl[ \{ x^{(\pi)}_{i,k} : i \in [m], k \in [n] \}
        \cup \{ y^{(\pi)}_{k,j} : j \in [m], k \in \calN_G(j) \} \Biggr].
    \]

    Let $\bfA$ be defined as in \cref{def: iterated cnf phi}.
    We define $\widetilde{\phi}_G(\bfA, \kappa)$ as the polynomial system
    \[
        \bigcup_{\pi \in \Q^{\le\kappa}} \bigcup_{i,j \in [m]} \left\{ \sum_{k \in \calN_G(j)} x^{(\pi)}_{i,k} y^{(\pi)}_{k,j} = A^{(\pi)}_{i,j} \right\}.
    \]
\end{definition}

In what follows we use the graph gadgets $P(G)$ and $H(G)$.

\begin{definition}[Graph $P(G)$]
    Let $m,n>0$ and let $G$ be an $m \times n$ bipartite graph.
    Given an adjacency matrix $M_G$ of $G$, define $P(G)$ as the $m \times (n+2)$ bipartite graph with the adjacency matrix
    \[
        M_{P(G)} = \begin{bmatrix}
            \begin{matrix}
                1 & 1 \\
                \vdots & \vdots \\
                1 & 1
            \end{matrix} &
            M_G
        \end{bmatrix}.
    \]
\end{definition}

\begin{definition}[Graph $H(G)$]
    Let $m,n>0$ and let $G$ be an $m \times n$ bipartite graph.
    Given an adjacency matrix $M_G$ of $G$, define $H(G)$ as the $m \times (8n+8)$ bipartite graph with the adjacency matrix
    \[
        M_{H(G)} = \underbrace{\begin{bmatrix}
            \begin{array}{*{16}c}
                M_G & \begin{matrix} 1 \\ \vdots \\ 1 \end{matrix} &
                M_G & \begin{matrix} 1 \\ \vdots \\ 1 \end{matrix} &
                M_G & \begin{matrix} 1 \\ \vdots \\ 1 \end{matrix} &
                M_G & \begin{matrix} 1 \\ \vdots \\ 1 \end{matrix} &
                M_G & \begin{matrix} 1 \\ \vdots \\ 1 \end{matrix} &
                M_G & \begin{matrix} 1 \\ \vdots \\ 1 \end{matrix} &
                M_G & \begin{matrix} 1 \\ \vdots \\ 1 \end{matrix} &
                M_G & \begin{matrix} 1 \\ \vdots \\ 1 \end{matrix}
            \end{array}
        \end{bmatrix}}_{8}.
    \]
\end{definition}

Finally, we will also need the following lower bound on the $\PC$ degree of $\FPHP(G)$.
\begin{proposition}[\cite{MiksaN24}]\label{prop: fphp degree bound}
    Let $G$ be an $(r,c)$-boundary expander with left-degree at most $\Deltamax$. Then any $\PCR_\F$ refutation of $\FPHP(G)$ requires degree strictly larger than $\frac{cr}{2\Deltamax}$.
\end{proposition}

The main result of this subsection is a lower bound on the size of $\PCR_{\F_2}$ refutations of $\psi_{P(H(G))}(\bfA, \kappa)$.

\begin{theorem}\label{thm: lower bound on iterated formula psi}
    Let $n>0$, $N=8n+8$, $m=16(N+2)$, $\kappa \le n^{\alpha}$ for some $\alpha < 1$, and let $G$ be an $m \times n$ $(n^\delta,d)$-lossless expander with parts $[m]$ and $[n]$, the minimum left-degree at least $\Deltamin = 50 \log n$, the maximum left-degree at most $\Deltamax = 150 \log n$, and $\delta > 4\alpha$.
    Let $\bfA = \{ A^{(\pi)} : \pi \in \Sigma^{\le\kappa} \}$ be a collection of $m \times m$ matrices as in \cref{def: iterated cnf psi}.
    Then any $\PCR_{\F_2}$ refutation of $\psi_{P(H(G))}(\bfA, \kappa)$ requires size $2^{\Omega({(n^\delta/\kappa)}^{1/3})}$.
\end{theorem}

The proof of this theorem is executed as follows.
\begin{enumerate}
    \item Assume, for the sake of contradiction, that $\psi_{P(H(G))}(\bfA, \kappa)$ has a refutation $\Pi$ of size $2^{\epsilon {(n^\delta/\kappa)}^{1/3}}$. Let $d_1$ be a certain threshold parameter with $d_1  = \Theta(n^\delta/\kappa)$.
    \item We construct a random substitution $\rho_1$ that turns $\Pi$ into a refutation $\Pi_1$ of $\phi_{H(G)}(\bfA_1, \kappa)$, where $\bfA_1 = \bfA \restriction \rho_1$. $\rho_1$ eliminates all terms $t$ from $\Pi$ that either have high degree in the $u$-variables or have high degree (exceeding $d_1$) in the \emph{positive} occurrences of the $z$-variables. Moreover, it substitutes all the $w$-variables with either constants or $x$- or $y$-variables.
    \item We proceed by constructing another random substitution $\rho_2$ that further eliminates all terms in $\Pi_1$ that mention many columns of $G$ (specifically, exceeding $d_1^{1/3}$) using $x$-, $y$- or $z$-variables. The resulting refutation $\Pi_2$ is a refutation of $\phi_{H(G')}(\bfA_2, \kappa)$, where $\bfA_2 = \bfA_1 \restriction \rho_2$ and $G'$ is still a sufficiently good expander.
    \item Then we construct a substitution $\rho_3$ that eliminates all terms of high degree (again, exceeding $d_1$) in $x$- and $y$-variables and negative occurrences in $z$-variables. The resulting refutation $\Pi_3$ is a refutation of $\widetilde{\phi}_{G'}(\bfA_3, \kappa)$, where $\bfA_3 = \bfA_2 \restriction \rho_3$.
    \item Finally, we reduce $\FPHP(G')$ to $\widetilde{\phi}_{G'}(\bfA_3, \kappa)$, while increasing the degree of a refutation by at most $\Theta(\kappa)$. Combining these degree bounds, we obtain a refutation $\Pi_4$ of $\FPHP(G')$ of degree $\Theta(d_1 \kappa)$, which contradicts \cref{prop: fphp degree bound} due to our choice of parameters.
\end{enumerate}

% In our applications, we assume the setting from \cref{lem:good expander exists} to simplify the computations.

% \sout{Given a term $t$ and $\theta_1, \ldots, \theta_k \in \{x,y,z,u,w\}$, let $t_{\theta_1,\ldots,\theta_k}$ be the projection of $t$ on the corresponding variables.} \slava{I want to get rid of this notation.}

Let $d$ be a threshold parameter. We begin by defining a restriction $\rho_1$ that w.h.p.~eliminates any term $t$ that either $U^{(\pi)}$-left-row-mentions or $U^{(\pi)}$-right-row-mentions at least $d$ rows, or positively-$Z^{(\pi)}$-left-row-mentions or positively-$Z^{(\pi)}$-right-row-mentions $d$ rows.
Moreover, $\rho_1$ is defined so that it substitutes all $w$-variables with either constants or $x$- or $y$-variables.
The construction uses the auxiliary matrices $\calB$ and $\calC$.
\begin{equation*}
    \calB = \begin{pmatrix}
        * & * & 0 & 0 \\
        0 & 0 & * & *
    \end{pmatrix},
    \quad
    \calC = \begin{pmatrix}
        * & 0 & * & 0 \\
        0 & * & 0 & *
    \end{pmatrix}.
\end{equation*}

Every row $\calB_i$ of $\calB$ and every row $\calC_j$ of $\calC$ satisfy the following properties:
\begin{itemize}
    \item there is exactly one $k \in [4]$ satisfying $\calB_{i,k} = \calC_{j,k} = *$;
    \item the remaining indices $\ell \in [4] \setminus \{k\}$ satisfy $\calB_{i,\ell} = 0$ or $\calC_{j,\ell} = 0$.
\end{itemize}
Moreover, each column of $\calB$ and $\calC$ contains one $0$-entry.

We will define $\rho_1$ only for graphs $P(G)$, i.e., those that have two all-ones columns in their adjacency matrix. Let $G$ be a bipartite graph with parts $[m]$ and $[n]$.

\begin{definition}
    Let $t$ be a term in the variables of $\psi_{P(G)}(\bfA, \kappa)$.
    We say that $i \in [m]$ is \emph{$U^{(\pi)}$-left-row-mentioned in $t$} if there are $\mu \in [4]$, $j \in [m]$ and a non-empty initial segment $\Sigma$ of $\calN_{P(G)}(j)$ with $\{1,2\} \subsetneq \Sigma$ such that $u^{(\pi, \mu)}_{i,j,\Sigma}$ or $\overline{u}^{(\pi, \mu)}_{i,j,\Sigma}$ appears in $t$.
    We say that $j \in [m]$ is \emph{$U^{(\pi)}$-right-row-mentioned in $t$} if there are $\mu \in [4]$, $i \in [m]$ and a non-empty initial segment $\Sigma$ of $\calN_{P(G)}(j)$ with $\{1,2\} \subsetneq \Sigma$ such that $u^{(\pi, \mu)}_{i,j,\Sigma}$ or $\overline{u}^{(\pi, \mu)}_{i,j,\Sigma}$ appears in $t$.
    We say that $i \in [m]$ is \emph{positively-$Z^{(\pi)}$-left-row-mentioned in $t$} if there are $\mu \in [4]$, $j \in [m]$ and $k \in \{3, \ldots, n+2\}$ such that $z^{(\pi, \mu)}_{i,j,k}$ appears in $t$.
    We say that $j \in [m]$ is \emph{positively-$Z^{(\pi)}$-right-row-mentioned in $t$} if there are $\mu \in [4]$, $i \in [m]$ and $k \in \{3, \ldots, n+2\}$ such that $z^{(\pi, \mu)}_{i,j,k}$ appears in $t$.
\end{definition}

We substitute the first two columns with constants; therefore, we do not consider them in the above definition.

\begin{definition}[Substitution $\rho_1$ for $\psi_{P(G)}(\bfA, \kappa)$]\label{def: substitution rho_1 for psi}
    Let $\kappa \ge 0$, $m, n > 0$ and let $G$ be an $m \times n$ bipartite graph with parts $[m]$ and $[n]$. 

    For every $\pi \in \Q^{\le\kappa}$, let $\{ \tilde{x}^{(\pi)}_{i,k} : i \in [m], k \in [n] \}$, $\{ \tilde{y}^{(\pi)}_{k,j} : j \in [m], k \in \calN_{G}(j) \}$, $\{ \tilde{z}^{(\pi)}_{i,j,k} : i, j \in [m], k \in \calN_{G}(j) \}$ and $\{ \tilde{u}^{(\pi)}_{i,j,\Sigma} : i,j \in [m], \Sigma \text{ is a non-empty initial segment of } \calN_{G}(j) \}$ be fresh variables.
    The random substitution $\rho_1$ for $\psi_{P(G)}(\bfA, \kappa)$ is defined as follows. For every $\pi \in \Q^{\le\kappa}$:
    \begin{enumerate}
        \item For each $i \in [m]$, independently and uniformly sample $b^{(\pi)}_i, c^{(\pi)}_i \in [2]$ and $\alpha^{(\pi)}_i, \beta^{(\pi)}_i \in \Q$.

        \item For each $\mu \in [4]$, $i \in [m]$, $k \in [n+2]$, set
        \[
            \rho_1(x^{(\pi,\mu)}_{i,k}) = \begin{cases}
                \alpha^{(\pi)}_i & \text{if } k = 1, \\
                1 & \text{if } k = 2, \\
                0 & \text{if } k \ge 3 \text{ and } \calB_{b^{(\pi)}_i,\mu} = 0, \\
                \tilde{x}^{(\pi)}_{i,k-2} & \text{if } k \ge 3 \text{ and } \calB_{b^{(\pi)}_i,\mu} = *.
            \end{cases}
        \]
        \item For each $\mu \in [4]$, $j \in [m]$, $k \in \calN_{P(G)}(j)$, set
        \[
            \rho_1(y^{(\pi,\mu)}_{k,j}) = \begin{cases}
                1 & \text{if } k = 1, \\
                \beta^{(\pi)}_j & \text{if } k = 2, \\
                0 & \text{if } k \ge 3 \text{ and } \calC_{c^{(\pi)}_j,\mu} = 0, \\
                \tilde{y}^{(\pi)}_{k-2,j} & \text{if } k \ge 3 \text{ and } \calC_{c^{(\pi)}_j,\mu} = *.
            \end{cases}
        \]
        \item For each $\mu \in [4]$, $i,j \in [m]$, $k \in \calN_{P(G)}(j)$, set
        \[
            \rho_1(z^{(\pi,\mu)}_{i,j,k}) = \begin{cases}
                \alpha^{(\pi)}_i & \text{if } k = 1, \\
                \beta^{(\pi)}_j & \text{if } k = 2, \\
                0 & \text{if } k \ge 3 \text{ and either } \calB_{b^{(\pi)}_i,\mu} = 0 \text{ or } \calC_{c^{(\pi)}_j,\mu} = 0, \\
                \tilde{z}^{(\pi)}_{i,j,k-2} & \text{if } k \ge 3 \text{ and } \calB_{b^{(\pi)}_i,\mu} = \calC_{c^{(\pi)}_j,\mu} = *.
            \end{cases}
        \]
        \item For each $\mu \in [4]$, $i,j \in [m]$ and every non-empty initial segment $\Sigma$ of $\calN_{P(G)}(j)$, let $\Sigma' = ( k-2 : k \in \Sigma, k > 2 )$ and set
        \[
            \rho_1(u^{(\pi,\mu)}_{i,j,\Sigma}) = \begin{cases}
                \alpha^{(\pi)}_i & \text{if } \Sigma = (1), \\
                \alpha^{(\pi)}_i + \beta^{(\pi)}_j & \text{if } \Sigma = (1, 2), \\
                \tilde{u}^{(\pi)}_{i,j,\Sigma'} & \text{if } \Sigma \setminus \{1,2\} \neq \varnothing, \calB_{b^{(\pi)}_i,\mu} = \calC_{c^{(\pi)}_j,\mu} = * \text{ and } \alpha^{(\pi)}_i = \beta^{(\pi)}_j, \\
                \overline{\tilde{u}}^{(\pi)}_{i,j,\Sigma'} & \text{if } \Sigma \setminus \{1,2\} \neq \varnothing, \calB_{b^{(\pi)}_i,\mu} = \calC_{c^{(\pi)}_j,\mu} = * \text{ and } \alpha^{(\pi)}_i \neq \beta^{(\pi)}_j, \\
                \alpha^{(\pi)}_i + \beta^{(\pi)}_j & \text{otherwise}.
            \end{cases}
        \]
        \item For each $\mu \in [4]$, $i,j \in [m]$, set
        \begin{align*}
            \rho_1(w^{(\pi)}_{i,j,0}) &= \rho_1(A^{(\pi)}_{i,j}), \\
            \rho_1(w^{(\pi)}_{i,j,1}) &= \begin{cases}
                \rho_1(A^{(\pi)}_{i,j}) & \text{if } b^{(\pi)}_i = 1, \\
                0 & \text{if } b^{(\pi)}_i = 2, \\
            \end{cases} \\
            \rho_1(w^{(\pi)}_{i,j,2}) &= \begin{cases}
                0 & \text{if } b^{(\pi)}_i = 1, \\
                \rho_1(A^{(\pi)}_{i,j}) & \text{if } b^{(\pi)}_i = 2. \\
            \end{cases}
        \end{align*}
        \item For any variable $\theta$ mapped to $0$ or $1$ by $\rho_1$, set $\rho_1(\overline{\theta}) = 1 - \rho_1(\theta)$.
        \item For any variable $\theta$ not mapped to a Boolean constant by $\rho_1$, set $\rho_1(\overline{\theta}) = \overline{\rho_1(\theta)}$.
    \end{enumerate}
\end{definition}

\begin{lemma}\label{lem: rho1 reduces u-degree}
    Let $\kappa \ge 0$, $m,n>0$ and let $G$ be an $m \times n$ bipartite graph with parts $[m]$ and $[n]$ and the maximum left-degree at most $\Deltamax$.
    Let $\bfA = \{ A^{(\pi)} : \pi \in \Sigma^{\le\kappa} \}$ be a collection of $m \times m$ matrices as in \cref{def: iterated cnf psi}.
    Let $t$ be a term in the variables of $\psi_{P(G)}(\bfA, \kappa)$ and $t_u$ be the subterm of $t$ that consists of all the $u$-variables of the form $u^{(\pi,\mu)}_{i,j,\Sigma}$ or $\overline{u}^{(\pi,\mu)}_{i,j,\Sigma}$ with $\{1,2\} \subsetneq \Sigma$ (that is, all variables that either $U^{(\pi)}$-left-row-mention or $U^{(\pi)}$-right-row-mention some indices from $[m]$).
    Similarly, let $t_{+z}$ be the subterm of $t$ consisting of all the \emph{positive} $z$-variables of the form $z^{(\pi,\mu)}_{i,j,k}$ with $k>2$ (such variables positively-$Z^{(\pi)}$-left-row-mention or positively-$Z^{(\pi)}$-right-row-mention some indices from $[m]$). Assume that either $\deg(t_u)$ or $\deg(t_{+z})$ is at least $d$.
    Then
    \[
        \Prb{t \restriction \rho_1 \neq 0} \le {\left(\sqrt{3/4}\right)}^{\sqrt{\frac{d}{4\Deltamax}}}.
    \]
\end{lemma}
\begin{proof}
    Assume that $\deg(t_u) \ge d$.
    For each $\pi \in \Q^{\le\kappa}$, let $I^{(\pi)}$ be the set of indices $i \in [m]$ that are $U^{(\pi)}$-left-row-mentioned in $t_u$, and let $J^{(\pi)}$ be the set of indices $j \in [m]$ that are $U^{(\pi)}$-right-row-mentioned in $t_u$.
    Let $D = \sum_{\pi \in \Q^{\le\kappa}} (\card{I^{(\pi)}}+\card{J^{(\pi)}})$.
    Then $d \le \deg(t_u) \le 4 D^2 \Deltamax$, which implies that $D \ge \sqrt{\frac{d}{4\Deltamax}}$.

    Let $t'$ be a minimal subterm of $t_u$ that for each $\pi \in \Q^{\le\kappa}$, $U^{(\pi)}$-left-row-mentions $I^{(\pi)}$ and $U^{(\pi)}$-right-row-mentions $J^{(\pi)}$.    
    If we think of the variables $u^{(\pi,\mu)}_{i,j,\Sigma}$ (or their negations) appearing in $t'$ as edges $((\pi,i),(\pi,j))$ in a bipartite graph, then $t'$ corresponds to a minimal edge cover of such a graph. This cover cannot contain a path of length three, thus its connected components are stars of the form $\prod_{(j,\mu,\Sigma) \in P} u^{(\pi_0,\mu)}_{i_0,j,\Sigma} \prod_{(j,\mu,\Sigma) \in N} \overline{u}^{(\pi_0,\mu)}_{i_0,j,\Sigma}$ or $\prod_{(i,\mu,\Sigma) \in P} u^{(\pi_0,\mu)}_{i,j_0,\Sigma} \prod_{(i,\mu,\Sigma) \in N} \overline{u}^{(\pi_0,\mu)}_{i,j_0,\Sigma}$, and each of these stars mentions its own disjoint set of indices.

    Assume that we have a star with $r$ rays, say $s = \prod_{(j,\mu,\Sigma) \in P} u^{(\pi_0,\mu)}_{i_0,j,\Sigma} \prod_{(j,\mu,\Sigma) \in N} \overline{u}^{(\pi_0,\mu)}_{i_0,j,\Sigma}$.
    For every $b \in [2]$ and $\alpha \in \Q$, the events
    \[
        \left( u^{(\pi_0,\mu)}_{i_0,j,\Sigma} \restriction \rho_1 \neq 0 \;\middle\vert\; b^{(\pi_0)}_{i_0} = b \text{ and } \alpha^{(\pi_0)}_{i_0} = \alpha \right), \text{ where } (j,\mu,\Sigma) \in P,
    \]
    and
    \[
        \left( \overline{u}^{(\pi_0,\mu)}_{i_0,j,\Sigma} \restriction \rho_1 \neq 0 \;\middle\vert\; b^{(\pi_0)}_{i_0} = b \text{ and } \alpha^{(\pi_0)}_{i_0} = \alpha \right), \text{ where } (j,\mu,\Sigma) \in N,
    \]
    are independent and each occurs with probability at most $3/4$.
    This is because when we fix the value of $b^{(\pi_0)}_{i_0}$ and $\alpha^{(\pi_0)}_{i_0}$, which determine the centre of the star $(\pi_0, i_0)$, the value of $u^{(\pi_0,\mu)}_{i_0,j,\Sigma}$ depends only on $\beta^{(\pi_0)}_j$ and $c^{(\pi_0)}_j$, which are independent for different $j$.
    Thus, all of them simultaneously occur with probability at most ${\left(3/4\right)}^r \le {\left(\sqrt{3/4}\right)}^{r+1}$.
    %\hanlin{Do we need to justify that different stars are independent as well?}
    %\slava{This follows from the fact that the indices are disjoint, so all star-shaped-subterms are independent. This is stated as ``each of these stars mentions its own disjoint set of indices'' above.}
    \hanlin{\sout{Actually I'm confused again about why the events inside each star is independent... I guess I'm just missing something... Also, maybe say one sentence about why the proof doesn't work for $t_{-z}$?}}
    \slava{\sout{Each index appears exactly in one variable that itself appears in exactly one star (we chose the minimal star-cover). If we fix the centre of the star to a fixed value, for each ray-index $j$ we set the block to zero with probability $1/2$ and thus $\rho_1$ assigns the corresponding $u$-value to $\alpha_{i_0} + \beta_{j}$. With probability $1/2$ the value of $\beta_j$ is correct here. This depends only on the row-type of $j$ and the value $\beta_j$, thus is independent from the other rows.}}
    \slava{\sout{For $t_{-z}$, the problem is that if we want to set a positive $z$-variable to zero we can set the related $x$-variable or $y$-variable to zero independently (either works, so if we fix the value of one of them, the other can be chosen arbitrarily). Thus, we can apply the same star-argument. However, if we want to set the negative occurrence to $0$, we need to set \textbf{both} $x$- and $y$-variables to one, thus it depends on two indices simultaneously.}}
    This implies that
    \[
        \Prb{t \restriction \rho_1 \neq 0} \le {\left(\sqrt{3/4}\right)}^{D} \le {\left(\sqrt{3/4}\right)}^{\sqrt{\frac{d}{4\Deltamax}}}.
    \]

    The case when $\deg(t_{+z}) \ge d$ is proved analogously: each $z^{(\pi)}_{i,j,k}$ occurring in $t_{+z}$ can be independently set to $0$ by choosing the appropriate values for $b^{(\pi)}_i$ (assigning $\rho_1(x^{(\pi)}_{i,k})=0$) or for $c^{(\pi)}_j$ (assigning $\rho_1(y^{(\pi)}_{k,j})=0$).
    Note that we cannot apply the same argument to the negative occurrences of $z$-variables, since they depend on both $x$- and $y$-variables simultaneously.
\end{proof}

% We define another polynomial system $\phi_G(\bfA, \kappa)$, which is analogous to $\psi^m_n(\bfA, \kappa, G)$.
The substitution $\rho_1$ transforms any refutation of $\psi_{P(G)}(\bfA, \kappa)$ into a refutation of $\phi_G(\bfA_1, \kappa)$, where $\bfA_1 = \bfA \restriction \rho_1$.

\begin{lemma}\label{lem: rho1 preserves refutations}
    Let $\kappa \ge 0$, $m,n>0$ and let $G$ be an $m \times n$ bipartite graph.
    Let $\bfA = \{ A^{(\pi)} : \pi \in \Sigma^{\le\kappa} \}$ be a collection of $m \times m$ matrices as in \cref{def: iterated cnf psi}. Let $\bfA_1 = \bfA \restriction \rho_1 = \{ A^{(\pi)} \restriction \rho_1 : A^{(\pi)} \in \bfA \}$.
    Then for every $\PCR_{\F_2}$ refutation $\Pi$ of $\psi_{P(G)}(\bfA, \kappa)$, $\Pi \restriction \rho_1$ is a refutation of $\phi_G(\bfA_1, \kappa)$ (modulo the axioms for twin variables $\theta + \overline{\theta} + 1$).
\end{lemma}
\begin{proof}
    Follows from the definition of $\rho_1$. Under $\rho_1$, the axioms of $\psi_{P(G)}(\bfA, \kappa)$ are either satisfied or become axioms of the same type for $\phi_G(\bfA_1, \kappa)$. Since $\rho_1$ sometimes substitutes variables with negated variables, we need to apply the twin variable axioms $\theta + \overline{\theta} + 1$ to replace such occurrences of $\overline{\theta}$ with $\theta+1$. This can lead to an insignificant increase in the refutation size, which we can safely ignore.
\end{proof}

Next, we define a random substitution $\rho_2$ that w.h.p.~turns any term $t$ that mentions many columns into $0$.
As with $\rho_1$, we only apply $\rho_2$ to graphs that have a specific structure, namely $H(G)$.

\begin{definition}
    Let $t$ be a term in the variables of $\phi_{H(G)}(\bfA, \kappa)$. Let $N = 8n+8$.
    Given $k \in [N]$, let $\eta(k) = ((k-1) \bmod (n+1))+1$ and $\nu(k) = \lfloor \frac{k-1}{n+1} \rfloor + 1$.

    We say that $k \in [n]$ is \emph{$X^{(\pi)}$-column-mentioned in $t$} if there are $i \in [m]$ and $\ell \in [N]$ with $\eta(\ell) = k$ such that $x^{(\pi)}_{i,\ell}$ or $\overline{x}^{(\pi)}_{i,\ell}$ appears in $t$.
    We say that $k \in [n]$ is \emph{${(Y^{(\pi)})}^T$-column-mentioned in $t$} if there are $j \in [m]$ and $\ell \in [N]$ with $\eta(\ell) = k$ such that $y^{(\pi)}_{\ell,j}$ or $\overline{y}^{(\pi)}_{\ell,j}$ appears in $t$.
    We say that $k \in [n]$ is \emph{$Z^{(\pi)}$-column-mentioned in $t$} if there are $i,j \in [m]$ and $\ell \in [N]$ with $\eta(\ell) = k$ such that $z^{(\pi)}_{i,j,\ell}$ or $\overline{z}^{(\pi)}_{i,j,\ell}$ appears in $t$.
\end{definition}

\begin{definition}[Substitution $\rho_2$]\label{def: substitution rho_2 for phi}
    % \slava{WAS : Let $\kappa \ge 0$, $n>0$, $N=8n+8$, $m=16(N+2)$ and let $G$ be an $m \times n$ $(n^\delta,d)$-lossless expander with parts $[m]$ and $[n]$, the minimum left-degree at least $\Deltamin = 50 \log n$ and the maximum left-degree at most $\Deltamax = 150 \log n$.}

    Let $\kappa \ge 0$, $m,n>0$, $N=8n+8$ and let $G$ be an $m \times n$ bipartite graph with parts $[m]$ and $[n]$.

    The random substitution $\rho_2$ for $\phi_{H(G)}(\bfA, \kappa)$ is defined as follows.
    \begin{enumerate}
        \item For each $k \in [n]$, independently and uniformly sample $c_k \in \{*, 1, 0\}$. Assign $c_{n+1} = *$.
        \item Let $\tilde{K} = \{ k \in [N] : c_{\eta(k)} \in \Q \}$ and let $\tilde{G} = G \setminus \{ k \in [n] : c_k \in \Q \}$. Note that $\calN_{H(\tilde{G})}(j) = \calN_{H(G)}(j) \setminus \tilde{K}$ for all $j \in [m]$.

        \item For every $\pi \in \Q^{\le\kappa}$:
        \begin{enumerate}
            \item Let $\{ \tilde{u}^{(\pi)}_{i,j,\Sigma} : i,j \in [m], \Sigma \text{ is a non-empty initial segment of } \calN_{H(\tilde{G})}(j) \}$ be fresh extension variables.

            \item For every $i \in [m]$, and $k \in [N]$, set
            \[
                \rho_2(x^{(\pi)}_{i,k}) = \begin{cases}
                    c_{\eta(k)} & \text{if } c_{\eta(k)} \in \Q, \\
                    x^{(\pi)}_{i,k} & \text{otherwise}.
                \end{cases}
            \]

            \item For every $j \in [m]$, and $k \in \calN_{H(G)}(j)$, set
            \[
                \rho_2(y^{(\pi)}_{k,j}) = \begin{cases}
                    c_{\eta(k)} & \text{if } c_{\eta(k)} \in \Q, \\
                    y^{(\pi)}_{k,j} & \text{otherwise}.
                \end{cases}
            \]

            \item For every $i,j \in [m]$, and $k \in \calN_{H(G)}(j)$, set
            \[
                \rho_2(z^{(\pi)}_{i,j,k}) = \begin{cases}
                    c_{\eta(k)} & \text{if } c_{\eta(k)} \in \Q, \\
                    z^{(\pi)}_{i,j,k} & \text{otherwise}.
                \end{cases}
            \]

            \item For every $i,j \in [m]$ and every non-empty initial segment $\Sigma$ of $\calN_{H(G)}(j)$, set
            \[
                \rho_2(u^{(\pi)}_{i,j,\Sigma}) = \begin{cases}
                    0 & \text{if } \Sigma \setminus \tilde{K} = \varnothing \text{ and } \card{\{ k \in \Sigma : c_{\eta(k)}=1 \}} \text{ is even}, \\
                    1 & \text{if } \Sigma \setminus \tilde{K} = \varnothing \text{ and } \card{\{ k \in \Sigma : c_{\eta(k)}=1 \}} \text{ is odd}, \\
                    \tilde{u}^{(\pi)}_{i,j,(\Sigma \setminus \tilde{K})} & \text{if } \Sigma \setminus \tilde{K} \neq \varnothing \text{ and } \card{\{ k \in \Sigma : c_{\eta(k)}=1 \}} \text{ is even}, \\
                    \overline{\tilde{u}}^{(\pi)}_{i,j,(\Sigma \setminus \tilde{K})} & \text{if } \Sigma \setminus \tilde{K} \neq \varnothing \text{ and } \card{\{ k \in \Sigma : c_{\eta(k)}=1 \}} \text{ is odd}.
                \end{cases}
            \]

            \item For any variable $\theta$ mapped to $0$ or $1$ by $\rho_2$, set $\rho_2(\overline{\theta}) = 1 - \rho_2(\theta)$.
            \item For any variable $\theta$ not mapped to a Boolean constant by $\rho_2$, set $\rho_2(\overline{\theta}) = \overline{\rho_2(\theta)}$.
        \end{enumerate}
    \end{enumerate}
\end{definition}

We show that under $\rho_2$ every term that mentions many columns becomes zero w.h.p.
\begin{lemma}\label{lem: rho2 reduces column degree}
    Let $\kappa \ge 0$, $m,n>0$ and let $G$ be an $m \times n$ bipartite graph with parts $[m]$ and $[n]$.
    Let $t$ be a term in the variables of $\phi_{H(G)}(\bfA, \kappa)$.
    Let $K$ be the set of all $k \in [n]$ that are $X^{(\pi)}$-column-mentioned, ${(Y^{(\pi)})}^T$-column-mentioned or $Z^{(\pi)}$-column-mentioned in $t$, for some $\pi \in \Q^{\le\kappa}$.
    Then $\Prb{t \restriction \rho_2 \neq 0} \le {(2/3)}^{\card{K}}$.
\end{lemma}
\begin{proof}
    For every $x$-, $y$- or $z$-variable $\theta$ appearing in $t$, $\Prb{\rho_2(\theta) \neq 0} \le 2/3$, which occurs independently for distinct columns. Thus, $(t \restriction \rho_2 \neq 0)$ occurs with probability at most ${(2/3)}^{\card{K}}$.
\end{proof}

When $G$ is an expander with sufficiently large minimum left-degree, the graph $\tilde{G}$ defined in \cref{def: substitution rho_2 for phi} remains a good expander due to \cref{lem:remains good expander}.
\begin{proposition}\label{prop: rho2 gives good expander}
    Let $\kappa \ge 0$, $n>0$, $N=8n+8$, $m=16(N+2)$ and let $G$ be an $m \times n$ $(n^\delta,d)$-lossless expander with parts $[m]$ and $[n]$, the minimum left-degree at least $\Deltamin = 50 \log n$ and the maximum left-degree at most $\Deltamax = 150 \log n$.
    Then with probability at least $1/2$, the graph $\tilde{G}$ from \cref{def: substitution rho_2 for phi} is an $(n^\delta, c)$-boundary expander, where $c = \Deltamin / 6 - d$.
\end{proposition}

We show that $\rho_2$ transforms any refutation of $\phi_{H(G)}(\bfA, \kappa)$ into a refutation of $\phi_{H(\tilde{G})}(\bfA_2, \kappa)$, where $\bfA_2 = \bfA \restriction \rho_2$.

\begin{lemma}\label{lem: rho2 preserves refutations}
    Let $\kappa \ge 0$, $m,n>0$ and let $G$ be an $m \times n$ bipartite graph with parts $[m]$ and $[n]$.
    Let $\bfA = \{ A^{(\pi)} : \pi \in \Sigma^{\le\kappa} \}$ be a collection of $m \times m$ matrices as in \cref{def: iterated cnf phi}.
    Let $\bfA_2 = \bfA \restriction \rho_2 = \{ A^{(\pi)} \restriction \rho_2 : A^{(\pi)} \in \bfA \}$.
    Then for every $\PCR_{\F_2}$ refutation $\Pi$ of $\phi_{H(G)}(\bfA, \kappa)$, $\Pi \restriction \rho_2$ is a refutation of $\phi_{H(\tilde{G})}(\bfA_2, \kappa)$.
\end{lemma}
\begin{proof}
    Can be verified straightforwardly from the definition of $\rho_2$, similarly to \cref{lem: column PCR2 iterated alg Rank ref remains ref}.
    % \slava{Write a short comment?}
\end{proof}

Now we define a random restriction $\rho_3$ that reduces the total \emph{degree} of the terms in any refutation of $\phi_{H(G)}(\bfA,\kappa)$ provided such a refutation has certain restrictions on its $u$-degree and column degree.
We need the following matrices, which we used in \cref{sec: iterability of Rank}.
\begin{equation*}
    \calD = \begin{pmatrix}
        1 & * & 1 & 0 & 0 & * & * & * \\
        0 & 1 & * & 1 & * & * & * & 0 \\
        * & * & 0 & * & 1 & 0 & 1 & * \\
        * & 0 & * & * & * & 1 & 0 & 1
    \end{pmatrix},
    \quad
    \calE = \begin{pmatrix}
        1 & 0 & 0 & 1 & * & * & * & * \\
        0 & * & 1 & * & 1 & * & 0 & * \\
        * & 1 & * & 0 & * & 0 & * & 1 \\
        * & * & * & * & 0 & 1 & 1 & 0
    \end{pmatrix}.
\end{equation*}

\begin{definition}[Substitution $\rho_3$]\label{def: substitution rho_3 for phi}
    Let $\kappa \ge 0$, $m,n>0$, $N = 8n+8$, and let $G$ be an $m \times n$ bipartite graph with parts $[m]$ and $[n]$.
    Given $k \in [N]$, let $\eta(k) = ((k-1) \bmod (n+1))+1$ and $\nu(k) = \lfloor \frac{k-1}{n+1} \rfloor + 1$.

    The substitution $\rho_3$ for $\phi_{H(G)}(\bfA,\kappa)$ is defined as follows.
    For every $\pi \in \Q^{\le\kappa}$:
    \begin{enumerate}
        \item For every $i \in [m]$ uniformly and independently sample $s^{(\pi)}_i \in [4]$ and $r^{(\pi)}_i \in [4]$.

        \item For each $i \in [m]$ and $k \in [N]$, set
        \[
            \rho_3(x^{(\pi)}_{i,k}) = \begin{cases}
                1 & \text{if } \eta(k) = n + 1 \text{ and } \calD_{s^{(\pi)}_i,\nu(k)} = 1, \\
                0 & \text{if } \eta(k) = n + 1 \text{ and } \calD_{s^{(\pi)}_i,\nu(k)} \neq 1, \\
                x^{(\pi)}_{i,\eta(k)} & \text{if } \eta(k) \neq n + 1 \text{ and } \calD_{s^{(\pi)}_i,\nu(k)} = *, \\
                \calD_{s^{(\pi)}_i,\nu(k)} & \text{if } \eta(k) \neq n + 1 \text{ and } \calD_{s^{(\pi)}_i,\nu(k)} \in \Q.
            \end{cases}
        \]

        \item For each $j \in [m]$ and $k \in \calN_{H(G)}(j)$, set
        \[
            \rho_3(y^{(\pi)}_{k,j}) = \begin{cases}
                \card{\calN_{G}(j)} \bmod 2 & \text{if } \eta(k) = n + 1 \text{ and } \calE_{r^{(\pi)}_j,\nu(k)} = 1, \\
                0 & \text{if } \eta(k) = n + 1 \text{ and } \calE_{r^{(\pi)}_j,\nu(k)} \neq 1, \\
                y^{(\pi)}_{\eta(k),j} & \text{if } \eta(k) \neq n + 1 \text{ and } \calE_{r^{(\pi)}_j,\nu(k)} = *, \\
                \calE_{r^{(\pi)}_j,\nu(k)} & \text{if } \eta(k) \neq n + 1 \text{ and } \calE_{r^{(\pi)}_j,\nu(k)} \in \Q.
            \end{cases}
        \]

        \item For each $i,j \in [m]$ and $k \in \calN_{H(G)}(j)$ set
        \[
            \rho_3(z^{(\pi)}_{i,j,k}) = \rho_3(x^{(\pi)}_{i,k}) \rho_3(y^{(\pi)}_{k,j}).
        \]

        \item For each $i,j \in [m]$ and every non-empty initial segment $\Sigma$ of $\calN_{H(G)}(j)$, set
        \[
            \rho_3(u^{(\pi)}_{i,j,\Sigma}) = \sum_{k \in \Sigma} \rho_3(z^{(\pi)}_{i,j,k}).
        \]

        \item For every variable $\theta$ assigned by $\rho_3$, set $\rho_3(\overline{\theta}) = 1 - \rho_3(\theta)$.
    \end{enumerate}
\end{definition}

Note that $\rho_3$ does not substitute any negated variables and expands the extension variables as polynomials. It can be trivially extended to a $\PC$ refutation. The \emph{size} of the resulting refutation can grow exponentially, but we will prove that its \emph{degree} becomes small~w.h.p.
For every refutation $\Pi$ of $\phi_{H(G)}(\bfA, \kappa)$, $\Pi \restriction \rho_3$ can be transformed into a refutation of $\widetilde{\phi}_G(\bfA_3, \kappa)$, where $\bfA_3 = \bfA \restriction \rho_3$.

In order to handle the negative occurrences of the $z$-variables, we need to define a transformation $R_d$. It is a linear mapping defined on the terms in the variables of $\phi_{H(G)}(\bfA, \kappa)$.
In the definitions below, when working with $\phi_{H(G)}$, we ignore all variables whose column $k$ satisfies $\eta(k)=n+1$. These variables are mapped to constants by $\rho_3$, and hence do not contribute to the degree after the restriction. Accordingly, all $z$-stars considered by $R_d$ are formed only from variables $\overline z^{(\pi)}_{i,j,k}$ with $\eta(k)\neq n+1$.
\begin{definition}
    Let $t$ be a term in the variables of $\phi_{H(G)}(\bfA, \kappa)$.
    We define $R_d(t)$ by \emph{contracting} every subterm of the form $\prod_{j \in J} \overline{z}^{(\pi)}_{i_0,j,k}$ or $\prod_{i \in I} \overline{z}^{(\pi)}_{i,j_0,k}$, where $\card{I},\card{J} \ge d$, using the following procedure.
    Let $t = t_1 t_2$, where $t_1$ contains only negated $z$-variables that do not mention columns $k$ with $\eta(k) = n + 1$ and $t_2$ contains the rest.
    Let
    \begin{align*}
        S_{\mathit{left}} &= \{ (\pi, i_0, \{ j \in [m] : \overline{z}^{(\pi)}_{i_0,j,k} \in \vars(t_1) \}, k) :
        \begin{aligned}[t]
            & \pi \in \Q^{\le\kappa}, i_0 \in [m], \\
            & k \in [N], \eta(k) \neq n + 1 \},
        \end{aligned} \\
        S_{\mathit{right}} &= \{ (\pi, \{ i \in [m] : \overline{z}^{(\pi)}_{i,j_0,k} \in \vars(t_1) \}, j_0, k) :
        \begin{aligned}[t]
            & \pi \in \Q^{\le\kappa}, j_0 \in [m], \\
            & k \in [N], \eta(k) \neq n + 1 \}.
        \end{aligned}
    \end{align*}
    Given a star $(\pi, i_0, J, k) \in S_{\mathit{left}}$ we can think of the set $J$ as the set of \emph{rays} of the star. Similarly, given a star $(\pi, I, j_0, k) \in S_{\mathit{right}}$ we can think of the set $I$ as the set of rays of the star. 
    Let $C$ be the set of quadruples $(\pi,i,j,k)$ with $\overline{z}^{(\pi)}_{i,j,k} \in \vars(t_1)$ that do not belong to any star in $S_{\mathit{left}} \cup S_{\mathit{right}}$ with at least $d$ rays. We then contract all stars with at least $d$ rays to a single variable (its centre). Formally, we define
    \[
        R_d(t) =
        \left(\prod_{\substack{(\pi,i_0,J,k) \in S_{\mathit{left}} \\ \card{J} \ge d}} \overline{x}^{(\pi)}_{i_0,k}\right)
        \cdot
        \left(\prod_{\substack{(\pi,I,j_0,k) \in S_{\mathit{right}} \\ \card{I} \ge d}} \overline{y}^{(\pi)}_{k,j_0}\right)
        \cdot
        \prod_{(\pi, i, j, k) \in C} \overline{z}^{(\pi)}_{i,j,k}
        \cdot
        t_2.\footnotemark
    \]
    % \begin{align*}
    %     t_{\mathit{left}} &=
    %     \left(\prod_{\substack{(\pi,i_0,J,k) \in S_{\mathit{left}} \\ \card{J} < d}} \prod_{j \in J} \overline{z}^{(\pi)}_{i_0,j,k}\right)
    %     \left(\prod_{\substack{(\pi,i_0,J,k) \in S_{\mathit{left}} \\ \card{J} \ge d}} \overline{x}^{(\pi)}_{i_0,k}\right), \\
    %     t_{\mathit{right}} &=
    %     \left(\prod_{\substack{(\pi,I,j_0,k) \in S_{\mathit{right}} \\ \card{I} < d}} \prod_{i \in I} \overline{z}^{(\pi)}_{i,j_0,k}\right)
    %     \left(\prod_{\substack{(\pi,I,j_0,k) \in S_{\mathit{right}} \\ \card{I} \ge d}} \overline{y}^{(\pi)}_{k,j_0}\right).
    % \end{align*}
    % Then
    % \[
    %     R_d(t) = t_{\mathit{left}} \cdot t_{\mathit{right}} \cdot t_2.\footnotemark
    % \]
    \footnotetext{As usual, products are understood modulo the Boolean axioms, so the resulting terms are multilinear.}
    We proceed by extending $R_d$ to polynomials linearly.
\end{definition}

When $d \ge 2$,\footnote{In our application, we choose $d$ to be even larger.} the mapping $R_d$ does not increase the degree of any term.
Indeed, suppose that $R_d$ inserted $q$ centres and removed $r$ rays (negated $z$-variables appearing in large stars). The sum of sizes of all contracted stars is at least $dq$, while every removed ray belongs to at most one contracted left-star and one contracted right-star. Therefore, we remove at least $dq/2$ rays, and hence $r \ge dq/2 \ge q$, so the degree does not increase.

The intuition behind $R_d$ is that every star with at least $d$ rays is contracted to a single variable, which corresponds to its centre. The following lemma shows that $\rho_3$ reduces the degree of any term $t$ with many $x$- and $y$-variables or many negative occurrences of $z$-variables after applying $R_d$.

\begin{lemma}\label{lem: rho3 reduces xyz-degree}
    Let $\kappa \ge 0$, $m,n>0$, $d, \tilde{d}>0$ and let $G$ be an $m \times n$ bipartite graph with parts $[m]$ and $[n]$.
    Let $t$ be a term in the variables of $\phi_{H(G)}(\bfA, \kappa)$.
    Let $K$ be the set of all $k \in [n]$ that are $X^{(\pi)}$-column-mentioned, ${(Y^{(\pi)})}^T$-column-mentioned or $Z^{(\pi)}$-column-mentioned in $t$, for some $\pi \in \Q^{\le\kappa}$.
    Let $t_{x,y}$ be the subterm of $R_{\tilde{d}}(t)$ consisting of all $x$- and $y$-variables appearing in $R_{\tilde{d}}(t)$, excluding those that mention columns $k$ with $\eta(k) = n+1$.
    Similarly, let $t_{-z}$ be the subterm of $R_{\tilde{d}}(t)$ consisting of all \emph{negative} occurrences of $z$-variables in $R_{\tilde{d}}(t)$, excluding those that mention columns $k$ with $\eta(k) = n+1$.
    Assume that $\deg(t_{x,y}) \ge d$ or $\deg(t_{-z}) \ge d$.
    Then
    \[
        \Prb{R_{\tilde{d}}(t) \restriction \rho_3 \neq 0} \le {\left(15/16\right)}^{\frac{d}{16\tilde{d} \card{K}}}.
    \]
\end{lemma}
\begin{proof}
    Suppose that $\deg(t_{x,y}) \ge d$. For every $\pi \in \Q^{\le\kappa}$, let
    \begin{align*}
        I_x^{(\pi)} &= \{ i \in [m] : \exists k \in [N] \text{ with $\eta(k) \neq n+1$ s.t. } x^{(\pi)}_{i,k} \text{ or } \overline{x}^{(\pi)}_{i,k} \text{ appears in } t_{x,y} \}, \\
        J_y^{(\pi)} &= \{ j \in [m] : \exists k \in [N] \text{ with $\eta(k) \neq n+1$ s.t. } y^{(\pi)}_{k,j} \text{ or } \overline{y}^{(\pi)}_{k,j} \text{ appears in } t_{x,y} \}.
    \end{align*}
    Then
    \[
        d \le \deg(t_{x,y}) \le 8 \card{K} \sum_{\pi \in \Q^{\le\kappa}} (\card{I_x^{(\pi)}} + \card{J_y^{(\pi)}}).
    \]
    For every $i \in I_x^{(\pi)}$ there is a variable in $t_{x,y}$ that is independently set to $0$ with probability $1/4$ by $\rho_3$. Similarly, for every $j \in J_y^{(\pi)}$, there is a variable in $t_{x,y}$ that is also set to $0$ by $\rho_3$ with the same probability. This implies that
    \[
        \Prb{t_{x,y} \restriction \rho_3 \neq 0} \le {\left(3/4\right)}^{\frac{d}{8 \card{K}}}.
    \]

    % Let ${(R_{\tilde{d}}(t))}_z = s_1 s_2$, where $s_1$ and $s_2$ consist of all the positive and negative occurrences of the $z$-variables in $R_{\tilde{d}}(t)$, respectively.

    % Suppose that $\deg(s_1) \ge \deg({(R_{\tilde{d}}(t))}_{x,y,z})/3$.
    % For every $\pi \in \Q^{\le\kappa}$, let
    % \begin{align*}
    %     I_z^{(\pi)} &= \{ i \in [m] : \exists j \in [m], k \in [n]\colon z^{(\pi)}_{i,j,k} \text{ or } \overline{z}^{(\pi)}_{i,j,k} \text{ appears in } t \}, \\
    %     J_z^{(\pi)} &= \{ j \in [m] : \exists i \in [m], k \in [n]\colon z^{(\pi)}_{i,j,k} \text{ or } \overline{z}^{(\pi)}_{i,j,k} \text{ appears in } t \}.
    % \end{align*}
    % Then
    % \[
    %     \deg(s_1) \le 8\Deltamax \sum_{\pi \in \Q^{\le\kappa}} (\card{I_z^{(\pi)}} + \card{J_z^{(\pi)}}) \le 16\Deltamax \sum_{\pi \in \Q^{\le\kappa}} \max\{\card{I_z^{(\pi)}}, \card{J_z^{(\pi)}}\}.
    % \]
    % The term $s_1$ contains at least $\sum_{\pi \in \Q^{\le\kappa}} \max\{\card{I_z^{(\pi)}}, \card{J_z^{(\pi)}}\}$ variables that are assigned independently by $\rho$, and each does not become zero with probability at most ${(3/4)}^2$.
    % It implies that
    % \[
    %     \Prb{s_1 \restriction \rho_3 \neq 0} \le {(3/4)}^{\frac{\deg({(R_{\tilde{d}}(t))}_{x,y,z})}{24\Deltamax}}.
    % \]

    % Finally, consider the remaining case $\deg(s_2) \ge \deg({(R_{\tilde{d}}(t))}_{x,y,z})/3$.
    Now suppose instead that $\deg(t_{-z}) \ge d$.
    We greedily identify a large set of variables in $t_{-z}$ that are independently set to $0$ by $\rho_3$ w.h.p. Initially set $S = \vars(t_{-z})$ and $T = \varnothing$.
    \begin{itemize}
        \item Pick any variable $\overline{z}^{(\pi)}_{i,j,k} \in S$ and add it to $T$.
        \item Remove all variables $\{ \overline{z}^{(\pi)}_{i,j',k'} : j' \in [m], \eta(k') \in K \} \cup \{ \overline{z}^{(\pi)}_{i',j,k'} : i' \in [m], \eta(k') \in K \}$ from $S$.
        \item Repeat until $S$ is empty.
    \end{itemize}
    Observe that each variable in $T$ mentions its own indices: for every $\pi \in \Q^{\le\kappa}$ and $\ell \in [m]$, there is at most one variable of the form $\overline{z}^{(\pi)}_{i,\ell,k}$ or $\overline{z}^{(\pi)}_{\ell,j,k}$. Thus, every variable from $T$ independently becomes zero under $\rho_3$ with probability at least $1/16$.
    It remains to show that the size of $T$ is sufficiently large.
    At each step we remove at most $16\tilde{d} \card{K}$ $z$-variables from $S$ due to the properties of $R_{\tilde{d}}$, therefore the size of $T$ is at least $\frac{d}{16\tilde{d} \card{K}}$.
    This implies that
    \[
        \Prb{t_{-z} \restriction \rho_3 \neq 0} \le {\left(15/16\right)}^{\frac{d}{16\tilde{d} \card{K}}}.
    \]
\end{proof}

We proceed by showing that for every $\PCR_{\F_2}$ refutation $\Pi$ of $\phi_{H(G)}(\bfA, \kappa)$ of degree $d$, $R_{\tilde{d}}(\Pi) \restriction \rho_3$ can be extended to a refutation of $\widetilde{\phi}_G(\bfA \restriction \rho_3, \kappa)$.

% It might be cleaner to prove instead: for every term $t$ if the property about large stars in $t$ holds, then $R_{\tilde{d}​(t) \restriction \rho_3 ​= t \restriction \rho_3​$.
% This is what we prove (with extra steps).
% Then for multiplication in particular, we have $R(t \cdot \theta) \restriction \rho_3 = t \cdot \theta \restriction \rho_3$ and $R(t) \cdot \theta \restriction \rho_3 = t \cdot \theta \restriction \rho_3$, so they are equal.
% But the current proof is also fine, just a bit more complicated (slightly more cases to consider).
\begin{lemma}\label{lem: rho3 eliminates stars correctly}
    Let $\kappa \ge 0$, $m,n>0$, $N=8n+8$, $\tilde{d}\ge 2$ and let $G$ be an $m \times n$ bipartite graph with parts $[m]$ and $[n]$.
    Let $\Pi$ be a $\PCR_{\F_2}$ refutation of $\phi_{H(G)}(\bfA, \kappa)$ of size at most $\frac{ {(4/3)}^{\tilde{d}} }{ 8m N 2^{\kappa} }$.

    With probability at least $1/2$ over the choice of $\rho_3$, the following holds:
    if $R_{\tilde{d}}(\Pi) \restriction \rho_3$ has degree at most $d$ for some $d \ge 0$, then it can be extended to a valid $\PC$ refutation $\tilde{\Pi}$ of $\widetilde{\phi}_G(\bfA \restriction \rho_3, \kappa)$ with degree at most $d+O(1)$.
\end{lemma}
\begin{proof}
    We restrict our attention only to the terms that do not mention columns $k$ with $\eta(k) = n + 1$, as these are mapped to constants by $\rho_3$.

    We begin by demonstrating the following property of $\rho_3$.
    \begin{claim}\label{clm: rho3 stars property}
        Let $\pi \in \Q^{\le\kappa}$ and $k \in [N]$ with $\eta(k) \neq n + 1$. For every $i \in [m]$ and $J \subseteq \{ \ell \in [m] : k \in \calN_{H(G)}(\ell) \}$,
        \[
            \Prb{\forall \ell \in J : \rho_3(y^{(\pi)}_{k,\ell}) \neq 1} \le {(3/4)}^{\card{J}}.
        \]
        Observe that if $\rho_3(y^{(\pi)}_{k,\ell}) = 1$, then $\rho_3(\overline{z}^{(\pi)}_{i,\ell,k}) = \rho_3(\overline{x}^{(\pi)}_{i,k})$ by the definition of $\rho_3$. Similarly, for every $j \in [m]$ with $k \in \calN_{H(G)}(j)$ and $I \subseteq [m]$,
        \[
            \Prb{\forall \ell \in I : \rho_3(x^{(\pi)}_{\ell,k}) \neq 1} \le {(3/4)}^{\card{I}}.
        \]
        As before, if $\rho_3(x^{(\pi)}_{\ell,k}) = 1$, then $\rho_3(\overline{z}^{(\pi)}_{\ell,j,k}) = \rho_3(\overline{y}^{(\pi)}_{k,j})$.
    \end{claim}
    \begin{claimproof}
        We will show the first statement, and the second is symmetric.
        The events $\rho_3(y^{(\pi)}_{k,\ell}) = 1$ for $\ell \in J$ are independent, since they depend on the independent random choices of $r^{(\pi)}_\ell$ for $\ell \in J$. Each event occurs with probability at least $1/4$, since $\calE$ contains at least one $1$ in every column.
        Now fix $s^{(\pi)}_i$ and thus $\rho_3(x^{(\pi)}_{i,k})$. Then $\rho_3(\overline{z}^{(\pi)}_{i,\ell,k}) = \rho_3(\overline{x}^{(\pi)}_{i,k})$ when $\rho_3(y^{(\pi)}_{k,\ell}) = 1$.
    \end{claimproof}

    There are at most $4m N 2^{\kappa} \card{\Pi}$ distinct maximal stars in $\Pi$: for every term $t$ choose a triple $(\pi,\ell,k)$ and consider two induced stars $\{ \overline{z}^{(\pi)}_{i,\ell,k} : i \in [m] \} \cap \vars(t)$ and $\{ \overline{z}^{(\pi)}_{\ell,j,k} : j \in [m] \} \cap \vars(t)$. Together with the union bound, the above claim shows that with probability at least $1/2$, the following holds for every term $t$ appearing in $\Pi$:
    \begin{itemize}
        \item For every $\pi \in \Q^{\le\kappa}$, $k \in [N]$ with $\eta(k) \neq n + 1$, $i \in [m]$, if the star $\{ \overline{z}^{(\pi)}_{i,j,k} : j \in [m] \} \cap \vars(t)$ has at least $\tilde{d}$ rays, then there exists $\tilde j\in[m]$ such that $\overline z^{(\pi)}_{i,\tilde j,k}$ appears in $t$ and
        \[
            \rho_3(y^{(\pi)}_{k,\tilde j})=1.
        \]
        \item For every $\pi \in \Q^{\le\kappa}$, $k \in [N]$ with $\eta(k) \neq n + 1$, $j \in [m]$, if the star $\{ \overline{z}^{(\pi)}_{i,j,k} : i \in [m] \} \cap \vars(t)$ has at least $\tilde{d}$ rays, then there exists $\tilde i\in[m]$ such that $\overline z^{(\pi)}_{\tilde i,j,k}$ appears in $t$ and
        \[
            \rho_3(x^{(\pi)}_{\tilde i,k})=1.
        \]
    \end{itemize}
    For the rest of the proof, fix such $\rho_3$.

    In what follows we omit the subscript $\tilde{d}$ and simply write $R$ to denote $R_{\tilde{d}}$.
    \begin{claim}\label{clm: good star property implies R(t)}
        For every term $t$ appearing in $\Pi$, we have
        \[
            R(t) \restriction \rho_3 = t \restriction \rho_3.
        \]
    \end{claim}
    \begin{claimproof}
        We use the fact the conditions from \cref{clm: rho3 stars property} hold for every term $t$ in $\Pi$ due to the choice of $\rho_3$.
        Let $t$ be a term in $\Pi$. $R$ leaves all variables that are not negated $z$-variables unchanged, so we only need to consider the negative occurrences of $z$-variables in $t$. Moreover, for a fixed $\pi \in \Q^{\le\kappa}$ and $k \in [N]$ with $\eta(k) \neq n + 1$, $R$ acts independently on the variable sets $\{ \overline{z}^{(\pi)}_{i,j,k} : i, j \in [m] \}$.

        Fix $\pi \in \Q^{\le\kappa}$ and $k \in [N]$ with $\eta(k) \neq n + 1$. Let $t'$ be the subterm of $t$ consisting of all negative occurrences of $z$-variables of the form $\overline{z}^{(\pi)}_{i,j,k}$ for some $i,j \in [m]$. It suffices to show that $R(t') \restriction \rho_3 = t' \restriction \rho_3$.
        Let $E = \{(i, j) : \overline{z}^{(\pi)}_{i,j,k} \text{ appears in } t'\}$.

        Let $C_\mathit{left}$ be the set of all centres $i \in [m]$ such that the left-star $(i, \{ j \in [m] : (i, j) \in E \})$ has at least $\tilde{d}$ rays, and let $C_\mathit{right}$ be the set of all centres $j \in [m]$ such that the right-star $(\{ i \in [m] : (i, j) \in E \}, j)$ has at least $\tilde{d}$ rays. By the choice of $\rho_3$, for every $i \in C_\mathit{left}$ there exists $j_i\in[m]$ such that $\overline z^{(\pi)}_{i,j_i,k}$ appears in $t'$ and $\rho_3(y^{(\pi)}_{k,j_i})=1$. Similarly, for every $j \in C_\mathit{right}$ there exists $i_j\in[m]$ such that $\overline z^{(\pi)}_{i_j,j,k}$ appears in $t'$ and $\rho_3(x^{(\pi)}_{i_j,k})=1$.

        Hence we have
        \[
            t' = \prod_{(i, j) \in E} \overline{z}^{(\pi)}_{i,j,k}
        \]
        and
        \begin{equation}\label{eq: R(t')}
            R(t') = \prod_{i \in C_\mathit{left}} \overline{x}^{(\pi)}_{i,k} \cdot \prod_{j \in C_\mathit{right}} \overline{y}^{(\pi)}_{k,j} \cdot \prod_{\substack{(i, j) \in E \\ i \notin C_\mathit{left},\,j \notin C_\mathit{right}}} \overline{z}^{(\pi)}_{i,j,k}.
        \end{equation}

        If for some $i \in C_\mathit{left}$ we have $\rho_3(x^{(\pi)}_{i,k}) = 1$, then $R(t') \restriction \rho_3 = 0$. Our choice of $j_i$ ensures that $\rho_3(\overline{z}^{(\pi)}_{i,j_i,k}) = 1 - \rho_3(x^{(\pi)}_{i,k}) \cdot \rho_3(y^{(\pi)}_{k,j_i}) = 0$, and hence $t' \restriction \rho_3 = 0$ as well. The same argument applies to every $j \in C_\mathit{right}$ with $\rho_3(y^{(\pi)}_{k,j}) = 1$. The claim is proved if either of these cases occurs.
        Otherwise, we have $j_i \notin C_\mathit{right}$ for every $i \in C_\mathit{left}$ and $i_j \notin C_\mathit{left}$ for every $j \in C_\mathit{right}$. We can equivalently rewrite $t'$ as
        \begin{equation}\label{eq: t' rewritten}
            t'
            = \prod_{i \in C_\mathit{left}} \prod_{j : (i, j) \in E} \overline{z}^{(\pi)}_{i,j,k}
            \cdot \prod_{j \in C_\mathit{right}} \prod_{\substack{i : (i, j) \in E \\ i \notin C_\mathit{left}}} \overline{z}^{(\pi)}_{i,j,k}
            \cdot \prod_{\substack{(i, j) \in E \\ i \notin C_\mathit{left},\, j \notin C_\mathit{right}}} \overline{z}^{(\pi)}_{i,j,k}.
        \end{equation}

        Fix $i \in C_\mathit{left}$. By definition of $\rho_3$, we have $\rho_3(\overline{z}^{(\pi)}_{i,j_i,k}) = \rho_3(\overline{x}^{(\pi)}_{i,k})$. Moreover, for every $j \in [m]$, we have 
        \begin{align*}
            \rho_3(\overline{z}^{(\pi)}_{i,j,k}) \cdot \rho_3(\overline{x}^{(\pi)}_{i,k}) &= (1 - \rho_3(x^{(\pi)}_{i,k}) \cdot \rho_3(y^{(\pi)}_{k,j})) \cdot (1 - \rho_3(x^{(\pi)}_{i,k})) \\
            &= 1 - \rho_3(x^{(\pi)}_{i,k}) \cdot \rho_3(y^{(\pi)}_{k,j}) - \rho_3(x^{(\pi)}_{i,k}) + \rho_3^2(x^{(\pi)}_{i,k}) \cdot \rho_3(y^{(\pi)}_{k,j}) \\
            &= 1 - \rho_3(x^{(\pi)}_{i,k}) = \rho_3(\overline{x}^{(\pi)}_{i,k})
        \end{align*}
        modulo the Boolean axioms.
        Thus,
        \[
            \prod_{j : (i, j) \in E} \rho_3(\overline{z}^{(\pi)}_{i,j,k}) = \rho_3(\overline{z}^{(\pi)}_{i,j_i,k}) \cdot \prod_{\substack{j : (i, j) \in E \\ j \neq j_i}} \rho_3(\overline{z}^{(\pi)}_{i,j,k}) = \rho_3(\overline{x}^{(\pi)}_{i,k}) \cdot \prod_{\substack{j : (i, j) \in E \\ j \neq j_i}} \rho_3(\overline{z}^{(\pi)}_{i,j,k}) = \rho_3(\overline{x}^{(\pi)}_{i,k}).
        \]

        Now fix $j \in C_\mathit{right}$. By definition of $\rho_3$, we have $\rho_3(\overline{z}^{(\pi)}_{i_j,j,k}) = \rho_3(\overline{y}^{(\pi)}_{k,j})$. Moreover, for every $i \in [m]$, we have $\rho_3(\overline{z}^{(\pi)}_{i,j,k}) \cdot \rho_3(\overline{y}^{(\pi)}_{k,j}) = \rho_3(\overline{y}^{(\pi)}_{k,j})$. Thus,
        \[
            \prod_{\substack{i : (i, j) \in E \\ i \notin C_\mathit{left}}} \rho_3(\overline{z}^{(\pi)}_{i,j,k}) = \rho_3(\overline{y}^{(\pi)}_{k,j}).
        \]

        It implies that equations \eqref{eq: R(t')} and \eqref{eq: t' rewritten} are the same under $\rho_3$ modulo the Boolean axioms, and the claim follows.
    \end{claimproof}

    Now we prove the lemma by induction on the proof lines in $\Pi$. Let $f$ be a proof line in $\Pi$.
    \begin{itemize}
        \item If $f$ is an axiom, then $R(f) = f$.
        After applying $\rho_3$, this axiom is either an identity, or a simple consequence of the axioms of $\widetilde{\phi}_G(\bfA \restriction \rho_3,\kappa)$, by the definition of $\rho_3$.

        \item If $f$ is obtained from polynomials $g$ and $h$ as $f = \alpha g + \beta h$, then clearly $R(f) = \alpha R(g) + \beta R(h)$ by linearity of $R$.

        \item The remaining case is multiplication by a variable $\theta$: $f = g \cdot \theta$.
        Applying \cref{clm: good star property implies R(t)} term-wise to $f$ and $g$, we obtain
        \[
            R(g) \restriction \rho_3 = g \restriction \rho_3
        \]
        and
        \[
            R(f) \restriction \rho_3 = f \restriction \rho_3 = (g \restriction \rho_3) \cdot (\theta \restriction \rho_3) = (R(g) \restriction \rho_3) \cdot (\theta \restriction \rho_3).
        \]
        Hence $R(f) \restriction \rho_3$ can be derived from $R(g) \restriction \rho_3$ by multiplying by $\theta \restriction \rho_3$, which is polynomial of degree at most $2$. This can be done in $\PC$ by multiplying by the monomials of $\theta \restriction \rho_3$ and taking linear combinations and can result in an increase of degree by at most a constant.
    \end{itemize}
    The absorption identities used in the multiplication step above are consequences of the Boolean axioms and can be derived in $\PC$ in constant degree. Hence the resulting derivation has degree at most $d+O(1)$.
\end{proof}

Finally, we show how $\widetilde{\phi}_{G}(\bfA, \kappa)$ can be reduced to $\FPHP(G)$.

\begin{definition}[Reduction $\tau$]\label{def: reduction tau for phi tilda}
    Let $\kappa \ge 0$, $m,n>0$ and let $G$ be an $m \times n$ bipartite graph with parts $[m]$ and $[n]$.
    Let $P = {(p_{i,k})}_{i \in [m], k \in \calN_{G}(i)}$ be fresh variables.\footnote{It is useful to think of $P$ as an $m \times n$ variable matrix, where $P_{i,k} = 0$ for all $i \in [m]$ and $k \notin \calN_{G}(i)$. Then $\tau(x^{(\pi)}_{i,k}) = {((A^{(\pi)} \restriction \tau) P)}_{i,k}$.}
    The reduction $\tau$ for $\widetilde{\phi}_{G}(\bfA, \kappa)$ is defined as follows.
    For every $\pi \in \Q^{\le\kappa}$ in decreasing order:
    \begin{enumerate}
        \item For each $i \in [m]$ and $k \in [n]$ set
        \[
            \tau(x^{(\pi)}_{i,k}) = \sum_{\substack{s \in [m] \\ \text{such that } k \in N_{G}(s)}} \tau(A^{(\pi)}_{i,s}) p_{s,k}.
        \]
        \item For each $j \in [m]$ and $k \in \calN_{G}(j)$ set
        \[
            \tau(y^{(\pi)}_{k,j}) = p_{j,k}.
        \]
    \end{enumerate}
\end{definition}

This definition is well-defined: $A^{(\pi)} \restriction \tau = A^{(\pi)}$ for every $\pi \in \Q^{\kappa}$, because $A^{(\pi)}$ contains only constants in these cases.

\begin{lemma}\label{lem: tau produces refutations of FPHP}
    Let $\kappa \ge 0$, $m,n>0$ and let $G$ be an $m \times n$ bipartite graph with parts $[m]$ and $[n]$ and the maximum left-degree $\Deltamax$.
    Let $\bfA = \{ A^{(\pi)} : \pi \in \Sigma^{\le\kappa} \}$ be a collection of $m \times m$ matrices as in \cref{def: iterated algebraic phi tilde}.
    Then for any $\PCR_{\F_2}$ refutation $\Pi$ of $\widetilde{\phi}_G(\bfA, \kappa)$ with degree $d$, $\Pi \restriction \tau$ can be transformed into a refutation of $\FPHP(G)$ with degree at most $\max(\Deltamax, d(\kappa+1))$.
\end{lemma}
\begin{proof}

It follows by induction on $\card{\pi}$ that for every $\pi \in \Q^{\le\kappa}$, $i \in [m]$, and $k \in [n]$, $\tau(x^{(\pi)}_{i,k})$ (and thus $\tau(\overline{x^{(\pi)}_{i,k}})$) is a polynomial of degree at most $\kappa-\card{\pi}+1$.

This implies that all polynomials from $\widetilde{\phi}_G(\bfA, \kappa) \restriction \tau$ can be easily derived from the axioms of $\FPHP(G)$.
\begin{claim}
    Every polynomial from $\widetilde{\phi}_G(\bfA, \kappa) \restriction \tau$ has a degree-$(\kappa+2)$ proof from $\FPHP(G)$.
\end{claim}
\begin{claimproof}
    As shown above, all the entries of $A^{(\pi)} \restriction \tau$ are polynomials of degree at most $\kappa - \card{\pi}$.

    Let $\pi \in \Sigma^{\le\kappa}$ and $i, j \in [m]$. The $(\pi,i,j)$th axiom of $\widetilde{\phi}_G(\bfA, \kappa)$ under $\tau$ transforms as follows:
    \begin{align*}
        \left( \sum_{k \in \calN_{G}(j)} x^{(\pi)}_{i,k} y^{(\pi)}_{k,j} + A^{(\pi)}_{i,j} \right) \restriction \tau &=
        \sum_{k \in \calN_{G}(j)} \left( \sum_{\substack{s \in [m] \\ \text{such that } k \in N_{G}(s)}} \tau(A^{(\pi)}_{i,s}) p_{s,k} \right) p_{j,k} + \tau(A^{(\pi)}_{i,j}) \\
        &= \sum_{s \in [m]} \tau(A^{(\pi)}_{i,s}) \left( \sum_{k \in \calN_{G}(j) \cap N_{G}(s)} p_{s,k} p_{j,k} \right) + \tau(A^{(\pi)}_{i,j}) \\
        &= \sum_{s \in [m] \setminus \{j\}} \tau(A^{(\pi)}_{i,s}) \left( \sum_{k \in \calN_{G}(j) \cap N_{G}(s)} p_{s,k} p_{j,k} \right) \\
        &\quad +
        \tau(A^{(\pi)}_{i,j}) \left( \sum_{k \in \calN_{G}(j)} p_{j,k} + 1 \right),
    \end{align*}
    which is trivially derivable in degree $\kappa-\card{\pi}+2$ from the hole axioms $p_{s,k} p_{j,k}$ and the polynomial
    \[
        \sum_{k \in \calN_{G}(j)} p_{j,k} + 1.
    \]
    Note that the latter is an immediate consequence of the pigeon axiom for pigeon $j$ together with its functional axioms, and thus has a derivation of degree at most $\Deltamax$ from $\FPHP(G)$.

    Thus, we can extend $\Pi \restriction \tau$ to a refutation of $\FPHP(G)$ of degree at most $\max(\Deltamax, d(\kappa+1))$.
\end{claimproof}

\end{proof}

Finally, we are ready to prove \cref{thm: lower bound on iterated formula psi}.

\begin{proof}[Proof of \cref{thm: lower bound on iterated formula psi}]

Let
\begin{align*}
    c &= \Deltamin/6-d, \\
    d_1 &= \frac{cn^\delta}{32 \Deltamax (\kappa + 1)}, \\
    d_2 &= d_1^{1/3} = {\left( \frac{cn^\delta}{32 \Deltamax (\kappa + 1)} \right)}^{1/3}.
\end{align*}
Note that $d_1 = \Theta(n^\delta / \kappa)$ and $d_2 = \Theta({(n^\delta / \kappa)}^{1/3})$ because of our choice of $\Deltamin$ and $\Deltamax$.

Let $\Pi$ be a $\PCR_{\F_2}$ refutation of $\psi_{P(H(G))}(\bfA, \kappa)$. Suppose, for the sake of contradiction, that $\card{\Pi} \le 2^{\epsilon d_2}$ for some small $\epsilon > 0$. We will fix the value of $\epsilon$ later.

We start by applying the restriction $\rho_1$ from \cref{def: substitution rho_1 for psi} to $\Pi$.
Let $\Pi_1 = \Pi \restriction \rho_1$.
By \cref{lem: rho1 reduces u-degree} and the union bound, there exists $\rho_1$ such that every term $t$ in $\Pi_1$ contains at most $d_1$ $u$-variables and at most $d_1$ positively occurring $z$-variables. This holds since the maximum left-degree of $H(G)$ is $8\Deltamax+8$ and the size of $\Pi$ is less than ${(4/3)}^{\sqrt{\frac{d_1}{4(8\Deltamax+8)}}} = 2^{\Theta({(\frac{d_1}{\log{n}})}^{1/2})}$ for a sufficiently small $\epsilon$.
By \cref{lem: rho1 preserves refutations}, $\Pi_1$ is a refutation of $\phi_{H(G)}(\bfA \restriction \rho_1, \kappa)$.
Since $\rho_1$ substitutes only constants and literals, the size and degree of $\Pi_1$ are at most the size and degree of $\Pi$, respectively.

We proceed by applying the restriction $\rho_2$ from \cref{def: substitution rho_2 for phi} to $\Pi_1$.
Let $\Pi_2 = \Pi_1 \restriction \rho_2$.
By \cref{lem: rho2 reduces column degree} and the union bound, with probability greater than $1/2$, for every term $t$ in $\Pi_2$ it holds that $t$ mentions at most $d_2$ columns using $x$-, $y$- or $z$-variables; that is, there is a set $K \subseteq [n]$ of size at most $d_2$ such that no $k \notin K$ is $X^{(\pi)}$-column-mentioned, ${(Y^{(\pi)})}^T$-column-mentioned or $Z^{(\pi)}$-column-mentioned in $t$.
This holds since the size of $\Pi_1$ is less than $\frac{1}{2} {(3/2)}^{d_2} = 2^{\Theta(d_2)}$ for a sufficiently small $\epsilon$.
% \hanlin{Why is it true that size of $\Pi_1$ is smaller than $2^{\Theta(d_2)}$? If I understand correctly, the size of $\Pi_1$ should be $2^{\Theta((\frac{d_1}{\log n})^{1/2})}$?}
% \slava{At the beginning of the theorem we say that the size of $\Pi$ is at most $2^{\epsilon d_2}$ and substitutions $\rho_1$ and $\rho_2$ do not drastically increase the size (there can be a blow-up by a polynomial factor in the worst case). When we apply $\rho_1$ we only need $\Pi$ to be smaller than $2^{O((\frac{d_1}{\log n})^{1/2})}$ though.}
By \cref{prop: rho2 gives good expander}, the graph $G'$ is an $(n^\delta, c)$-boundary expander with probability at least $1/2$.
Thus, by the union bound, there is $\rho_2$ for which both of these events occur.
By \cref{lem: rho2 preserves refutations}, $\Pi_2$ is a refutation of $\phi_{H(G')}(\bfA \restriction \rho_1 \restriction \rho_2, \kappa)$.
Since $\rho_2$ substitutes only constants and literals, the size and degree of $\Pi_2$ are at most the size and degree of $\Pi_1$ (and thus also of $\Pi$), respectively.

Next, we apply the restriction $\rho_3$ from \cref{def: substitution rho_3 for phi} to $\Pi_2$.
Let $\tilde{d} = d_2$ and $\Pi' = R_{d_2}(\Pi_2) \restriction \rho_3$.
By \cref{lem: rho3 reduces xyz-degree} and the union bound, with probability larger than $1/2$, every term $t$ of $R_{d_2}(\Pi_2)$ satisfying $t\restriction\rho_3\neq0$ contains at most $d_1$ $x$- or $y$-variables and at most $d_1$ negative $z$-variables.
Together with the bounds obtained after $\rho_1$, this implies that every term of $\Pi'$ has degree at most $8d_1$.
This is because $\card{K}$ is less than $d_2$ due to $\rho_2$, and the size of $\Pi_2$ is less than $\frac{1}{2}{(16/15)}^{\frac{d_1}{16d_2^2}} = 2^{\Theta(d_2)}$ for a sufficiently small $\epsilon$.

Furthermore, since $\delta>4\alpha$, we have $d_2 = \omega(\kappa+\log(mN))$. Hence, by choosing
$\epsilon$ sufficiently small, the size assumption
\[
    \card{\Pi_2} \le \frac{(4/3)^{d_2}}{8mN2^\kappa}
\]
of \cref{lem: rho3 eliminates stars correctly} is satisfied.
Thus, $\Pi'$ can be extended to a refutation $\Pi_3$ of $\widetilde{\phi}_{G'}(\bfA \restriction \rho_1 \restriction \rho_2 \restriction \rho_3, \kappa)$ of degree at most $8d_1 + O(1)$.
Thus, by the union bound, there is $\rho_3$ for which both of the above events happen and $\Pi_3$ is a valid refutation of $\widetilde{\phi}_{G'}(\bfA \restriction \rho_1 \restriction \rho_2 \restriction \rho_3, \kappa)$ of degree at most $8d_1+O(1)$.

Finally, we apply the reduction $\tau$ from \cref{def: reduction tau for phi tilda} to $\Pi_3$.
% Let $\Pi_4$ be a $\PC$ refutation obtained by extending $\Pi_3 \restriction \tau$ to a valid refutation (since we replace some variables with polynomials, we need to simulate some of the multiplication rules).
By \cref{lem: tau produces refutations of FPHP}, $\Pi_3 \restriction \tau$ can be extended to a refutation $\Pi_4$ of $\FPHP(G')$. The degree of $\Pi_4$ is at most $8(d_1+O(1))(\kappa+1) = \frac{cn^\delta}{4\Deltamax}+O(\kappa) \le \frac{cn^\delta}{2\Deltamax}$. This contradicts the degree bound from \cref{prop: fphp degree bound}.
\end{proof}

%!TEX root = main.tex

\section{Proof Complexity Generators for \texorpdfstring{$\SA$}{SA}}\label{sec:generator sa}

In this section, we will state our results for a stronger formula $\simpleBTRank^m_n(A)$, which is similar to $\BTRank^m_n(A)$ except that it does not use $z$-variables. Formally, it is defined as follows.

\begin{definition}[The CNF encoding $\simpleBTRank^m_n(A)$]
    Assume $m > n$ are positive integers and $A \in \Q^{m \times m}$ is an $m$ by $m$ Boolean matrix.
    The (simple) bamboo-tree CNF encoding, denoted $\simpleBTRank^m_n(A)$, uses input variables $x_{i,k}, y_{k,i}: i \in [m], k \in [n]$ together with extension variables $u_{i,j,k} : i,j \in [m], k \in [n]$, and consists of the following axioms. 

    \textbf{Output Axioms:} For every $i,j \in [m]$, the clause $u_{i,j,n}$ if $A_{i,j} = 1$ and the clause $\neg u_{i,j,n}$ if $A_{i,j} = 0$.

    \textbf{Summation Base Axioms:} For every $i,j \in [m]$, $u_{i,j,1} = x_{i,1} y_{1,j}$, encoded by the clauses $u_{i,j,1} \lor \neg x_{i,1} \lor \neg y_{1,j}$, $x_{i,1} \lor \neg u_{i,j,1}$ and $y_{1,j} \lor \neg u_{i,j,1}$.

    \textbf{Summation Axioms:} For every $i,j \in [m]$, and $k\in \{2,\ldots,n\}$, $u_{i,j,k} = u_{i, j, k-1} + x_{i,k}y_{k,j}$, encoded by the following six clauses:
        \begin{align*}
            & x_{i,k} \lor \neg u_{i,j,k-1} \lor u_{i,j,k}, \\
            & x_{i,k} \lor u_{i,j,k-1} \lor \neg u_{i,j,k}, \\
            & y_{k,j} \lor \neg u_{i,j,k-1} \lor u_{i,j,k}, \\
            & y_{k,j} \lor u_{i,j,k-1} \lor \neg u_{i,j,k}, \\
            & \neg x_{i,k} \lor \neg y_{k,j}  \lor \neg u_{i,j,k-1} \lor \neg u_{i,j,k}, \\
            & \neg x_{i,k} \lor \neg y_{k,j}  \lor u_{i,j,k-1} \lor u_{i,j,k}. 
        \end{align*}
\end{definition}

In what follows, it will be useful to additionally enforce that each row of $X$ and each column of $Y$ have odd weight. Let $m > n$ be positive integers and consider a matrix $A \in \Q^{m \times m}$. We extend $A$ to an $(m+1) \times (m+1)$ matrix $\tilde{A}$ by adding a new column and row. For every $i, j \in [m+1]$ we define
\begin{equation}\label{eq: sa: matrix tildeA}
    \tilde{A}_{i,j} = \begin{cases}
        A_{i,j} & i\in[m] \text{ and } j\in[m], \\
        n \bmod 2 & \text{if } i=m+1 \text{ and } j=m+1, \\
        1 & \text{if } i=m+1 \text{ and } j\in[m], \\
        1 & \text{if } i\in[m] \text{ and } j=m+1.
    \end{cases}
\end{equation}
Next, we construct a partial restriction $\sigma$, that we apply to $\simpleBTRank^{m+1}_n(\tilde{A})$ that fixes the last ($(m+1)$th) row of $X$ and the last ($(m+1)$th) column of $Y$ to the all-ones vector. Formally, for every $k \in [n]$, we have $\sigma(x_{m+1,k}) = \sigma(y_{k,m+1}) = 1$, $\sigma(\overline{x}_{m+1,k}) = \sigma(\overline{y}_{k,m+1}) = 0$, $\sigma(u_{m+1,m+1,k}) = (k \bmod 2)$ and $\sigma(\overline{u}_{m+1,m+1,k}) = 1 - (k \bmod 2)$. The variables of the form $u_{m+1,j,k}$, where $j \in [m]$ and $k \in [n]$, and $u_{i,m+1,k}$, where $i \in [m]$ and $k \in [n]$, (and their negated variants) are left unassigned. We denote by ${\simpleBTRank'}^m_n(A)$ the restricted formula $\simpleBTRank^{m+1}_n(\tilde{A}) \restriction \sigma$. Observe that ${\simpleBTRank'}^m_n(A)$ is weaker than $\simpleBTRank^m_n(A)$ since it contains additional restrictions encoding $\sum_{k \in [n]} x_{i,k} = 1$, for all $i \in [m]$, and $\sum_{k \in [n]} y_{k,j} = 1$, for all $j \in [m]$, using the extra $u$-variables $u_{i,j,k}$ with either $i=m+1$ or $j=m+1$, and (augmented) Output, Summation Base and Summation Axioms.

% We will also consider a substitution instance of $\simpleBTRank$, denoted as $\simpleBTRank'$. \michal{The new formula should be properly defined including the specification of its parameters, i.e. to make it explicit whether $\simpleBTRank^m_n(A)$ after the substitution is denoted still as $\simpleBTRank'^m_n(A)$ or  with some other $\tilde{m}$ or $\tilde{A}$.} In this instance, the last row of $X$ and the last column of $Y$ are substituted with ones. Specifically, for every $k \in [n]$, we assign $x_{m,k} = y_{k,m} = 1$ and $u_{m,m,k} = k \bmod 2$. \michal{How about the negations of these variables? The partial assignment $\rho_{I,J}$ does not assign them values.}

Our plan for proving a lower bound for $\simpleBTRank^m_n(A)$ is to reduce it to ${\simpleBTRank'}^{m}_{\tilde{n}}(A)$ for some $\tilde{n} = \Theta(n)$, in such a way that any small $\SA$ refutation is transformed into one of a small row degree (see \cref{def:row degree sa}), and then to establish an explicit row degree lower bound for the latter formula. We start with the latter.
 
\subsection{Row Degree Lower Bound}

We start by proving a row degree bound on ${\simpleBTRank'}^{m}_{n}(A)$.

\begin{definition}
    Let $t$ be a term in the variables of ${\simpleBTRank}'^m_n(A)$.
    We say that $i \in [m]$ is \emph{$X$-row-mentioned in $t$} if there is $k \in [n]$ such that $x_{i,k}$ or $\overline{x}_{i,k}$ appears in $t$.
    We say that $j \in [m]$ is \emph{$Y^T$-row-mentioned in $t$} if there is $k \in [n]$ such that $y_{k,j}$ or $\overline{y}_{k,j}$ appears in $t$.
    We say that $i \in [m]$ is \emph{$U$-left-row-mentioned in $t$} if there are $j \in [m+1]$ and $k \in [n]$ such that $u_{i,j,k}$ or $\overline{u}_{i,j,k}$ appears in $t$.
    We say that $j \in [m]$ is \emph{$U$-right-row-mentioned in $t$} if there are $i \in [m+1]$ and $k \in [n]$ such that $u_{i,j,k}$ or $\overline{u}_{i,j,k}$ appears in $t$.
\end{definition}

\begin{definition}\label{def:row degree sa}[Row degree]
    Let $t$ be a term in the variables of ${\simpleBTRank}'^m_n(A)$.
    Let $I$ be the set of all $i \in [m]$ that are either $X$-row-mentioned or $U$-left-row-mentioned in $t$.
    Similarly, let $J$ be the set of all $j \in [m]$ that are either $Y^T$-row-mentioned or $U$-right-row-mentioned in $t$.
    We say that $t$ \emph{mentions} rows $(I,J)$.
    The \emph{row degree} of $t$ is defined as the sum $\card{I} + \card{J}$.
\end{definition}

Lower bounds on the $\SA$ degree are typically established by constructing a \emph{pseudoexpectation}, which is a linear mapping that satisfies the following properties: it maps all low-degree multilinear polynomials to $\bbR$, vanishes on multiples of the axioms, and is non-zero on the constant-$1$ polynomial. We present such a mapping for multilinear terms with small row degree.

\begin{lemma}\label{lem: SA BTRank rowe degree bound}
    For every $m > n$ and every $A \in \Q^{m \times m}$, every $\SA$ refutation of ${\simpleBTRank}'^m_n(A)$ requires row degree $n-1$.
\end{lemma}

Before proving the lemma, we construct a distribution of the pairs of matrices which will be used in the proof.
\begin{definition}[The random family $\calD_{I,J}$ for ${\simpleBTRank'}^m_n(A)$]\label{def: sa segree random family of matrices}
    Let $I, J \subseteq [m]$, $d = \card{I} + \card{J} \le n-2$ and recall the definition of the matrix $\tilde{A}$ from \eqref{eq: sa: matrix tildeA}.
    We define $\calD_{I,J}$ as the uniform distribution on the pairs of full-rank matrices $M \in \Q^{(I \cup \{m+1\}) \times [n]}$ and $N \in \Q^{[n] \times (J \cup \{m+1\})}$ that satisfy $MN = \tilde{A}_{I \cup \{m+1\},J \cup \{m+1\}}$ and for every $k \in [n]$, $M_{m+1,k} = N_{k,m+1} = 1$.
\end{definition}

Based on $\calD_{I,J}$ we construct a random assignment $\rho_{I,J}$.
\begin{definition}[The random assignment $\rho_{I,J}$ for ${\simpleBTRank}'^m_n(A)$]\label{def: sa degree random assignment}
    Let $I, J \subseteq [m]$ and $d = \card{I} + \card{J} \le n-2$. We define a random partial assignment $\rho_{I,J}$ as follows.
    \begin{enumerate}
        \item Sample a pair of matrices $(M, N)$ according to $\calD_{I,J}$.
        \item Assign $\rho_{I,J}(X_{I,[n]}) = M_{I,[n]}$, meaning that for every $i \in I$ and $k \in [n]$, set $\rho_{I,J}(x_{i,k}) = M_{i,k}$. Similarly, assign $\rho_{I,J}(Y_{[n],J}) = N_{[n],J}$: for every $j \in J$ and $k \in [n]$, set $\rho_{I,J}(y_{k,j}) = N_{k,j}$.
        \item For every $(i,j) \in ((I \cup \{m+1\}) \times (J \cup \{m+1\}) \setminus \{(m+1,m+1)\})$, and $k \in [n]$, set 
        \[
            \rho_{I,J}(u_{i,j,k}) = \left( \sum_{\ell=1}^k M_{i,\ell} N_{\ell,j} \bmod 2\right).
        \]
        \item For every variable $\theta$ assigned by $\rho_{I,J}$, set $\rho_{I,J}(\overline{\theta}) = 1 - \rho_{I,J}(\theta)$.
    \end{enumerate}
\end{definition}

A key property of $\rho_{I,J}$ is that it remains consistent when restricted to smaller sets of rows. Let $I, J \subseteq [m]$ with $\card{I} + \card{J} \le n-2$ and consider any two subsets $I' \subseteq I$ and $J' \subseteq J$. Then, for every term $t$ that mentions rows $(I',J')$, we have
\begin{equation}\label{eq: sa degree property}
    \Prb{\rho_{I',J'}(t) = 1} = \Prb{\rho_{I,J}(t) = 1}.
\end{equation}
\Iddo{put here the condition for technical section in intro.}
% We show this by describing a random process that iteratively samples $\rho_{I,J}$.
We show this by induction on $\card{I \setminus I'} + \card{J \setminus J'}$.
The base case is trivial and the inductive step follows from the following lemma.
\begin{lemma}
    Let $m > n$ be positive integers and let $A \in \Q^{m \times m}$ be a Boolean matrix.
    Let $I, J \subseteq [m]$ with $\card{I} + \card{J} < n-2$, $i_0 \in [m] \setminus I$ and $j_0 \in [m] \setminus J$. Then for every term $t$ that mentions rows $(I,J)$ we have
    \[
        \Prb{\rho_{I,J}(t) = 1} = \Prb{\rho_{I \cup \{i_0\},J}(t) = 1} = \Prb{\rho_{I,J \cup \{j_0\}}(t) = 1}.
    \]
\end{lemma}
\begin{proof}
    By symmetry, it is sufficient to show
    \[
        \Prb{\rho_{I,J}(t) = 1} = \Prb{\rho_{I,J \cup \{j_0\}}(t) = 1}.
    \]
    Let $\calD'$ be the following distribution.
    \begin{enumerate}
        \item\label{item: sa sampling 1} Sample a pair of matrices $(M, N)$ according to $\calD_{I,J}$.
        \item\label{item: sa sampling 2} Choose uniformly at random a vector $v \in \Q^n$ such that $Mv = \tilde{A}_{I \cup \{m+1\},j_0}$ and the matrix $N' = \begin{bmatrix} N & v \end{bmatrix}$ has full rank (that is, $\rank(N') = \card{J}+2$).
        \item Output the pair $(M, N')$.
    \end{enumerate}
    Showing that $\calD'$ is the same as $\calD_{I,J\cup \{j_0\}}$ concludes the proof.

    Fix a pair of matrices $(M,N)$ resulting from sampling from $\calD_{I,J}$ in \cref{item: sa sampling 1}.
    The number of choices of a vector $v \in \Q^n$ that satisfy $Mv = \tilde{A}_{I \cup \{m+1\},j_0}$ is $2^{n-\card{I}-1}$. This is because $M$ has full rank. Given such $v$, the rank of the matrix $N' = \begin{bmatrix} N & v \end{bmatrix}$ is either $\card{J}+1$ or $\card{J}+2$. The former happens if and only if $v$ is linearly dependent with the columns of $N$. That is, there exists a vector $\mu \in \Q^{J \cup \{m+1\}}$ with $N\mu = v$. Moreover, each such $\mu$ corresponds uniquely to $v$ since the rank of $N$ is full.
    
    Substituting $N\mu = v$ into $Mv = \tilde{A}_{I \cup \{m+1\},j_0}$, we conclude that $\mu$ must satisfy $MN\mu = \tilde{A}_{I \cup \{m+1\},j_0}$, which is equivalent to $\tilde{A}_{I \cup \{m+1\},J \cup \{m+1\}} \mu = \tilde{A}_{I \cup \{m+1\},j_0}$ due to our choice of $(M,N)$. Let $\nu$ be the number of vectors $\mu$ satisfying this last equality. Observe that $\nu \le 2^{\card{J}+1}$ because of the dimension of $\mu$.

    Thus, there are $\nu \le 2^{\card{J}+1}$ vectors $v$ satisfying $Mv = \tilde{A}_{I \cup \{m+1\},j_0}$ and $\rank(N') = \card{J}+1$. For the remaining $2^{n-\card{I}-1} - \nu > 0$ vectors, we have $\rank(N') = \card{J}+2$. This implies that the definition of $\calD'$ is correct (that is, the vectors $v$ from \cref{item: sa sampling 2} exist). Moreover, the number of choices of such $v$ -- that is, $2^{n-\card{I}-1} - \nu$ -- is independent of the initial choice of $(M, N)$ and is determined by $A$, $I$, $J$, and $j_0$.

    Every pair of matrices in $\calD_{I,J\cup\{j_0\}}$ has a unique restriction to $\calD_{I,J}$, and every pair of matrices in $\calD_{I,J}$ has the same number $2^{n-\card{I}-1}-\nu$ of admissible extensions. Therefore, $\calD'$ is the uniform distribution $\calD_{I,J\cup\{j_0\}}$.

    Observe that this also implies that each distribution $\calD_{I,J}$ is non-empty: clearly $\calD_{\varnothing,\varnothing}$ is non-empty and the above induction step shows that if $\calD_{I,J}$ is non-empty, then so are $\calD_{I \cup \{i_0\},J}$ and $\calD_{I,J \cup \{j_0\}}$.
\end{proof}

\begin{proof}[Proof of \cref{lem: SA BTRank rowe degree bound}]

Let $S_{n-2}$ be the vector space of all multilinear polynomials in the variables of ${\simpleBTRank}'^m_n(A)$ with row degree at most $n-2$.
That is, $S_{n-2} \subseteq \bbR[\vec{x},\vec{y},\vec{u}] / (\vec{x}^{\,2}-\vec{x},\vec{y}^{\,2}-\vec{y},\vec{u}^{\,2}-\vec{u})$.
We construct a linear mapping
\[
    R\colon S_{n-2} \to \bbR
\]
that satisfies the following properties.
\begin{itemize}
    \item $R(1) = 1$.
    \item $R(t) \geq 0$ for every term $t$ with row degree at most $n-2$.
    \item For every axiom $\ax$ and every term $t$ such that $\ax \cdot t$ has row degree at most $n-2$, $R(\ax \cdot t) = 0$.
\end{itemize}
This mapping rules out the existence of a refutation of ${\simpleBTRank}'^m_n(A)$ with row degree at most $n-2$. Indeed, in any $\SA$ refutation of ${\simpleBTRank}'^m_n(A)$, the left-hand side becomes non-negative under $R$, while the right-hand side evaluates to $R(-1) = -1$, leading to a contradiction.

Let $t$ be a term in the variables of ${\simpleBTRank}'^m_n(A)$ with row degree at most $n-2$; namely, it mentions rows $(I,J)$ with $\card{I} + \card{J} \le n-2$. We define $R(t) = \Prb{\rho_{I,J}(t) = 1}$.
Note that $\rho_{I,J}$ assigns values to all the variables in $t$, which implies that $\rho_{I,J}(t) \in \Q$.
We then extend $R$ to all multilinear polynomials of row degree at most $n-2$ linearly.

It immediately follows from the definition that $R(1) = 1$ and $R(t) \geq 0$ on all terms of row degree at most $n-2$.

% \michal{``We show .. `the' pair.." does not make much sense, and it is unclear which matrices are being referred to by the definite article. E.g. first, the property of $R$ that we need can be stated. Then, we can say that in order to prove it, we define a sampling that gives the uniform distribution in point 1 above. See also a different comment below that suggests to first treat $\rho$ in a separate lemma and to derive the properties of $R$ afterwards.}

\eqref{eq: sa degree property} simplifies verifying the last properties of $R$. Let $t$ be a term and $\ax$ an axiom of ${\simpleBTRank}'^m_n(A)$.
Suppose the polynomial $t \cdot \ax$ mentions rows $(I,J)$ with $\card{I} + \card{J} \le n-2$. Expanding $t \cdot \ax$ as a sum of monomials $\sum_{\ell=1}^q \alpha_\ell t_\ell$, where each monomial $t_\ell$ mentions rows $(I_\ell,J_\ell)$ with $I_\ell \subseteq I$ and $J_\ell \subseteq J$, we have:
\begin{align*}
    R(t \cdot \ax) &= R\parens*{\sum_{\ell=1}^q \alpha_\ell t_\ell} = \sum_{\ell=1}^q \alpha_\ell R(t_\ell) = \sum_{\ell=1}^q \alpha_\ell \Prb{\rho_{I_\ell, J_\ell}(t_\ell) = 1} = \sum_{\ell=1}^q \alpha_\ell \Prb{\rho_{I, J}(t_\ell) = 1} \\
    &= \sum_{\ell=1}^q \alpha_\ell \Exp[\rho_{I, J}(t_\ell)] = \Exp\brackets*{\sum_{\ell=1}^q \alpha_\ell \rho_{I, J}(t_\ell)} = \Exp[\rho_{I, J}(t \cdot \ax)] = 0.
\end{align*}
The last equality holds because $t \cdot \ax$ evaluates to zero under \emph{every} choice of matrices $M$ and $N$ in $\rho_{I,J}$: the $x$- and $y$-variables are assigned values from $M$ and $N$, respectively, and the $u$-variables are assigned the corresponding partial inner products.
Since $MN = \tilde{A}_{I \cup \{m+1\},J \cup \{m+1\}}$, every relevant axiom of ${\simpleBTRank}'^m_n(A)$ is satisfied, and thus $t \cdot \ax$ evaluates to zero.
\end{proof}

\subsection{Size to Row Degree Reduction}

We define a random substitution that transforms a refutation of $\simpleBTRank^m_n(A)$ into a refutation of ${\simpleBTRank'}^m_{\tilde{n}}(A)$ (for some $\tilde{n} = \Theta(n)$) while decreasing the row degree of the terms appearing in the refutation. Recall the matrices $\calD$ and $\calE$ (they were already defined at the beginning of \cref{sec: iterability of Rank}).  
\begin{equation*}
    \calD = \begin{pmatrix}
        1 & * & 1 & 0 & 0 & * & * & * \\
        0 & 1 & * & 1 & * & * & * & 0 \\
        * & * & 0 & * & 1 & 0 & 1 & * \\
        * & 0 & * & * & * & 1 & 0 & 1
    \end{pmatrix},
    \quad
    \calE = \begin{pmatrix}
        1 & 0 & 0 & 1 & * & * & * & * \\
        0 & * & 1 & * & 1 & * & 0 & * \\
        * & 1 & * & 0 & * & 0 & * & 1 \\
        * & * & * & * & 0 & 1 & 1 & 0
    \end{pmatrix}.
\end{equation*}

The following definition is similar to \cref{def:first_restriction_BTRank}.
\begin{definition}[Random substitution for $\simpleBTRank^m_n(A)$]\label{def: random sub for simpleBTRank}
    Assume $m > n \ge 20$ and $(n-4)$ is divisible by $16$ and let $\tilde{n} = (n-4)/8$. Let $A \in \Q^{m \times m}$ be a Boolean matrix.
    % We define a random restriction $\rho$ for $\simpleBTRank^m_n(A)$.
    Consider fresh variables
    $\{\tilde{x}_{i,k} : i \in [m], k \in [\tilde{n}]\}$,
    $\{\tilde{y}_{k,j} : j \in [m], k \in [\tilde{n}]\}$ and
    $\{\tilde{u}_{i,j,k} : (i,j) \in ([m+1] \times [m+1]) \setminus \{(m+1,m+1)\}, k \in [\tilde{n}]\}$, which serve as the variables of ${\simpleBTRank'}^m_{\tilde{n}}$.
    The random substitution $\rho$ is defined as follows.
    
    For each $i \in [m]$, we sample its \emph{row type} $s_i \in [4]$ (which corresponds to a row of $\calD$) and a random bit $\alpha_i \in \Q$. Then we set the first and $(n-1)$th bits of $X_i$ (the $i$th row of the matrix $X$ from $\simpleBTRank^m_n(A)$) to $\alpha_i$ and the second and $n$th bits of $X_i$ to $1$. The remaining $(n-4)$ bits are partitioned into eight equal consecutive parts (corresponding to the columns of $\calD$). That is, given $k \in \{3, \ldots, n-2\}$, let $\eta(k) = ((k-3) \bmod \tilde{n}) + 1$ and $\nu(k) = \lfloor \frac{k-3}{\tilde{n}} \rfloor + 1$. Then $\nu(k)$ gives the index of the part $k$ belongs to and $\eta(k)$ gives its index within this part.
    Each variable $x_{i,k}$ will be replaced with either a constant or $\tilde{x}_{i,\eta(k)}$ according to the value of $\calD_{s_i,\nu(k)}$.
    
    Similarly, for each $j \in [m]$, we choose its row type $r_j \in [4]$ (corresponding to a row of $\calE$) and $\beta_j \in \Q$. Then the first and $(n-1)$th bits of $Y_j$ (the $j$th column of $Y$) are set to $1$, the second and $n$th bits to $\beta_j$ and the rest are set to a constant or $\tilde{y}_{\eta(k),j}$ according to the value of $\calE_{r_j,\nu(k)}$.

    Then the value of the $u$-variables is defined consistently with the above assignments.

    Formally, this is defined by the following random process.
    \begin{enumerate}
        \item For each $i \in [m]$, uniformly and independently sample $s_i, r_i \in [4]$ and $\alpha_i, \beta_i \in \Q$.
        \item For each $i \in [m]$ and $k \in [n]$, set
        \[
            \rho(x_{i,k}) = \begin{cases}
                \alpha_i & \text{if } k \in \{1,n-1\}, \\
                1 & \text{if } k \in \{2,n\}, \\
                \tilde{x}_{i,\eta(k)} & \text{if } k \in \{3, \ldots, n-2\} \text{ and } \calD_{s_i,\nu(k)} = *, \\
                \calD_{s_i,\nu(k)} & \text{if } k \in \{3, \ldots, n-2\} \text{ and } \calD_{s_i,\nu(k)} \in \Q.
            \end{cases}
        \]
        \item For each $j \in [m]$ and $k \in [n]$, set
        \[
            \rho(y_{k,j}) = \begin{cases}
                1 & \text{if } k \in \{1,n-1\}, \\
                \beta_j & \text{if } k \in \{2,n\}, \\
                \tilde{y}_{\eta(k),j} & \text{if } k \in \{3, \ldots, n-2\} \text{ and } \calE_{r_j,\nu(k)} = *, \\
                \calE_{r_j,\nu(k)} & \text{if } k \in \{3, \ldots, n-2\} \text{ and } \calE_{r_j,\nu(k)} \in \Q.
            \end{cases}
        \]
        \item
        For each $i,j \in [m]$ and $k \in [n]$, we define $\rho(u_{i,j,k})$ according to $\sum_{\ell=1}^{k} \rho(x_{i,\ell}) \rho(y_{\ell,j}) \bmod 2$.

        For every $b \in [8]$, define
        \[
            \gamma_{i,j,b} = \begin{cases}
                A_{i, j} & \text{if } (\calD_{s_i,b}, \calE_{r_j,b}) = (*, *), \\
                1 & \text{if } (\calD_{s_i,b}, \calE_{r_j,b}) = (*, 1) \text{ or } (1, *), \\
                0 & \text{otherwise},
            \end{cases}
        \]
        and
        \[
            p_{i,j,b} = \alpha_i + \beta_j +\sum_{b'=1}^{b-1} \gamma_{i,j,b'} \bmod 2.
        \]
        % Given $k \in \{3, \ldots, n-2\}$, let $p_{i,j,k} = \sum_{\ell=1}^{(\nu(k)-1)\tilde{n}+2} \rho(x_{i,\ell}) \rho(y_{\ell,j}) \bmod 2$ simplified under the assumptions $\sum_{\ell=1}^{\tilde{n}} \tilde{x}_{i,\ell} \tilde{y}_{\ell,j} = A_{i,j} \bmod 2$, $\sum_{\ell=1}^{\tilde{n}} \tilde{x}_{i,\ell} = 1 \bmod 2$ and $\sum_{\ell=1}^{\tilde{n}} \tilde{y}_{\ell,j} = 1 \bmod 2$. Note that $p_{i,j,k} \in \Q$.

        Then for each $i,j \in [m]$ and $k \in [n]$, set
        \[
            \rho(u_{i,j,k}) = \begin{cases}
                \alpha_i & \text{if } k = 1, \\
                \alpha_i + \beta_j & \text{if } k = 2, \\
                A_{i,j} + \beta_j & \text{if } k = n-1, \\
                A_{i,j} & \text{if } k = n, \\
                \tilde{u}_{i,j,\eta(k)}^{1-p_{i,j,\nu(k)}} & \text{if } k \in \{3, \ldots, n-2\}, \calD_{s_i,\nu(k)} = \calE_{r_j,\nu(k)} = *, \\
                \tilde{u}_{i,m+1,\eta(k)}^{1-p_{i,j,\nu(k)}} & \text{if } k \in \{3, \ldots, n-2\}, \calD_{s_i,\nu(k)} = *, \calE_{r_j,\nu(k)} = 1, \\
                \tilde{u}_{m+1,j,\eta(k)}^{1-p_{i,j,\nu(k)}} & \text{if } k \in \{3, \ldots, n-2\}, \calD_{s_i,\nu(k)} = 1, \calE_{r_j,\nu(k)} = *, \\
                p_{i,j,\nu(k)} + (\eta(k) \bmod 2) & \text{if } k \in \{3, \ldots, n-2\}, \calD_{s_i,\nu(k)} = \calE_{r_j,\nu(k)} = 1, \\
                p_{i,j,\nu(k)} & \text{if } k \in \{3, \ldots, n-2\}, \calD_{s_i,\nu(k)} = 0 \text{ or } \calE_{r_j,\nu(k)} = 0.
            \end{cases}
        \]
    \end{enumerate}
\end{definition}

We proceed by showing that $\rho$ eliminates every term in the variables of $\simpleBTRank^m_n(A)$ of high row degree.
\begin{lemma}\label{lem: SA BTRank eliminates terms whp}
    Assume $m > n \ge 20$, $(n-4)$ is divisible by $16$ and $A \in \Q^{m \times m}$ is a Boolean matrix. Let $t$ be a term in the variables of $\simpleBTRank^m_n(A)$ and $t'$ be any subterm of $t$ such that $t'$ does not mention columns $\{1,2,n-1,n\}$. Let $(I,J) \subseteq [m] \times [m]$ be the rows that $t'$ mentions. Then $\Prb{t \restriction \rho \neq 0} \le {(\sqrt{7/8})}^{\card{I}+\card{J}}$.
\end{lemma}
\begin{proof}
    Consider the bipartite graph $G$ with parts $I$ and $J$ and edges
    \[
        \{ (i,j) : \exists k \in [n] \text{ such that } u_{i,j,k} \text{ or } \overline{u}_{i,j,k} \text{ appears in } t' \}.
    \]

    Let $t''$ be a minimal subterm of $t'$ that mentions rows $(I,J)$. In terms of $G$, the term $t''$ corresponds to one of its minimal edge covers. Such a cover does not contain any paths of length three, thus its connected components are
    \begin{itemize}
        \item isolated vertices $x_{i,k}$, $\overline{x}_{i,k}$, $y_{k,j}$ or $\overline{y}_{k,j}$,
        \item stars $\prod_{(j,k) \in P} u_{i_0,j,k} \prod_{(j,k) \in N} \overline{u}_{i_0,j,k}$ and $\prod_{(i,k) \in P} u_{i,j_0,k} \prod_{(i,k) \in N} \overline{u}_{i,j_0,k}$.
    \end{itemize}
    The minimality of $t''$ ensures that each of these components mentions its own disjoint set of indices.

    In the first case, we have $\Prb{\rho(x_{i,k}) \neq 0}, \Prb{\rho(\overline{x}_{i,k}) \neq 0}, \Prb{\rho(y_{k,j}) \neq 0}, \Prb{\rho(\overline{y}_{k,j}) \neq 0} \le 3/4$ (because each column of $\calD$ and $\calE$ contains one $0$ and one $1$).

    In the second case, assume that we have a star with $r$ rays, say $S = \prod_{(j,k) \in P} u_{i_0,j,k} \prod_{(j,k) \in N} \overline{u}_{i_0,j,k}$ where $\card{P} + \card{N} = r$. This star $U$-left-row-mentions index $i_0 \in I$ and $U$-right-row-mentions $r$ indices $\{ j : \exists k \text{ such that } (j,k) \in P \sqcup N \} \subseteq J$.
    For every $s \in [4]$ and $\alpha \in \Q$, the events
    \[
        ( u_{i_0,j,k} \restriction \rho \neq 0 \mid s_{i_0} = s, \alpha_{i_0} = \alpha ), \text{ where } (j,k) \in P,
    \]
    and
    \[
        ( \overline{u}_{i_0,j,k} \restriction \rho \neq 0 \mid s_{i_0} = s, \alpha_{i_0} = \alpha ), \text{ where } (j,k) \in N,
    \]
    are independent and each occur with probability at most $7/8$.
    Thus, all of them simultaneously occur with probability at most ${(7/8)}^r \le {(\sqrt{7/8})}^{r+1}$.
    It implies that
    \[
        \Prb{t' \restriction \rho \neq 0} \le \Prb{t'' \restriction \rho \neq 0} \le {\left(\sqrt{7/8}\right)}^{\card{I}+\card{J}}.\qedhere
    \]
\end{proof}

Next, we show that the random substitution $\rho$ transforms any $\SA$ refutation of $\simpleBTRank^m_n(A)$ into a refutation of ${\simpleBTRank'}^m_{(n-4)/8}(A)$.
\begin{lemma}\label{lem: SA BTRank ref remains ref}
    Assume $m > n \ge 20$, $(n-4)$ is divisible by $16$ and $A \in \Q^{m \times m}$ is a Boolean matrix.
    Then for every $\SA$ refutation $\Pi$ of $\simpleBTRank^m_n(A)$, $\Pi \restriction \rho$ can be converted into a $\SA$ refutation of ${\simpleBTRank}'^{m}_{(n-4)/8}(A)$ without increasing its row degree.
\end{lemma}
\begin{proof}
    Observe that due to our choice of $n$, $\tilde{n} = (n-4)/8$ is even.
    Recall the properties of $\calD$ and $\calE$:
    \begin{itemize}
        \item there is an odd number of indices $k \in [8]$ such that $\calD_{i,k} = \calE_{j,k} = *$;
        \item there is an even number of indices $k \in [8]$ such that  $\calD_{i,k} = *$ and $\calE_{j,k} = 1$;
        \item there is an even number of indices $k \in [8]$ such that  $\calD_{i,k} = 1$ and $\calE_{j,k} = *$.
    \end{itemize}
    In particular, these properties ensure that the auxiliary values $p_{i,j,b}$ and $\gamma_{i,j,b}$ are defined consistently. In particular, we have $\sum_{b=1}^8 \gamma_{i,j,b} = A_{i,j} \bmod 2$. This implies that $p_{i,j,8} = \alpha_i + \beta_j + A_{i,j} + \gamma_{i,j,8} \bmod 2$.

    Each Output Axiom is satisfied by $\rho$ since it always assigns $\rho(u_{i,j,n}) = A_{i,j}$.

    Similarly, every Summation Base Axiom is also satisfied by the definition of $\rho$.

    % \michal{Crucial facts used in the proof, such as ``2 divides $(n-4)/8$'' or ``every pair of rows, one from $\mathcal{D}$ and one from $\mathcal{E}$, share a star in exactly three or one columns and have a star meeting a 1 in exactly zero or four columns'' should be stated explicitly.}

    % Every Output Axiom $u_{i,j,n}^{A_{i,j}}$ is either satisfied, or becomes \michal{It cannot turn into these forms, it is always satisfied, as $\rho(u_{i,j,n}) = A_{i,j}$ always.}the Output Axiom of the form $\tilde{u}_{i,j,\tilde{n}}^{A_{i,j}}$, $\tilde{u}_{i,m+1,\tilde{n}}$ or $\tilde{u}_{j,m+1,\tilde{n}}$.

    % Every set of Summation Base Axioms encoding $u_{i,j,1} = x_{i,1} y_{1,j}$ is either satisfied or \michal{Again, this particular axiom is always satisfied, by the definition of $\rho$ at the index $k=1$. It should be stated what axioms are mapped by $\rho$ to the new Output and Summation Base axioms.}encodes the statement of the form $\tilde{u}_{i,j,1} = \tilde{x}_{i,1} \tilde{y}_{1,j}$, $\tilde{u}_{i,m+1,1} = \tilde{x}_{i,1}$ or $\tilde{u}_{m+1,j,1} = \tilde{y}_{1,j}$.

    Now consider a Summation Axiom. For readability, we reason about the parity relation $u_{i,j,k} = u_{i,j,k-1} + x_{i,k} y_{k,j} \bmod 2$ semantically instead of the CNF that encodes it. Each such equation involves only constantly many variables, thus the choice of encoding does not affect the argument: all the derivations below can be stated as constant-size $\SA$ derivations in the CNF encoding without mentioning any additional rows.
    For brevity, we omit the $\bmod 2$ in the analysis below, but it is understood that all equalities are modulo $2$.
    When $k \in \{2,n-1,n\}$, we have the following cases.
    \begin{itemize}
        \item For $k=2$, the axiom is satisfied since it becomes $\alpha_i + \beta_j = \alpha_i + 1 \cdot \beta_j$.
        \item For $k=n$, the axiom is satisfied since it becomes $A_{i,j} = (A_{i,j} + \beta_j) + 1 \cdot \beta_j$.
        \item For $k = n-1$, we have $A_{i,j} + \beta_j = \rho(u_{i,j,n-2}) + \alpha_i \cdot 1$.
        If $(\calD_{s_i,8}, \calE_{r_j,8}) = (*,*)$, then $\gamma_{i,j,8} = A_{i,j}$ and $p_{i,j,8} = \alpha_i + \beta_j \bmod 2$. In this case, $\rho(u_{i,j,n-2}) = \tilde{u}_{i,j,\tilde{n}}^{1 - p_{i,j,8}}$. Thus we have
        \[
            A_{i,j} + \beta_j = \tilde{u}_{i,j,\tilde{n}}^{1 - (\alpha_i + \beta_j \bmod 2)} + \alpha_i \cdot 1,
        \]
        which is a consequence of the Output Axiom $\tilde{u}_{i,j,\tilde{n}} = A_{i,j}$.
        Similarly, if $(\calD_{s_i,8}, \calE_{r_j,8}) \in \{(*,1), (1,*)\}$, we have a consequence of the Output Axiom $\tilde{u}_{i,m+1,\tilde{n}} = 1$ or $\tilde{u}_{m+1,j,\tilde{n}} = 1$, respectively.
        Otherwise, $\rho(u_{i,j,n-2})$ is a constant and the axiom is satisfied.
    \end{itemize}
    The remaining case is $k \in \{3, \ldots, n-2\}$.
    We have two options depending on the value of $\eta(k)$.

    Suppose that $\eta(k) = 1$.
    \begin{itemize}
        \item If $k=3$, then either the axiom is satisfied or it becomes a Summation Base Axiom for rows $(i,j)$, $(i,m+1)$ or $(m+1,j)$ (depending on the values of $\calD_{s_i,\nu(k)}$ and $\calE_{r_j,\nu(k)}$).
        \item If $k>3$, we have $\nu(k) > 1$ and $\eta(k-1) = \tilde{n}$. Then either the axiom is satisfied or it becomes a simple consequence of the Output Axiom involving $\rho(u_{i,j,k-1})$ and the Summation Base Axiom for rows $(i,j)$, $(i,m+1)$ or $(m+1,j)$ (depending on the values of $\calD_{s_i,\nu(k)}$ and $\calE_{r_j,\nu(k)}$).
    \end{itemize}

    Now assume that $\eta(k) > 1$. Then $\nu(k) = \nu(k-1)$ and $\eta(k-1) = \eta(k) - 1$.
    Then the axiom is either satisfied, or it becomes a Summation Axiom involving rows $(i,j)$, $(i,m+1)$ or $(m+1,j)$ (depending on the values of $\calD_{s_i,\nu(k)}$ and $\calE_{r_j,\nu(k)}$).

    Moreover, applying $\rho$ to the conical junta in the original refutation yields another conical junta, since $\rho$ maps variables to variables or constants.

    It follows from this case analysis that the image of every axiom of $\simpleBTRank^m_n(A)$ is either satisfied, an axiom of ${\simpleBTRank'}^m_{\tilde n}(A)$, or has a short $\SA$ derivation from the relevant Summation Base Axiom and Output Axiom. These derivations involve only the rows mentioned by the original axiom (and possibly $m+1$, which does not contribute to the row degree) and therefore do not increase row degree. Replacing every restricted source axiom by the corresponding derivation converts $\Pi\restriction\rho$ into a $\SA$ refutation of ${\simpleBTRank'}^m_{\tilde n}(A)$.

    % \begin{itemize}
        % \item either satisfied,
        % \item encodes the statement of the form $\tilde{u}_{i,j,\eta(k)} = \tilde{u}_{i,j,\eta(k)-1} + \tilde{x}_{i,\eta(k)} \tilde{y}_{\eta(k),j}$, $\tilde{u}_{i,m+1,\eta(k)} = \tilde{u}_{i,j,\eta(k)-1} + \tilde{x}_{i,\eta(k)}$
        % \michal{These polynomials mixing $\tilde{u}_{i,m+1}$ and $\tilde{u}_{i,j}$ are not axioms of $\simpleBTRank'^m_n(A)$. They occur only for specific values of $k$, and it should be then stated that they are derivable from the axioms by a small proof.}
        % or  $\tilde{u}_{m+1,j,\eta(k)} = \tilde{u}_{i,j,\eta(k)-1} + \tilde{y}_{\eta(k),j}$, or
        % \michal{This should be checked because other expressions also arise, e.g. for $s_i=1, r_j=4, k=\tilde{n}+3, \alpha_i=1, \beta_j=0$ we get that $\rho$ maps $u_{i,j,k} = u_{i,j,k-1} + x_{i,k}y_{k,j}$ to $\tilde{u}^1_{i,j,1} = \tilde{u}^0_{m+1,j,\tilde{n}} + \tilde{x}_{i,1}\tilde{y}_{1,j}$, which is a consequence of the axioms rather than an axiom per se.}
        % \item becomes an Output Axiom encoding $\tilde{u}_{i,j,\tilde{n}} = A^{i,j}$, $\tilde{u}_{i,m+1,\tilde{n}} = 1$ o
        % of the form $\tilde{u}_{i,j,\tilde{n}}^{A_{i,j}}$, $\tilde{u}_{i,m+1,\tilde{n}}$ or $\tilde{u}_{j,m+1,\tilde{n}}$.
        % \michal{Also, it should be stated whether this is a polynomial or a literal (CNF), because the sign depends on this; in particular, in the lines above similar expressions are treated as polynomials, which corresponds to $\tilde{u}^0_{m+1,j,\tilde{n}}$ here rather than $\tilde{u}_{m+1,j,\tilde{n}}$, as $\tilde{A}_{m+1,j} = 1$.}
    % \end{itemize}

\end{proof}

\begin{theorem}\label{thm: SA lower bound on simpleBTRank}
    Assume $m > n \ge 20$, $(n-4)$ is divisible by $16$ and $A \in \Q^{m \times m}$ is a Boolean matrix.
    Then any $\SA$ refutation of $\simpleBTRank^m_n(A)$ requires size ${(\sqrt{8/7})}^{(n-4)/8-1}$.
\end{theorem}
\begin{proof}
    Let $\tilde{n} = (n-4)/8$ and suppose, for the sake of contradiction, that there exists a refutation $\Pi$ of $\simpleBTRank^m_n(A)$ with size $\card{\Pi} < {(\sqrt{8/7})}^{\tilde{n}-1}$.
    By the union bound and \cref{lem: SA BTRank eliminates terms whp}, there exists $\rho$ such that the row degree of any term in $\Pi \restriction \rho$ is at most $\tilde{n}-2$.
    By \cref{lem: SA BTRank ref remains ref}, $\Pi \restriction \rho$ can be converted into a refutation $\Pi'$ of ${\simpleBTRank}'^{m}_{\tilde{n}}(A)$ of row degree at most $\tilde{n}-2$. This contradicts the row degree lower bound $\tilde{n}-1$ by \cref{lem: SA BTRank rowe degree bound}.
\end{proof}

%!TEX root = main.tex

\section{Structural Results}\label{sec: amplification}

\subsection{Amplification of the Rank Principle}

In this section, we prove an amplification theorem for the weak rank principle, namely that the proof complexity of $\Rank^m_n$ is robust for different choices of parameters $(m, n)$, under low-degree algebraic reductions and assuming that $m$ and $n$ are not too close. The proof is an adaptation of known amplification theorems for weak pigeonhole principles~\cite{PWW88, Thapen-PhD, Jerabek04}. This result should be regarded as a basic property of the weak rank principle that essentially follows from previous constructions.
%rather than a novel or technically sophisticated contribution. 
Nonetheless, it plays a foundational role in later sections, and we include it here for completeness. Moreover, this result provides another aspect in which the weak rank principle behaves similarly to the weak pigeonhole principles.

\begin{restatable}{theorem}{ThmAmplificationOfWRank}\label{cor: robustness of wRank wrt parameters}
        Let $m > 2n$ be positive integers, $\eps > 0$, and $d = O(\eps^{-1}\log (m/n))$, then there is a degree-$d$ algebraic reduction
        \[
                \Rank^{(1+\eps)n}_n \le^\alg_d \Rank^m_n.
        \]
\end{restatable}

We prove \autoref{cor: robustness of wRank wrt parameters} in two steps. First, we reduce $\Rank^{2n}_n$ to $\Rank^m_n$ for arbitrarily large $m$.

\begin{theorem}\label{thm: stretching wRank 1}
        Let $m > 2n$ be positive integers and $d = \lceil \log(m/n)\rceil$. Then there is a degree-$d$ algebraic reduction
        \[
                \Rank^{2n}_n \le^\alg_d \Rank^m_n.
        \]
\end{theorem}
\begin{proof}
        Let $X \in \F^{2n \times n}$ and $Y \in \F^{n\times 2n}$ be the variables of $\Rank^{2n}_n$. For every positive integer $i$, we will build a reduction from $\Rank^{2n}_n$ to $\Rank^{2^i n}_n$ inductively. Let $X^{(i)} \in \F^{2^i n\times n}$ and $Y^{(i)}\in \F^{n\times 2^i n}$ be the variables of $\Rank^{2^i n}_n$. Abusing notation, we will also specify our reduction $R_i : \F^{4n^2} \to \F^{2^{i+1}n^2}$ by defining the following two matrices $X^{(i)} \in \F^{2^i n\times n}$ and $Y^{(i)} \in \F^{n\times 2^i n}$ whose entries are polynomials over $(X, Y)$.
        
        This is done inductively. First, $X^{(1)} = X$ and $Y^{(1)} = Y$. For each $i > 1$:
        \[
                \begin{array}{cc}
                        X^{(i+1)} = \underbrace{\begin{pmatrix}
                                X^{(i)} & 0\\
                                0 & X^{(i)}
                        \end{pmatrix}}_{2^{i+1} n\times 2n}
                        \cdot
                        \underbrace{X}_{2n\times n},
                        &
                        Y^{(i+1)} = \underbrace{Y}_{n\times 2n}
                        \cdot
                        \underbrace{\begin{pmatrix}
                                Y^{(i)} & 0\\
                                0 & Y^{(i)}
                        \end{pmatrix}}_{2n\times 2^{i+1}n}.
                \end{array}
        \]
        The idea underlying our reduction is that if $XY = I_{2n}$, then $X^{(i)}Y^{(i)} = I_{2^i n}$ for every $i\ge 1$. This is easily seen by an induction on $i$:
        \begin{align*}
                X^{(i+1)}Y^{(i+1)} =&\, \begin{pmatrix}X^{(i)}&0\\0&X^{(i)}\end{pmatrix}XY\begin{pmatrix}Y^{(i)}&0\\0&Y^{(i)}\end{pmatrix}\\
                =&\, \begin{pmatrix}X^{(i)}&0\\0&X^{(i)}\end{pmatrix}\begin{pmatrix}Y^{(i)}&0\\0&Y^{(i)}\end{pmatrix}\\
                =&\, I_{2^{i+1}n}.
        \end{align*}

        It is easy to see that each entry of $X^{(i)}$ ($Y^{(i)}$ respectively) is a degree-$i$ polynomial over $X$ ($Y$ respectively). Now we prove that each axiom of $\Rank^{2^i n}_n$, when composed with our reduction $R$, admits a degree-$2i$ $\NS$-proof from the axioms of $\Rank^{2n}_n$. Again, this is via an induction on $i$. The case when $i = 1$ is trivial; assuming this is true for $\Rank^{2^i n}_n$, we prove this for $\Rank^{2^{i+1} n}_n$:
        \begin{align*}
                 &\, X^{(i+1)}Y^{(i+1)} - I_{2^{i+1}n}\\
                =&\, \begin{pmatrix}X^{(i)}&0\\0&X^{(i)}\end{pmatrix}XY\begin{pmatrix}Y^{(i)}&0\\0&Y^{(i)}\end{pmatrix} - I_{2^{i+1}n}\\
                =&\, \begin{pmatrix}X^{(i)}&0\\0&X^{(i)}\end{pmatrix}\begin{pmatrix}Y^{(i)}&0\\0&Y^{(i)}\end{pmatrix} + \begin{pmatrix}X^{(i)}&0\\0&X^{(i)}\end{pmatrix}(XY - I_{2n})\begin{pmatrix}Y^{(i)}&0\\0&Y^{(i)}\end{pmatrix} - I_{2^{i+1}n}\\
                =&\, \underbrace{\begin{pmatrix}X^{(i)}Y^{(i)} - I_{2^in}&0\\0&X^{(i)}Y^{(i)} - I_{2^in}\end{pmatrix}}_{\text{(Part I)}} + \underbrace{\begin{pmatrix}X^{(i)}&0\\0&X^{(i)}\end{pmatrix}(XY - I_{2n})\begin{pmatrix}Y^{(i)}&0\\0&Y^{(i)}\end{pmatrix}}_{\text{(Part II)}}.
        \end{align*}
        By our induction hypothesis, each entry of (Part I) admits a degree-$2i$ $\NS$-proof from the axioms of $\Rank^{2n}_n$. Since the axioms of $\Rank^{2n}_n$ only express $XY = I_{2n}$, and each $X^{(i)}$ and $Y^{(i)}$ are degree-$i$ polynomials over $(X, Y)$, it follows that each entry of (Part II) also admits a degree-$2(i+1)$ $\NS$-proof from the axioms of $\Rank^{2n}_n$. This proves our claim for $i+1$.

        Now we wrap up the proof. Let $k = \lceil \log (m/n)\rceil$ be the smallest integer such that $2^k n \ge m$. Let $X' \in \F^{m\times n}$ and $Y' \in \F^{n\times m}$ be the variables of $\Rank^{m}_n$. Given $(X, Y)$, our reduction sets $X'$ to be the first $m$ rows of $X^{(k)}$ and sets $Y'$ to be the first $m$ columns of $Y^{(k)}$. It is easy to see that the reduction can be computed in degree $k$ and that each axiom of $\Rank^{m}_n$ admits a degree-$2k$ $\NS$-proof from the axioms of $\Rank^{2n}_n$.
\end{proof}

Now we reduce $\Rank_n^{(1+\eps)n}$ to $\Rank_n^{2n}$. The proof idea is very similar to that of \autoref{thm: stretching wRank 1} but we include the full proof here for completeness.

%\slava{Do we need to essentially repeat the proof the theorem above again? It takes one page, but the underlying idea is the same.} \hanlin{Yes but the constructions are sort of different. What should we do if we don't write a full proof of this theorem?}\slava{We can define $X^{(i)}$ and $Y^{(i)}$, but omit the proof of $X^{(i)}Y^{(i)} = I$ and just say that it's analogous to the one above.}\Iddo{I\ am in favour of keeping the full proof of both constructions; We can simply say that the idea is the same and we add the proof for completeness.}

\begin{theorem}\label{thm: stretching wRank 2}
        Let $n$ be a positive integer, $\eps > 0$ such that $\eps n$ is an integer, and $d = \lceil 1/\eps\rceil$. Then there is a degree-$d$ algebraic reduction
        \[
                \Rank^{(1+\eps)n}_n \le_d^\alg \Rank^{2n}_n.
        \]
\end{theorem}
\begin{proof}
        Let $X \in \F^{(1+\eps)n \times n}$ and $Y \in \F^{n\times (1+\eps)n}$ be the variables of $\Rank^{(1+\eps)n}_n$. For every positive integer $i$, we will build a reduction from $\Rank^{(1+\eps)n}_n$ to $\Rank^{(1+i\eps)n}_n$ inductively. Let $X^{(i)} \in \F^{(1+i\eps)n \times n}$ and $Y^{(i)} \in \F^{n\times (1+i\eps)n}$ be the variables of $\Rank^{(1+i\eps)n}_n$. Abusing notation, we will also specify our reduction $R : \F^{2(1+\eps)n^2} \to \F^{2(1+i\eps)n^2}$ by defining the following two matrices $X^{(i)}$ and $Y^{(i)}$ whose entries are polynomials over $(X, Y)$.

        This is done inductively. First, $X^{(1)} = X$ and $Y^{(1)} = Y$. For each $i > 1$:
        \[
                \begin{array}{cc}
                        X^{(i+1)} = \underbrace{\begin{pmatrix}X^{(i)} & 0 \\ 0 & I_{\eps n}\end{pmatrix}}_{(1+(i+1)\eps)n\times (1+\eps)n}\cdot \underbrace{X}_{(1+\eps) n \times n},
                        &
                        Y^{(i+1)} = \underbrace{Y}_{n\times (1+\eps)n}\cdot \underbrace{\begin{pmatrix}Y^{(i)} & 0 \\ 0 & I_{\eps n}\end{pmatrix}}_{(1+\eps)n\times (1+(i+1)\eps) n}.
                \end{array}
        \]

        For the sake of intuition, let us see that if $XY = I_{(1+\eps) n}$, then $X^{(i)}Y^{(i)} = I_{(1+i\eps)n}$ for every $i\ge 1$. This is via an induction over $i$:
        \begin{align*}
                X^{(i+1)}Y^{(i+1)} =&\, \begin{pmatrix}X^{(i)} & 0 \\ 0 & I_{\eps n}\end{pmatrix} X Y \begin{pmatrix}Y^{(i)} & 0 \\ 0 & I_{\eps n}\end{pmatrix}\\
                =&\, \begin{pmatrix}X^{(i)} & 0 \\ 0 & I_{\eps n}\end{pmatrix} \begin{pmatrix}Y^{(i)} & 0 \\ 0 & I_{\eps n}\end{pmatrix}\\
                =&\, \begin{pmatrix}I_{(1+i\eps) n} & 0 \\ 0 & I_{\eps n}\end{pmatrix}\\
                =&\, I_{(1+(i+1)\eps) n}.
        \end{align*}

        It is easy to see that each entry of $X^{(i)}$ ($Y^{(i)}$ respectively) is a degree-$i$ polynomial over $X$ ($Y$ respectively). Now we prove that each axiom of $\Rank^{(1+i\eps)n}_n$, when composed with our reduction $R$, admits a degree-$2i$ $\NS$-proof from the axioms of $\Rank^{(1+\eps)n}_n$. Again, this is via an induction on $i$. The case that $i = 1$ is trivial; assuming this is true for $\Rank^{(1+i\eps)n}_n$, we prove this for $\Rank^{(1+(i+1)\eps)n}_n$:
        \begin{align*}
                &\, X^{(i+1)}Y^{(i+1)} - I_{(1+(i+1)\eps)n}\\
                =&\, \begin{pmatrix}X^{(i)} & 0 \\ 0 & I_{\eps n}\end{pmatrix} X Y \begin{pmatrix}Y^{(i)} & 0 \\ 0 & I_{\eps n}\end{pmatrix} - I_{(1+(i+1)\eps) n}\\
                =&\, \begin{pmatrix}X^{(i)} & 0 \\ 0 & I_{\eps n}\end{pmatrix} (XY - I_{(1+\eps)n}) \begin{pmatrix}Y^{(i)} & 0 \\ 0 & I_{\eps n}\end{pmatrix} + \begin{pmatrix}X^{(i)} & 0 \\ 0 & I_{\eps n}\end{pmatrix}\begin{pmatrix}Y^{(i)} & 0 \\ 0 & I_{\eps n}\end{pmatrix} - I_{(1+(i+1)\eps) n}\\
                =&\, \underbrace{\begin{pmatrix}X^{(i)} & 0 \\ 0 & I_{\eps n}\end{pmatrix} (XY - I_{(1+\eps)n}) \begin{pmatrix}Y^{(i)} & 0 \\ 0 & I_{\eps n}\end{pmatrix}}_{\text{(Part I)}} + \underbrace{\begin{pmatrix}X^{(i)}Y^{(i)} - I_{(1+i\eps) n} & 0 \\ 0 & 0\end{pmatrix}}_{\text{(Part II)}}
        \end{align*}
        By our induction hypothesis, each entry of (Part II) admits a degree-$2i$ $\NS$-proof from the axioms of $\Rank^{(1+\eps)n}_n$. Since the axioms of $\Rank^{(1+\eps)n}_n$ only express $XY = I_{(1+\eps n)}$, and each $X^{(i)}$ and $Y^{(i)}$ are degree-$i$ polynomials over $(X, Y)$, it follows that each entry of (Part I) also admits a degree-$2(i+1)$ $\NS$-proof from the axioms of $\Rank^{(1+\eps)n}_n$. This proves our claim for $i+1$.

        Now we wrap up the proof. Let $k = \lceil 1/\eps\rceil$ be the smallest integer such that $(1+k\eps)n \ge 2n$. Let $X' \in \F^{2n\times n}$ and $Y' \in \F^{n\times 2n}$ be the variables of $\Rank^{2n}_n$. Given $(X, Y)$, our reduction sets $X'$ to be the first $2n$ rows of $X^{(k)}$ and sets $Y'$ to be the first $2n$ columns of $Y^{(k)}$. It is easy to see that the reduction can be computed in degree $k$ and that each axiom of $\Rank^{2n}_n$ admits a degree-$2k$ proof from the axioms of $\Rank^{(1+\eps)n}_n$.
\end{proof}

The takeaway message from the above two reductions is that the proof complexity of weak rank formulas $\Rank^{m}_n$ are robust (in the regime when $m$ and $n$ are not \emph{too close}), in the sense that they can be reduced to each other via low-degree reductions.

\ThmAmplificationOfWRank*
\begin{proof}[Proof Sketch]
    This follows from \autoref{thm: stretching wRank 1}, \autoref{thm: stretching wRank 2}, and \autoref{fact: composition of algebraic reductions}.
\end{proof}

\subsection{Low-Degree Reductions for WRank in Strong Proof Systems}

For sufficiently strong proof systems closed under low-degree reductions, we show that it is enough to prove the hardness of WRank with $A=I$ to get a generator (namely, hardness of WRank for every $A$). We demonstrate this simple property for the bamboo-tree encoding $\simpleBTRank^m_n(A)$.

% We need the following folklore result.
% \begin{proposition}[Full rank factorisation; folklore, see, e.g.,~\cite{piziak1999full}]\label{thm: full rank factorisation}
%     Let $A$ be an $m \times \ell$ matrix of rank $r$ over $\F$. For every $k \ge r$, there exist matrices $B$ and $C$ of dimensions $m \times k$ and $k \times \ell$, respectively, such that $BC = A$.
% \end{proposition}

We show that the formula $\simpleBTRank^m_n(A)$ can be reduced to $\simpleBTRank^r_n(I_r)$, where $r>n$, by a degree-$1$ reduction.
The main idea consists of substitution $X'=AX$ and $Y'=Y$ into the statement $X'Y'=A$, thus obtaining $AXY=A$, which is derivable from $XY=I_r$.

\newcommand{\hist}{u}
\newcommand\OutAx{\mathsf{OutAx}}
\newcommand\MultBaseAx{\mathsf{MultBaseAx}}
\newcommand\MultAx{\mathsf{MultAx}}

\begin{lemma}\label{lem:degree-1 reduction for CNF rank formula}
    Let $A \in \Q^{m\times m}$ be an arbitrary matrix of rank $r$ with $r > n$. Then there is a degree-$1$ reduction from $\simpleBTRank^m_n(I_m)$ to $\simpleBTRank^m_n(A)$.\footnote{It is possible to construct a reduction from $\simpleBTRank^r_n(I_r)$: that would require factorising $A$ into $BC$ for some $B \in \F^{m\times r}$ and $C \in \F^{r\times m}$, and then substituting $X' = BX$ and $Y' = CY$. Here we present a simpler reduction.}
\end{lemma}
\begin{proof}
    Let the variables of $\simpleBTRank^m_n(I_m)$ be $X = {(x_{i,k})}_{i\in[m],k\in[n]}$, $Y = {(y_{k,j})}_{k\in[n],j\in[m]}$, and $U = {(u_{i,j,k})}_{i,j\in[m],k\in[n]}$, and the variables of $\simpleBTRank^m_n(A)$ be $X' = {(x'_{i,k})}_{i\in[m],k\in[n]}$, $Y' = {(y'_{k,j})}_{k\in[n],j\in[m]}$, and $U' = {(u'_{i,j,k})}_{i,j\in[m],k\in[n]}$.
    Here we are going to slightly abuse the notation of degree-$1$ reductions: we are going to construct the reduction $R$ that substitutes $(X',Y',U')$ with degree-$1$ polynomials in $(X,Y,U)$, but the axioms of $\simpleBTRank^m_n(A) \circ R$ will have \emph{constant degree} $\NS$-proofs from the axioms of $\simpleBTRank^m_n(I_m)$. Since the axioms of $\simpleBTRank$ are themselves have constant degrees, this does not pose any problems and such a reduction is sufficient for our applications.

    % From \cref{thm: full rank factorisation} there exist two matrices $B \in \F_2^{m \times r}$ and $C \in \F_2^{r \times \ell}$ such that $BC=A$.
    We can consider the following degree-$1$ reduction $R$ that computes $(X', Y', E')$ from $(X, Y, E)$:
    \begin{itemize}
        \item for every $i\in [m], k\in [n]$, $x'_{i, k} = x_{i,k}$;
        \item for every $k\in [n], j\in [m]$, $y'_{k, j} = \sum_{b \in [r]} y_{k, b} A_{b, j}$;
        \item for every $i,j\in [m]$ and $k\in[n]$, $\hist'_{i, j, k} = \sum_{b \in [r]} A_{b,j} \hist_{i, b, k}$.
    \end{itemize}
    % However, the proof that this reduction reduces $\simpleBTRank^r_n(I_r)$ to $\simpleBTRank^m_n(A)$ is quite verbose and technical. Instead, without sacrificing any ideas, we can consider two reductions $R^1$ and $R^2$ such that $R = R^1 \circ R^2$.

    % Consider the variables $X' \in \F_2^{r\times n}$, $Y'' \in \F_2^{n\times m}$, and $U'' \in \F_2^{r\times m \times n}$.
    
    % Consider the degree-$1$ reduction $R^1$ that computes $(X'', Y'', E'')$ from $(X', Y', E')$:
    % \begin{itemize}
    %     \item for every $i\in [m], k\in [n]$, $x''_{i, k} = \sum_{a \in [r]} B_{i, a} x'_{a, k}$;
    %     \item for every $k\in [n], j\in [m]$, $y''_{k, j} = y'_{k, j}$;
    %     \item for every $i\in [m]$, $j\in[m]$ and $k\in[n]$, $\hist''_{i, j, k} = \sum_{a \in [r]} B_{i,a} \hist'_{a, j, k}$,
    % \end{itemize}
    % and the degree-$1$ reduction $R^2$ that computes $(X', Y', E')$ from $(X, Y, E)$:
    % \begin{itemize}
    %     \item for every $i\in [r], k\in [n]$, $x'_{i, k} = x_{i, k}$;
    %     \item for every $k\in [n], j\in [m]$, $y'_{k, j} = \sum_{b \in [r]} y_{k, b} C_{b, j}$;
    %     \item for every $i\in [r]$, $j\in[m]$ and $k\in[n]$, $\hist'_{i, j, k} = \sum_{b \in [r]} C_{b,j} \hist_{i, b, k}$.
    % \end{itemize}
    % It is immediately clear that $R^1 \circ R^2 = R$.
    We show that $\simpleBTRank^n_n(I_n)$ reduces to $\simpleBTRank^m_n(A)$ under $R$.

    % Now we need to show that for each axiom $\ax$ of $\simpleBTRank^m_n(A)$, the polynomial $\ax \circ R$ has a small $\NS$-proof from the clauses of $\simpleBTRank^r_n(I_r)$.
    % We are going to show that for each axiom $\ax$ of $\simpleBTRank^r_n \circ R^2$, the polynomial $\ax \circ R^2$ has a small $\NS$-proof from the clauses of $\simpleBTRank^r_n(I_r)$.

    Recall that in algebraic proof systems, we encode a clause $C = \bigvee_{i \in P} v_i \lor \bigvee_{i \in N} \lnot v_i$ as the polynomial $\prod_{i \in P} (1 + v_i) \prod_{i \in N} v_i$.
    We give the following names to the axioms in $\simpleBTRank^m_n(I_m)$:
    \begin{align*}
        \OutAx_{i,j} &= \delta_{i,j}+\hist_{i,j,n}, &i,j\in [m],\\
        \MultBaseAx_{i,j} &= \hist_{i,j,0},&i,j\in [m],\\
        \MultAx^1_{i,j,k} &= (1 + x_{i,k}) \hist_{i,j,k-1} (1+\hist_{i,j,k}),&i,j\in [m],k\in[n], \\
        \MultAx^2_{i,j,k} &= (1 + x_{i,k}) (1+\hist_{i,j,k-1}) \hist_{i,j,k},&i,j\in [m],k\in[n], \\
        \MultAx^3_{i,j,k} &= (1 + y_{k,j}) \hist_{i,j,k-1} (1+\hist_{i,j,k}),&i,j\in [m],k\in[n], \\
        \MultAx^4_{i,j,k} &= (1 + y_{k,j}) (1+\hist_{i,j,k-1}) \hist_{i,j,k},&i,j\in [m],k\in[n], \\
        \MultAx^5_{i,j,k} &= x_{i,k} y_{k,j} \hist_{i,j,k-1} \hist_{i,j,k};&i,j\in [m],k\in[n], \\
        \MultAx^6_{i,j,k} &= x_{i,k} y_{k,j} (1+\hist_{i,j,k-1}) (1+\hist_{i,j,k}),&i,j\in [m],k\in[n].
    \end{align*}
    To simplify the proofs, we also consider the following three types of auxiliary polynomials:
    \begin{align*}
        \MultAx^7_{i,j,k} &= (1 + x_{i,k}) (\hist_{i,j,k-1} + \hist_{i,j,k}),&i,j\in [m],k\in[n], \\
        \MultAx^8_{i,j,k} &= (1 + y_{k,j}) (\hist_{i,j,k-1} + \hist_{i,j,k}),&i,j\in [m],k\in[n], \\
        \MultAx^9_{i,j,k} &= x_{i,k} y_{k,j} (1 + \hist_{i,j,k-1} + \hist_{i,j,k}),&i,j\in [m],k\in[n].
    \end{align*}
    It is easy to see that for all $i,j\in[m]$ and $k\in [n]$ (over characteristic $2$):
    \begin{align*}
        \MultAx^7_{i,j,k} &= \MultAx^1_{i,j,k} + \MultAx^2_{i,j,k}, \\
        \MultAx^8_{i,j,k} &= \MultAx^3_{i,j,k} + \MultAx^4_{i,j,k}, \\
        \MultAx^9_{i,j,k} &= \MultAx^5_{i,j,k} + \MultAx^6_{i,j,k}.
    \end{align*}

    % \slava{The correct proof strategy (presentation-wise): define two restrictions that substitute the $B$ and $C$ parts separately. This can be shown in reasonable space. Then notice that $R$ is their combination $R^1 \circ R^2$ and make a comment that degree-bound is a bit better (6 and 7) if carefully analysed.}

    Let $\ax$ be any polynomial from $\simpleBTRank^m_n(A)$. There are the following cases:
    \begin{itemize}
        \item \underline{\textbf{Output Axioms:}} Suppose $\ax = A_{i,j} + \hist'_{i,j,n}$, where $i,j\in[m]$. Then
        \begin{align*}
            \ax \circ R &= A_{i,j} + \sum_{b\in [r]} A_{b,j} \hist_{i,b,n}\\
            &= \sum_{b\in [r]} A_{b,j} ((\delta_{i,b} + \hist_{i,b,n}) + \delta_{i,b}) + A_{i,j}\\
            &= \sum_{b\in [r]} A_{b,j} \OutAx_{i,b},
        \end{align*}
        which is a linear combination of the axioms of $\simpleBTRank^r_n(I_r)$.

        \item \underline{\textbf{Multiplication Base Axioms:}} Suppose $\ax = \hist'_{i,j,0}$, where $i,j\in[m]$.
        Then
        \[
            \ax \circ R = \sum_{b\in[r]} A_{b,j} \hist_{i,b,0} = \sum_{b\in[r]} A_{b,j} \MultBaseAx_{i,b},
        \]
        which is a linear combination of the axioms of $\simpleBTRank^r_n(I_r)$.

        \item \underline{\textbf{Multiplication Axioms:}}

        \begin{enumerate}
            \item\label{item: main item for CNF ERank}
            Suppose $\ax = (1 + x'_{i,k}) \hist'_{i,j,k-1} (1+\hist'_{i,j,k})$, where $i,j\in[m]$, and $k \in[n]$.
            Then
            \begin{align*}
                \ax \circ R &= (1+x_{i,k}) (\sum_{b_1 \in [m]} A_{b_1, j} \hist_{i,b_1,k-1}) (1+\sum_{b_2 \in [m]} A_{b_2, j} \hist_{i,b_2,k}) \\
                &= \sum_{b_1, b_2 \in [m]} A_{b_1,j} A_{b_2,j} \hist_{i,b_1,k-1} \MultAx^7_{i,b_2,k},
            \end{align*}
            which is a degree-4 NS proof of $\ax \circ R$ from $\simpleBTRank^m_n(I_m)$.
            % \begin{align*}
            %     \ax \circ R =& (1+\sum_{a_1 \in [m]} B_{i,a_1} x_{a_1,k}) (\sum_{a_2,b_2 \in [m]} B_{i,a_2} C_{b_2,j} \hist_{a_2,b_2,k-1}) (1+\sum_{a_3, b_3 \in [m]} B_{i,a_3} C_{b_3,j} \hist_{a_3,b_3,k}) \\
            %     =& \sum_{\substack{a_2, b_2 \in [m] \\ a_3, b_3 \in [m]}} B_{i,a_2}C_{b_2,j} B_{i,a_3}C_{b_3,j} \cdot \hist_{a_2,b_2,k-1} \MultAx^7_{a_3,b_3,k} \\
            %     &+ \sum_{\substack{a_2, b_2 \in [m] \\ a_3, b_3 \in [m] \\ a_1 \in [m] \setminus \{a_3\}}} B_{i,a_1} B_{i,a_2}C_{b_2,j} B_{i,a_3}C_{b_3,j} \cdot x_{a_1,k} \hist_{a_2,b_2,k-1} (\MultAx^9_{a_3,b_3,k} \\
            %     &~~~~~~~~~~~~~~~~+ \MultAx^8_{a_3,b_3,k} + y_{k,b_3} \MultAx^7_{a_3,b_3,k}),
            % \end{align*}
            % which is a degree-$6$ $\NS$-proof of $\ax \circ R$ from $\simpleBTRank^m_n(I_m)$.
            % \slava{This looks ugly.}
            This polynomial identity can be verified by induction on $m$.

%             \slava{This is a polynomial identity that can be verified by just comparing the monomial coefficients.}
% \Iddo{What are you suggesting to do?}\slava{I hoped someone can independently verify this and maybe suggest a good way of showing this? Induction on $m$ should work, but this belongs to an appendix.}

            \item Suppose $\ax = (1 + x'_{i,k}) (1+\hist'_{i,j,k-1}) \hist'_{i,j,k}$, where $i,j\in[m]$, and $k \in[n]$.
            Then $\ax \circ R$ has a degree-$4$ $\NS$-proof from $\simpleBTRank^r_n(I_r)$ due to symmetry: simply exchange $\hist_{a,b,k-1}$ and $\hist_{a,b,k}$ in \cref{item: main item for CNF ERank}.

            \item\label{item: additional item for CNF ERank} Suppose $\ax = (1 + y'_{k,j}) \hist'_{i,j,k-1} (1+\hist'_{i,j,k})$, where $i,j\in [m]$ and $k \in[n]$.
            % Then $\ax \circ R$ has a degree-$4$ $\NS$-proof from $\simpleBTRank^r_n(I_r)$ due to symmetry: exchange $x$ and $y$ variables (and also replace $B$ with $C$ appropriately) in \cref{item: main item for CNF ERank}.
            Then
            \begin{align*}
                \ax \circ R^2 =&~ (1+\sum_{b_3 \in [m]} A_{b_3,j} y_{k,b_3}) (\sum_{b_1 \in [m]} A_{b_1, j} \hist_{i,b_1,k-1}) (1+\sum_{b_2 \in [m]} A_{b_2, j} e_{i,b_2,k}) \\
                =&~ \sum_{b_1, b_2 \in [m]} A_{b_1,j} A_{b_2,j} \hist_{i,b_1,k-1} \MultAx^8_{i,b_2,k} \\
                &+ \sum_{\substack{b_1,b_2,b_3\in[m] \\ b_2 \neq b_3}} A_{b_1,j} A_{b_2,j} A_{b_3,j} y_{k,b_3} \hist_{i,b_1,k-1} (\MultAx^7_{i,b_2,k} \\
                &~~~~~~~~~~~~~~~~+ \MultAx^9_{i,b_2,k} + x_{i,k} \MultAx^8_{i,b_2,k}),
            \end{align*}
            which is a degree-$6$ NS proof of $\ax \circ R$ from $\simpleBTRank^m_n(I_m)$.

            \item Suppose $\ax = (1 + y'_{k,j}) (1+\hist'_{i,j,k-1}) \hist'_{i,j,k}$, where $i,j\in[m]$, and $k \in[n]$.
            Then $\ax \circ R$ has a degree-$6$ $\NS$-proof from $\simpleBTRank^m_n(I_m)$ due to symmetry: simply exchange $\hist_{a,b,k-1}$ and $\hist_{a,b,k}$ in \cref{item: additional item for CNF ERank}.

            \item Suppose $\ax = x'_{i,k} y'_{k,j} \hist'_{i,j,k-1} \hist'_{i,j,k}$, where $i,j\in [m]$ and $k \in[n]$.
            % Then $\ax \circ R^2$ again has a degree-$7$ $\NS$-proof from $\simpleBTRank^r_n(I_r)$, which is analogous to \cref{item: main item for CNF ERank}.
            Then
            \begin{align*}
                \ax \circ R =&~ x_{i,k} (\sum_{b_3 \in [m]} A_{b_3,j} y_{k,b_3}) (\sum_{b_1 \in [m]} A_{b_1, j} \hist_{i,b_1,k-1}) (\sum_{b_2 \in [m]} A_{b_2, j} \hist_{i,b_2,k}) \\
                =&~ \sum_{b_1, b_2 \in [m]} A_{b_1,j} A_{b_2,j} \hist_{i,b_1,k-1} \MultAx^9_{i,b_2,k} \\
                &+ \sum_{\substack{b_1,b_2,b_3\in[r] \\ b_2 \neq b_3}} A_{b_1,j} A_{b_2,j} A_{b_3,j} y_{k,b_3} \hist_{i,b_1,k-1} (\MultAx^9_{i,b_2,k} \\
                &~~~~~~~~~~~~~~~~+ x_{i,k} \MultAx^8_{i,b_2,k}),
            \end{align*}
            which is a degree-$6$ NS proof of $\ax \circ R$ from $\simpleBTRank^r_n(I_r)$.

            \item Suppose $\ax = x'_{i,k} y'_{k,j} (1+\hist'_{i,j,k-1}) (1+\hist'_{i,j,k})$, where $i,j\in [m]$ and $k \in[n]$.
            % Then $\ax \circ R^2$ again has a degree-$7$ $\NS$-proof from $\simpleBTRank^r_n(I_r)$, which is analogous to \cref{item: main item for CNF ERank}.
            Then
            \begin{align*}
                \ax \circ R =&~ x_{i,k} (\sum_{b_3 \in [r]} A_{b_3,j} y_{k,b_3}) (\sum_{b_1 \in [r]} A_{b_1, j} \hist_{i,b_1,k-1}) (\sum_{b_2 \in [m]} A_{b_2, j} \hist_{i,b_2,k}) \\
                =&~ \sum_{b_2 \in [r]} A_{b_2,j} \MultAx^9_{i,b_2,k} \\
                &+ \sum_{\substack{b_2,b_3\in[r] \\ b_2 \neq b_3}} A_{b_2,j} A_{b_3,j} y_{k,b_3} (\MultAx^9_{i,b_2,k} + x_{i,k} \MultAx^8_{i,b_2,k}) \\
                &+ (x'_{i,k} y'_{k,j} \hist'_{i,j,k-1} \hist'_{i,j,k}) \circ R,
            \end{align*}
            which is a degree-$6$ NS proof of $\ax \circ R$ from $\simpleBTRank^r_n(I_r)$.\qedhere
        \end{enumerate}
    \end{itemize}
\end{proof}

%!TEX root = main.tex

\section{Hardness of Proving Circuit Lower Bounds}\label{sec: hardness of CLB from rank}

In this section, we show that there is a low-degree reduction from the weak rank principles to circuit lower bound statements. In other words, for every algebraic proof system $\calP$ closed under low-degree reductions, if $\calP$ cannot prove the weak rank principle, then $\calP$ cannot prove circuit lower bounds for the truth table of any function. We also unconditionally show the hardness of such circuit lower bound statements for $\PCR_{\F_2}$. Although $\PCR_{\F_2}$, when measured by size, does not seem to be closed under low-degree reductions, we still manage to prove the hardness of circuit lower bound statements for $\PCR_{\F_2}$ by reducing them to an appropriate iterated variant of the weak rank principle.

\begin{remark}[{Comparison with~\cite{Razb98, Razborov04}}]
    Razborov showed that low-degree polynomial calculus proofs~\cite{Razb98} and short resolution proofs~\cite{Razborov04} cannot prove any circuit lower bounds. Both proofs proceed by first showing that (variants of) weak pigeonhole principles are hard for the underlying proof system and then reducing the weak pigeonhole principle to the circuit lower bound statements. Intuitively, these weak proof systems cannot prove circuit lower bounds because they cannot \emph{count}.

    Complementing Razborov's results above, in this section we show that there is a low-degree reduction from the rank principles to the circuit lower bound statements. Intuitively, weak proof systems that cannot reason about \emph{linear algebra} cannot prove circuit lower bounds, either. Our result demonstrates yet another aspect in which the (weak) rank principles behave as an algebraic analogue of the (weak) pigeonhole principle.
\end{remark}

%\hanlin{Maybe a comparison with Razborov98 and Razborov04. Razborov 98 is like: Since PC cannot \emph{count}, PC cannot prove circuit lower bounds. We replace ``count'' with something slightly less trivial --- if P cannot reason about matrix rank, then P cannot prove circuit lower bounds. We can say something like ``this opens up new ways of showing unprovability of circuit lower bounds''}

Roughly speaking, our arguments uses \emph{non-commutative algebraic branching programs} (ncABPs) as a proxy. Nisan~\cite{Nis91} showed that the $\ncABP$ complexity of every non-commutative homogeneous polynomial $f$ is \emph{characterised} by the rank of certain matrix associated with $f$. Hence, if $\calP$ cannot prove the weak rank principles, then $\calP$ cannot prove any lower bound on the $\ncABP$ complexity of non-commutative homogeneous polynomials. In \autoref{sec: ncABP lb}, we formalise the above intuition as a proof complexity reduction from the weak rank principles to polynomial equations expressing $\ncABP$ lower bound. Then, in \autoref{sec: unprovability of Boolean circuit lb}, we show that the unprovability of $\ncABP$ lower bounds imply the unprovability of lower bounds for general Boolean circuits.

\subsection{Lower Bounds for Non-commutative Algebraic Branching Programs}\label{sec: ncABP lb}

Fix an underlying field $\F$. We consider non-commutative polynomials over the variable set $X = \{x_1, x_2, \dots, x_n\}$. Denote
\[X^k = \{x_{i_1}x_{i_2}\dots x_{i_k}: i_1, i_2, \dots, i_k\in[n]\}\]
as the set of degree-$k$ non-commutative monomials.

\paragraph{Formalisation of ncABP lower bounds.} Let $f$ be a non-commutative homogeneous polynomial of degree $d$ over the variable set $X = \{x_1, x_2, \dots, x_n\}$. Let $\vec{r} = (r_0, r_1, \dots, r_d)$ where $r_0 = r_d = 1$, we now define $\lb_\ncABP(f, \vec{r})$, which is a set of polynomial equations expressing that $f$ cannot be computed by a non-commutative ABP where the $i$-th layer has $r_i$ nodes.
\begin{itemize}
    \item [\textbf{(Variables)}] For every layer $1\le i\le d$ and every variable $x_j$ ($1\le j\le n$), there is a matrix $M_{i, x_j} \in \F^{r_{i-1} \times r_i}$ among the variables of $\lb_\ncABP(f, \vec{r})$. We will also denote $M_i$ as the $r_{i-1}\times r_i$ matrix where the $(u, v)$-th entry is the (formal) linear form $\sum_{j\in [n]}M_{i, x_j}[u, v]x_j$. Note that here, $M_{i, x_j}[u, v]$ is a variable of the polynomial equation system $\lb_\ncABP(f, \vec{r})$ and $x_j$ is a formal variable of $f$.
    \item [\textbf{(Equations)}] For every degree-$d$ non-commutative monomial $m \in X^d$, $\lb_\ncABP(f, \vec{r})$ contains a degree-$d$ equation stating that the $m$-th coefficient of $f$ is equal to the $m$-th coefficient of $f'$, where $f'$ is the non-commutative polynomial computed by the ABP. Suppose that $m = x_{j_1}x_{j_2} \dots x_{j_d}$, then this equation can be written as
    \[\prod_{i=1}^d M_{i, x_{j_i}} - \coeff[f, m] = 0.\]
    (Note that $\prod_{i=1}^d M_{i, x_{j_i}}$ is a $1\times 1$ matrix over $\F$, i.e., a scalar in $\F$.)
\end{itemize}

\begin{theorem}\label{thm: reduction from Rank to ncABP lb}
    Let $f$ be any non-commutative homogeneous polynomial of degree $d$ over the variable set $X = \{x_1, x_2, \dots, x_n\}$ and let $N = n^d$. Let $r\in \N$ and denote $\vec{r} = (1, \underbrace{r, r, \dots, r}_{(d-1)\text{ $r$'s}}, 1)$.
    
    Then there is a degree-$2$ reduction from $\Rank^N_r$ to $\lb_\ncABP(f, \vec{r})$.
\end{theorem}
\def\bf{\mathsf{bf}}
\begin{proof}
    %The idea is simple: Let $M_1, M_2, \dots, M_d$ denote the matrices describing an ncABP, where each $M_i$ is of dimension $N\times N$ (except that $M_1$ has dimension $1\times N$ and $M_d$ has dimension $N\times 1$). If there \emph{were} a rank-$r$ decomposition of $I_N$, i.e., $I_N = XY$ where $X$ is an $N\times r$ matrix and $Y$ is an $r\times N$ matrix, then we also have
    %\[M_1\cdot M_2\cdot M_3\cdot \ldots \cdot M_{d-1} \cdot M_d = (M_1X) \cdot (YM_2X) \cdot (YM_3X) \cdot\ldots \cdot (YM_{d-1}X)\cdot (YM_d),\]
    %which describes a width-$r$ ncABP computing the same non-commutative polynomial.
    Identify $[N]$ with $X^d$, i.e., the set of non-commutative degree-$d$ monomials over $X$. Consider what we call the \emph{brute-force ncABP} computing $f$: this is a ncABP consisting of $d+1$ layers, with each internal layer of width $N$. The ncABP consists of $N$  disjoint (with respect to nodes, except for the source and sink nodes) paths, where each such path from the root to the sink computes a single non-commutative monomial. More precisely, we  can describe the brute-force ncABP in terms of matrices whose entries are linear forms over $X$ as follows:
    \begin{itemize}
        \item For every $m\in X^d$, let $M_1^\bf[1, m] = \coeff[f, m]x_1$ where $x_1$ is the first variable in $m$.
        \item For every $2\le i < d$ and every $m\in X^d$, let $M_i^\bf[m, m] = x_i$ where $x_i$ is the $i$-th variable in $m$; for every $m, m'\in X^d$ with $m\ne m'$, let $M_i^\bf[m, m'] = 0$.
        \item For every $m\in X^d$, let $M_d^\bf[m, 1] = x_d$ where $x_d$ is the $d$-th variable in $m$.
    \end{itemize}
     
    Let $X \in \F^{N\times r}$ and $Y \in \F^{r\times N}$ be the variables of $\Rank^N_r$ and $(M_1, M_2, \dots, M_d)$ be the variables of $\lb_\ncABP(f, \vec{r})$. Consider the following degree-$2$ mapping $R$ that maps $(X, Y)$ into:
    \begin{itemize}
        \item $M_1 = M_1^\bf\cdot X$, $M_d = Y\cdot M_d^\bf$;
        
        (What this actually means is that $M_{1, x_j} = M_{1, x_j}^\bf\cdot X$ and $M_{d, x_j} = Y\cdot M_{d, x_j}^\bf$ for every $x_j\in X$.)
        \item $M_i = Y\cdot M_i^\bf\cdot X$ for every $2\le i < d$.
        
        (Similarly, what this actually means is that $M_{i, x_j} = Y\cdot M_{i, x_j}^\bf \cdot X$ for every $x_j \in X$.)
    \end{itemize}
    We prove this is a valid reduction from $\Rank^N_r$ to $\lb_\ncABP(f, \vec{r})$. Towards this end fix any monomial $m = x_{j_1}x_{j_2}\dots x_{j_d} \in X^d$ and consider the polynomial corresponding to $m$:
    \[p_m = \prod_{i=1}^d M_{i, x_{j_i}} - \coeff[f, m].\]
    Then, $p_m\circ R$ is the following polynomial over $(X, Y)$:
    \[p_m\circ R = M_{1, x_1}^\bf (XY) M_{2, x_2}^\bf (XY) M_{3, x_3}^\bf \dots M_{d-1, x_{d-1}}^\bf (XY) M_{d, x_d}^\bf - \coeff[f, m].\]
    On the other hand, the following is immediate from the definitions of $M_i^\bf$:
    \[M_{1, x_1}^\bf M_{2, x_2}^\bf M_{3, x_3}^\bf \dots M_{d-1, x_{d-1}}^\bf M_{d, x_d}^\bf - \coeff[f, m] = 0.\]
    Now we have the following telescoping sum:
    \begin{align}
        p_m\circ R=&\, M_{1, x_1}^\bf M_{2, x_2}^\bf \dots M_{d, x_d}^\bf - \coeff[f, m]\nonumber\\ 
        &\, +M_{1, x_1}^\bf(XY-I)M_{2, x_2}^\bf \dots M_{d, x_d}^\bf\nonumber\\
        &\, +M_{1, x_1}^\bf(XY) M_{2, x_2}^\bf (XY-I)M_{3, x_3}^\bf \dots M_{d, x_d}^\bf\nonumber\\
        &\, +\dots\nonumber\\
        &\, +M_{1, x_1}^\bf (XY) M_{2, x_2}^\bf (XY) M_{3, x_3}^\bf \dots (XY-I)M_{d, x_d}^\bf\nonumber\\
        =&\, \sum_{i=1}^{d-1}{\sf Prefix}_i (XY-I){\sf Suffix}_{i+1},\label{eq: pm o R}
    \end{align}
    where ${\sf Prefix}_i = M_{1, x_1}^\bf (XY) M_{2, x_2}^\bf (XY) \dots (XY) M_{i, x_i}^\bf$ and ${\sf Suffix}_i = M_{i, x_i}^\bf M_{i+1, x_{i+1}}^\bf \dots M_{d, x_d}^\bf$. It is easy to see that \eqref{eq: pm o R} is of the form $\sum_{u, v \in [N]}\mathsf{axiom}_{u, v} p_{u, v}$, where $\mathsf{axiom}_{u, v}$ is the $(u, v)$-th entry of $(XY-I)$, and each $p_{u, v}$ is a polynomial over $(X, Y)$ of degree at most $2d-2$.
\end{proof}

As a corollary, there is a degree-$O(\log(n^d/r))$ reduction from $\Rank^{2r}_r$ to $\lb_\ncABP(f, \vec{r})$.

\subsection{Boolean Circuit Lower Bounds}\label{sec: unprovability of Boolean circuit lb}
We show that there is a degree-$O(n)$ algebraic reduction from polynomial equations expressing ncABP lower bounds to polynomial equations expressing Boolean circuit lower bounds. This implies that for every proof system $\calP$ closed under low-degree algebraic reductions, if $\calP$ cannot prove the weak rank principle, then it also cannot prove circuit lower bounds for any Boolean function. (Note that in our formalisation, the circuit lower bound statement has size $2^{O(n)}$, hence a reduction of degree $O(n)$ can be seen as \emph{low degree}.)

Let $f: \{0, 1\}^n \to \{0, 1\}$ denote a Boolean function (represented by its length-$2^n$ truth table), $s\in\N$, and $\lb(f, s)$ denote the polynomial equation system expressing that the circuit complexity of $f$ is greater than $s$.

\begin{theorem}\label{thm: hardness of boolean circuit lower bounds}
    For every $f:\{0, 1\}^n \to \{0, 1\}$, there is a non-commutative polynomial $p_f: \F^2 \to \F$ of degree $n$ such that for every parameter $r$, if we let $\vec{r} = (1, \underbrace{r, r, \dots, r}_{(n-1)\text{ }r\text{'s}}, 1)$, then %\vspace{-2em}
    \[\lb_\ncABP(p_f, \vec{r})\le^{O(n)}_\alg\lb(f, s)\text{ where }s = O(nr^2).\]
\end{theorem}

\paragraph{Formalisation of $\lb(f, s)$.} We now recall the precise  definition of $\lb(f, s)$ from~\cite{Razb98,Razborov04,Razb15-annals}.

First, we list all variables of $\lb(f, s)$ along with their intended meaning. (In what follows, $a\in\{0, 1\}^n$ represents the input of the circuit, $v, v'\in[s]$ represent gates of the circuit, $v' < v$, $b\in\{1, 2\}$, and $i\in[n]$ represents a variable.)

\def\Fanin{\mathrm{Fanin}}
\def\Type{\mathrm{Type}}
\def\InputType{\mathrm{InputType}}
\def\InputVar{\mathrm{InputVar}}
\def\INPUTVAR{\mathrm{INPUTVAR}}
\def\InputNode{\mathrm{InputNode}}
\def\INPUTNODE{\mathrm{INPUTNODE}}

\begin{longtable}{|c|p{.74\textwidth}|} 
    \hline
    \textbf{Name} & \textbf{Intended Meaning}\\
    \hline
    $y_{av}$ & The Boolean value computed at the computational node $v$ on the input $a$\\
    \hline
    $y_{abv}$ & The value on $a$ brought to $v$ by the $b$-th input to $v$\\
    \hline
    $\Fanin(v)$ & This is $0$ if $v$ is a NOT gate and $1$ if $v$ is an AND gate or an OR gate\\
    \hline
    $\Type(v)$ & When $\Fanin(v) = 1$, this is $0$ if $v$ is an AND gate and $1$ if $v$ is an OR gate\\
    \hline
    $\InputType_b(v)$ & This is $0$ if the $b$-th input to $v$ is a constant or a variable and $1$ if it is one of the previous computational gates\\
    \hline
    $\InputType'_b(v)$ & When $\InputType_b(v) = 0$, this is $0$ if the $b$-th input to $v$ is a constant, and is $1$ if it is a variable\\
    \hline
    $\InputType''_b(v)$ & When $\InputType_b(v) = \InputType'_b(v) = 0$, this equals the $b$-th input to $v$\\
    \hline
    $\InputVar_b(v, i)$ & When $\InputType_b(v) = 0$ and $\InputType'_b(v) = 1$, this is $1$ if and only if the $b$-th input to $v$ is the $i$-th variable\\
    \hline
    $\INPUTVAR_b(v, i)$ & Equals $\bigvee_{i'\le i}\InputVar_b(v, i')$, introduced to keep bottom fan-in bounded\\
    \hline
    $\InputNode_b(v, v')$ & When $\InputType_b(v) = 1$, this is $1$ if and only if the $b$-th input to $v$ is the previous gate $v'$\\
    \hline
    $\INPUTNODE_b(v, v')$ & Analogous to $\INPUTVAR_b(v, i)$ (i.e., equals $\bigvee_{v'' \le v'}\InputNode_b(v, v'')$).\\
    \hline
    \caption{Variables of $\lb(f, s)$}
    \label{tab: variables of bool lb}
\end{longtable}

Then, $\lb(f, s)$ consists of the following clauses (we can translate them into low-degree polynomials in the standard way):
\begin{enumerate}[({Group }1)]
    \item If the $b$-th input of $v$ is a constant, then this constant is equal to $y_{abv}$: \label{item: boolclb case1}
    \begin{itemize}
        \item $\lnot\InputType_b(v) \land \lnot\InputType'_b(v) \to (y_{abv} = \InputType''_b(v))$;
    \end{itemize}

    \item If the $b$-th input of $v$ is a variable, then there is a unique $i$ such that it is the $i$-th variable:\label{item: boolclb case2}
    \begin{itemize}
        \item $\lnot\InputType_b(v) \land \InputType'_b(v) \to \lnot(\InputVar_b(v, i)\land \InputVar_b(v, i'))$ where $i\ne i'$;
        \item $\lnot\InputType_b(v) \land \InputType'_b(v) \to (\INPUTVAR_b(v, i) = \INPUTVAR_b(v, i-1)\lor \InputVar_b(v, i))$ where we define $\INPUTVAR_b(v, 0) = 0$;
        \item $\lnot\InputType_b(v) \land \InputType'_b(v) \to \INPUTVAR_b(v, n)$;
        \item $\lnot\InputType_b(v) \land \InputType'_b(v) \land \InputVar_b(v, i) \to y_{abv} = a_i$;
    \end{itemize}

    \item If the $b$-th input of $v$ is a previous gate, then there is a unique $v' < v$ such that it is the $v'$-th gate: \label{item: boolclb case3}
    \begin{itemize}
        \item $\InputType_b(v) \to \lnot(\InputNode_b(v, v')\land \InputNode_b(v, v''))$ where $v'\ne v''$;
        \item $\InputType_b(v) \to (\INPUTNODE_b(v, v') = \INPUTNODE_b(v, v'-1)\lor \InputNode_b(v, v'))$ where we define $\INPUTNODE_b(v, 0) = 0$;
        \item $\InputType_b(v) \to \INPUTNODE_b(v, v-1)$;
        \item $\InputType_b(v) \land \InputNode_b(v, v') \to y_{abv} = y_{av'}$;
    \end{itemize}
    
    \item Every gate operates as expected: \label{item: boolclb case4}
    \begin{itemize}
        \item $\lnot\Fanin(v)\to (y_{av} = \lnot y_{a1v})$;
        \item $\Fanin(v) \land \lnot\Type(v)\to (y_{av} = y_{a1v} \land y_{a2v})$;
        \item $\Fanin(v) \land \Type(v)\to (y_{av} = y_{a1v} \lor y_{a2v})$;
    \end{itemize}

    \item The circuit computes $f$: \label{item: boolclb case5}
    \begin{itemize}
        \item $y_{as} = f(a)$.
    \end{itemize}
\end{enumerate}

\paragraph{The reduction.} We first provide an outline of the reduction. Given any Boolean function $f: \{0, 1\}^n \to \{0, 1\}$, let $p_f: \F^2 \to \F$ be the degree-$n$ non-commutative polynomial over the variables $z_0, z_1$ such that:
\begin{itemize}
    \item For every $x\in\{0, 1\}^n$, the coefficient of the monomial $z_{x_1}z_{x_2} \dots z_{x_n}$ is equal to $f(x)$. (Here we treat $f(x) \in \{0, 1\}$ as an element in $\F$.)
\end{itemize}
Suppose that $p_f$ can be computed by a small ncABP $P$. For every degree-$n$ non-commutative monomial $m$, let $\Coeff(P, m)$ denote the coefficient of $m$ in the non-commutative polynomial computed by $P$ (this non-commutative polynomial should be equal to $p_f$). Then $\Coeff$ can be computed by a small Boolean circuit. Plugging $P$ into $\Coeff$ gives us a small circuit for $f$. %I'm not super sure if this would be a good proof complexity reduction...

In more details, let $\vec{r} = (r_0, r_1, \dots, r_n)$ where $r_0 = r_d = 1$ and $r_i = r$ for every $1 \le i < d$. Let $M_1, \dots, M_n$ denote the matrices representing the ncABP $P$, where each $M_i$ is of dimension $r_{i-1}\times r_i$. Note that each entry of each $M_i$ is a linear form over (the non-commutative variables) $z_0$ and $z_1$, hence we can write $M_i = M_{i, z_0}z_0 + M_{i, z_1}z_1$ where each $M_{i, z_b} \in \F^{r_{i-1}\times r_i}$. Every $x \in \{0, 1\}^n$ corresponds to the non-commutative monomial $m_x = z_{x_1}z_{x_2} \dots z_{x_n}$, and the coefficient of $m_x$ is
\begin{equation}
    \prod_{i=1}^n M_{i, z_{x_i}}.\label{eq: coefficient of m_x}
\end{equation}

Let $\Coeff(\{M_{i, z_b}\}, x)$ denote the circuit that given $\{M_{i, z_b}\}$ and $x$ as inputs, outputs the value of \eqref{eq: coefficient of m_x}. The size of $\Coeff$ is at most $s = O(nr^2)$. Moreover, it is possible to design a circuit implementation of $\Coeff$ such that for every internal gate $g$ and every $x\in \{0, 1\}^n$, there is an $\F$-polynomial $q_{g, x}$ of degree $O(n)$ over the inputs $\{M_{i, z_b}\}$ that computes the value of $g$.

Now we are ready to specify the reduction from $\lb_\ncABP(p_f, \vec{r})$ to $\lb(f, s)$, where $s = O(nr^2)$. Let $\{M_{i, z_b}\}$ be the inputs of $\lb_\ncABP(p_f, \vec{r})$. Consider the circuit $C(x) = \Coeff(\{M_{i, z_b}\}, x)$ that takes $x\in \{0, 1\}^n$ as an input. The variables
\begin{align}
    &\Fanin(v), \Type(v), \InputType_b(v), \InputType'_b(v),\nonumber\\
    &\InputVar_b(v, i), \INPUTVAR_b(v, i), \InputNode_b(v, v'), \INPUTNODE_b(v, v')\label{eq: circuit structure variables}
\end{align}
are fixed constants that reflect the wiring of $C$. Note that the circuit $\Coeff$ takes $\{M_{i, z_b}\}$ as inputs, but the circuit $C$ treats $\{M_{i, z_b}\}$ as hardwired constants. Therefore, some variables of the form $\InputType''_b(v)$ are projections over $\{M_{i, z_b}\}$. Finally, the variables $y_{av}$ and $y_{abv}$ are internal values on the computation $C(a)$; all of them can be computed by a degree-$O(n)$ polynomial over $\{M_{i, z_b}\}$ (i.e., $q_{g, a}$ for the corresponding gate $g$).

Next we verify that every clause in $\lb(f, s)$ (composed with our reduction) admits a degree-$O(n)$ $\NS$-proof from $\lb_\ncABP(p_f, \vec{r})$. In fact this essentially follows from definition:
\begin{itemize}
    \item \groupref{item: boolclb case1} is true since we assign both $y_{abv}$ and $\InputType_b''(v)$ the same projection over $\{M_{i, z_b}\}$.
    \item \groupref{item: boolclb case2} and \groupref{item: boolclb case3} only contains variables in \eqref{eq: circuit structure variables} (which are constants under our reduction). They assert that the circuit has a valid structure; since our circuit $C$ indeed has a valid structure, they are trivially true.
    
    The only two exceptions are
    \begin{align*}
        \lnot\InputType_b(v) \land \InputType'_b(v) \land \InputVar_b(v, i) \to y_{abv} = a_i;&~\text{and}\\
        \InputType_b(v) \land \InputNode_b(v, v') \to y_{abv} = y_{av'}.
    \end{align*}
    as they contain not only variables in \eqref{eq: circuit structure variables} but also variables of the form $y_{abv}$ and $y_{av'}$. But they are also trivially true by our definitions of $y_{abv}$ and $y_{av'}$.

    \item \groupref{item: boolclb case4} is true because the polynomials $q_{g, x}$ correctly captures the values being computed at the gate $g$ on input $x$.
    \item \groupref{item: boolclb case5} asserts that $y_{as} = f(a)$. This is exactly the $a$-th axiom in $\lb_\ncABP(p_f, \vec{r})$ and hence trivially admits a degree-$O(n)$ $\NS$-proof from $\lb_\ncABP(p_f, \vec{r})$.
\end{itemize}

\begin{corollary}\label{cor: hardness of proving circuit lower bounds from rank principles}
    For every $f: \{0, 1\}^n \to \{0, 1\}$ and every parameter $r$, if the circuit complexity of $f$ is greater than $s = O(nr^2)$, then there is a degree-$O(n^2)$ algebraic reduction from $\Rank^{2r}_r$ to $\lb(f, s)$.
\end{corollary}
\begin{proof}
    Let $\vec{r} = (1, \underbrace{r, r, \dots, r}_{(n-1)\text{ }r\text{'s}}, 1)$ and $p_f: \F^2\to \F$ be the non-commutative polynomial guaranteed by \autoref{thm: hardness of boolean circuit lower bounds}, then there is a degree-$O(n)$ reduction from $\lb_\ncABP(p_f, \vec{r})$ to $\lb(f, s)$. By \autoref{thm: reduction from Rank to ncABP lb}, there is a degree-$2$ reduction from $\Rank_r^{2^n}$ to $\lb_\ncABP(p_f, \vec{r})$. By \autoref{thm: stretching wRank 1}, there is a degree-$O(n)$ reduction from $\Rank_r^{2r}$ to $\Rank_r^{2^n}$. Composing these three reductions (via \autoref{fact: composition of algebraic reductions}) gives us a degree-$O(n^2)$ reduction from $\Rank_r^{2r}$ to $\lb(f, s)$.
\end{proof}

%As a corollary, there is a degree-$O(n^2)$ reduction from $\Rank^{2r}_r$ to $\lb(f, O(nr^2))$.
%\slava{Again, refer to \autoref{cor: robustness of wRank wrt parameters}?}
%\hanlin{put it together, from rank principle to unprovability of circuit lb.}

\begin{remark}\label{remark: unprovability of NC2 lower bounds}
    The circuit for $\Coeff$ can be computed in $\NC^2$ (since it computes an iterated matrix multiplication over $\F$). Hence, the above argument also gives a reduction from the weak rank principle to sentences expressing $\NC^2$ circuit lower bounds. In particular, let $\lb_{\NC^2}(f, s)$ denote the collection of polynomial equations encoding that $f$ cannot be computed by a circuit of size $s$ and depth $\log^2 s$ (we do not attempt to formally define $\lb_{\NC^2}(f, s)$ here). The same argument above shows that for every Boolean function $f$ and every parameter $r$, letting $s = \poly(n, r)$, there is a degree-$O(n^2)$ algebraic reduction from $\Rank_r^{2r}$ to $\lb_{\NC^2}(f, s)$. Hence, if the rank principle is hard for some proof system $\calP$ (that is closed under low-degree reductions), then $\calP$ cannot prove any lower bound against $\NC^2$ circuits as well.
\end{remark}

%\paragraph{Better generators from weak rank principles.} We observe that the above proof actually implies a proof complexity generator of larger stretch from the weak rank principles. \hanlin{Talk about how the low-rank generator has stretch at most $n\mapsto n^2$.}

%In particular, consider the 

%\subsection{Algebraic Circuit Lower Bounds}
%\paragraph{Formalization of $\lb_\alg(f, s)$.} \hanlin{TODO: needs the universal circuits}

%!TEX root = main.tex

\subsection{Lower Bounds for \texorpdfstring{$\PCR_{\F_2}$}{PCR\_F2}}\label{sec: iterability gives ckt lbs}

We apply the iterated rank principles from \cref{sec: iterability of Rank} to obtain $\PCR_{\F_2}$ size lower bounds for $\lb_{\ncABP}(f,\vec{r})$ and $\lb^{\oplus}(f,s)$. The latter is a variant of $\lb(f,s)$ that allows binary $\oplus$ gates.
The first lower bound uses the algebraic iterated rank
principle, whereas the second uses its bamboo-tree encoding.

%!TEX root = main.tex

\subsubsection{Lower Bounds for Non-commutative ABPs}\label{sec: LB-ncABP}

We start by describing the formula that encodes the existence of small non-commutative algebraic branching programs for a given non-commutative polynomial $f$.

\begin{definition}[Lower bound formula for ncABPs]
    Let $n \ge 2$, $d \ge 2$, and $r \ge 1$ be integers.
    Let $f(\vec{z}) = f(z_1, z_2, \ldots, z_n)$ be a non-commutative homogeneous polynomial over $\F_2$ of degree $d$ in $n$ variables. Let $\vec{r} = (r_0, r_1, \ldots, r_d)$ with $r_0 = r_d = 1$.
    For every $\Delta \in [d]$ and $k \in [n]$, let $M_{\Delta,z_k}$ be a variable matrix of dimension $r_{\Delta-1} \times r_\Delta$.
    Let $\calM_{\Delta}$ be the set of all degree-$\Delta$ non-commutative monomials in $z_1, \ldots, z_n$. For every $\Delta \in [d]$ and $m \in \calM_{\Delta}$, let $P_{\Delta,m}$ be a $1\times r_\Delta$ row vector of variables.

    The lower bound formula $\ncABP(f, \vec{r})$ is defined as the following polynomial system:
    \begin{align}
        \label{eq: ncabp base}& P_{1,z_k} = M_{1,z_k} & k \in [n], \\
        \label{eq: ncabp step}& P_{\Delta, m z_k} = P_{\Delta-1, m} M_{\Delta,z_k} & k \in [n], \Delta \in \{2, \ldots, d\}, m \in \calM_{\Delta-1}, \\
        \label{eq: ncabp output}& P_{d,m} = \coeff(f, m) & m \in \calM_d.
    \end{align}
\end{definition}

The above formula encodes the existence of a non-commutative ABP of depth $d+1$ and width $\vec{r}$. The $\Delta$th layer, where $\Delta \in [d]$, computes the row vector of homogeneous
non-commutative polynomials, represented by their coefficient vectors:
$$
    \sum_{m \in \calM_\Delta} P_{\Delta,m} m.
$$
Edge labels between consecutive layers of the program are encoded by the matrices $M_{\Delta,z_k}$. The last set of equations \eqref{eq: ncabp output} ensures that the program computes the given polynomial $f$.

We show that any satisfying assignment to the formula $\IRank^m_n(\bfA, \kappa, \Sigma)$ (for a suitable choice of parameters) gives a construction of a polynomial-size $\ncABP$ circuit for any non-commutative polynomial.

\begin{theorem}\label{thm: ncABP lb for PCR2}
    Let $n \ge 2$, $d \ge 2$, and $r \ge 1$ be integers.
    Let $f(\vec{z}) = f(z_1, z_2, \ldots, z_n)$ be a non-commutative homogeneous polynomial over $\F_2$ of degree $d$ in $n$ variables. Let $\vec{r} = (1, \underbrace{r, r, \ldots, r}_{d-1}, 1)$.
    
    Then any $\PCR_{\F_2}$ refutation of $\ncABP(f, \vec{r})$ requires size $2^{\Omega(\sqrt{r/d})}$.
\end{theorem}
\begin{proof}

We will construct a reduction to $\ncABP(f, \vec{r})$ from $\IRank^{nr}_r(\bfA, d-1, [n])$ for some particular family of matrices $\bfA$.
Let $\Pi$ be a $\PCR_{\F_2}$ refutation of $\ncABP(f, \vec{r})$.

For every $\pi = (j_1, \ldots, j_{d-1}) \in {[n]}^{d-1}$ define $A^{(\pi)}$ as any $nr \times nr$ Boolean constant matrix satisfying
\[
    A^{(\pi)}_{1,(j_d-1)r + 1} = \coeff(f, z_{j_1} \cdots z_{j_{d-1}} \cdot z_{j_d})
\]
for all $j_d \in [n]$. The remaining entries of $A^{(\pi)}$ can be filled arbitrarily.

For every $\pi \in {[n]}^{< d-1}$, define $A^{(\pi)}$ as an $nr \times nr$ block matrix
\[
    A^{(\pi)} = \begin{bmatrix}
        X^{(\pi * 1)} & X^{(\pi * 2)} & \cdots & X^{(\pi * n)}
    \end{bmatrix}.
\]

Let $\bfA = \{ A^{(\pi)} : \pi \in {[n]}^{\le d-1} \}$.

Also split the matrix $Y$ into $n$ consecutive blocks of dimension $r\times r$:
\[
    Y = \begin{bmatrix}
        Y^{(1)} & Y^{(2)} & \cdots & Y^{(n)}
    \end{bmatrix}.
\]

Given $\pi \in {[n]}^{\le d-1}$, $X^{(\pi)}_1$ denotes the first \emph{row} of the matrix $X^{(\pi)}$.
Similarly, given $k \in [n]$, $Y^{(k)}_1$ denotes the first \emph{column} of the matrix $Y^{(k)}$.
Now define the reduction $\tau$ as follows.
\begin{align*}
    & \tau(P_{\Delta,m}) = X^{(j_1, \ldots, j_\Delta)}_1 & \Delta \in [d-1], m = z_{j_1} \cdots z_{j_\Delta} \in \calM_\Delta, \\
    & \tau(P_{d,m}) = A^{(j_1, \ldots, j_{d-1})}_{1,(j_d-1)r + 1} & m = z_{j_1} \cdots z_{j_d} \in \calM_d, \\
    & \tau(M_{1,z_k}) = X^{(k)}_1 & k \in [n], \\
    & \tau(M_{\Delta,z_k}) = Y^{(k)} & \Delta \in \{2, \ldots, d-1\}, k \in [n], \\
    & \tau(M_{d,z_k}) = Y^{(k)}_1 & k \in [n].
\end{align*}

For every $k\in[n]$, $\tau$ assigns $P_{1, z_k}$ to $X^{(k)}_1$ and $M_{1, z_k}$ to $X^{(k)}_1$, which satisfies \eqref{eq: ncabp base}.

For every $m = z_{j_1} \cdots z_{j_d} \in \calM_d$, we have
\[
    A^{(j_1, \ldots, j_{d-1})}_{1,(j_d-1)r + 1} = \coeff(f, z_{j_1} \cdots z_{j_d}),
\]
which satisfies \eqref{eq: ncabp output}.

Now consider the axioms \eqref{eq: ncabp step}. For every $\Delta \in \{2, \ldots, d-1\}$, $m = z_{j_1} \cdots z_{j_{\Delta-1}} \in \calM_{\Delta-1}$ and $j_\Delta \in [n]$, under $\tau$ the axiom becomes
\[
    X^{(j_1, \ldots, j_\Delta)}_1 = X^{(j_1, \ldots, j_{\Delta-1})}_1 Y^{(j_\Delta)}.
\]
Similarly, for every $m = z_{j_1} \cdots z_{j_{d-1}} \in \calM_{d-1}$ and $j_d \in [n]$, we have
\[
    A^{(j_1, \ldots, j_{d-1})}_{1,(j_d-1)r + 1} = X^{(j_1, \ldots, j_{d-1})}_1 Y^{(j_d)}_1.
\]
In both cases, these polynomial equations are among the axioms of $\IRank^{nr}_r(\bfA, d-1, [n])$.

Hence every axiom of $\ncABP(f, \vec{r}) \restriction \tau$ is either satisfied or becomes an axiom of $\IRank^{nr}_r(\bfA, d-1, [n])$, and $\Pi \restriction \tau$ is a refutation of $\IRank^{nr}_r(\bfA, d-1, [n])$.
By \cref{thm: lower bound for iterated algebraic irank}, this implies that the size of $\Pi$ is $2^{\Omega(\sqrt{r/d})}$.
\end{proof}

Observe that the size of $\ncABP(f, \vec{r})$ is polynomial in $n^d d r^2$, thus the bound becomes non-trivial when $r = \omega(d^3 \log^2 n)$.

%!TEX root = main.tex

\subsubsection{Lower Bounds for Boolean Circuits}

As in~\cite{Razb15-annals}, we consider Boolean circuits in the basis $\lnot, \land, \lor, \oplus$, where the last three connectives are binary.

\begin{definition}[Formula $\lb^{\oplus}(f, s)$]
    Let $n$ and $s$ be positive integers, and let $f\colon \Q^n \to \Q$ be a Boolean function.
    The formula $\lb^{\oplus}(f, s)$ is defined similarly to $\lb(f, s)$ except that it also allows binary $\oplus$ connectives.
\end{definition}

We show that if, for a particular choice of parameters, the formula $\psi_{P(H(G))}(\bfA, \kappa)$ is satisfiable, we can construct a polynomial-size circuit for any Boolean function.

\begin{theorem}\label{thm: boolean lb for PCR2}
    There exists $\alpha > 0$ such that for every sufficiently large integer $n$, every integer $s \ge n^5$, and every Boolean function $f\colon \Q^n \to \Q$, any $\PCR_{\F_2}$ refutation of $\lb^{\oplus}(f, s)$ requires size $2^{\Omega({(s^{\alpha/5} / n)}^{1/3})}$.
\end{theorem}
\begin{proof}

Let $r = s^{1/5} \ge n$ and $q = (r-10)/8$. Note that $q=\Theta(r)=\Theta(s^{1/5})$.

Choose $\delta,d > 0$ from \cref{lem:good expander exists} such that there exists a $(16r) \times q$ $(q^\delta, d)$-lossless expander $G$ with the appropriate degree bounds and denote $G' = P(H(G))$.
Then fix any $0 < \alpha < \delta/4$.
If $q^\alpha < n$, the lower bound is trivial, so we can assume that $q^\alpha \ge n$. This allows us to apply \cref{thm: lower bound on iterated formula psi} later.

We will reduce $\psi_{P(H(G))}(\bfA, n)$ to $\lb^{\oplus}(f, s)$ for some particular $\bfA$.
Let $\Pi$ be a $\PCR_{\F_2}$ refutation of $\lb^{\oplus}(f, s)$.

Define $A_1$ and $A_0$ as $16r \times 16r$ matrices as follows:
\[
    A_1 = \begin{pmatrix}
        1 & 0 & 0 & \cdots & 0 \\
        0 & 1 & 0 & \cdots & 0 \\
        0 & 0 & 1 & \cdots & 0 \\
        \vdots & \vdots & \vdots & \ddots & \vdots \\
        0 & 0 & 0 & \cdots & 1
    \end{pmatrix}
    \text{ and }
    A_0 = \begin{pmatrix}
        0 & 1 & 0 & \cdots & 0 \\
        1 & 0 & 0 & \cdots & 0 \\
        0 & 0 & 1 & \cdots & 0 \\
        \vdots & \vdots & \vdots & \ddots & \vdots \\
        0 & 0 & 0 & \cdots & 1
    \end{pmatrix}.
\]
The important property of these matrices is that ${(A_1)_{1,1}} = 1$ and ${(A_0)_{1,1}} = 0$.
For every $\pi \in \Q^n$, define $A^{(\pi)} = A_{f(\pi)}$.

For every $\pi \in \Q^{\le n}$, we define block matrices $X^{(\pi)}$ and $Y^{(\pi)}$ as follows:
\[
    X^{(\pi)} = \begin{bmatrix}
        X^{(\pi,1)} & X^{(\pi,2)} & X^{(\pi,3)} & X^{(\pi,4)}
    \end{bmatrix},
    \text{ and }
    Y^{(\pi)} = \begin{bmatrix}
        Y^{(\pi,1)} \\ Y^{(\pi,2)} \\ Y^{(\pi,3)} \\ Y^{(\pi,4)}
    \end{bmatrix}.
\]
Here, for every $\mu \in [4]$, $k \in [r]$, and $j \in [16r]$ with $k\notin\calN_{G'}(j)$, the $(k,j)$th entry of $Y^{(\pi,\mu)}$ is interpreted as $0$.
Then for every $\pi \in \Q^{< n}$, $A^{(\pi)}$ is a $16r \times 16r$ block matrix
\[
    A^{(\pi)} = \begin{bmatrix}
        X^{(\pi * 0)} & X^{(\pi * 1)} & {Y^{(\pi * 0)}}^T & {Y^{(\pi * 1)}}^T
    \end{bmatrix}.
\]

Let
\[
    \bfA=\{A^{(\pi)}:\pi\in\Q^{\le n}\}.
\]

We construct the circuit $\calC(v_1, \ldots, v_n)$ that computes $f$, provided that $\psi_{P(H(G))}(\bfA, n)$ is satisfied. The circuit uses gates that correspond to the extension variables of $\psi_{P(H(G))}(\bfA, n)$.
\begin{enumerate}
    \item For every $\mu \in [4]$, $i \in [16r]$, $k \in [r]$:
    \[
        g_x[0, \mu, i, k] = x^{((),\mu)}_{i,k}.
    \]

    \item For every $\ell \in [n]$, $\mu \in [4]$, $i \in [16r]$, $k \in [r]$:
    \[
        g_x[\ell, \mu, i, k] = (\lnot v_\ell \land g_w[\ell-1, i, (\mu-1)r + k, 0]) \lor (v_\ell \land g_w[\ell-1, i, (\mu-1)r + k + 4r, 0]).
    \]

    \item For every $\mu \in [4]$, $j \in [16r]$, $k \in \calN_{G'}(j)$:
    \[
        g_y[0, \mu, k, j] = y^{((),\mu)}_{k,j}.
    \]

    \item For every $\ell \in [n]$, $\mu \in [4]$, $j \in [16r]$, $k \in \calN_{G'}(j)$:
    \[
        g_y[\ell, \mu, k, j] = (\lnot v_\ell \land g_w[\ell-1, j, (\mu-1)r + k + 8r, 0]) \lor (v_\ell \land g_w[\ell-1, j, (\mu-1)r + k + 12r, 0]).
    \]

    \item For every $\ell \in \{0, \ldots, n\}$, $\mu \in [4]$, $i,j \in [16r]$, $k \in \calN_{G'}(j)$:
    \[
        g_z[\ell, \mu, i, j, k] = g_x[\ell, \mu, i, k] \land g_y[\ell, \mu, k, j].
    \]

    \item For every $\ell \in \{0, \ldots, n\}$, $\mu \in [4]$, $i,j \in [16r]$:
    \[
        g_u[\ell, \mu, i, j, (1)] = g_z[\ell, \mu, i, j, 1].
    \]

    \item For every $\ell \in \{0, \ldots, n\}$, $\mu \in [4]$, $i,j \in [16r]$ and every initial segment $\Sigma \cup (k)$ of $\calN_{G'}(j)$ with $\Sigma \neq \varnothing$:
    \[
        g_u[\ell, \mu, i, j, \Sigma \cup (k)] = g_u[\ell, \mu, i, j, \Sigma] \oplus g_z[\ell, \mu, i, j, k].
    \]

    \item For every $\ell \in \{0, \ldots, n\}$, $i,j \in [16r]$:
    \begin{align*}
        g_w[\ell, i, j, 1] &= g_u[\ell, 1, i, j, \calN_{G'}(j)] \oplus g_u[\ell, 2, i, j, \calN_{G'}(j)], \\
        g_w[\ell, i, j, 2] &= g_u[\ell, 3, i, j, \calN_{G'}(j)] \oplus g_u[\ell, 4, i, j, \calN_{G'}(j)], \\
        g_w[\ell, i, j, 0] &= g_w[\ell, i, j, 1] \oplus g_w[\ell, i, j, 2].
    \end{align*}
\end{enumerate}
The output of the circuit is $g_w[n, 1, 1, 0]$. The size of the circuit is $O(nr^3)$, which is $o(s)$ for large enough $n$. It follows from the construction that for every $\pi \in \Q^n$, $\calC(\pi) = {A^{(\pi)}}_{1,1} = f(\pi)$.

We plug $\calC(v_1,\ldots,v_n)$ into $\lb^{\oplus}(f,s)$. Under the resulting substitution $\tau$, every axiom encoding the circuit structure is satisfied. The remaining gate axioms become the corresponding Binary AND, Summation Base, Summation, or Output Axioms of $\psi_{G'}(\bfA,n)$. For every input $\pi\in\Q^n$, the circuit-output axiom becomes
\[
    g_w[n,1,1,0]=A^{(\pi)}_{1,1}=f(\pi).
\]
Thus, every axiom of $\lb^{\oplus}(f,s)\restriction\tau$ is either an identity or an axiom of $\psi_{G'}(\bfA,n)$. Consequently, $\Pi\restriction\tau$ is a refutation of $\psi_{G'}(\bfA,n)$. By \cref{thm: lower bound on iterated formula psi}, this implies that $\lb^{\oplus}(f, s)$ requires a refutation of size $2^{\Omega({(q^\delta / n)}^{1/3})} = 2^{\Omega({(s^{\delta/5} / n)}^{1/3})} \ge 2^{\Omega({(s^{\alpha/5} / n)}^{1/3})}$.
\end{proof}

Observe that the size of $\lb^{\oplus}(f, s)$ is $\poly(2^n, s)$. Thus, the bound becomes non-trivial when $s = \omega(n^{20/\alpha})$.

%!TEX root = main.tex

\section{Proving Circuit Lower Bounds via the Weak Rank Principle}\label{sec: Smolensky from weak rank principle}

We show that the weak rank principle, when added as an axiom to  a weak formal theory corresponding to $\AC^0[p]$, is capable of proving lower bounds against $\AC^0[p]$ circuits by formalising Smolensky's lower bound~\cite{Smolensky87}.
%Below already appeared in the introduction
%Fix an odd prime $p \le O(1)$, Smolensky~\cite{Smolensky87} showed that any depth-$d$ $\AC^0[p]$ circuit requires size $2^{\Omega(n^{1/2d})}$ to compute the $\MOD_2$ function.

%This section examines the ``reasoning power'' of this proof carefully. Our main conclusion is that to prove Smolensky's lower bounds, it suffices to use basic operations over $\AC^0[p]$ circuits and \emph{one invocation of the weak rank principle} over $2^{O(n)}\times 2^{O(n)}$ matrices. More formally, we consider the theory $\overline{\V^0(p)} + \wRank_p$ (defined in \autoref{sec: def of V0(3) + wRank}) and show that this theory proves Smolensky's lower bounds. We think that the contribution of this section is mostly conceptual: it demonstrates that (even the weak version of) rank principles are powerful enough to prove lower bounds against $\AC^0[p]$ circuits.

%Previously, M\"uller and Pich~\cite{MullerP20} formalised Smolensky's $\AC^0[p]$ lower bounds in \Jerabek's theory $\mathbf{APC}_1$~\cite{Jerabek04, Jer07}. The difference between our formalisation and theirs will be discussed in \autoref{remark: comparison with Muller-Pich}, at the end of \autoref{sec: formalisations of smolensky}.

\paragraph{Roadmap of \autoref{sec: Smolensky from weak rank principle}.} Smolensky's proof consists of two parts: a correlation bound against low-degree polynomials (i.e., any low-degree polynomial over $\F_p$ has to make many errors when computing $\prod_{i=1}^n x_i$) and the ``polynomial method'' (i.e., every small $\AC^0[p]$ circuit is well-approximated by a low-degree polynomial over $\F_p$). Hence we arrange this section as follows:
\begin{itemize}
    \item \autoref{sec: smolensky's lower bound} provides an exposition of Smolensky's proof. The exposition is equivalent to the original proof, but it will be presented in a way that only involves simple manipulations computable in $\AC^0[p]$ and the weak rank principles.
    
    In \autoref{sec: correlation lb against low-deg polys} we prove the correlation bound against low-degree $\F_p$-polynomials. This is the main step where the weak rank principle is involved. In \autoref{sec: polynomial method} we show that small $\AC^0[p]$ circuits can be approximated by low-degree $\F_p$ polynomials, and put everything together to finish the circuit lower bound.
    \item In \autoref{sec: formlisation} we formalise the lower bound in $\overline{\V^0(p)} + \wRank_p$.
    
    We start with a description of the relevant theories ($\overline{\V^0(p)}$ and $\overline{\V^0(p)} + \wRank_p$) in \autoref{sec: def of V0(3) + wRank}. The main formalisation will appear in \autoref{sec: formalisations of smolensky}. Finally, \autoref{sec: proof of ECC in BA} contains a postponed proof of a technical theorem related to error-correcting codes.
\end{itemize}

\Iddo{Bar theory suspected to be conservative over non-bar theory.}

\Iddo{Development of theories of algebra: Gaisin; Extending to get some algbic reasoning for well studied theories; whereas people considered stronger extension. Ours is a first step in linear algebra.}

\Iddo{Open problem: Compare to SC04: LAP conjectured to prove the DET idens and hence also strong Rank princinple; So probably stronger than our theories, which are WRank (weak means 2n-->n).}

It is  recommended to read \autoref{sec: smolensky's lower bound} before delving into \autoref{sec: formlisation} for the following reasons. First, to make the invocation of weak rank principles explicit, the proofs in \autoref{sec: smolensky's lower bound} are presented in a somewhat different way than it is usually presented in textbooks (e.g.,~\cite{AB09}); it would be beneficial to be familiarised with the proof before formalising it. Second, \autoref{sec: smolensky's lower bound} contains important definitions (such as the matrices on which the weak rank principle is invoked) that the formalisation in \autoref{sec: formlisation} needs to reason about.

\subsection{Smolensky's Lower Bound}\label{sec: smolensky's lower bound}

\subsubsection{Correlation Bound Against Low-Degree Polynomials}\label{sec: correlation lb against low-deg polys}

We need the following correlation bound against low-degree polynomials.
\begin{theorem}\label{thm: corr bounds for low-deg poly}
    Let $P: \F_p^n\to \F_p$ be a polynomial of degree at most $0.1\sqrt{n}$. Let $S = \{x \in \{-1, 1\}^n: P(x) = \prod_{i=1}^n x_i\}$, then $|S| \le 0.99\cdot 2^n$.
\end{theorem}
\begin{proof}[Proof Ideas]
    The idea is that every function $f: S \to \F_p$ is equivalent to a polynomial of degree at most $d = n/2 + 0.1\sqrt{n}$ over $S$. In fact, let $T\subseteq[n]$ and $m = \prod_{i\in T}x_i$ be a multilinear monomial. If $|T| = \deg(m) \le n/2$, then $m$ is trivially equivalent to a polynomial of degree at most $d$ (i.e., $m$ itself). If $|T| = \deg(m) > n/2$, then the equality 
    \begin{equation}\label{eq: making monomials low degree}
        m \equiv \mleft(\prod_{i\in [n]\setminus T}x_i\mright) \cdot P(x)
    \end{equation}
    still holds over $S$, and the RHS of \eqref{eq: making monomials low degree} is a polynomial of degree at most $d$. Hence every multilinear monomial $m$ is equivalent to a polynomial of degree at most $d$ over $S$. Since $S\subseteq\{-1, 1\}^n$, every function $f: S\to \F_p$ can be written as a linear combination of multilinear monomials, hence is also equivalent to a polynomial of degree at most $d$.
    
    Given this, it follows that there are at most 
    \[p^{\sum_{i=0}^d\binom{n}{i}} \le p^{0.99\cdot 2^n}\]
    different functions $f: S\to \F_p$, indicating that $|S| \le 0.99\cdot 2^n$.
\end{proof}

To aid our formalisations later, we need to explicitly construct the matrices over which the weak rank principle is used (instead of using concepts such as \emph{dimensions of vector spaces}). We now rephrase the above proof so that the invocation of the weak rank principle becomes clear.

%We now rephrase the above proof in the following way so that it becomes clear that it only involves some easy manipulation plus one invocation of the weak rank principle. %As our formalisations crucially depend on the proof below, we recommend going through (i.e., not skipping) this proof.

\def\low{{\sf low}}
\def\high{{\sf high}}

\begin{theorem}\label{thm: corr bounds using weak rank principle}
    Let $S\subseteq \{-1, 1\}^n$, $|S| \ge 0.999\cdot 2^n$, $P: S\to\F_p$ be any polynomial of degree at most $0.1\sqrt{n}$. Then there exists some $z\in S$ such that $P(z)\ne \prod_{i=1}^n z_i$.
\end{theorem}
\begin{proof}
    Given $S\subseteq \{-1, 1\}^n$, we construct the following matrices $A\in \F_p^{|S|\times r}$ and $B\in \F_p^{r\times |S|}$ where $r = 0.99\cdot 2^n$. We identify $[r]$ (i.e., columns of $A$ and rows of $B$) with monomials $m$ of degree at most $d = n/2 + 0.1\sqrt{n}$. Now, for an input $x\in S$ and a monomial $m\in [r]$:
    \begin{itemize}
        \item Define $A[x, m] = m(x)$ to be the value of $m$ on input $x$. Note that $A$ can be seen as a linear mapping for \emph{polynomial evaluation}: Let $\vec{e} \in \F_p^r$ and think of $\vec{e}$ as the coefficient list of a polynomial $P$ with degree at most $d$. (That is, for every monomial $m$ of degree at most $d$, $\vec{e}_m$ is the coefficient of $m$ in $P$.) Then $A\vec{e} \in \F_p^S$ is exactly the list of values of $P$ over $S$.

        \item Define $B[m, x]$ to be the sum of the following values:
        \begin{enumerate}
            \item if $\deg(m)\le n/2$, we add $(\frac{p+1}{2})^n m(x)$ to $B[m, x]$;
            \item for every monomial $m'$ that appears in $P$ with coefficient $\alpha_{m'}$, if the degree of $\tilde{m} = m'\cdot m\cdot \prod_{i\in [n]}x_i$ is greater than $n/2$, we add $(\frac{p+1}{2})^n \cdot \alpha_{m'}\cdot \tilde{m}(x)$ to $B[m, x]$.
        \end{enumerate}
    \end{itemize}

    The intended meaning of $B$ is the following. It was shown in \autoref{thm: corr bounds for low-deg poly} that every function $f: S\to \F_p$ is equivalent to a polynomial of degree at most $d = n/2 + 0.1\sqrt{n}$. Let $\vec{v}_f \in \F_p^S$ be the vector representing the value list of $f$, then $B\vec{v}_f \in \F_p^r$ is exactly the coefficient list of this polynomial.

    To be more precise, we introduce the following notation. Let $\calM$ denote the set of all multilinear monomials on variables $z_1, \dots, z_n$, $\calM_{\le n/2}$ and $\calM_{>n/2}$ denote the subset of monomials with degree $\le n/2$ and $>n/2$ respectively. Let $f: S\to \F_p$ be a function, then it can be expressed as the polynomial
    \[f(z) = \sum_{m\in \calM}\lambda_{f, m}m(z)\]
    where $\lambda_{f, m} = (\frac{p+1}{2})^n \sum_{x\in S}f(x)m(x)$.

    To see this, notice that
        \begin{align*}
            \sum_{m\in \calM}\lambda_{f, m}m(z) = &\,\sum_{m\in\calM}\mleft(\frac{p+1}{2}\mright)^n\sum_{x\in S}f(x)m(x)m(z)\\
            =&\, \mleft(\frac{p+1}{2}\mright)^n\sum_{x\in S}f(x)\sum_{m\in \calM}m(x)m(z)\\
            =&\, \mleft(\frac{p+1}{2}\mright)^n\sum_{x\in S}f(x)\cdot \delta_{x, z}\cdot 2^n\\
            =&\, f(z). & (\text{over }\F_p)
        \end{align*}

        Given any function $f:S\to \F_p$, denote
    \[f_\low(z) = \sum_{m\in \calM_{\le n/2}}\lambda_{f, m}m(z)\quad\text{and}\quad f_\high(z) = \sum_{m\in \calM_{>n/2}}\lambda_{f, m}m(z).\]
    
    The following claim expresses the ``intended meaning of $B$'' mentioned above:

    \begin{claim}\label{claim: meaning of B}
        Let $f: S\to \F_p$ be any function, $\vec{v}_f\in \F_p^S$ be the list of values of $f$ on $S$, and $\hat{f}$ denote the polynomial whose coefficient list is $B\cdot \vec{v}_f$. Then for every $z\in S$,
        \[\hat{f}(z) = f_\low(z) + f_\high(z) \cdot P(z)\cdot \prod_{i=1}^n z_i.\]
    \end{claim}
    \begin{claimproof}
        We have that
        \begin{align}
             &\, f_\low(z) + f_\high(z)\cdot P(z)\cdot \prod_{i=1}^n z_i\nonumber\\
            =&\, \sum_{m\in \calM_{\le n/2}}\lambda_{f, m}m(z) + \sum_{\tilde{m}\in \calM_{>n/2}}\lambda_{f, \tilde{m}}\tilde{m}(z)P(z)\prod_{i=1}^n z_i\nonumber\\
            =&\, (\frac{p+1}{2})^n\sum_{x\in S}\mleft(\underbrace{\sum_{m\in \calM_{\le n/2}}f(x)m(x)m(z)}_{\text{(Part I)}} + \underbrace{\sum_{\tilde{m}\in \calM_{>n/2}}f(x)\tilde{m}(x)\tilde{m}(z)P(z)\prod_{i=1}^n z_i}_{\text{(Part II)}}\mright).\label{eq: flow + fhigh * diff}
        \end{align}
        Let $x\in S$, $m\in \calM$, consider the coefficient of $f(x)m(z)$ in \eqref{eq: flow + fhigh * diff}. If $\deg(m) \le n/2$ then the contribution of (Part I) is $(\frac{p+1}{2})^nm(x)$; otherwise it is $0$. For every monomial $m'$ of $P$ with coefficient $\alpha_{m'}$, if $\tilde{m} = m\cdot m'\cdot \prod_{i=1}^n z_i$ has degree $>n/2$, then it contributes to (Part II) a coefficient of $(\frac{p+1}{2})^n\tilde{m}(x)\alpha_{m'}$; otherwise it contributes nothing. It follows from the definition of $B$ that
        \[\eqref{eq: flow + fhigh * diff} = \sum_{x\in S}\sum_{m\in \calM}B[m, x]f(x)m(z),\]
        which by definition is equal to $\hat{f}(z)$.
    \end{claimproof}

    We also claim that any witness that $A\cdot B$ is not the identity matrix gives rise to a bad input on which $P$ and $\prod_{i=1}^n z_i$ differ:
    \begin{claim}\label{claim: witness of wRank implies a bad input}
        If $z, x\in S$ and $(AB)[z, x]\ne \delta_{z, x}$, then $P(z) \ne \prod_{i=1}^n z_i$.
    \end{claim}
    \begin{claimproof}
        Let $f: \{-1, 1\}^n \to \F_p$ be the function that evaluates to $1$ on input $x$ and evaluates to $0$ on every input in $\{-1, 1\}^n \setminus\{x\}$; note that $\vec{1}_x$ is exactly the ``value list'' vector of $f$. By \autoref{claim: meaning of B}, $B\cdot \vec{1}_x \in \F_p^r$ is the coefficient list of the polynomial $f'(z) = f_\low(z) + f_\high(z) \cdot P(z)\cdot \prod_{i=1}^n z_i$. Note that $f(z) = f_\low(z) + f_\high(z) = \delta_{z, x}$. However, $(AB)[z, x] = f'(z) \ne \delta_{z, x}$. This means that $P(z) \ne \prod_{i=1}^n z_i$.
    \end{claimproof}

    Since $|S| \ge 0.999\cdot 2^n \gg r$, it follows from the weak rank principle that there exists $z, x\in S$ such that $(AB)[z, x]\ne \delta_{z, x}$. This implies that $P(z) \ne \prod_{i=1}^n z_i$ as shown by \autoref{claim: witness of wRank implies a bad input}.
\end{proof}

The matrices $A$ and $B$ in the above proof will be crucial for our formalisation, hence we reiterate their definitions here. (Note that $A$ does not depend on the polynomial $P$ but $B$ does.)

\begin{definition}\label{def: matrices A and B}
    Let $N = 2^n$, $r = 0.99\cdot 2^n$, and $P: \{-1, 1\}^n \to \F_p$ be a polynomial of degree at most $0.1\sqrt{n}$. Identify $[N]$ with $\{-1, 1\}^n$ and $[r]$ with the set of monomials of degree at most $d = n/2 + 0.1\sqrt{n}$. Define:
    \begin{itemize}
        \item $A$ to be the $N\times r$ matrix such that $A[x, m] = m(x)$ for every $x\in\{-1, 1\}^n$ and $m \in \calM_{\le d}$.
        \item $B(P)$ to be the $r\times N$ matrix such that for every $x\in \{-1, 1\}^n$ and $m\in \calM_{\le d}$, $B[m, x]$ is the sum of the following values. (1) If $\deg(m) \le n/2$, we add $(\frac{p+1}{2})^n m(x)$ to $B[m, x]$; (2) for every monomial $m'$ that appears in $P$ with coefficient $\alpha_{m'}$, if the degree of $\tilde{m} = m'\cdot m\cdot \prod_{i\in[n]}x_i$ is greater than $n/2$, then we add $(\frac{p+1}{2})^n\cdot \alpha_{m'}\cdot \tilde{m}(x)$ to $B[m, x]$.
    \end{itemize}
\end{definition}

\subsubsection{Approximating \texorpdfstring{$\AC^0[p]$}{AC0[p]} Circuits by Low-Degree Polynomials}\label{sec: polynomial method}

\def\rand{\mathsf{rand}}
The next ingredient in the proof is the \emph{low-degree} approximation of $\AC^0[p]$ circuits~\cite{Razborov87}: For every parameter $\Delta$ and every $\AC^0[p]$ circuit $C$ of depth $d$ and size $s$, there is a polynomial $P: \F_p^n\to \F_p$ of degree at most $(p\Delta)^d$ that agrees with $C$ on at least a $(1-s\cdot 0.9^\Delta)$ fraction of inputs. Here we treat $\{0, 1\}$ as a subset of $\F_p$ and we guarantee that for every $x \in \{0, 1\}^n$, $P(x) \in \{0, 1\}$. We also assume that $C$ only consists of $\NOT$ gates, $\MOD_p$ gates, and $\OR$ gates ($\AND$ gates can be simulated by $\NOT$ and $\OR$ gates using De Morgan's law), and the definition of $\MOD_p$ gates is
\[\MOD_p(x_1, x_2, \dots, x_n) = \begin{cases}0&\text{if }\sum_{i=1}^n x_i \equiv 0\pmod p;\\1 & \text{otherwise}.\end{cases}\]

\paragraph{Defining the low-degree polynomials.} In fact, the Razborov--Smolensky proof provides us a stronger guarantee: There is a \emph{distribution $\calP$ of low-degree polynomials} such that for every input $x\in\{0, 1\}^n$, with high probability over $P\gets \calP$, we have $P(x) = C(x)$. Using error-correcting codes, we can make $\calP$ the uniform distribution over a support of size $L = 2^{O(n)}$.\footnote{As our formalisation in \autoref{sec: formlisation} will use the weak rank principles over matrices of dimension $\poly(L, 2^n)$, it is crucial that $L$ is not too large.} That is, there is a list of $L$ polynomials $P_{C, 1}, P_{C, 2}, \dots, P_{C, L}$ of degree at most $(p\Delta)^d$ such that for every $x\in\{0, 1\}^n$,
\[\Pr_{j\gets [L]}[P_{C, j}(x) \ne C(x)] \le s\cdot 0.9^\Delta.\]

%For some $\ell = \poly(s)$, let $\rand \in [\ell]$ be the random bits ``used'' in the proof, and $P_{C, \rand}$ be the degree-$(2\Delta)^d$ polynomial constructed in the proof. Then
%\[\Pr_{\rand\gets \{0, 1\}^\ell, x\gets \{0, 1\}^n}[P_{C, \rand}(x) \ne C(x)] \le s/2^\Delta.\]
%For the sake of our formalisations, it would be helpful to define the polynomial $P_{C, \rand}$ explicitly.

\def\IP{\mathrm{IP}}

In particular, let $\ell\le\poly(t)$, $\Enc: \F_p^t \to \F_p^\ell$ be an $\F_p$-linear error-correcting code such that for every $\vec{x}\in \F_p^t\setminus \{0^t\}$, at least a $0.1$ fraction of indices $i\in [\ell]$ satisfy that $\Enc(\vec{x})_i \ne 0$. Let $\vec{j} = (j_1, j_2, \dots, j_\Delta)$ where each $j_i\in [\ell]$. We will have a polynomial $P_{C, \vec{j}}$ for each such $\vec{j}$, hence there are $L = \ell^\Delta$ polynomials in total. We define $P_{C, \vec{j}}$ inductively over the structure of $C$:
\begin{enumerate}
    \item If the top gate of $C$ is a $\NOT$ gate, then $C = \lnot C'$ for some circuit $C'$. We define $P_{C, \vec{j}} = 1-P_{C', \vec{j}}$.
    \item If the top gate of $C$ is a $\MOD_p$ gate, then $C = \MOD_p(C_1, C_2, \dots, C_t)$ for some $t\in\N$ and sub-circuits $C_1, C_2, \dots, C_t$. We define
    \[P_{C, \vec{j}} = (\sum_{i=1}^t P_{C_i, \vec{j}})^{p-1}.\]
    The correctness of $P_{C, \vec{j}}$ follows from Fermat's little theorem.
    \item If the top gate of $C$ is an $\OR$ gate, then $C = \OR(C_1, C_2, \dots, C_t)$ for some $t\in\N$ and sub-circuits $C_1, C_2, \dots, C_t$. The idea here is to approximate the $\OR$ function using the $j_1$-th, $\dots$, $j_\Delta$-th outputs of $\Enc$: let $H_C(x) = (C_1(x), \dots, C_t(x))$, we ``think'' that $C(x)$ outputs $1$ if there exists some $i\in[\Delta]$ such that $\Enc(H_C(x))_{j_i} \not\equiv 0\pmod p$.
    
    In particular, let $\OR: \F_p^\Delta\to \F_p$ be any degree-$\Delta$ polynomial over $\F_p$ that computes the $\OR$ function over $\Delta$ Boolean inputs; for example, one could take $\OR(x_1, \dots, x_\Delta) = 1-\prod_{i=1}^\Delta(1-x_i)$. We define
    \[P_{C, \vec{j}} = \OR\mleft((\Enc(H_C(x))_{j_1})^{p-1}, (\Enc(H_C(x))_{j_2})^{p-1}, \dots, (\Enc(H_C(x))_{j_\Delta})^{p-1}\mright).\]
\end{enumerate}
It can be seen from the above inductive proof that the degree of $P_{C, \vec{j}}$ is at most $(p\Delta)^d$.

\def\Bad{\mathsf{Bad}}
\paragraph{Correctness of $P_{C, \vec{j}}$.} Next we show that for a random $\vec{j}$, there are only a fraction of $0.9^\Delta\cdot s$ inputs $x \in \{0, 1\}^n$ such that $P_{C, \vec{j}}(x) \ne C(x)$. In fact, let $\Bad$ denote the set of $(x, \vec{j})$ such that $P_{C, \vec{j}}(x) \ne C(x)$, and let $K = 2^nL\cdot 0.9^\Delta\cdot s$, then $|\Bad| \le K$. Moreover, this inequality can be ``feasibly witnessed'' by a pair of functions
\[\phi: \Bad\to[K]\quad\text{and}\quad\psi: [K]\to \Bad\]
such that for every $(x, \vec{j}) \in \Bad$, $\psi(\phi(x, \vec{j})) = (x, \vec{j})$. (Clearly, the existence of such functions $(\phi, \psi)$ implies that $|\Bad| \le K$. The complexity of computing $\phi$ and $\psi$ will be crucial our formalisation and will be discussed in \autoref{sec: formalisations of smolensky}.)

We identify $[K]$ with $\{0, 1\}^n \times [s] \times [0.9^\Delta\cdot L]$. That is, each element in $[K]$ is treated as a tuple $(x, g, c)$ where $x\in\{0, 1\}^n$ is an input, $g\in [s]$ is the index of some $\OR$ gate in $C$, and $c\in [0.9^\Delta\cdot L]$ is the ``compressed'' representation of some $\vec{j} \in [L] = [\ell]^\Delta$ that introduced an error on the $g$-th gate of $C$ when computing $C(x)$.

Now, let $(x, \vec{j})\in\Bad$. Let $g$ be the bottommost gate of $C$ that introduced an error; that is, if we denote by $C^g$ the sub-circuit computed by $g$, then $g$ is the bottommost gate such that $C^g(x) \ne P_{C^g, \vec{j}}(x)$. Suppose that $C^g$ is the $\OR$ of sub-circuits $C_1, C_2, \dots, C_t$. Then we have:
\begin{itemize}
    \item $H_g(x) = (C_1(x), C_2(x), \dots, C_t(x))$ is not the all-zero vector, but for every $k\in [\Delta]$ we have $\Enc(H_g(x))_{j_k} = 0$.
\end{itemize}
For every fixed $x$ and $g$, the probability over a random $j\gets[\ell]$ that $H_g(x) \ne \vec{0}$ but $\Enc(H_g(x))_j = 0$ is at most $0.9$; hence the probability over a random $\vec{j}$ that the above bullet happens is at most $0.9^\Delta$. It follows that given $x$ and $g$, every $\vec{j}$ that makes the above bullet happens can be encoded by a number $c\in [0.9^\Delta\cdot L]$. We then define $\phi(x, \vec{j}) = (x, g, c)$. It is easy to define the inverse function $\psi$ such that $\psi(x, g, c) = (x, \vec{j})$ so that for every $(x, \vec{j}) \in \Bad$, we have that $\psi(\phi(x, \vec{j})) = (x, \vec{j})$.

%Given $(x, \rand) \in \Bad$, let $g$ be the bottommost gate on which an error was introduced, and $c \in [2^{\ell - \Delta}]$ be the compressed representation of $\rand$ given $(x, g)$, then we define $\phi(x, \rand) = (x, g, c)$. Similarly, given $(x, g, c) \in [K]$, we can recover the ``$c$-th'' bad randomness $\rand$ that introduces an error on the $g$-th gate of $C$ when computing $C(x)$, and we define $\psi(x, g, c) = (x, \rand)$. It is easy to see that for every $(x, \rand) \in \Bad$, we have $\psi(\phi(x, \rand)) = (x, \rand)$.

\paragraph{Basis change.} A minor but annoying issue is that \autoref{thm: corr bounds using weak rank principle} deals with low-degree polynomials $P: \{-1, 1\}^n \to \{-1, 1\}$ but our $P_{C, \vec{j}}$ are polynomials from $\{0, 1\}^n$ to $\{0, 1\}$. Observe that the linear map $\tau(x) = 1-2x$ maps $0$ to $1$ and maps $1$ to $-1$, and its inverse map is $\tau^{-1}(x) = \frac{p-1}{2}(x-1)$. We define
\[Q_{C, \vec{j}}(x_1, \dots, x_n) = \tau(P_{C, \vec{j}}(\tau^{-1}(x_1), \dots, \tau^{-1}(x_n))).\]
Clearly, the degree of $Q_{C, \vec{j}}$ is the same as the degree of $P_{C, \vec{j}}$. Moreover, for every $x\in \{0, 1\}^n$, we have that $P_{C, \vec{j}}(x) = \MOD_2(x)$ if and only if $Q_{C, \vec{j}}(\tau(x)) = \prod_{i=1}^n \tau(x_i)$. (Hereafter, for $x\in\F_p^n$, $\tau(x)$ is a vector in $\F_p^n$ obtained by applying $\tau$ to each coordinate of $x$ individually.) %(Hereafter, for $x\in \F_p^n$ and $c\in\F_p$, we denote by $x+\vec{c}$ the vector in $\F_p^n$ whose $i$-th entry is equal to $x_i + c$ for each $i\in[n]$.)

\paragraph{Proof of the main lower bound.} Now we are ready to prove the main lower bound against $\AC^0[p]$ circuits. Let $C: \{0, 1\}^n \to \{0, 1\}$ be an $\AC^0[p]$ circuit that purportedly computes $\MOD_2$. A na\"ive attempt is to find a good $\vec{j}$, construct the matrices $A\in \F_p^{2^n\times 0.99\cdot 2^n}$ and $B = B(Q_{C, \vec{j}})\in \F_p^{0.99\cdot 2^n\times 2^n}$ (see \autoref{def: matrices A and B}), use the weak rank principle to find inputs $z, x\in\{-1, 1\}^n$ such that $(AB)[z, x] \ne \delta_{z, x}$, and conclude that $C(\tau^{-1}(z))\ne \MOD_2(\tau^{-1}(z))$. However, it may be the case that $(\tau^{-1}(z), \vec{j})\in \Bad$, which does not give us the desired conclusion.

To remedy this, we put the matrices $A$ and $B(Q_{C, \vec{j}})$ together for every possible $\vec{j}$ (there are $L$ many possibilities) in a block-diagonal manner. This results in a giant matrix multiplication of dimension $(2^n L)\times (0.99\cdot 2^n L)\times (2^n L)$. We still need to handle the entries in $\Bad$. However, since $|\Bad| \ll 0.01\cdot 2^n L$, we can afford to increase the middle dimension (i.e., $0.99\cdot 2^n L$) by $|\Bad|$ to ``correct'' all the values related to entries in $\Bad$.

\begin{theorem}\label{thm: smolensky main}
    Let $C$ be an $\AC^0[p]$ circuit of depth $d$ and size $s < 2^{n^{1/2d}/(20p)}$, then there exists an input $x\in \{0, 1\}^n$ such that $C(x) \ne \MOD_2(x)$.
\end{theorem}
\begin{proof}
    We set the following parameters:
    \begin{table}[H]
        \centering
        \begin{tabular}[b]{|c|c|}
            \hline
            {\textbf{Value}} & {\textbf{Intended meaning}}\\
            \hline
            $\Delta = \frac{n^{1/2d}}{10p}$ & \makecell{Number of ``samples'' taken in $\calS$.\\Equiv.~the degree of the ``polynomial approximation'' of $\OR$}\\
            \hline
            $\ell = \poly(s)$ & Output length of $\Enc$\\
            \hline
            $L = \ell^\Delta \le \exp(n)$ & Number of polynomials in $\calP$\\
            \hline
            $N = 2^n L$ & Dimension of the weak rank principle\\
            \hline
            $K = 2^n L \cdot 0.6^\Delta \cdot s \le 0.001N$ & $|\Bad|$\\
            \hline
            $R = 0.99\cdot 2^n L + K \le 0.991N$ & Rank parameter of the weak rank principle\\
            \hline
        \end{tabular}
    \end{table}
    We construct the following two matrices $\hat{A}$ and $\hat{B}$ where $\hat{A} \in \F_p^{N\times R}$ and $\hat{B} \in \F_p^{R\times N}$. It would be convenient to identify $[N]$ with $\{-1, 1\}^n \times [L]$, so every element in $[N]$ can be parsed as $(x, \vec{j})$ where $x\in \{-1, 1\}^n$ and $\vec{j}\in[\ell]^\Delta$. Similarly, we identify $[R]$ with the disjoint union of $[0.99\cdot 2^n]\times [\ell]^\Delta$ and $[K]$. Recall from the proof of \autoref{thm: corr bounds using weak rank principle} that we also identified $[0.99\cdot 2^n]$ with monomials of degree at most $d = n/2 + 0.1\sqrt{n}$.

    Let $(z, \vec{j})\in [N]$ and $a\in [R]$, we define $\hat{A}[(z, \vec{j}), a]$ as follows:\begin{itemize}
        \item Suppose $a = (m, \vec{j}') \in [0.99\cdot 2^n]\times [\ell]^\Delta$. If $\vec{j} \ne \vec{j}'$ then $\hat{A}[(z, \vec{j}), a] = 0$; otherwise $\hat{A}[(z, \vec{j}), a] = m(z)$.
        \item Suppose $a \in [K]$. If $(\tau^{-1}(z), \vec{j}) \in \Bad$ and $a = \phi(\tau^{-1}(z), \vec{j})$ then $\hat{A}[(z, \vec{j}), a] = 1$; otherwise $\hat{A}[(z, \vec{j}), a] = 0$.
    \end{itemize}
    Similarly, let $a\in[R]$ and $(x, \vec{j}) \in [N]$, we define $\hat{B}[a, (x, \vec{j})]$ as follows:\begin{itemize}
        \item Suppose $a = (m, \vec{j}') \in [0.99\cdot 2^n]\times [\ell]^\Delta$. If $\vec{j} = \vec{j}'$ then $\hat{B}[a, (x, \vec{j})] = B(Q_{C, \vec{j}})[a, x]$ (recall the definition of $B(Q_{C, \vec{j}})$ in \autoref{def: matrices A and B}); otherwise $\hat{B}[a, (x, \vec{j})] = 0$.
        \item Suppose $a \in [K]$. Let $(\tau^{-1}(z), \vec{j}') = \psi(a)$, we define
        \[\hat{B}[a, (x, \vec{j})] = \delta_{(z, \vec{j}'), (x, \vec{j})} - \sum_{m \in [0.99\cdot 2^n]}\hat{A}[(z, \vec{j}'), (m, \vec{j}')]\cdot \hat{B}[(m, \vec{j}'), (x, \vec{j})].\]
        (The intention is to set $\hat{B}[a, (x, \vec{j})]$ so that $(\hat{A}\cdot \hat{B})[(z, \vec{j}'), (x, \vec{j})] = \delta_{(z, \vec{j}'), (x, \vec{j})}$.)
    \end{itemize}

    From the above definition it is clear that for every $(\tau^{-1}(z), \vec{j}')\in \Bad$ and every $(x, \vec{j}) \in [N]$, we have $(\hat{A}\hat{B})[(z, \vec{j}'), (x, \vec{j})] = \delta_{(z, \vec{j}'), (x, \vec{j})}$. On the other hand, the weak rank principle implies that there exists $(z, \vec{j}'), (x, \vec{j}) \in [N]$ such that $(\hat{A}\hat{B})[(z, \vec{j}'), (x, \vec{j})] \ne \delta_{(z, \vec{j}'), (x, \vec{j})}$. It is easy to see that $\vec{j} = \vec{j}'$ as otherwise $(\hat{A}\hat{B})[(z, \vec{j}'), (x, \vec{j})] = 0 = \delta_{(z, \vec{j}'), (x, \vec{j})}$. Then we can see that $(A\cdot B(Q_{C, \vec{j}}))[z, x] \ne \delta_{z, x}$. It follows from \autoref{claim: witness of wRank implies a bad input} that $Q_{C, \vec{j}}(z)\ne \prod_{i=1}^n z_i$, hence $P_{C, \vec{j}}(\tau^{-1}(z)) \ne \MOD_2(\tau^{-1}(z))$. However, since $(\tau^{-1}(z), \vec{j}')\not \in\Bad$, we have that $P_{C, \vec{j}}(\tau^{-1}(z)) = C(\tau^{-1}(z))$. Hence $C(\tau^{-1}(z)) \ne \MOD_2(\tau^{-1}(z))$ and $C$ fails to compute the $\MOD_2$ function.
\end{proof}

\subsection{Formalisation in \texorpdfstring{$\overline{\V^0(p)} + \wRank_p$}{V0(p) + wRank p}}\label{sec: formlisation}

\subsubsection{The Theory \texorpdfstring{$\overline{\V^0(p)} + \wRank_p$}{V0(p) + wRank p}}\label{sec: def of V0(3) + wRank}

We start with a brief introduction to the theories $\V^0(p)$, $\overline{\V^0(p)}$, and $\overline{\V^0(p)} + \wRank_p$. A more comprehensive treatment for the theories $\V^0(p)$ and $\overline{\V^0(p)}$ can be found in Cook and Nguyen's textbook~\cite{CN10}. The theory $\overline{\V^0(p)} + \wRank_p$ is introduced and studied for the first time in this work.

We consider \emph{two-sorted} logic: the first sort are natural numbers (i.e., indices) and the second sort are bit strings (or more precisely, finite sets of natural numbers whose characteristic vectors are bit strings). In this context, we usually use capital letters ($X, Y, Z$) to denote strings and lower case letters ($x, y, z$) to denote numbers; however our notation does not strictly follow this convention and we list the exceptions in \autoref{remark: numbers vs strings}.

The vocabulary for two-sorted logic is $\calL^2_A$, which contains symbols
\[0, 1, +, \cdot, |\cdot|; =_1, =_2, \le, {\in}.\]
Here, $=_1$ and $=_2$ denotes equality over numbers and strings respectively; it is usually clear in the context whether numbers or strings are being compared, so we often omit the subscripts $1$ and $2$ and simply write $=$ for convenience. We will also use $X(t)$ or $X_t$ to denote ``$t\in X$''. These symbols are governed by a set of ``basic axioms'' called $2$-$\BASIC$ (see~\cite[Figure 2]{CN10}).

Let $\Phi$ be a set of formulas, then we can define the following \emph{axiom schema}:
\begin{itemize}
    \item $\Phi$-$\COMP$ (comprehension): $\exists X\le y\forall z<y~(X(z) \leftrightarrow \varphi(z))$, where $\varphi(z)$ is any formula in $\Phi$ and $X$ does not occur free in $\varphi(z)$.
    \item $\Phi$-$\IND$ (number induction): $(\varphi(0) \land \forall x~(\varphi(x)\to \varphi(x+1)))\to \forall z~\varphi(z)$, where $\varphi(z)$ is any formula in $\Phi$.
    \item $\Phi$-$\MIN$ (number minimisation): $\varphi(y)\implies (\exists x\le y ~(\varphi(x) \land \lnot\exists z<x~\varphi(z)))$, where $\varphi(z)$ is any formula in $\Phi$.
\end{itemize}

Let $\calL$ be a two-sorted vocabulary that contains $\calL^2_A$, then $\Sigma^B_0(\calL)$ is the set of $\calL$-formulas whose only quantifiers are bounded number quantifiers (there can be free string variables).

Let $m$ be a constant,\footnote{Here we mainly consider the case that $m = p$ is a prime, but it also makes perfect sense to consider the theory $\V^0(m)$ when $m$ is a composite number.} the theory $\V^0(m)$ has vocabulary $\calL^2_A$ and is axiomatised by $2$-$\BASIC$, $\Sigma^B_0(\calL^2_A)$-$\COMP$, and an axiom defining the function $\MOD_m$; see~\cite[Section IX.4.6]{CN10} for details. %\hanlin{I actually don't understand what this means in~\cite{CN10-book}. Why don't they use (266)-(269) directly?}

We actually work with the theory $\overline{\V^0(m)}$ instead of $\V^0(m)$. This theory has vocabulary $\calL_{\FAC^0(m)}$ that contains a symbol for every function in $\FAC^0(m)$ ($\FAC^0(m)$ is the class of (possibly non-Boolean) \emph{functions} computable in $\AC^0(m)$). It proves $\Sigma_0^B(\calL_{\FAC^0(m)})$-$\COMP$, $\Sigma_0^B(\calL_{\FAC^0(m)})$-$\IND$, and $\Sigma_0^B(\calL_{\FAC^0(m)})$-$\MIN$: for example, for every $\FAC^0(m)$ function $f$ that maps a string $X$ to the string $f(X)$, we can use the number minimisation axiom inside $\overline{\V^0(m)}$ to conclude that if $f(X)$ is not the all-zero string, then there exists the smallest $i$ such that $f(X)_i = 1$. Again, we refer to~\cite[Section IX.4.6]{CN10} for more details.

Although $\overline{\V^0(m)}$ ``contains much more features'' than $\V^0(m)$, it is a conservative extension of $\V^0(m)$, i.e., every formula in the vocabulary $\calL^2_A$ that can be proved in $\overline{\V^0(m)}$ can be proved in $\V^0(m)$ as well. Hence it is without loss of generality to work with $\overline{\V^0(m)}$ instead of $\V^0(m)$.

%It is axiomatised by $2$-$\BASIC$, $\Sigma_0^B$-$\COMP$ \hanlin{Why not $\Sigma_0^B(\calL)$-$\COMP$?}, defining axioms for $mod'_m$ (\cite[(266)-(269)]{CN10-book}), and (86) for each function $F_{\varphi(z), t}$.

%It is more convenient to work with $\overline{\V^0(m)}$ as it contains a function symbol for every $\FAC^0(m)$ function and it is a universal conservative extension of $\V^0(m)$.

%A function is in $\FAC^0(m)$ iff it is represented by a symbol in $\calL_{\FAC^0(m)}$. A relation is in $\AC^0(m)$ iff it is represented by an open (or a $\Sigma^B_0$) formula of $\calL_{\FAC^0(m)}$.

%$\Sigma_0^B(\calL)$-$\COMP$, $\Sigma_0^B(\calL)$-$\IND$, and $\Sigma_0^B(\calL)$-$\MIN$ are available in $\overline{\V^0(m)}$ (where $ = \calL_{\FAC^0(m)}$).

%\hanlin{Can we define wRank over the vocabulary $\calL^2_A$? Is $\V^0(p) + \wRank_p(\text{functions over }\calL^2_A)$ equivalent to $\overline{\V^0(p)} + \wRank_p(\text{strings})$?}

Finally, let $m = p$ be a prime. The \emph{weak rank principle} over $\F_p$, denoted as $\wRank_p$, is the following axiom. Let $n$ be a number, $A \in \F_p^{2n\times n}$ and $B \in \F_p^{n\times 2n}$ be matrices coded as strings, then there exists $x, y\in [2n]$ such that
\[\sum_{z\in [n]}a_{x, z}b_{z, y} \not\equiv \delta_{x, y} \pmod p.\]
(Note that $\sum_{z\in [n]}a_{x, z}b_{z, y} \bmod p$ can be expressed as a function in $\FAC^0[p]$, hence $\wRank_p$ is indeed a universal sentence in the language $\calL_{\FAC^0(p)}$.) For every constant $\eps > 0$, one can use a similar argument as \autoref{thm: stretching wRank 2} to show that $\overline{\V^0(p)} + \wRank_p$ proves the following: Let $n$ be a number, $A\in \F_p^{(1+\eps)n\times n}$ and $B\in \F_p^{n\times (1+\eps)n}$ be two matrices encoded as strings, then $AB\ne I_{(1+\eps)n}$. In other words, ``$\rank(I_{2n}) > n$'' and ``$\rank(I_{(1+\eps)n}) > n$'' are equivalent, and replacing the constant $2$ in the definition of $\wRank_p$ with other constants greater than $1$ does not affect the strength of the theory $\overline{\V^0(p)} + \wRank_p$. In fact, our formalisation will use ``$\rank(I_{(1+\eps)n}) > n$'' for some small constant $\eps > 0$.

\subsubsection{Smolensky's Lower Bound in \texorpdfstring{$\overline{\V^0(p)} + \wRank_p$}{V0(p) + wRank p}}\label{sec: formalisations of smolensky}

Fix a constant $d\in\N$ (in the meta-theory). Let $n\in \Log$ (that is, we assume $2^n$ is a ``number''), $s = 2^{n^{1/2d}/(20p)}$, and $C$ be a string that codes an $\AC^0[p]$ circuit of depth $d$ and size $s$. Let $\Eval_d(C, x)$ denote the circuit evaluation function on depth-$d$ $\AC^0[p]$ circuits, where $x\in [2^n]$ is a number that denotes the input of $C$. Since $\Eval_d$ is in $\AC^0[p]$, it is in the language of $\overline{\V^0(p)}$. Also, since $x$ is a number, we can talk about $\MOD_2(x)$ in $\overline{\V^0(p)}$. Our goal is to prove that
\[\exists x~\Eval_d(C, x) \ne \MOD_2(x).\]

\begin{remark}\label{remark: numbers vs strings}
    It should be clear from the context whether a variable is of the number sort or the string sort. Most of the time we use capital letters ($X, Y, Z$) to denote strings and use lower-case letters ($x, y, z$) to denote numbers. However, one exception is that $L, N, K, R$ (in the proof of \autoref{thm: smolensky main}) are numbers.
\end{remark}

%Before formalising the proof of \autoref{thm: smolensky main}, we need to show that $\overline{\V^0(3)}$ can reason about the following concepts.

%\paragraph{Polynomials.} Let $\calM$ denote the set of multilinear monomials over $x_1, \dots, x_n$, $f: \calM \to \F_p$ be a polynomial represented as the list of coefficients (which is a string $F$). Then there is an $\FAC^0[p]$ function that given $F$ and a number $x\in\F_p^n$, computes $f(x)$. Similarly, let $Z\in \F_p^{2^n}$ be a string that consists of the value list of a multilinear polynomial $f$ over $\{-1, 1\}^n$, then there is an $\FAC^0[p]$ function that given a string $Z$ and a monomial $m\in \calM$ coded as an integer, computes the coefficient of $m$ in $f$ (recall that this is equal to $(\frac{p+1}{2})^n\sum_{x\in \{-1, 1\}^n}f(x)m(x)$). In what follows, for polynomials $P$ over the domain $\{-1, 1\}^n$, we will not specify whether $P$ is represented by a list of values or a list of coefficients, as one can convert between these two representations in $\FAC^0[p]$.

%\hanlin{Let's only use coefficients to represent a polynomial.}

\paragraph{The matrix $B(P)$ (\autoref{def: matrices A and B}).} It is clear from the definition that there is an $\FAC^0[p]$ function that given $(m, x)$ and the list of coefficients in $P$, outputs the value of $B(P)[m, x]$. %It is important that the following claim is provable in $\overline{\V^0(3)}$: %Moreover, the following version of \autoref{claim: meaning of B} can be proved in $\overline{\V^0(3)}$:

%\begin{claim}[$\overline{\V^0(3)}\vdash:$]
%    Let $n\in\Log$, $\vec{v}_f \in \F_3^{2^n}$ be a string encoding the list of values of a function $f: \{-1, 1\}^n \to \F_3$. Let $f_\low$ and $f_\high$ be the degree-$(\le n/2)$ and degree-$(>n/2)$ parts of $f$ respectively (which can be computed by $\FAC^0[3]$ functions). Let $\hat{f}$ denote the polynomial whose coefficient list is $B(P)\cdot \vec{v}_f$. Then for every $z\in\{-1, 1\}^n$,
%    \[\hat{f}(z) = f_\low(z) + f_\high(z) \cdot P(z)\cdot \prod_{i=1}^n z_i.\]
%\end{claim}
\begin{claim}[$\overline{\V^0(p)}\vdash:$]\label{claim: invoking smaller rank principle}
    Let $z, x\in \{-1, 1\}^n$ such that $(A\cdot B(P))[z, x] \ne \delta_{z, x}$. Then $P(z) \ne \prod_{i=1}^n z_i$.
\end{claim}
\begin{proof}[Proof Sketch]
    The proofs of both \autoref{claim: meaning of B} and \autoref{claim: witness of wRank implies a bad input} can be carried out in $\overline{\V^0(p)}$.
\end{proof}

\def\Val{\mathsf{Val}}
\def\Coeff{\mathsf{Coeff}}
\paragraph{The polynomials $P_{C, \vec{j}}$ and $Q_{C, \vec{j}}$.} For each gate $g$, let $C^g$ denote the sub-circuit computed by $g$, we represent the polynomials $P_{C^g, \vec{j}}$ and $Q_{C^g, \vec{j}}$ by their list of coefficients. There is an $\FAC^0[p]$ circuit that given the description of $C$, an index $g$, and a list of vectors $\vec{j}$, outputs the coefficient list of $P_{C^g, \vec{j}}$. (Note that since $\vec{j} \in [\ell]^\Delta$ and $\ell^\Delta = L$ is a number in the theory, $\vec{j}$ can be represented as a single number.) The coefficient list of $Q_{C, \vec{j}}(x) = P_{C, \vec{j}}(x-\vec{1}) + 1$ can be computed in $\FAC^0[p]$ as well, If $C^g$ is an $\AC^0[p]$ circuit of depth $d$, then the degree of $P_{C^g, \vec{j}}$ is at most $(p\Delta)^d$ (and we only consider monomials up to this degree in our representation).

\paragraph{A good error-correcting code.} Before formalising the low-degree approximations of $\AC^0[p]$ circuits, we need the following $\F_p$-linear error-correcting code whose correctness can be proved feasibly. In fact, we need to feasibly prove the following formalisation of ``every non-zero codeword of $\Enc$ has constant relative Hamming weight'':

\begin{restatable}{theorem}{ThmECCInBA}\label{thm: error-correcting code in VAC0(3)}
    Suppose that $\log^2 s\in \Log$. For some $\ell \le \poly(s)$, there is an $\F_p$-linear code $\Enc: \F_p^s \to \F_p^{\ell}$ and functions $\tilde{\phi}(X, i_1, i), \tilde{\psi}(X, i_1, c)$ computable in $\FAC^0[p]$ such that $\overline{\V^0(p)}$ proves the following. For every string $X\in \F_p^s$, every numbers $i_1\in[s]$ and $i \in [\ell]$, if $X_{i_1} \ne 0$ but $\Enc(X)_i = 0$, then:
    \begin{itemize}
        \item $\tilde{\phi}(X, i_1, i)$ returns a number in $[0.9\ell]$, and
        \item $\tilde{\psi}(X, i_1, \tilde{\phi}(X, i_1, i)) = i$.
    \end{itemize}
\end{restatable}

Note that the above statement implies that the relative Hamming weight of $\Enc(X)$ is at least $0.1$. Roughly speaking, the error-correcting code is the concatenation of Reed--Muller code and Hadamard code; the proof proceeds by formalising the local (unique-)decoding algorithm of the Reed--Muller code (see e.g.,~\cite[Section 19.4.2]{AB09}). We postpone the proof to \autoref{sec: proof of ECC in BA}.

\paragraph{The functions $\phi, \psi$.} We need to show that $\phi$ and $\psi$ are computable in $\FAC^0[p]$. Given a string $C$ representing our $\AC^0[p]$ circuit and two numbers $x$, $\vec{j}$, we can compute $\phi(x, \vec{j})$ as follows:
\begin{enumerate}
    \item First, find the bottommost gate $g$ for which $C^g(x) \ne P_{C^g, \vec{j}}(x)$.
    
    For every $g$, we can compute the indicator of whether $C^g(x) = P_{C^g, \vec{j}}(x)$ in $\AC^0[p]$. Then we use an $\FAC^0$ circuit to compute the number $g$, which is the index of the bottommost gate on which the indicator is False. (If there is no such gate then we set $g = \bot$.)
    
    \item Let $H_g(x) = (C_1(x), \dots, C_t(x))$ where $C_1, \dots, C_t$ are input gates of $g$. For each $k\in [\Delta]$, we have that $\Enc(H_g(x))_{j_k} = 0$. However, there exists an index $j'\in[t]$ such that $C_{j'}(x) = 1$, and the smallest such index can be computed in $\FAC^0$ given $H_g(x)$. Hence we compute $c_k = \tilde{\phi}(H_g(x), j', j_k)\in [0.9 \ell]$ using \autoref{thm: error-correcting code in VAC0(3)}. This gives us a vector $\vec{c} = (c_1, \dots, c_\Delta)$. (Note that $\ell^\Delta = L$ is a number in the theory, hence $\vec{c}$ can be represented as a single number.) We output $\phi(x, \vec{j}) = (x, g, \vec{c})$.
\end{enumerate}

Similarly, one can compute $\psi$ in $\FAC^0[p]$ by invoking the function $\tilde{\psi}$ in \autoref{thm: error-correcting code in VAC0(3)}. Namely, given a string $C$ representing our $\AC^0[p]$ circuit and three numbers $x, g, \vec{c}$, we can compute $\psi(x, g, \vec{c})$ as follows. First, let $C_1, \dots, C_t$ be the input gates of $g$, we can compute $H_g(x) = (C_1(x), \dots, C_t(x))$ in $\FAC^0[p]$. We can also compute the smallest index $j'$ such that $C_{j'}(x) = 1$ in $\FAC^0$ given $H_g(x)$. Then, for each $K\in[\Delta]$, we let $j_k = \tilde{\psi}(H_g(x), j', c_k)$. We return $\psi(x, g, \vec{c}) = (x, \vec{j})$.

\begin{claim}[$\overline{\V^0(p)}\vdash:$]\label{claim: (phi, psi) is a good inj surj pair}
    Suppose that $\log^2 s \in \Log$. For every $x\in\{0, 1\}^n$ and $\vec{j}\in [\ell]^\Delta$, if $P_{C, \vec{j}}(x) \ne C(x)$, then $\psi(\phi(x, \vec{j})) = (x, \vec{j})$.
\end{claim}
\begin{proof}[Proof Sketch]
    Since $P_{C, \vec{j}}(x) \ne C(x)$, it follows from $\Sigma_0^B(\calL_{\FAC^0(p)})\text{-}\MIN$ that there exists a bottommost gate $g$ for which $C^g(x) \ne P_{C^g, \vec{j}}(x)$. It follows that $H_g(x)_{j'} \ne 0$ but $\Enc(H_g(x))_{j_k} = 0$ for every $k \in [\Delta]$. Let $c_k = \tilde{\phi}(H_g(x), j', j_k)$, then $\phi(x, \vec{j}) = (x, g, \vec{c})$. Let $\psi(x, g, \vec{c}) = (x, \vec{j}')$, then for every $k\in [\Delta]$, 
    \[j'_k = \tilde{\psi}(H_g(x), j', c_k) = \tilde{\psi}(H_g(x), j', \phi(H_g(x), j', j_k)) = j_k\]
    where the last equality follows from \autoref{thm: error-correcting code in VAC0(3)}. Hence $\psi(\phi(x, \vec{j})) = (x, \vec{j})$.
\end{proof}

\paragraph{The matrices $\hat{A}, \hat{B}$} as defined in the proof of \autoref{thm: smolensky main}. It is easy to see that these matrices can be computed in $\FAC^0[p]$. That is, there is an $\FAC^0[p]$ function that takes a string $C$ and numbers $z, \vec{j}, a$ as inputs and outputs the value of $\hat{A}[(z, \vec{j}), a]$, and the analogous statement holds for $\hat{B}$.

\begin{claim}[$\overline{\V^0(p)}\vdash:$]\label{claim: properties of hat A and hat B}
    Let $z, x\in \{-1, 1\}^n$, $\vec{j}, \vec{j}'\in [\ell]^\Delta$, then:
    \begin{compactenum}[\normalfont(a)]
        \item If $\vec{j} \ne \vec{j}'$, then $(\hat{A}\hat{B})[(z, \vec{j}), (x, \vec{j}')] = 0$.\label{item: can only make error on the same j}
        \item If $\psi(\phi(z-\vec{1}, \vec{j})) = (z-\vec{1}, \vec{j})$, then $(\hat{A}\hat{B})[(z, \vec{j}), (x, \vec{j})] = \delta_{(z, \vec{j}), (x, \vec{j}))}$.\label{item: getting rid of Bad}
        \item If $\psi(\phi(z-\vec{1}, \vec{j}))\ne(z-\vec{1}, \vec{j})$, then $(\hat{A}\hat{B})[(z, \vec{j}), (x, \vec{j})] = (A\cdot B(Q))[z, x]$.\label{item: reduce to the smaller matrix}
    \end{compactenum}
\end{claim}
\begin{proof}[Proof Sketch]
    These follow easily from the definitions.
\end{proof}

\paragraph{Putting it together.} Fix a constant $d\in\N$ (in the meta-language), we have:

\begin{theorem}[$\overline{\V^0(p)} + \wRank_p\vdash:$]
\label{thm:main-smolensk-in-BA}
    Let $n\in\Log$, $s = 2^{n^{1/2d}/(20p)}$, and $C$ be a string that codes an $\AC^0[p]$ circuit of depth $d$ and size $s$. Then there exists an input $x\in\{0, 1\}^n$ (coded as a number) such that $\Eval_d(C, x) \ne \MOD_2(x)$.
\end{theorem}
\begin{proof}
    Consider the matrices $\hat{A}$ and $\hat{B}$ as defined above. By $\wRank_p$, there exists $(z, \vec{j}'), (x, \vec{j}) \in [N]$ such that $(\hat{A} \hat{B})[(z, \vec{j}'), (x, \vec{j})] \ne \delta_{(z, \vec{j}'), (x, \vec{j})}$. Then:
    \begin{compactitem}
        \item By \autoref{claim: properties of hat A and hat B} (\autoref{item: can only make error on the same j}), $\vec{j} = \vec{j}'$;
        \item By \autoref{claim: properties of hat A and hat B} (\autoref{item: getting rid of Bad}), $\psi(\phi(z - \vec{1}, \vec{j})) \ne (z - \vec{1}, \vec{j})$;
        \item By \autoref{claim: properties of hat A and hat B} (\autoref{item: reduce to the smaller matrix}), $(AB)[(z, \vec{j}), (x, \vec{j})]$ = $(A\cdot B(Q_{C, \vec{j}}))[z, x] \ne \delta_{z, x}$.
        \item By \autoref{claim: invoking smaller rank principle}, $Q_{C, \vec{j}}(z) \ne \prod_{i=1}^n z_i$, hence $P_{C, \vec{j}}(z-\vec{1}) \ne \MOD_2(z-\vec{1})$;
        \item By \autoref{claim: (phi, psi) is a good inj surj pair} (and note that it is indeed the case that $\log^2 s \in \Log$), $P_{C, \vec{j}}(z-\vec{1}) = C(z-\vec{1})$. Hence $C(z-\vec{1}) \ne \MOD_2(z-\vec{1})$.\qedhere
    \end{compactitem}
\end{proof}

\Iddo{ I moved the remark her to the intro as it is important
}
\subsubsection{Proof of \texorpdfstring{\autoref{thm: error-correcting code in VAC0(3)}}{Theorem 9.10}}\label{sec: proof of ECC in BA}
\ThmECCInBA*
\def\Had{\mathrm{Had}}
\begin{proof}
    Let $n = O(\log s/\log\log s)$, $h = \lceil s^{1/n}\rceil$, $q$ be the smallest power of $p$ that is greater than $10nh$. Note that $q\le \log^2 s\in\Log$, hence $\overline{\V^0(p)}$ can reason about $\F_q$ by brute force. (For example, $\overline{\V^0(p)}$ can prove the existence of an irreducible degree-$\log_p q$ polynomial over $\F_p$. See e.g.,~\cite[Section 4.3]{Jer05-PhD}.) Let $H$ be the set of lexicographically first $h$ elements in $\F_q$. We interpret $X\in \F_p^s\subseteq \F_q^s$ as a function $X: H^n \to \F_q$. Then there is a unique polynomial $\tilde{X}: \F_q^n \to \F_q$ of individual degree at most $h$ that is consistent with $X$ on $H^n$. Moreover, one can compute $\tilde{X}(\vec{v})$ using an $\FAC^0[p]$ circuit (in fact, an $\F_q$-linear transformation) over the string input $X$ and the number input $\vec{v}$.
    
    We now concatenate the above Reed--Muller code with the Hadamard code. Treat $\F_q$ as the $\log_p q$-dimensional vector space over $\F_p$, then each $\alpha \in \F_q$ is encoded as a vector $\Had(\alpha)\in \F_p^q$ defined as follows. For each $u\in \F_q$, we interpret both $\alpha$ and $u$ as vectors in $\F_p^{\log_p q}$, and the $u$-th element of $\Had(\alpha)$ is equal to ${\rm IP}(\alpha, u) = \sum_{j=1}^{\log_p q}\alpha_j u_j$. Again, since $q\in\Log$, $\overline{\V^0(p)}$ can reason about the Hadamard code using brute force.
    
    Define $\Enc(X)$ be the concatenation of $\Had(\tilde{X}(\vec{v}))$ for each $\vec{v}\in \F_q^n$. That is, the length of $\Enc(X)$ is $\ell = q^{n+1} \le \poly(s)$. Given an index $i\in [\ell]$, we interpret $i$ as a pair of vectors $(\vec{v} \in \F_q^n, u \in \F_q = \F_p^{\log_p q})$, and define $\Enc(X)_i = {\rm IP}(\tilde{X}(v), u)$.

    The local decoder of the Reed--Muller code requires the following equivalence class over ``directions'' in $\F_q^n$: two non-zero vectors $\vec{v}_1, \vec{v}_2 \in \F_q^n$ are equivalent if and only if there is a non-zero scalar $\alpha \in \F_q$ such that $\vec{v}_1 = \alpha \vec{v}_2$. Let $\vec{v}$ be a non-zero vector in $\F_q^n$ and define $[\vec{v}]$ as the following vector: let $i\in [n]$ be the first index such that $v_i \ne 0$, and for each $j\in [n]$, the $j$-th element of $[\vec{v}]$ is equal to $v_i^{-1}v_j$ (over $\F_q$). Again, $\overline{\V^0(p)}$ can prove (by brute force) that $v_1$ and $v_2$ are equivalent if and only if $[v_1] = [v_2]$, and that there are $q^{n-1} + q^{n-2} + \dots + q + 1 \le 2q^{n-1}$ equivalence classes.
    
    Let $X \in \F_p^s$ be a non-zero string, $i_1 \in [s]$ be an index such that $X_{i_1} = 1$, and $i\in [\ell]$ be an index such that $\Enc(X)_i = 0$. We interpret $i$ as a pair of vectors $(\vec{v}\in \F_q^n, u \in \F_q)$. To compute $\hat{\phi}(X, i_1, i)$, we consider the following two cases:\begin{itemize}
        \item If $\tilde{X}(\vec{v}) \ne 0$, then we consider the Hadamard code. In particular, $u$ is among the vectors in $\F_q$ for which ${\rm IP}(\tilde{X}(\vec{v}), u) = 0$. There are at most $q/p$ such elements, hence we let $\tilde{\phi}(X, i_1, i) = (\textsf{Hadamard}, \vec{v}, k)$ where $u$ is the $k$-th smallest element in $\F_q$ such that ${\rm IP}(\tilde{X}(\vec{v}), u) = 0$.
        \item If $\tilde{X}(\vec{v}) = 0$, then we consider the Reed--Muller code. In particular, consider the directional vector $\vec{w} = \vec{v} - i_1$ (where we interpret $i_1 \in [s]\subseteq \F_q^n$ and the subtraction happens in $\F_q^n$). The unique line determined by $\vec{v}$ and $i_1$ is $\ell(t) = t\cdot [\vec{w}] + i_1$. Composing $\tilde{X}$ with $\ell$ gives us a univariate polynomial $f(t) = \tilde{X}(t\cdot [\vec{w}] + i_1)$ of degree at most $nh$. Since $f(0)\ne 0$, this polynomial has at most $nh$ roots. We define $\tilde{\phi}(X, i_1, i) = (\textsf{Reed--Muller}, [\vec{w}], k, u)$ where $\vec{v}$ corresponds to the $k$-th root of $f$. (Recall that $k$ can be found by brute force.)
    \end{itemize}
    Now we can see that $\tilde{\phi}(X, i_1, i)$ either returns $(\textsf{Hadamard}, \vec{v}, k)$ where $\vec{v}\in \F_q^n, k\in[q/p]$ or returns $(\textsf{Reed--Muller}, [\vec{w}], k, u)$ where $[\vec{v}] \in [2q^{n-1}]$, $k\in [q/10]$, $u\in \F_q$. It follows that the size of $\mathrm{Range}(\tilde{\phi})$ is at most $q^{n+1}/p + q^{n+1}/5 \le 0.9\ell$.

    We can similarly define the function $\tilde{\psi}(X, i_1, c)$:
    \begin{itemize}
        \item If $c$ is of the form $(\textsf{Hadamard}, \vec{v}, k)$, then we compute the $k$-th smallest $u\in \F_q$ such that ${\rm IP}(\tilde{X}(\vec{v}), u) = 0$ and output $(\vec{v}, u)$.
        \item If $c$ is of the form $(\textsf{Reed--Muller}, [\vec{w}], k, u)$, then we compute the $k$-th root $\vec{v}$ of the degree-$nh$ polynomial $f(t) = \tilde{X}(t\cdot [\vec{w}] + i_1)$ and output $(\vec{v}, \vec{u})$.
    \end{itemize}
    It is easy to see that $\psi(X, i_1, \phi(X, i_1, i)) = i$ for every $i$ such that $\Enc(X)_i = 0$.
\end{proof}

\appendix
%!TEX root = main.tex

\section{Expander Properties}\label{sec:missing proofs}

The proofs here mostly mimic the corresponding proofs from~\cite{Razb15-annals}.

\goodexpanderlemma*
\begin{proof}
    The proof essentially follows~\cite{Razb15-annals} but with a slightly different choice of parameters.
    We construct a bipartite graph $G$ with parts $[m]$ and $[n]$.
    For every $i \in [m]$ and $k \in [n]$, we add an edge between $i$ and $k$ with probability $\frac{100 \log n}{n}$. Let $M_G = {(a_{i,j})}_{i \in [m], k \in [n]}$ be the adjacency matrix of such graph.

    % By the Chernoff bound,\Iddo{Better to have a reference and specify the specific formulation of the bound we use. I did it below in the sequel proof.  } for each $i \in [m]$,

    For each $i \in [m]$, the left-degree of $i$ is the sum of $n$ independent variables $a_{i,k} : k \in [n]$, with expectation $\mu = 100\log{n}$. By \cref{thm:chernoff bound}, applied with $\delta=1/2$,
    \[
        \Pr\Bigg[\Big\lvert \sum_{k \in [n]} a_{i,k} - 100\log n \Big\rvert > 50 \log n\Bigg] \le 2e^{-100\log{n}/12} = o(1/n).
    \]
    By the union bound, these events happen simultaneously for all $i \in [m]$ with probability $1-o(1)$.

    Now let $r = n^\delta$.
    \begin{claim}
        If   property~\eqref{eq:lossless expander property} is violated for some set $I \subseteq [m]$ of size $l \le r$, then there exist a set $K \subseteq [n]$ of size $ld/2$ such that the rectangle $I \times K$ contains at least $ld$ ones in $M_G$.
    \end{claim}
    \begin{claimproof}
        Let $K = \bigcup_{i \in I} \calN_G(i) \setminus \boundary_G(I)$, i.e., it contains all non-unique neighbours of $I$. In particular, every column of the rectangle $I \times K$ contains at least two ones. Thus, if $\card{K} \ge ld/2$, we can take any $K' \subseteq K$ of size exactly $ld/2$.
        Now assume that $\card{K} < ld/2$. The number of ones in $I \times K$ is at least
        \[
            \sum_{i \in I} \card{\calN_G(i) \setminus \boundary_G(I)} = \sum_{i \in {I}} \card{\calN_G(i)} - \card{\boundary_G(I)},
        \]
        since each $k \in \boundary_G(I)$ is contained in exactly one $\calN_G(i)$. Thus, since~\eqref{eq:lossless expander property} is violated for $I$, this is further lower bounded by $d \card{I} = ld$. To conclude the proof, we can arbitrarily enlarge $K$ to have size exactly $ld/2$.
    \end{claimproof}

    Now we will bound the probability of the existence of such $I \times K$. For every $\ell \in [r]$, there is at most ${(Cn)}^{\ell}$ choices of $I$, $n^{\ell d/2}$ choices of $K$, and ${(\ell d)}^{2\ell d}$ choices of the positions of ones in $I \times K$. Thus, the probability that there exists such a rectangle $I \times K$ with $\card{I} = \ell \le r = n^\delta$ is at most
    \[
        {(Cn)}^{\ell} n^{\ell d/2} {(\ell d)}^{2\ell d} {\left( \frac{100 \log n}{n} \right)}^{\ell d} \le
        n^{2\ell} {\left( \frac{{n^{4\delta} d^4 (100 \log n)}^2}{n} \right)}^{\ell d/2} \le
        n^{-2\ell}
    \]
    for small enough $\delta$ and large enough $n$ and $d$. Observing that $\sum_{\ell=1}^r n^{-2\ell} = o(1)$, the proof follows.
\end{proof}

\remainsgoodexpanderclaim*
\begin{proof}
    Let $G' = G \setminus K$.
    For every $i \in L$, let $M_i \subseteq \calN_G(i)$ be any fixed subset of size exactly $\Deltamin$.
    Hence $\card{M_i \setminus K}$ is the sum of $\Deltamin$ independent indicator variables, each of which is $1$ with probability $1/3$. The expectation of this sum is $\mu = \Exp[\card{M_i \setminus K}] = \Deltamin/3$.
    By \cref{thm:chernoff bound}, applied with $\delta=1/2$,
    \[
        \Pr[\card{M_i \setminus K} < \Deltamin/6] \le e^{-\Deltamin/24} = o(1/n),
    \]
    since $\Deltamin = 50\log n$.

    By the union bound, the left-degree of $G'$ is at least $\Deltamin/6$ with probability $1-o(1)$.

    The remaining part of the proof repeats the argument from~\cite{Razb15-annals}. Let $I \subseteq L$ with $\card{I} \le n^\delta$. For every $k \in R$, let $\gamma_k = \card{\{ i \in I : k \in \calN_G(i) \}}$.
    Using the bound from~\eqref{eq:lossless expander property}, it is easy to see that
    \[
        \sum_{\substack{k \in R \\ \gamma_k \ge 2}} \gamma_k = \sum_{i \in I} \card{\calN_G(i)} - \card{\boundary_G(I)} \le d \card{I}.
    \]

    Now let $\gamma_k' = \card{\{ i \in I : k \in \calN_{G'}(i) \}}$. Clearly, $\gamma_k' \le \gamma_k$ for all $k$, since $\gamma'_k = 0$ for all $k \in K$.
    This implies that
    \[
        \sum_{i \in I} \card{\calN_{G'}(i)} - \card{\boundary_{G'}(I)} = \sum_{\substack{k \in R \\ \gamma_k' \ge 2}} \gamma_k' \le
        \sum_{\substack{k \in R \\ \gamma_k \ge 2}} \gamma_k \le d \card{I}.\qedhere
    \]
\end{proof}

\section*{Acknowledgment}
We thank the anonymous reviewers for helpful comments that improved the exposition of this work. 

\paragraph{AI usage notice.} AI tools played no role in conducting the research, developing or drafting the proofs, or producing the scientific content in this work. Their use was limited to basic copy-editing and assistance in preparing some figures.

\small 
\bibliographystyle{plain}
\bibliography{main}

% TO BE REMOVED:
\ifnum\IddoSys=0
\newpage
\listoffixmes
\fi

%  \newpage
%  \pagestyle{empty}
% \begin{center}\small{--- Page left blank for ECCC stamp ---}\end{center}

\end{document}